\RequirePackage{booktabs}
\documentclass[sn-mathphys-num]{sn-jnl}

\usepackage{graphicx}%
\usepackage{multirow}%
\usepackage{enumerate}
\usepackage{enumitem}
\usepackage{amsmath,amssymb,amsfonts}%
\usepackage{amsthm}%
\usepackage{mathrsfs}%
\usepackage{braket}
\usepackage[title]{appendix}%
\usepackage{xcolor}%
\usepackage{textcomp}%
\usepackage{manyfoot}%
\usepackage{booktabs}%
\usepackage{algorithm}%
\usepackage{algorithmicx}%
\usepackage{algpseudocode}%
\usepackage{listings}%
\usepackage{xcolor}
\usepackage[normalem]{ulem}
\usepackage{tikz}
\usepackage{array}
\usepackage{longtable}
\usepackage{multirow}
\usepackage{hhline}
\usepackage{dsfont}
\usepackage{setspace}
\usepackage{titlesec}
\usepackage{titletoc}
\hypersetup{
  hypertexnames = false
}

\titlecontents{section}[0em]
{\addvspace{1em}\bfseries}
{\phantomsection\hyperlink{section::\arabic{section}}{\thecontentslabel}\ }
{\hspace*{-0em}}
{\titlerule*[3pt]{}\bfseries\contentspage}
\titlecontents{subsection}[0em]
{\addvspace{.2em}}
{\phantomsection\hyperlink{subsection::\arabic{section}.\arabic{subsection}}{\thecontentslabel}\ }
{\hspace*{-0em}}
{\titlerule*[3pt]{.}\contentspage}
    
\newcommand*\emptycirc[1][1ex]{\raisebox{-.4ex}{\tikz\draw[line width=.75pt] (0, 0) circle (#1);}} 
\newcommand*\fullcirc[1][1ex]{\raisebox{-.4ex}{\tikz\draw[fill] (0, 0) circle (#1);}} 
\newcommand*\greycirc[1][1.0125ex]{\raisebox{-.4ex}{\tikz\draw[lightgray] (0, 0) circle (#1);}} 

\newcommand*\halfcirc[1][1ex]{%
  \raisebox{-.4ex}{%
  \begin{tikzpicture}
  \draw[fill] (0,0)-- (90:#1) arc (90:0:#1) arc (360:270:#1) -- cycle ;
  \draw[line width=.75pt] (0,0) circle (#1);
  \end{tikzpicture}}%
}
\newcommand*\hatchcirc[1][1ex]{%
  \raisebox{-.4ex}{%
  \begin{tikzpicture}
  \draw[line width=.75pt] (0,0) circle (#1);
  \draw[line width=.75pt] (0.66ex, -.75ex) -- (0.66ex, .75ex);
  \draw[line width=.75pt] (0.33ex, -.94ex) -- (0.33ex, .94ex);
  \draw[line width=.75pt] (0ex, -1ex) -- (0ex, 1ex);
  \draw[line width=.75pt] (-0.33ex, -.94ex) -- (-0.33ex, .94ex);
  \draw[line width=.75pt] (-0.66ex, -.75ex) -- (-0.66ex, .75ex);
  \end{tikzpicture}}%
}

\newcommand*\screen[1][1ex]{%
	\raisebox{-.2ex}{%
	\begin{tikzpicture}
	\draw[line width=.5pt] (-1.2ex, -.75ex) -- (1.2ex, -.75ex);
	\draw[line width=.5pt] (-1.2ex, 1ex) -- (1.2ex, 1ex);
	\draw[line width=.5pt] (-1.4ex, -.55ex) -- (-1.4ex, .8ex);
	\draw[line width=.5pt] (1.4ex, -.55ex) -- (1.4ex, .8ex);
	\draw[line width=.5pt] (-1.2ex, 1ex) arc (90:180:0.2ex);
	\draw[line width=.5pt] (-1.4ex, -.55ex) arc (180:270:0.2ex);
	\draw[line width=.5pt] (1.2ex, -.75ex) arc (270:360:0.2ex);
	\draw[line width=.5pt] (1.4ex, .8ex) arc (0:90:0.2ex);
	\draw[line width=.5pt] (-1.15ex, -.6ex) -- (.55ex, .85ex);
	\draw[line width=.5pt] (-.85ex, -.6ex) -- (.85ex, .85ex);
	\draw[line width=.5pt] (-.55ex, -.6ex) -- (1.15ex, .85ex);
	\end{tikzpicture}}%
}

\newcommand{\beginsupplement}{%
        \setcounter{table}{0}
        \renewcommand{\thetable}{S\arabic{table}}%
        \setcounter{figure}{0}
        \renewcommand{\thefigure}{S\arabic{figure}}%
        \setcounter{section}{0}
        \renewcommand{\thesection}{S\arabic{section}}%
        \setcounter{equation}{0}
        \renewcommand{\theequation}{S\arabic{equation}}%
     }

\theoremstyle{thmstyleone}%
\theoremstyle{thmstyletwo}%
\usepackage{pifont}
\newcommand{\cmark}{\ding{51}}%
\newcommand{\xmark}{\ding{55}}%

\theoremstyle{thmstylethree}%

\begin{document}

\title[Article Title]{BEADS: A canonical visualization of quantum states for applications in quantum information processing}


\author[1,2]{\fnm{Dennis} \sur{Huber}}\email{dennis.huber@tum.de}

\author*[1,2]{\fnm{Steffen J.} \sur{Glaser}}\email{glaser@tum.de}

\affil[1]{\orgdiv{TUM School of Natural Sciences}, \orgname{Technical University of Munich}, \orgaddress{\street{Lichtenbergstrasse 4}, \city{Garching}, \postcode{85747}, \state{Bavaria}, \country{Germany}}}

\affil[2]{\orgname{Munich Center for Quantum Science and Technology (MCQST)}, \orgaddress{\city{M\"unchen}, \postcode{80799}, \state{Bavaria}, \country{Munich}}}


\abstract{We introduce a generalized phase-space representation of qubit systems called the BEADS representation which makes it possible to visualize arbitrary quantum states in an intuitive and an easy to grasp way. Our representation is exact, bijective, and general. It bridges the gap between the highly abstract mathematical description of quantum mechanical phenomena and the mission to convey them to non-specialists in terms of meaningful pictures and tangible models. Several levels of simplifications can be chosen, e.g., when using the BEADS representation in the communication of quantum mechanics to the general public. In particular, this visualization has predictive power in contrast to simple metaphors such as Schr\"odinger's cat.}




\maketitle

\section{Introduction}\label{Introduction}

The field of quantum science and technology is rapidly developing \cite{Acin, NielsenChuang} which creates the necessity to educate scientists beyond physics, engineers, technicians, high school students, and the general public. However, quantum physics is notoriously difficult to convey: "Studying the quantum world is not intuitive, and many researchers have difficulty communicating the weirdness of quantum physics because it's not relatable to everyday experiences.(...) Because quantum processes are extremely difficult to visualize; the processes become distant and foreign" \cite{Castleberry21}. Similarly, J. Audretsch stated: "What makes the understanding of quantum theory so difficult is its lack of images. In its framework, there is no intuition evoked by memorable images and metaphors, but only the power of mathematical formulation" (translated from \cite{Audretsch12}). In fact, this problem has been around for about a century, which is vividly expressed by the following editorial introductory note \cite{BohrComment} to an article by Niels Bohr \cite{Bohr} published in Nature in 1928: "It must be confessed that the new quantum mechanics is far from satisfying the requirements of the layman who seeks to clothe his conceptions in figurative language. Indeed, its originators probably hold that such a symbolic representation is inherently impossible. It is earnestly to be hoped that this is not their last word on the subject, and that they may yet be successful in expressing the quantum postulate in picturesque form" \cite{BohrComment}.

In the following, we will present a powerful visualization technique that is able to provide meaningful pictures, animations as well as tangible models of abstract quantum phenomena. We hasten to add that we do not claim that it will be suitable for all types of quantum phenomena, yet it is applicable to the important field of quantum information processing where typically finite-dimensional quantum systems are studied in a non-relativistic setting. In particular, the proposed visualization can be applied (but is not limited) to systems consisting of coupled quantum bits (qubits), i.e., quantum mechanical two-level system such as electron or nuclear spins-$\frac{1}{2}$, a subset of the internal states of trapped ions or atoms, or superconducting quantum bits on a chip. In this scenario, the qubits are individually addressable, and their states can be individually measured. Due to the coupling between the qubits, a system of $N$ qubits has $2^N$ basis states and can be in highly entangled nonclassical superpositions of these basis states. This exponential growth of the available state space is a source of the power of quantum information and at the same time makes it impossible in practice to represent the full information hidden in a general quantum state by classical means if the quantum system consists of more than a few tens of qubits. This appears to dash any hope to be able to find a useful and informationally complete visualization of quantum states. However, as we will illustrate in the following, the availability of such a visualization for few quantum bits is extremely useful both in education and in research. In fact, for simplicity and practical convenience we will limit our examples to systems of less than four qubits. It is important to note that this is not due to fundamental limitations of the approach which can be pushed to much larger systems -- albeit at the cost of higher required computational power for the classical simulations and at the expense of higher complexity of the visualization which is a natural and necessary consequence of the exponential increase of the state space with the number of qubits.

At this point, it is instructive to point out another highly successful visualization of quantum systems. For the case of a single quantum bit (such as the spin of an electron), an extremely useful visualization is the Bloch vector \cite{Bloch, FeynmanBloch}. This is a three-dimensional vector with real components which can be easily visualized as an arrow in 3D space. The ability to visualize and to imagine the motion of the Bloch vector is enabling scientists to understand and to design highly nontrivial experimental protocols in magnetic resonance spectroscopy and imaging \cite{ErnstBodenhausen}. Unfortunately, this simple vector picture alone is not able to capture the full information encoded in the quantum state of two or more qubits. The \textit{BEADS} representation which we will introduce in the following can be seen as a natural (albeit perhaps not obvious) generalization of the Bloch vector representation to arbitrary multi-qubit systems. 

Note that the Bloch vector picture is able to represent the state of a quantum bit realized in a superconducting quantum chip and can be regarded as a physical model of the (superconducting) qubit. However, it is important to understand that physical models are in general not simply scaled versions (as, e.g., a model airplane) of another physical object. What is represented by the Bloch vector is not an upscaled version of the part of the chip hosting the qubit (or of a photon) but an equivalent mathematical representation of its abstract quantum state $\ket{\psi}$ (or its density operator $\rho$) that can be visualized as a three-dimensional arrow. Similarly, the BEADS representation is able to capture the information of the quantum state $\ket{\psi}$ for multi-qubit systems but should not be misunderstood as a magnified view of the qubits which may, e.g., be arrays of cold atoms, photons or superconducting circuits. Moreover, our representation is not a mechanistic model of quantum computing and is thus not capable of performing quantum computation on its own.
The scientific questions that we will address in detail in the following are

\begin{enumerate}[label=\alph*)]
	\item What are desirable features of graphical representations of the state of a system of a set of individually addressable and detectable qubits in quantum technology applications, such as quantum computing and quantum control?
	\item Can these desirable features be simultaneously realized in a canonical graphical representation?
	\item Can such a canonical graphical representation be rigorously formulated?
	\item What are the strengths and potential limitations of such a representation?
\end{enumerate}

\noindent
In this paper, we answer these questions constructively by providing the conceptual and mathematical description of the BEADS representation and by demonstrating its useful properties by illustrative examples. We have implemented an easy to use software package and created tangible models both which are being used in different educational formats (ranging from the physics quantum science and technology master program of the Technical University of Munich and the Ludwig Maximilian University of Munich, to advanced trainings for physics teachers, workshops for high-school students in the PhotonLab of the Max-Planck-Institute for  Quantum Optics (MPQ), and outreach activities for the general public at the Deutsches Museum in Munich and at science events.

Although the detailed mathematical and physical basis of the BEADS representation is non-trivial, due to its intuitiveness, it is possible to take full advantage of this visualization approach without having to understand the intricacies of the underlying mapping. For simplicity, we will focus our results on pure states. However, the BEADS representation is also suited to visualize arbitrary mixed states which will be addressed in future work.

\noindent
In order to provide a gentle introduction to the BEADS representation, we have structured the Results section in two main parts:
In the first part ("A light introduction to BEADS"), we introduce the BEADS representation in a phenomenological way, similarly to how a teacher might use it to visually introduce the properties of quantum bits, the standard computational basis states, the concepts of superpositions and entanglement, and the effect of quantum gates based on simple rules and recipes how to interpret the BEADS visualization. The advantages of the BEADS representation are illustrated in comparison with alternative visualizations techniques.

In the second part, we provide the theoretical background of the mapping and detail links to standard quantum concepts, such as expectation values of projective measurements, correlation functions and entanglement. Also, relations to already existing visualization approaches are discussed. Technical details and additional illustrative examples are provided in the supplementary material.

\section{Results}\label{Results}
\subsection{A light introduction to BEADS}\label{LightInt}
We start with a light, non-technical introduction to the \textit{Quantum BEADS representation} (or for short, the \textit{BEADS} representation) to fix ideas and get a feel of this unique representation of abstract quantum states (in parentheses, we sparingly add auxiliary information for experts). The BEADS representation provides a cohesive and seamless approach to visualize both classical and quantum information in a way which is simple, quantitative, and complete. 

In general, the BEADS representation consists of one or more colored spheres, called \textit{beads}. There are two principal kinds of beads: \textit{Q-Beads} that are associated with the state of individual physical qubits and additional correlation function beads, in particular \textit{E-Beads} (representing entanglement-based correlation functions in case of pure states). Before turning to the important correlation function beads and their relation to entanglement, let us first look at Q-Beads and their connection with the well-known classical bit states.

\subsubsection*{Classical bits}\label{CB}
A classical bit can only be in one of two possible states: either it has the value 0 or the value 1. We can represent these two states graphically, e.g., by a red and green colored square (called a \textit{color patch}), as illustrated on the left side of Fig.~\ref{Figure:Fig1}. A concrete tangible model of a classical bit could be a card with a red ($= 0$) and a green ($= 1$) side. We deliberately avoid representing bit states by the two sides of a coin, i.e., a \textit{circular disc} with different colors in order to clearly distinguish them from 2D images of spherical quantum beads, which of course also look disc-shaped.

\subsubsection*{Qubits}\label{QB}
\begin{figure}[H]
\centering
\includegraphics[width=.90\textwidth]{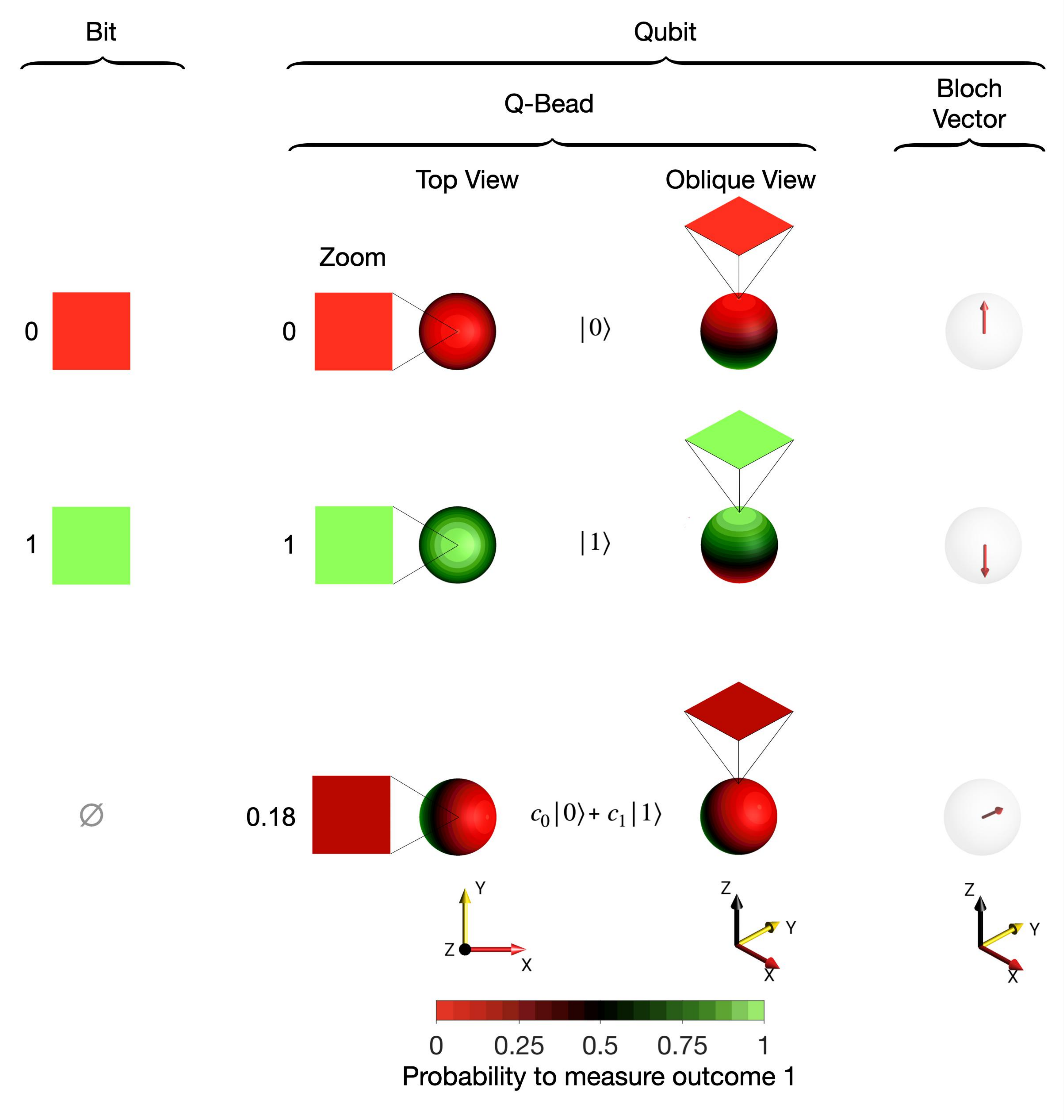}
\caption{\label{Figure:Fig1}Bit and corresponding qubit representations. The first two rows show visualizations of the two possible states 0 and 1 of a classical bit and of the \textit{computational basis states} $\ket{0}$ and $\ket{1}$ of a qubit. The bit values 0 and 1 are graphically represented by a red and a green square-shaped color patch, respectively.  In the BEADS representation, a qubit state is represented as a bipolar Q-Bead with a red and a green pole, shown from two different perspectives (a \textit{top view} and an \textit{oblique view}, see the different orientations of the coordinate system indicated below.) In addition, a zoomed view of the Q-Bead's north pole is shown as a square-shaped patch, whose color indicates the probability $p_1$ to obtain the outcome 1 if a measurement is performed along the z-axis (see color bar).  For computational basis states, there is a one-to-one correspondence between this color patch and the color patch representation of the corresponding bit states displayed on the left. In the third row, an example of the Q-Bead representation of a \textit{superposition state} $c_0\ket{0}+c_1\ket{1}$ is shown which has \textit{no classical equivalent}. In the right column, the equivalent Bloch vector representations of the qubit states are shown for comparison.}
\end{figure}

In the BEADS representation, a single qubit is represented by a \textit{sphere} with a red and an antipodal green pole, called a \textit{Q-Bead}, see Fig.~\ref{Figure:Fig1} (for a pure qubit state, the vector connecting the center of the sphere and the red pole is identical to the well-known Bloch vector). With the help of the color scale at the bottom of Fig.~\ref{Figure:Fig1}, the colors of a Q-Bead can be mapped to a value in the range between 0 and 1. Physically, this value corresponds to directional measurement probabilities, as explained in more detail below.  

\subsubsection*{Computational basis states}\label{CBS}
The computational basis states (with standard symbols $\ket{0}$ and $\ket{1}$) of a qubit correspond to Q-Bead orientations where either the red or the green pole is facing upwards (along the positive z-axis), see Fig.~\ref{Figure:Fig1}. Note that for these two special qubit states, a zoomed top view of the Q-Beads uppermost point (referred to as the Q-Bead's \textit{north pole}) looks exactly like the red or green \textit{color patches} that we have used to represent the classical bit values 0 and 1, cf. first two rows of Fig.~\ref{Figure:Fig1}. This provides a seamless transition between the graphical representations of qubits in computational basis states and classical bit states.

\subsubsection*{Superposition states}\label{SPS}
Whereas a classical bit is constrained to be in one of the two states 0 or 1, a qubit is not constrained to be in one of the two computational basis states $\ket{0}$ or $\ket{1}$. In fact, it can be in one of an infinite number of different \textit{superposition states} of the form $c_0\ket{0}+c_1\ket{1}$  (where the coefficients $c_0$ and $c_1$ are complex numbers satisfying the condition $|c_0|^2+|c_1|^2=1$). Every superposition state of this form can be uniquely mapped to a three-dimensional orientation of the Q-Bead (this mapping is identical to the standard mapping from an arbitrary pure qubit state to the Bloch vector orientation \cite{NielsenChuang}). For example, the Q-Bead shown in the third row of Fig.~\ref{Figure:Fig1} is oriented such that its red pole is tilted slightly away from the z-direction resulting in a dark-red zoomed color patch at the north pole, corresponding in this example to a numerical value of about 0.18.

\subsubsection*{Measurements}\label{Meas}

In order to better understand the color scheme introduced in Fig.~\ref{Figure:Fig1}, it is helpful to briefly review standard measurements (projective measurements) of a qubit and their similarities and differences to the measurement of a classical bit that is in state 0 or 1. For a classical bit, it is always possible to measure its value without altering it. When measuring a qubit along the z-axis, the same is only true if it is in one of the two computational basis states, where measuring the state $\ket{0}$ gives outcome 0 and measuring the state $\ket{1}$ gives outcome 1 without altering the respective states.

For \textit{superposition states}, a measurement in the z-direction still provides an outcome that can only be 0 or 1. However, even if the superposition state is precisely known, it is impossible to predict the outcome of an individual measurement with certainty. Instead, we can only predict the \textit{probabilities} $p_0$ and $p_1$ with which the outcome 0 or 1 will be obtained. (In fact, since only one of the two outcomes 0 or 1 is possible, it is sufficient to know only, e.g., $p_1$, because $p_0$ is simply given by $p_0 = 1 - p_1$). Furthermore, the initial superposition state \textit{is} modified by the measurement: If outcome 0 or 1 is measured, the qubit \textit{is} found in the corresponding computational basis state $\ket{0}$ or $\ket{1}$, respectively, after the measurement.

\begin{figure}[H]
\centering
\includegraphics[width=.8\textwidth]{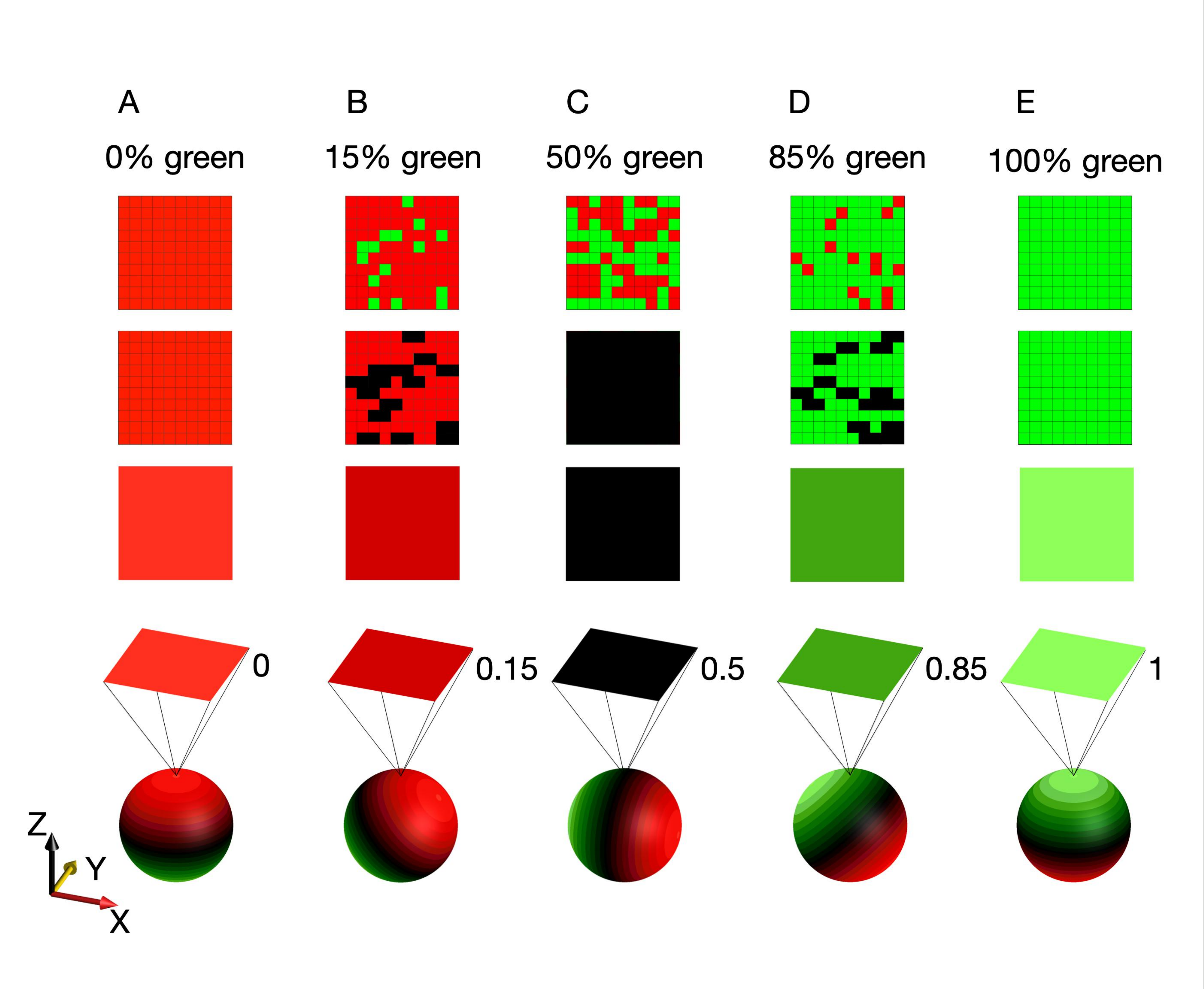}
\caption{\label{Figure:Fig2}Schematic relation between measurement statistics and the chosen color scheme of Q-Beads~\screen. Columns A-E correspond to five different qubit states $\psi=c_0\ket{0}+c_1\ket{1}$ with coefficients ($c_0$, $c_1$) of (1, 0) in (A), (0.92, 0.4) in (B), (0.71, 0.71) in (C), (0.4, 0.85) in (D), and (0, 1) in (E). In each column, the pixel matrix in the first row represents the statistics of 100 measurements along the z-axis, assuming that before each measurement the initial state corresponding to the respective column has been prepared (red and green pixels correspond to measurement outcomes 0 and 1). The second row shows the effect if arbitrary pairs of pixels with complementary colors are colored black. Blending the colors of the pixel matrices results in the uniformly colored color patches shown in the third row. The fourth row illustrates that these color patches correspond to the zoomed color patches at the upper point (north pole) for each of the Q-Bead representation of the five different qubit states prepared in columns A-E. The \screen$\ $ symbol at the end of the caption title indicates that this visualization is available as a dynamic simulation in the supplementary video.}
\end{figure}

Fig.~\ref{Figure:Fig2} (A)-(E) schematically illustrates for five exemplary qubit states how the color scale used for Q-Beads can be related to the stochastic measurement outcomes in experiments where the measurement direction was chosen to be along the positive z-axis. For each of these states, the following (simulated) steps were repeated 100 times: After preparing a qubit in the initial state $\psi$, a measurement of the qubit along the z-axis is performed. For each case, the 100 individual measurement outcomes were sequentially recorded and visualized in a 10-by-10 pixel matrix (shown in the first row) by coloring a pixel \textit{red} or \textit{green} if the outcome was 0 or 1, respectively \cite{DurHeusler}. Note that (in the limit of an infinite number of repeated experiments) the percentage of \textit{green} pixels in the top row for each case simply corresponds to the probability $p_1$ to obtain the measurement outcome 1 (indicated next to the zoomed color patches in the bottom of Fig.~\ref{Figure:Fig2}). We arrive at the color scheme for representing the value of $p_1$ on the surface of a Q-Bead in two simple steps: First, we select arbitrary pairs of pixels with complementary colors (red and green) in each matrix and color them black, indicating that for each pair the outcomes average to 0.5, see the resulting matrices in the second row of Fig.~\ref{Figure:Fig2}. Second, each of the five matrices is uniformly colored by blending the colors of all its pixels, resulting in the five color patches shown in the third row of Fig.~\ref{Figure:Fig2}. 

Note that these blended color patches match the color of the zoomed color patch at the upper point (north pole) for each of the five Q-Beads, see the fourth row of Fig.~\ref{Figure:Fig2}. Here, the value of $p_1$ given next to each color patch was directly calculated for each of the five considered states $\psi=c_0\ket{0}+c_1\ket{1}$) using Born's rule \cite{NielsenChuang} with the formula  $p_1=|c_1|^2$.

The previous discussion of measurements along the z-axis can be generalized for measurements along an arbitrary measurement axis: The color of each surface point of a Q-Bead indicates the probability $p_1(r)$ to measure the outcome 1 if the measurement is performed in the direction defined by the vector $r$ connecting the center of the Q-Bead and this surface point. Hence, the Q-Bead representation provides a convenient and compact way to display the probability $p_1(r)$ for all possible measurement directions $r$. (An alternative but equivalent interpretation of Q-Bead colors in terms of expectation values of measurement operators will be discussed later.)  

\subsubsection*{Multi-qubit states}\label{MQS}
In a multi-qubit system, each qubit $Q_k$ is represented by a corresponding Q-Bead. In the absence of entanglement, the Q-Beads of a multi-qubit pure state look just like the Q-Beads of the single-qubit states shown in Fig.~\ref{Figure:Fig1}, except for possible different orientations. However, in the presence of entanglement, the (reduced density operators of) individual qubits are in mixed (i.e., non-pure) single-qubit states, which is represented by a reduced brightness of the corresponding Q-Beads. In particular, in the case of totally mixed reduced single-qubit density operators, the corresponding Q-Beads are completely black, indicating that in any measurement direction, there is a 50:50 chance to measure the outcome 0 or 1. Even more importantly, in the BEADS representation the presence of entanglement in a qubit system is clearly visualized by the presence of additional entanglement-related correlation function beads (\textit{E-Beads}), as illustrated at the bottom of Fig.~\ref{Figure:Fig3}. As discussed in the following, it is rather straightforward to experimentally entangle qubits if suitable quantum gates can be applied.

\subsubsection*{Classical gates and quantum gates}
In classical computing, bit states are manipulated by (logical) gate operations. In quantum computing, the role of classical gates is taken over by (unitary) quantum gates, which are always reversible. (Although not all classical gates are reversible, it is not hard to construct a reversible classical analog of any classical gate \cite{NielsenChuang}). The only non-trivial reversible classical gate with a single input bit and output bit is the NOT gate which simply flips the bit state from 0 to 1 or vice versa (see the first table in the top row of Fig.~\ref{Figure:Fig3}). In contrast, there is an infinite number of possible single-qubit quantum gates. They all correspond to well-defined, simple three-dimensional rotations of the Q-Beads, which can be vividly illustrated using the BEADS representation (see section~\ref{sec:BEADSGates} of the supplementary material and the supplementary video). The effect of a quantum NOT gate (a.k.a. X-gate) is to rotate a Q-Bead by $180^\circ$ about the x-axis, which also flips the input states $\ket{0}$ and  $\ket{1}$ (second table). However, if the Q-Bead is initially oriented along the x-axis, it is clear that the $180^\circ$ x-rotation by the NOT gate cannot change its orientation (see third table). 

\subsubsection*{Entanglement and nonclassical correlations}
\begin{figure}[H]
\centering
\includegraphics[width=.86\textwidth]{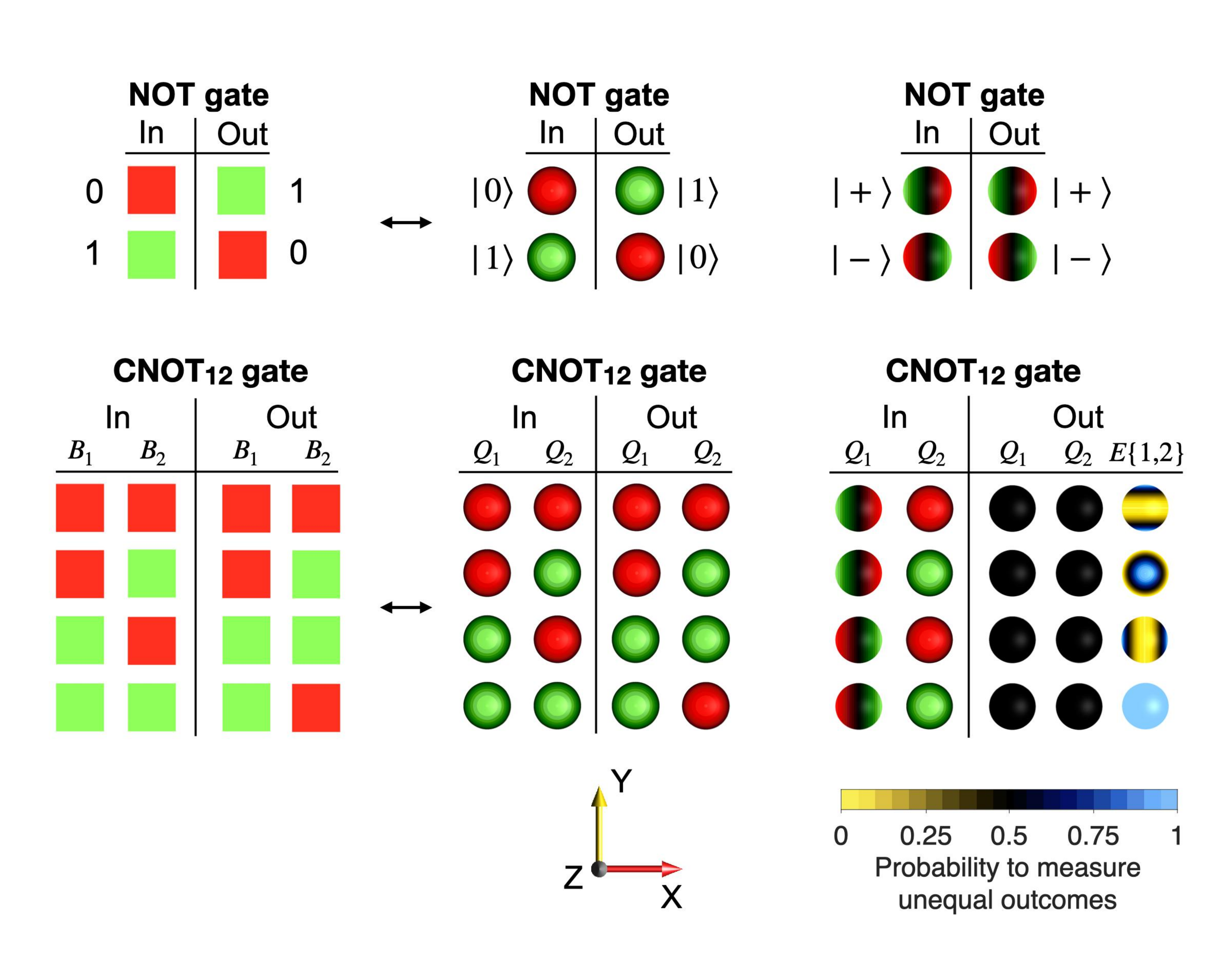}
\caption{\label{Figure:Fig3}Effect of the NOT and CNOT gates. Truth tables for the NOT gate (top) and the CNOT gate (bottom) for classical input states and output states are shown graphically on the left using red and green color patches as introduced in Fig.~\ref{Figure:Fig1}. The effect of the quantum versions of the NOT and controlled NOT (CNOT) gates is shown applied to computational basis states (center) or to selected superposition states (right) using the BEADS representation. In the center tables, the input qubit states are the computational basis states $\ket{0}$ and $\ket{2}$ for the NOT gate (top) and the two-qubit computational basis states $\ket{00}$, $\ket{01}$, $\ket{10}$, and $\ket{11}$ for the CNOT gate (bottom).  In the top right table, the input qubit states for the NOT gate are the superposition states $\ket{+}=1/\sqrt{2}(\ket{0}+\ket{1})$ and $\ket{-}=1/\sqrt{2}(\ket{0}-\ket{1})$. The table at the bottom right shows the results if the CNOT gate is applied to the two-qubit states $\ket{+0}$, $\ket{+1}$, $\ket{-0}$, or $\ket{-1}$, where the first qubit is in one of the superposition state $\ket{+}$ or $\ket{-}$ while the second qubit is in one of the basis states $\ket{0}$ or $\ket{1}$. In this case, entangled states are created which are not fully characterized by the output Q-Beads $Q_1^{out}$ and $Q_2^{out}$ but also require the E-Bead (short for entanglement-related correlation function bead) $E\{1,2\}$. With the help of the yellow-blue color scheme (see bottom right), each surface point of the E-Bead quantitatively indicates the probability to measure \textit{unequal} outcomes based on nonclassical correlations, if measured in this direction.}
\end{figure}

In the second row of Fig.~\ref{Figure:Fig3}, we see the effect of the \textit{controlled}-NOT (or CNOT) gate. It plays an important role in quantum information processing, because it is able to create \textit{entangled} multi-qubit states. This results in intriguing nonclassical correlations between the outcomes of measurements performed on the individual qubits, even if they are so far apart that no communication between them is possible during the measurement process. Before looking at the creation of entanglement and its corresponding visualizations in the BEADS representation, let us briefly see the effect of a CNOT gate on each of the four possible states 00, 01, 10, or 11 of two classical input bits, see the truth table at the lower left of Fig.~\ref{Figure:Fig3}. The state of the output bit $B_2^{out}$ depends uniquely on both input states $B_1^{in}$ and $B_2^{in}$: \textit{If} $B_1^{in}=0$ (red) then $B_2^{out}=B_2^{in}$, but \textit{if} $B_1^{in}=1$ (green) then $B_2^{out}=\text{NOT}\:B_2^{in}$. Thus, the state $B_1^{in}$ of the first input bit \textit{controls} whether or not the state $B_2^{in}$ must be flipped by a NOT gate to obtain the state $B_2^{out}$. Note that in all cases $B_1^{out}=B_1^{in}$ and the only purpose of $B_1^{out}$ is to make the gate reversible, so it is possible to construct a corresponding quantum version.

If the input state of the analogous quantum CNOT gate is one of the four two-qubit \textit{computational basis states} $\ket{00}$, $\ket{01}$, $\ket{10}$, or $\ket{11}$, the resulting table is congruent to the classical truth table of the CNOT gate (compare the second and first tables). However, things become much more interesting if the control qubit $Q_1$ is initially in a \textit{superposition state}. For example, in the third table, the first qubit is initially prepared in one of the superposition states $\ket{+}=1/\sqrt{2}(\ket{0}+\ket{1})$ or $\ket{-}=1/\sqrt{2}(\ket{0}-\ket{1})$, which correspond to the red pole of the corresponding Q-Bead pointing along the positive or negative x-direction, respectively. In this case, each of the two-qubit input states $\ket{+0}$, $\ket{+1}$,$\ket{-0}$, or $\ket{-1}$  is transformed into one of the four fully entangled \textit{Bell states} by the CNOT gate \cite{NielsenChuang} (for more details on this transformation, see supplementary section~\ref{sec:BellStates}). The BEADS representation of the resulting Bell state is schematically shown in the \textit{three} output columns of the third table in the lower row of Fig.~\ref{Figure:Fig3}: The Q-Beads $Q_1$ and $Q_2$ of the output states are completely black, that is, in any possible measurement direction, the measurement outcomes 0 or 1 have equal likelihood. However, the presence of the non-zero entanglement bead $E\{1,2\}$ indicates that for most measurement directions the measurement outcomes are at least partially nonclassically correlated. The yellow-blue color scheme (see color bar at the bottom right of Fig.~\ref{Figure:Fig3}) indicates the probability to measure \textit{unequal} outcomes if measured in the same (but arbitrary) direction. For example, the E-Bead shown in the first row of the table is bright yellow along the x-axis and the z-axis, indicating that in this case, the probability to measure different outcomes is 0, i.e., although the individual measurement outcomes for the two qubits are unpredictable, they are never different, that is, they are always identical! Conversely, the E-Bead is bright blue along the y-axis indicating that the probability to measure different outcomes is 1 such that we can predict that the measurement outcomes will be different in this case with certainty. The E-Beads corresponding to the first three Bell states are identical and only differ in their \textit{orientation}. However, the E-Bead for the state (called the \textit{singlet state} \cite{NielsenChuang}) obtained in the fourth row of the table is clearly of a different type and cannot be obtained by simply rotating one of the other Bell states. It is entirely blue, indicating that we can predict that for any chosen common measurement direction, the outcomes for two qubits will be \textit{different} with certainty, independent of their spatial separation. For more details on Bell states and other characteristic entangled two-qubit and three-qubit states, see sections~\ref{2Q}, \ref{3Q}, and supplementary sections~\ref{sec:PartEnt} and~\ref{sec:BellStates}.

\subsection{Theoretical background}\label{sec:BEADSTheoBackground}
From a mathematical theoretical point of view, the BEADS representation and its visualization is a generalized phase-space representation \cite{Curtright14,Ferrie11,Gosson17,Koczor23,Schleich01,Schroeck96,Weinbub18,Zachos05}. In particular, it belongs to the class of $s$-parametrized representations, which includes the well-known Wigner $W$ \cite{Wigner,Dowling94}, the Husimi $Q$ \cite{Husimi,Agarwal,Auzinsh09,Budker} and the Glauber $P$ \cite{Cahill69} representations, see \cite{Koczor20} for a unified description of $s$-parametrized phase-space representations of states in finite- or infinite-dimensional Hilbert spaces. More specifically, the BEADS representation is a new member of the family of \textit{continuous} phase space representations of \textit{finite}-dimensional quantum systems, which reflect inherent (rotational and permutational) symmetries of coupled two-level systems (also known as qubits or as spins $S = 1/2$, such as electrons or protons) or of coupled $d$-level systems with $d > 2$ (also known as qudits or spins $S > 1/2$ with individual Hilbert-space dimensions $d = 2S + 1$) \cite{Brif98, Brif99, RundleEveritt, RundleEveritt19, Tilma16}. These types of phase-space representations are closely related to the seminal paper by Stratonovich \cite{Stratonovich}. Related work in nuclear magnetic resonance (NMR) \cite{Philp05} can also be traced back to at least the work of Pines et al. \cite{Pines} where (albeit without a formal map) selected density operator terms of a spin-1 particle and their symmetry properties were depicted using spherical harmonics.

Based on a mapping between spherical tensor basis operators and a set of corresponding spherical functions, several different multi-qubit phase-space representations have been developed which differ in the used criteria to define the spherical tensor operators and which lead to different groupings of the spherical functions. The subclass of so-called multipole representations \cite{Jessen, Sanctuary85, Budker2, DROPS} group the spherical tensor operator primarily according to their angular momentum properties but not with respect to the number and the identity of involved spins (qubits or qudits). Therefore, multipole representations are not ideally suited to represent states in quantum computations which require the ability to control and to detect qubits individually \cite{DiVincenzo}. 

This problem was solved by the construction of the so-called LISA basis \cite{DROPS}, which consists of tensor operators with defined \underline{li}nearity (which corresponds to the \textit{number} of involved qubits), \underline{s}ubsystem (the actual set or subset of involved qubits), and \underline{a}uxiliary criteria (permutation symmetry and fractional parentage), for more details and explicit matrix representations of the LISA basis for up to six qubits as well as for two coupled qudits, see \cite{DROPS, LeinerDROPS}. Based on the LISA basis, the DROPS representation of arbitrary operators was introduced \cite{DROPS}.

The DROPS representation is a generalized Wigner function that is based on the decomposition of the operator of interest (e.g., the density operator representing the pure or mixed state of a system of qubits) in the LISA basis $T_{j,m}^{(\ell)}$. The resulting expansion coefficients $c_{j,m}^{(\ell)}$ are directly used to calculate corresponding linear combinations of spherical harmonics $Y_{j,m}$ to create a set of spherical functions, called droplet functions (for details, see Eq.~\ref{eq:DROPSMapping} in section~\ref{GenMap}). This mapping indeed provides a unique visualization with many favorable properties, such as natural transformations under non-selective rotations and permutations, visualization of symmetry properties, characteristics patterns depending on the coherence order of a droplet function and the immediate identification of the spins or qubits that are involved in a certain droplet function). We also implemented the DROPS representation in the SpinDrops app, which makes it possible to visualize and interactively control the dynamics of multi-spin states related to NMR applications \cite{SpinDrops}. It is also noteworthy that it is possible to experimentally reconstruct droplet functions using a scanning-based tomography approach based on measured expectation values of rotated longitudinal LISA basis operators, as experimentally demonstrated on NMR-based quantum processors and on superconducting quantum computers \cite{LeinerWQST, LeinerWQPT, DevraTomog23}.

However, the DROPS representation also has a number of shortcomings: The values of the spherical droplet functions in a given direction $r$ are \textit{proportional} to measurable expectation values of longitudinal tensor operators but these operators are more or less complicated combinations of simple Cartesian Pauli product operators. However, it would be highly desirable to have a representation that can easily be directly interpreted in terms of expectation values of simple Cartesian Pauli product operators or in terms of measurement probabilities of interest. Furthermore, the originally proposed DROPS representation provided only partial information about the kind of entanglement that is present in a multi-qubit state. For the special case of a two-qubit system, the concurrence $C$, which is an important entanglement measure, can be expressed as a function of the maximum value $f_{max}^{\{k\}}$ of the linear droplets corresponding to the two qubits: $C=\sqrt{1-\frac{16\pi}{3}\left(f_{max}^{\{k\}}\right)^2}$ \cite{DROPS}. However, it would be highly desirable to see also in systems consisting of three or more qubits not only whether entanglement exists but also what kind of entanglement-based nonclassical correlation functions occur with respect to measurements.

Here, we present the novel BEADS representation, which is also based on the LISA operator basis. The development of the BEADS representation was inspired by the fact that the Wigner representation of a single qudit can be transformed to the corresponding Husimi representation by applying \textit{different} scaling factors to spherical harmonics of different rank $j$ \cite{Koczor20}. We therefore investigated the possibility to translate the DROPS representation of multi-qubit systems (which can be seen as a generalized Wigner representation) to a representation that is akin to a generalized Husimi representation. This problem was in particular interesting, because in the case of a single qudit, the value of the spherical Husimi representation is identical to experimentally important measurement probabilities up to a simple overall scaling factor (see appendix~\ref{sec:BEADSHusimi} for further details), and it would be desirable for the generalized Husimi representation to also have such a highly useful simple and intuitive interpretation. Indeed, this turns out to be the case as will be discussed in the following. 

Furthermore, in the BEADS representation we also included the possibility to efficiently separate total correlation functions of pure states, that is, expectation values of (multiqubit) measurement operators into what will be referred to as compound correlation functions and entanglement-based connected correlation functions based on the work of Schlienz and Mahler \cite{Mahler95, Mahler96, Mahler97} and related papers \cite{Tran17, Kumar09}. This allows to draw conclusions about the correlation and entanglement properties of a state by characterizing contributions of the visualized correlation functions. A similar approach has also been applied in \cite{Hazzard1, Hazzard2} to visualize an approximation of two-point connected spin correlation functions.

\subsection{A canonical representation of quantum states}\label{Preview}
We designed the BEADS representation such that, while providing a unique complete visualization of quantum systems, measurement outcomes can be predicted, and entanglement and symmetry properties can be identified visually with no or little experience. This contrasts our approach from other representations which also provide faithful visualizations of multiqubit systems yet require a deep understanding to extract relevant information or do not even allow to access the discussed aspects with reasonable effort.

\begin{figure}[H]
\centering
\includegraphics[width=.9\textwidth]{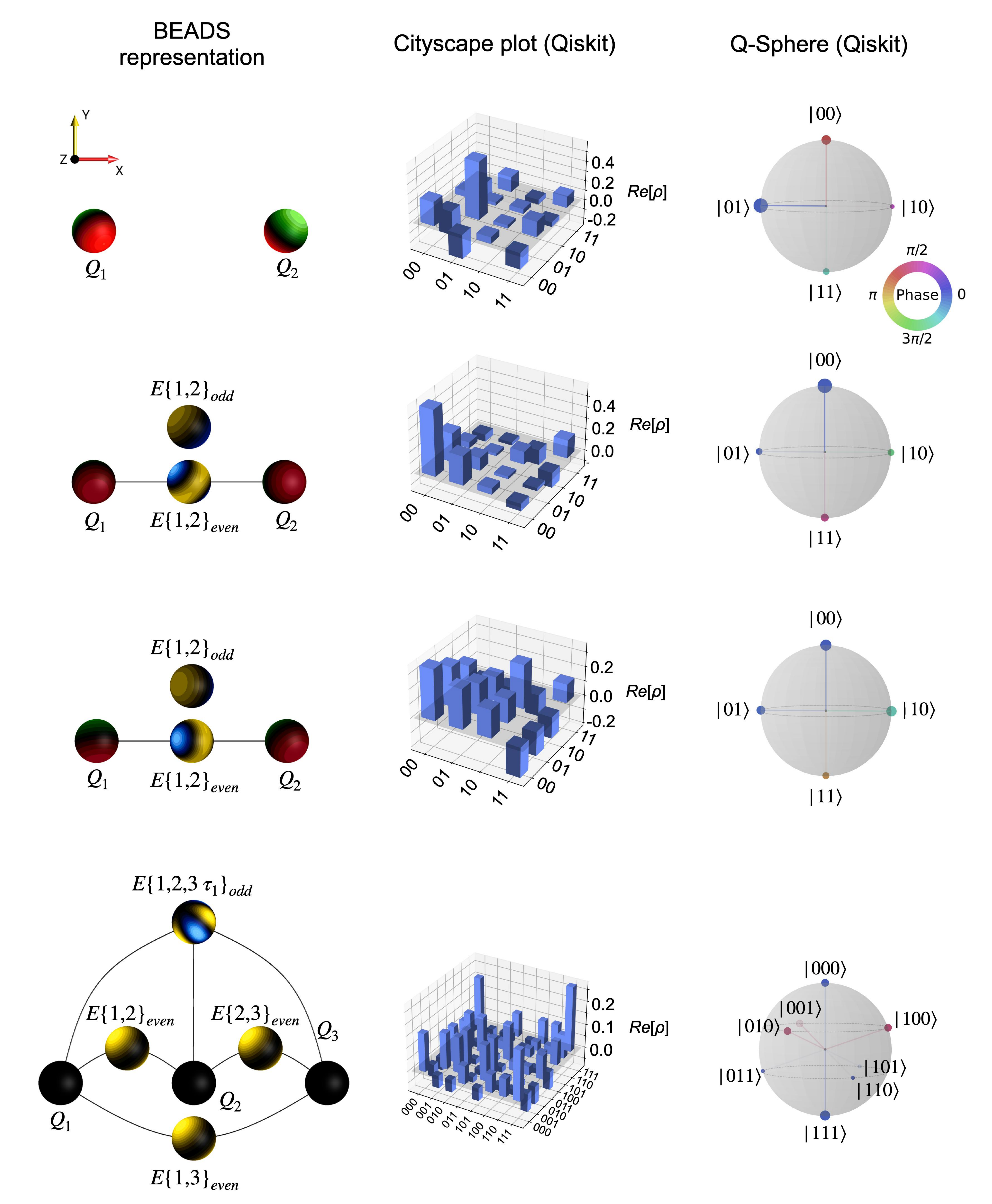}
\caption{\label{Figure:Fig4}BEADS, Cityscape, and Q-Sphere representation of several quantum states. The visualized examples comprise a fully separable state (top), two partially entangled two-qubit states which only differ by a global $45^\circ$-rotation around the x-axis (center), and a maximally entangled three-qubit state which is the GHZ state rotated by a $-60^\circ$-rotation around the xy-bisecting axis (bottom). The visualized states (read from top to bottom) are explicitly written out in Fig.~\ref{fig:FigComparison} of supplementary section~\ref{sec:BEADSComp}.}
\end{figure}

Before providing an in-depth physical analysis of the BEADS representation, we showcase the capability of our approach by comparing BEADS visualizations of selected non-trivial states with other representations including the Qiskit (real part) cityscape density operator plot \cite{Qiskit} and the Q-Sphere state vector visualization \cite{Qiskit} in Fig.~\ref{Figure:Fig4}. Further visualizations of the same states in terms of full cityscape plots, dimensional circle notations \cite{DCN}, and standard state vector expressions are given in section~\ref{sec:BEADSComp} of the supplementary material.

In the BEADS representation, one can directly see whether a state is entangled or separable. The topmost example in Fig.~\ref{Figure:Fig4} corresponds to a fully separable two-qubit product state. The BEADS visualization is only composed of Q-Beads, thus indicating in a simple way that there is no correlation and thus no entanglement between the qubits. In contrast, using the cityscape plot and Q-Sphere, for a non-expert, it is virtually impossible to make an immediate statement on the entanglement properties of the visualized state. Even experts will require some time to see if a state is entangled or not.

In the second and third state visualizations, we observe both, E-Beads and colored Q-Beads. Hence, we find the corresponding states to be (partially) entangled.  Moreover, both states have a antisymmetric component with respect to permutation represented by the $E\{1,2\}_{\textit{odd}}$ E-Bead. Again, all of this does not become directly evident in the alternative representations. Even more important though, only when comparing the two BEADS visualizations, it can be readily seen that both states only differ by a global $45^\circ$-rotation around the x-axis. 

The bottom three-qubit BEADS representation reveals a highly symmetric state. Later it will be discussed that this is the well-known GHZ state up to a global $-60^\circ$-rotation around the $xy$-bisecting axis. While the symmetry properties can also be assumed from the Q-Sphere in this case, the discussed small difference by a simple rotation compared to the GHZ state cannot be determined without advanced knowledge. The cityscape plot again barely provides any useful insights.

\subsection{A physical interpretation of BEADS: Two-qubit systems in pure states}\label{2Q}
We will now discuss the BEADS representation on a more detailed physical conceptual level and draw connections to underlying standard concepts of quantum mechanics.
An illustrative example of two-qubit entanglement can be found in states of Schmidt form which are two-qubit pure states that depend on a single angular parameter, the Schmidt angle $\theta$ \cite{Wharton16}. In Fig.~\ref{Figure:Fig5} the visualizations correspond to states of the form $\ket{\psi_\theta}=\cos{(\theta/2)}\ket{00}+\sin{(\theta/2)}\ket{11}$ where $\theta\in\left\{0, \pi/8, \pi/4, 3\pi/8, \pi/2\right\}$. The corresponding two-qubit BEADS representations are composed of multiple beads, the two Q-Beads $Q_1$ and $Q_2$, and the E-Bead (entanglement-based connected correlation function bead) $E\{1,2\}_{\textit{even}}$ which is located on a line connecting the Q-Beads indicating that this correlation is between the corresponding qubits. 

\begin{figure}[H]
\centering
\includegraphics[width=.8\textwidth]{FigSchmidtMain.pdf}
\caption{\label{Figure:Fig5}BEADS representations of Schmidt form states $\ket{\psi_\theta}=\cos{(\theta/2)}\ket{00}+\sin{(\theta/2)}\ket{11}$. A selection of states for different Schmidt angles $\theta$ is shown in oblique and top view. The E-Bead $E\{1,2\}_{\textit{even}}$ displays permutation symmetric two-qubit entanglement-based connected correlation coefficients. $\ket{\psi_0}$ corresponds to a fully separable state (no E-Bead) whereas $\ket{\psi_{\pi/2}}$ is maximally entangled which can be seen from black Q-Beads and maximally intense E-Bead colors. The remaining states are partially entangled. The color scales at the bottom serve as a reference for interpretation which can be made in terms of correlation coefficients, i.e., expectation values or related probabilities. The black lines connecting the Q-Beads serve as auxiliary entanglement indicators and their thickness is proportional to the norm of the underlying density operator component visualized by $E\{1,2\}_{\textit{even}}$.}
\end{figure}

As was briefly discussed in section~\ref{LightInt}, Q-Beads correspond to the physical qubits in a system. From a mathematical-physical perspective, Q-Beads are a visualization of the reduced density operators. Recall that the reduced density operator of a qubit is the partial trace of the density operator $\rho=\ket{\psi}\bra{\psi}$ \cite{NielsenChuang} with respect to all remaining qubits in the system. In addition, Q-Beads only map expectation values $\langle\sigma_k\rangle$ of single-qubit Pauli matrices $\sigma_k$ corresponding to the $k$-th qubit of a system and thus, they are equivalent to the individual Bloch vectors \cite{Bloch, FeynmanBloch}. As will be shown later, the previously introduced interpretation of Q-Beads in terms of probabilities is related to these expectation values and both can be bijectively transformed into each other. Since the reduced density operator of a qubit is pure only if the qubit is not entangled with any other qubit, we can use spherical function values of a Q-Bead (given in terms of colors) as a first indicator of entanglement. Specifically, the \textit{brightness} of the Q-Bead colors is maximum (Bloch vector of length 1) if a pure state is separable and minimum (Bloch vector of length 0, all-black Q-Beads) if the state is maximally entangled \cite{DROPS}. The latter arises due to the single-qubit density matrices being maximally mixed for which $\langle\sigma_k\rangle=0$.\newline

\begin{enumerate}[label=\textbf{Result \arabic*:}, itemindent=1cm, labelindent=1cm]
\item \label{res:BEADS1} Q-Beads $Q_k$ are a direct visualization of the reduced density operator corresponding to the $k$-th qubit of a system. The Q-Bead colors and brightness directly correspond to expectation values of single-qubit Pauli matrices $\sigma_k$. The brightness indicates whether a pure state is separable (maximum brightness), partially entangled (reduced brightness) or maximally entangled (fully black Q-Beads).\newline
\end{enumerate}

\noindent
Indeed, for the states shown in Fig.~\ref{Figure:Fig5}, the Q-Bead colors turn darker for increasing Schmidt angles $\theta$ which corresponds to increasing entanglement. Moreover, according to the discussed relations, $\ket{\psi_0}$ represents a fully separable state whereas $\ket{\psi_{\pi/2}}$ is maximally entangled.

E-Beads, e.g., $E\{1,2\}_{\textit{even}}$, are a direct visualization of entanglement-based connected correlation functions (see section~\ref{sec:BEADSCorrel}), i.e., nonclassical correlations of measurement outcomes. Two-qubit BEADS representations further comprise a second connected correlation function component $E\{1,2\}_{\textit{odd}}$, which is zero for states of the introduced Schmidt form though, and hence, is omitted in Fig.~\ref{Figure:Fig5}. 

Here, $E\{1,2\}_{\textit{even}}$ represents the fully permutation symmetric bilinear connected correlation component in a two-qubit system, whereas $E\{1,2\}_{\textit{odd}}$ corresponds to permutation antisymmetric bilinear connected correlation function components (see section~\ref{BEADSMap} for further details). Here, the subscript labels \textit{even} and \textit{odd} denote the point inversion symmetry of the E-Bead which allows us to tell for any point on the bead surface whether the corresponding correlation coefficient value at the antipodal point is identical (even symmetry) or of opposite sign (odd symmetry).

In Fig.~\ref{Figure:Fig5}, we observe that the correlation coefficients, that is, the correlation function values displayed by $E\{1,2\}_{\textit{even}}$ increase in magnitude and thus the color brightness rises for increasing values of $\theta$, oppositely to what was discussed for Q-Beads, which coincides with the increase of entanglement in the system. In the maximally entangled case $\theta=\pi/2$, the color brightness of $E\{1,2\}_{\textit{even}}$ is maximum. In contrast, for $\theta=0$, there is no E-Bead (and thus no entanglement) and the state must consequently be separable. Note that in this case, the E-Bead, which here is black in all spatial directions, is omitted. This is a subtlety in design which was chosen to minimize the complexity of the representation. However, black Q-Beads are not omitted since these correspond to the state of physical objects, the qubits.\newline

\begin{enumerate}[start = 2, label=\textbf{Result \arabic*:}, itemindent=1cm, labelindent=1cm]
\item \label{res:BEADS2} If all E-Beads in the BEADS representation of a pure state are black in all spatial directions, the state is a product state.\newline
\end{enumerate}

\begin{enumerate}[start = 3, label=\textbf{Result \arabic*:}, itemindent=1cm, labelindent=1cm]
\item \label{res:BEADS3} $n$-linear E-Beads are a direct visualization of $n$-partite direction-dependent connected correlation functions which for pure quantum states occur exclusively in the presence of entanglement \cite{Mahler95}.\newline
\end{enumerate}

\noindent
At this point, we conclude that in Fig.~\ref{Figure:Fig5} states corresponding to Schmidt angles $0<\theta<\pi/2$ are partially entangled. For such cases, the total correlation coefficient for a given observable is not merely given in terms of connected correlation coefficients. Instead, they are formed as a sum of genuine $n$-qubit connected correlation coefficients and what we refer to as \textit{compound} correlation coefficients. The latter correspond to products of single-qubit expectation values and/or lower-order connected correlation coefficients. Thus, compound correlation coefficients always represent redundant information. They can be removed from a density operator by applying \textit{Ursell} functions \cite{Mahler95} (see section~\ref{sec:BEADSCorrel}). It is also possible to visualize compound correlation coefficients by so-called \textit{C-Beads}, cf. Fig.~\ref{Figure:Fig6}.\newline

\begin{enumerate}[start = 4, label=\textbf{Result \arabic*:}, itemindent=1cm, labelindent=1cm]
\item \label{res:BEADS4} Compound correlation coefficients correspond to products of single-qubit expectation values and/or lower-order connected correlation coefficients. They are visualized by C-Beads (\textit{compound} correlation function beads).\newline
\end{enumerate}

\noindent
Note that from a quantum information theoretical perspective, a pure state is only entangled and thus considered "correlated" if it has non-zero \textit{covariance} \cite{Schilling} which manifests in non-zero \textit{connected} correlation coefficients represented by corresponding E-Beads in the BEADS visualization. This contrasts correlation functions from the more general concept of correlation in information theory \cite{Schilling2}. Importantly, the presence of non-zero \textit{compound} correlation coefficients in BEADS representations of pure states is not an indicator of the correlation of a state. In fact, pure product states, i.e., states of the form $\ket{\psi}=\bigotimes_k \ket{\psi_k}$, where $\ket{\psi_k}$ are single-qubit states, strictly contain only compound correlation coefficients and have zero covariance. They are thus considered \textit{uncorrelated} in quantum information theory. Nonetheless, the information encoded by C-Beads still manifests in \textit{correlations of measurement outcomes} when associated qubits are measured.

\begin{figure}[H]
\centering
\includegraphics[width=.7\textwidth]{SchmidtTotCoeff.pdf}
\caption{\label{Figure:Fig6}T-Bead representations and comparison of correlation function beads for states of Schmidt form $\ket{\psi_\theta}=\cos{(\theta/2)}\ket{00}+\sin{(\theta/2)}\ket{11}$. States for different Schmidt angles $\theta$ are shown in oblique view using T-Beads to visualize the total correlation function between both qubits (top). An extended color scheme is used to represent the ratio of connected ($E$) to compound ($C$) correlation coefficient components displayed by $\phi_{corr}= \text{atan}(E/C)$. The total correlation function is formed by combination of the compound correlation function ($C\{1,2\}_{\textit{even}}$) and the entanglement-based connected correlation function ($E\{1,2\}_{\textit{even}}$) all of which are visualized by the corresponding beads for the given states in the bottom table. Single-qubit expectation values and correlation coefficients along the z-axis are indicated numerically.}
\end{figure}

\noindent
In the BEADS representation, \textit{total} correlation coefficients, which directly correspond to expectation values of multipartite measurement operators (products of local observables), are visualized by so-called T-Beads for which we propose an extended color scheme, cf. Fig.~\ref{Figure:Fig6}, which characterizes the ratio of connected to compound correlation coefficients in each spatial direction.\newline

\begin{enumerate}[start = 5, label=\textbf{Result \arabic*:}, itemindent=1cm, labelindent=1cm]
\item \label{res:BEADS5} In case of pure states, total correlation coefficients are sums of entanglement-based connected and compound correlation coefficients which can be visualized by T-Beads (total correlation function beads).\newline
\end{enumerate}

\noindent
Indeed, total correlation coefficients for a given set of observables, and thus T-Beads, are key for predicting outcomes of projective measurements. E-Beads, while being suitable to characterize and quantify the entanglement in a system of qubits, are instead not sufficient to be used in measurement predictions except for maximally entangled pure states where the total correlation function is solely given by the connected correlation function, (such as for the entangled Bell states shown in Fig.~\ref{Figure:Fig3}). Fully separable states represent a special case, where the outcomes of individual qubits are independent of each other and thus, can be predicted from the local single-qubit information encoded in Q-Beads only. For the visualizations in Fig.~\ref{Figure:Fig5} these findings imply that a direct outcome prediction is only feasible for $\theta=0$ and $\theta=\pi/2$ whereas it is possible for all T-Bead visualizations in Fig.~\ref{Figure:Fig6}.\newline

\begin{enumerate}[start = 6, label=\textbf{Result \arabic*:}, itemindent=1cm, labelindent=1cm]
\item \label{res:BEADS6} In the BEADS representation, measurement outcomes can be predicted from Q-Beads and total correlations (T-Beads). For maximally entangled states, total correlation functions are equivalent to connected correlation functions and thus, E-Beads can be used. Correlation coefficients of product states trivially correspond to products of single-qubit expectation values and outcome predictions can be made using Q-Beads only.\newline
\end{enumerate}

\noindent
Recall that the color of Q-Beads and E-Beads (for maximally entangled states) along arbitrary directions was shown to correspond to measurement probabilities which can be deduced from color scales. We have previously introduced these color scales in Fig.~\ref{Figure:Fig1} and Fig.~\ref{Figure:Fig3} using simple descriptions which were specifically simplified and tailored to the discussed examples. Indeed, the interpretation of colors can be unified to so-called bit parity probabilities $p$ which specify the likelihood of measuring an odd number of qubits in the "down" states, that is, the eigenstate corresponding to eigenvalue $-1$ \cite{NielsenChuang} of the local observable $\sigma_{kr}=\vec{r}_k\cdot\vec{\sigma}_k=r_x\sigma_{kx}+r_y\sigma_{ky}+r_z\sigma_{kz}$ for the $k$-th qubit in the corresponding measurement direction $\vec{r}_k$. In case of a single qubit, this simply corresponds to the probability of measuring the qubit in the down state (see Fig.~\ref{Figure:Fig1}). 

As was briefly indicated for Q-Beads, an alternative and generally applicable interpretation of the utilized color scales is given in terms of expectation values. Indeed, the color of Q-Beads and fully symmetric T-Beads (e.g., $T\{1,2\}_{\textit{even}}$) along direction $\vec{r}$ then directly corresponds to expectation values $\langle O \rangle=\left\langle \prod _{k \in \mathcal{G}} \sigma_{k\vec{r}} \right\rangle$ where $\mathcal{G}$ is the set of all qubits involved in the bead, respectively. $\langle O \rangle$ adopts values in the range between $-1$ and $1$ and directly corresponds to single-qubit expectation values (Q-Beads) or total correlation coefficients (T-Beads). It is related to the bit parity probability \textit{p} by

\begin{equation}\label{eq:BPP}
p=\frac{1-\langle O \rangle}{2}.
\end{equation}

\noindent
\noindent
Full correlation of measurement outcomes corresponds to a expectation value (total correlation coefficient) $\langle O \rangle = 1$ and to a bit parity probability of $p = 0$ and is equivalent to measuring an even number of qubits in the down state of the associated measurement basis whereas $\langle O \rangle = -1$ (or $p = 1$) represents full anticorrelation, i.e., measuring an odd number of down states, see, e.g., Fig.~\ref{Figure:Fig6}.

Note that for T-Beads corresponding to other permutation symmetries (e.g., $T\{1,2\}_{\textit{odd}}$), $\langle O \rangle$ generally corresponds to expectation values of joint measurements, i.e., $O$ is expressed as a linear combination of Pauli product operators which does not correspond to a single projective measurement. While bit parity probabilities can still be calculated according to Eq.~\ref{eq:BPP} for these beads, they have no direct physical interpretation. Further, one should not mistakenly conclude that outcomes of asymmetric measurements, i.e., measurements where qubits are measured in different directions, can be merely predicted from beads which do not represent fully permutation symmetric components. In fact, predictions of asymmetric measurements require more elaborate methods that involve the analysis of symmetric and antisymmetric T-Beads alike which are explained in supplementary sections~\ref{sec:BEADSAsym} and~\ref{sec:BEADSBellTest}.\newline

\begin{enumerate}[start = 7, label=\textbf{Result \arabic*:}, itemindent=1cm, labelindent=1cm]
\item \label{res:BEADS7} In the BEADS representation, colors can be consistently interpreted as expectation values. T-Beads corresponding to full permutation symmetric components can equivalently be interpreted in terms of bit parity probabilities.\newline
\item \label{res:BEADS8} The correlation coefficient (expectation value) and the related bit parity probability associated with a symmetric measurement of $n$ qubits in a set $\mathcal{G}$ is represented by the color of the fully permutation symmetric $n$-linear T-Bead corresponding to $\mathcal{G}$ along the measurement direction $\vec{r}$.\newline
\end{enumerate}

\noindent
Applying these findings to the maximally entangled state ($\theta=\pi/2$) shown in Fig.~\ref{Figure:Fig5} and Fig.~\ref{Figure:Fig6}, we find full positive correlation of measurement outcomes and hence, a bit parity probability of $p = 0$ for symmetric measurements along the z- or x-axes (or any other direction in the xz-plane) can be found whereas the state is fully anticorrelated along the y-axis. Black regions indicate uncorrelated measurement outcomes in the corresponding directions, i.e., all possible outcomes are obtained statistically with equal probabilities. Indeed, the state $\ket{\psi_{\pi/2}}$ is the Bell state $\ket{\Phi^+}=1/\sqrt{2}(\ket{00}+\ket{11})$ for which correlations of outcomes along Cartesian coordinate axes are well-known but not directly evident for any other direction and must usually be calculated individually. In contrast, the BEADS representation provides a holistic view of the correlation coefficients along all possible measurement directions.

Moreover, the BEADS representation is symmetry-adapted. Besides permutational symmetries, it also captures spatial symmetries, e.g., rotational, point reflection, and reflection symmetries through a plane. These are directly evident from the types and surface patterns of beads, respectively. For the Bell state $\ket{\psi_{\pi/2}}=\ket{\Phi^+}$, Fig.~\ref{Figure:Fig5} and Fig.~\ref{Figure:Fig6} makes it immediately clear that the state is fully axially symmetric with respect to rotations around the y-axis and has a two-fold rotational symmetry around any axis in the zx-plane. An overview of symmetries in the BEADS representation is given in section~\ref{BEADSMap}.\newline

\begin{enumerate}[start = 9, label=\textbf{Result \arabic*:}, itemindent=1cm, labelindent=1cm]
\item \label{res:BEADS9} Spatial symmetries of expectation values, i.e., rotational, point reflection, and reflection symmetries through a plane of a quantum states are directly evident in the BEADS representation.
\end{enumerate}

\subsection{Three-qubit entanglement}\label{3Q}

Based on the concepts introduced in the preceding chapter we now visit the most prominent examples of three-qubit entanglement which are the GHZ state $\ket{\text{GHZ}}=1/\sqrt{2}(\ket{000}+\ket{111})$ \cite{GHZBeyondBell, Mermin90} and the W state $\ket{\text{W}} = 1/\sqrt{3}(\ket{001} + \ket{010} + \ket{100})$ \cite{Dur00} and the standard protocols to generate the states, both of which are shown in Fig.~\ref{Figure:Fig7}. Here, we introduce BEADS-augmented circuits which use standard circuit notation to outline the quantum circuit and display the Q-Bead $Q_k$ corresponding to the $k$-th qubit on the corresponding wire after every operation step. Additionally, we also show E-Beads (or alternatively T- or C-Beads) on additional grey lines (the lighter color indicating the virtual nature of the associated correlations displayed information). All beads are visualized in top view. This allows us to monitor the entire system evolution during the circuit on a compact visual level. Note that we use most significant bit first qubit numbering to define our quantum circuits which is standard in most quantum information textbooks (but different to what is used, e.g., in Qiskit\cite{Qiskit}). For simplicity, only fully permutation symmetric correlation beads are shown in this chapter which is sufficient to provide a complete visualization of the GHZ and W states.

The GHZ state \cite{GHZBeyondBell, Mermin90} is generated by first creating a Bell pair $\ket{\Phi^+}$ between the first and second qubit. This is achieved by applying a Hadamard gate on the first qubit represented by $Q_1$ in Fig.~\ref{Figure:Fig7} followed by a CNOT$_{12}$ operation which acts according to what was introduced in Fig.~\ref{Figure:Fig3}. Indeed, we can directly see that the Hadamard gate transforms the initial computational basis $\ket{0}$ state of the first qubit to the equal superposition state ($\ket{+}=1/\sqrt{2}(\ket{0}+\ket{1}$). The equal superposition can be recognized by the black color of $Q_1$ along the z-axis which corresponds to a 50\% probability of measuring $\ket{1}$. Importantly, while the circuit visualization is sufficient to understand the overall effect of the Hadamard gate, it does not reveal the dynamics which occur when the gate is acting on the system and which is vital to obtain a deeper understanding of quantum operations. Indeed, the BEADS representation is also particularly useful to visualize such dynamics. Among various single- and multi-qubit examples, the dynamics of the Hadamard gate are visualized in section~\ref{sec:BEADSGates} of the supplementary material and the animations in the supplementary video which reveal a $180^\circ$-rotation around the xz-bisecting axis.\newline

\begin{figure}[H]
\centering
\includegraphics[width=1.0\textwidth]{FigGHZW.pdf}
\caption{\label{Figure:Fig7}BEADS-augmented circuit representations of the three-qubit GHZ and W state generation \screen. Top views of beads representing the system state before and after gates in circuits to generate the GHZ state (top) and the W state (bottom) starting with an initial state $\ket{000}$ are shown on the left. Oblique view visualizations of the resulting final states are displayed on the right. Using the BEADS representation, it is obvious that the GHZ state is maximally entangled whereas the W state is not. Moreover, the W state is totally axially symmetric with respect to rotations around the z-axis while the GHZ has a threefold rotational symmetry around the same axis which can be deduced from the surface pattern of the trilinear E-Bead $E\{1,2,3\ \tau_1\}_{\textit{odd}}$. Lines connecting individual beads serve as auxiliary entanglement indicators.}
\vspace{-.3cm}
\end{figure}

\begin{enumerate}[start = 10, label=\textbf{Result \arabic*:}, itemindent=1.2cm, labelindent=1.2cm]
\item \label{res:BEADS10} The BEADS representation can be used to show the state of a quantum system after a gate but is also applicable to visualize system dynamics, e.g., during the action of a quantum gate.\newline
\end{enumerate}

\noindent
After the following CNOT$_{12}$ operation, only subsystem $\{1,2\}$ is maximally entangled (forming a Bell pair $\ket{\Phi^+}$) while the third qubit remains separable. These first two quantum operations further reveal the principal transformations of Q-Beads: rotations and scaling, the latter of which is observed as a change of brightness. Local unitary transformations, that is, single-qubit gates act on a Q-Bead simply by rotating it whereas multiqubit gates can induce both rotations and scaling.\newline

\begin{enumerate}[start = 11, label=\textbf{Result \arabic*:}, itemindent=1.2cm, labelindent=1.2cm]
\item \label{res:BEADS11} Quantum gates transform Q-Beads in terms of rotations and scaling of brightness.\newline
\end{enumerate}

\noindent
When a CNOT$_{23}$ (or a CNOT$_{13}$) is applied subsequently, the maximally entangled three-qubit state, the GHZ state is generated. The GHZ state is fully characterized by four E-Beads: three E-Beads corresponding to symmetric two-qubit correlations $E\{k,l\}_{\textit{even}}$, where $\{k,l\}\in\{\{1,2\},\{1,3\},\{2,3\}\}$ denotes the possible bilinear subsystems, and the trilinear E-Bead \linebreak[4]$E\{1,2,3\ \tau_1\}_{\textit{odd}}$. Following Result~\hyperref[res:BEADS3]{3}, the latter shows non-zero fully symmetric connected correlation coefficients which here arise due to genuine three-qubit entanglement. 

Observing the action of the last gate in the circuit on $E\{1,2\}_{\textit{even}}$, it becomes clear that E-Beads (and any other type of correlation function bead, i.e., T-Beads and C-Beads) are not only rotated or scaled but can undergo a third type of transformation which is termed morphing and which corresponds to a change in surface pattern independent of changes in brightness. Morphing arises due to beads involving spherical harmonics of higher ranks ($j > 1$) which do not exclusively transform in terms of simple rotations. In general, local selective rotations, that is, local gates applied to a subset of the qubits involved in a correlation component can cause morphing, whereas global rotations (i.e., the same local operation being applied to all qubits associated with a correlation function bead) induce simple rotations of the bead. Multiqubit gates can induce any of the possible transformations. Further examples of transformations of E-Beads are shown in section~\ref{sec:BEADSGates} of the supplementary material.\newline

\begin{enumerate}[start = 12, label=\textbf{Result \arabic*:}, itemindent=1.2cm, labelindent=1.2cm]
\item \label{res:BEADS12} Quantum gates act on correlation function beads (i.e., E-Beads, T-Beads, and C-Beads) in terms of rotation, scaling, and/or morphing.\newline
\end{enumerate}

\noindent
Since the resulting BEADS visualization only comprises fully permutation symmetric E-Beads, one can directly tell that the GHZ state is totally symmetric with respect to permutations of qubits, and it is immediately clear from the BEADS representation in Fig.~\ref{Figure:Fig7} that any bilinear correlation component is invariant under rotations around the z-axis. In contrast, the $E\{1,2,3\ \tau_1\}_{\textit{odd}}$ bead displays a threefold rotational symmetry with respect to global rotations around the z-axis, i.e., any $120^\circ$-z-rotation of all three qubits in the system, leaves the entire state invariant. Moreover, we immediately see that outcomes of the state are fully correlated for measurements along the z-axis of each two-qubit pair, yet the symmetric tripartite z-correlation coefficient is zero (see color scales in Fig.~\ref{Figure:Fig5} for reference). Given the observed bipartite correlation coefficients, this implies that if any qubit is measured along the z-axis, all qubits adopt the same state. For instance, if the first qubit is measured in state $\ket{0}$, the correlation coefficients represented by $E\{1,2\}_{\textit{even}}$ and $E\{1,3\}_{\textit{even}}$ imply that the second and third qubit will also be in state $\ket{0}$, and thus, the GHZ collapses to $\ket{000}$ overall.

In the BEADS representation, we can see that for the GHZ state measurement outcomes are in fact fully trilinearly correlated for symmetric local measurements along the x-axis, and in directions $(90^\circ, 120^\circ)$ and $(90^\circ, -120^\circ)$ given in terms of spherical coordinates. In contrast, the state is fully trilinearly anticorrelated along the -x-axis and directions $(90^\circ, 60^\circ)$ and $(90^\circ, -60^\circ)$.

Using the BEADS representation, the W state \cite{Dur00}, which is well-known to be fundamentally different from the GHZ state, can be identified to \textit{not} be maximally entangled due to non-zero Q-Beads ($\sigma_z$-expectation value along z-axis $\langle \sigma_z \rangle=1/3$). Based on the resulting BEADS representation we can directly tell the obtained W state to be fully axially symmetric with respect to arbitrary rotations about the z-axis. As the W state is not maximally entangled, it is necessary to look at the corresponding total correlation BEADS representation to understand its behavior under measurements (cf. supplementary section S14 for an exemplary analysis).

Another important class of entangled states are graph states \cite{HeinGraph04, HeinGraph06}. These states can be transformed into each other following well-known rules and represent a natural choice in measurement-based quantum computing. Using the BEADS representation, it is easily possible to illustrate graph state transformation rules and to get insights into the topologies and symmetries of the underlying correlation functions. We provide examples of such graph states and associated transformations in supplementary section~\ref{sec:BEADSGraphStates} for the interested reader.

In the following, we show how key features of quantum algorithms can be analyzed using the BEADS-augmented circuit representation, which provides a compact visualization of the quantum states obtained by each quantum gate, in particular, the generation and manipulation of entanglement. This is illustrated for Grover's algorithm and the teleportation protocol.

\subsection{Grover's algorithm}\label{Grover}
Grover's algorithm \cite{Grover} is also known as quantum search algorithm. It is one of the most prominent examples of quantum algorithms and it is specifically designed for efficiently searching solutions to a black box function, i.e., computational input strings for which the function yields a specific output (e.g., 1). In Fig.~\ref{Figure:Fig8}, we provide BEADS visualizations of two- and three-qubit Grover searches with a single solution, respectively, and show how our representation provides an overview of the algorithm.

The top two-qubit circuit \cite{NielsenChuang} in Fig.~\ref{Figure:Fig8} highlights the individual steps of the algorithm. During the Grover iteration U$_\text{G}$, maximum genuine two-qubit entanglement is created. It is immediately clear that the generated entangled states which have permutation-antisymmetric components $E\{1,2\}_{\textit{odd}}$ are different to the maximally entangled Bell states (see Fig.~\ref{Figure:Fig3}).

The resulting output state of U$_{\text{G}}$ corresponds to a computational basis state $\ket{01}$, the sought solution in this example, which can be directly deduced from the Q-Beads orientations, i.e., the visible color in the center of the upper hemisphere that is visible in top view. Based on the perfect alignment parallel to the measurement direction and vanishing E-Beads, it is clear that a single application of U$_{\text{G}}$ is sufficient to obtain the solution with certainty in a two-qubit single-solution Grover algorithm. Here, entanglement is created during the Grover iteration U$_{\text{G}}$ by the phase oracle U$_{\omega}$ and subsequently eliminated in the diffusion gate U$_{\text{S}}$ by the U$_{\text{0}}$ operation. 

The bottom circuit in Fig.~\ref{Figure:Fig8} corresponds to a three-qubit Grover quantum search, again for a case with only one solution (here, 011). Here, Grover iterations U$_G$ are summarized in single blocks and for simplicity we only show fully symmetric E-Beads. It is well-known, how the number of required iterations depends on the number of qubits and solutions \cite{Boyer98}, and the BEADS-augmented circuit makes it clear that unlike in the two-qubit case a single iteration does not yield the solution with certainty. By examining the Q-Beads after each iteration, the likelihood to measure the correct solution is visible. Using the BEADS representation, one can then directly see that the correct solution is almost, yet not precisely ($p = 78.13\%$) achieved after two iterations. The subsequent iterations worsen the performance. After six iterations the correct solution is almost found with certainty ($p = 99.98\%$) when the qubits are measured. We provide an extended analysis of Grover's algorithm using the BEADS representation in supplementary section~\ref{sec:BEADSGroverSupp}.

\begin{figure}[H]
\centering
\includegraphics[width=1\textwidth]{FigGrover.pdf}
\caption{\label{Figure:Fig8}BEADS-augmented circuit representation of Grover's algorithm \screen. Grover?s quantum search algorithm \cite{Grover} is visualized for two qubits (top) and three qubits (bottom). Using the BEADS representation, steps, where entanglement is maximum or minimum, can be readily identified and one can directly read off the state of interest based on the Q-Bead colors: 01 (red, green) for the two-qubit case on top and 011 (red, green, green) for the three-qubit case shown at the bottom. For three qubits, visualizing the outcomes of each Grover iteration U$_G$ reveals where a solution can be measured with high probability. The measurement probabilities for the shown measurement outcomes are displayed above the final states.}
\vspace{-.3cm}
\end{figure}

Based on the properties and capabilities of different hardware platforms, a compilation of quantum gates results in different control sequences, the effects of which can be visually compared to the action of the ideal target gates. An experimental implementation of a CNOT gate is visualized and explained in supplementary section~\ref{sec:BEADSNMRDS} using the BEADS representation. Furthermore, in realistic simulations the effects of gate imperfections can be visualized in the BEADS-augmented circuit representation.\newline

\begin{enumerate}[start = 13, label=\textbf{Result \arabic*:}, itemindent=1.2cm, labelindent=1.2cm]
\item \label{res:BEADS13} The BEADS representation can be used to analyze idealized gates or entire quantum algorithms which further can be compared with differently compiled experimental implementations. The effect of experimental imperfections can be visualized. The BEADS representation provides an overview over the evolution of entanglement-based connected correlation coefficients, the effects of gates, and measurements.\newline
\end{enumerate}

\subsection{Quantum teleportation}\label{QuTel}
\begin{figure}[H]
\centering
\includegraphics[width=1.0\textwidth]{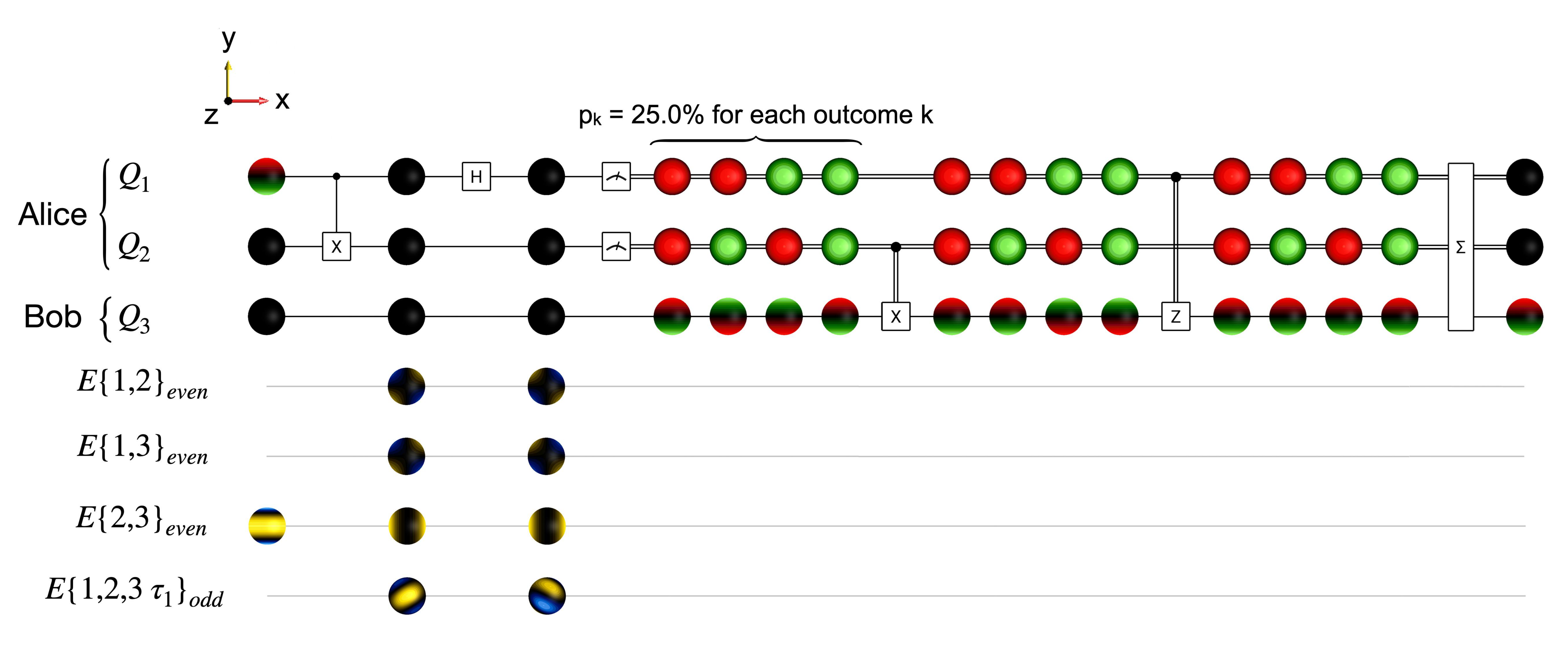}
\caption{\label{Figure:Fig9}BEADS representation of the quantum teleportation protocol \screen. The state ${\ket{R}=1/\sqrt{2}\left(\ket{0}+i\ket{1}\right)}$ is teleported from $Q_1$ onto $Q_3$. In this case, the protocol first involves entangling $Q_1$ with the Bell pair formed by $Q_2$ and $Q_3$. The three-qubit entanglement is then manipulated by a local Hadamard gate, which causes morphing of the affected E-Beads. Subsequently, the first and second qubits are measured in the computational basis. Together with the preceding gates, this implements a Bell measurement. The effect of the classically controlled correction gates indicated by double wires can be directly seen. In the last step, the equivalent mixed state representation is provided that comprises all possible outcomes.}
\vspace{-.3cm}
\end{figure}

In quantum information, quantum teleportation is a fundamental protocol which is specifically applied to transfer the state of a qubit onto another qubit by using entanglement and classical communication \cite{Teleportation}. For instance, we visualize the teleportation of the state $\ket{R}=1/\sqrt{2}\left(\ket{0}+i\ket{1}\right)$ using the BEADS representation in Fig.~\ref{Figure:Fig9}. As the intermediate two-qubit measurement is symmetric (both qubits are measured along the z-axis), we simplify the representation by just showing fully permutation symmetric E-Beads. A non-simplified BEADS visualization of the provided circuit is given among further teleportation examples in supplementary section~\ref{sec:BEADSTelSupp}.

We consider a system of three qubits where $Q_1$ and $Q_2$ are located in Alice's lab and $Q_3$ is located in Bob's remote lab. $Q_1$ is prepared in the state $\ket{R}$ and the remaining qubits form a Bell pair $\ket{\Phi^+}$ which has been previously generated as was shown in Fig.~\ref{Figure:Fig3}. The goal is to transfer the unknown state of $Q_1$ to $Q_3$. First, a Bell measurement is performed by Alice on the first two qubits. This is implemented by a CNOT$_{12}$ gate and a Hadamard gate H$_1$ followed by a measurement in the computational basis. 
Indeed, it is clearly visible that the initial CNOT$_{12}$ operation maximally entangles the first qubit with the already entangled EPR pair creating genuine three-qubit entanglement and non-zero connected two-qubit correlation coefficients between $Q_1$ and $Q_2$, and $Q_1$ and $Q_3$. As was discussed before, the following Hadamard gate morphs the E-Beads which are associated with the first qubit, that is, $E\{1,2\}_{\textit{even}}$, $E\{1,3\}_{\textit{even}}$, and $E\{1,2,3\ \tau_1\}_{\textit{odd}}$. For the subsequent measurement, the outcomes can be predicted based on the colors of the Q-Beads $Q_1$, $Q_2$, and the $E\{1,2\}_{\textit{even}}$ along the measurement direction (z-axis). Since all these beads are black, i.e., the probability to measure $\ket{1}$ for the individual qubits is $50\%$ and the zz-correlation coefficient is zero (i.e., the bit parity probability is $50\%$ as well), we expect any of the four possible outcomes with a probability of 25\% (fully uncorrelated outcomes). Indeed, we observe this constellation of outcomes in Fig.~\ref{Figure:Fig9} where each possible outcome is visualized as a beads column after the measurement. 

Note that from a global perspective the measurement outcomes are fully zzy-correlated. As was briefly mentioned before, such asymmetric correlations can also be predicted based on the BEADS representation. We provide several approaches that use the complete BEADS picture to deduce asymmetric correlations as applied examples in the context of Bell's theorem in supplementary section~\ref{sec:BEADSBellTest}.\newline

\begin{enumerate}[start = 14, label=\textbf{Result \arabic*:}, itemindent=1.2cm, labelindent=1.2cm]
\item \label{res:BEADS14} Comparing possible measurement outcomes in the BEADS representation allows to identify unobvious correlations of measurement outcomes on a visual level.\newline
\end{enumerate}

\noindent
If Alice sends the actually obtained measurement results to Bob through classical communication, the teleportation protocol requires Bob to apply an X-gate to $Q_3$ if the measurement outcome of $Q_2$ is 1 (corresponding to a classically controlled X-gate) followed by a Z-gate on $Q_3$ if the measurement outcome of $Q_1$ is 1 (corresponding to a classically controlled Z-gate).

These corrections finally ensure that the correct state is encoded on $Q_3$, and one can visually monitor but also directly predict the required corrections on $Q_3$ for every possible outcome in the BEADS picture.

\noindent

An alternative way to view the system state after a measurement in the BEADS representation, is given in terms of a corresponding mixed state. This is achieved by forming the weighted sum of the density operators of all possible outcome states, the outcome probabilities serving as weights, respectively. Despite the BEADS representation being especially useful to visualize pure states, it can also be used for representing mixed states. In Fig.~\ref{Figure:Fig9}, an operation $\Sigma$ is included which does not correspond to a gate or measurement, but which symbolizes the discussed mixed state formation, and we visualize the resulting mixed state as the last step in the circuit. Indeed, $Q_1$ and $Q_2$ are now fully black which corresponds to maximally mixed reduced states of both qubits, yet $Q_3$ retains the desired state. We provide an analogous visualization which directly uses a mixed state picture after the measurement in supplementary section~\ref{sec:BEADSTelSupp}.\newline

\begin{enumerate}[start = 15, label=\textbf{Result \arabic*:}, itemindent=1.2cm, labelindent=1.2cm]
\item \label{res:BEADS15} The BEADS representation is suitable to visualize pure and mixed states.\newline
\end{enumerate}

\noindent
At this point, it shall be emphasized that the BEADS representation is generally well-suited to visualize and understand quantum information protocols such as quantum teleportation. For instance, the ability to visualize entanglement-related connected correlation functions in the BEADS representation makes it an optimal choice to investigate entanglement swapping \cite{EntanglementSwapping1, EntanglementSwapping} where entanglement between two distant qubits is created by generating four-qubit entanglement using two Bell pairs. We refer the interested reader to supplementary section~\ref{sec:BEADSEntanglementSwapping} for BEADS visualization of the entanglement swapping protocol. 

\section{Discussion}\label{Discussion}
The BEADS representation introduced in this article provides a general, accurate, complete, and easy to interpret visualization of quantum states which is highly adaptive. It can be tailored to only show relevant information of interest, thus enabling the user to choose a suitable balance between completeness of the representation and simplicity of the visualized information depending on the application or intended recipients. The representation is based on the decomposition of an arbitrary operator into LISA tensor operators \cite{DROPS, LeinerDROPS} that can be directly mapped onto spherical harmonics (see section~\ref{BEADSMap}), which was previously introduced for the DROPS representation \cite{DROPS}. For increasing dimensions of a Hilbert space, i.e., for systems consisting of an increasing number of qubits, the number of required LISA operators grows exponentially. 

However, due to the various possibilities to simplify the BEADS representation (cf. section~\ref{sec:BEADSVariants}), this growth can be counteracted to some extent. We extend the DROPS mapping by introducing non-trivial scaling factors (see section~\ref{BEADSMap}) providing a novel phase-space representation that directly visualizes correlation coefficients (expectation values) of interest for systems of individually addressable particles and which we deem particularly useful for the visualization of pure or mixed quantum states represented by the density operator. With this, the achieved representation has significantly enhanced applicability. Visually, the BEADS and DROPS representations may appear the same at first glance. Hence, a detailed comparison is provided in Fig.~\ref{fig:FigDiscussionDROPS} which demonstrates important differences between the representations. DROPS, which are defined such that the integral over the spherical functions is normalized, can be described as a generalized Wigner function. While the DROPS representation can reveal symmetry properties of a visualized operator or reveal basic rotations, compared to BEADS, it does not provide direct quantitative information on entanglement-related correlations, or in general, correlations and hence it is significantly less useful for understanding gates or algorithms in quantum information.

Using selected examples, we demonstrated applications of the BEADS representation for single-qubit, two-qubit, or systems consisting of three or more qubits. Indeed, the visualization is not even restricted to the visualization of qubit system states and can be extended to the visualization of states of arbitrary qudit systems as well.

In the BEADS representation, Q-Beads are a representation of the reduced single-qubit density operators and provide a direct visualization of the spatial orientation of qubit states, rotations of qubit states, direction-dependent correlation coefficients (expectation values), and measurement probabilities. Although Q-Beads are essentially equivalent to the Bloch vector, the alternative visualization approach by spherical functions notably facilitates an interpretation of expectation values and probabilities in arbitrary directions (see appendix~\ref{sec:BEADSBloch} for further details).

The BEADS representation further gives a direct and intuitive visualization of correlation functions. Different types of beads allow to visualize entanglement-based connected (E-Beads), total (T-Beads), and compound correlation functions (C-Beads) between all or subsets of qubits in a system and provide corresponding direction-dependent correlation coefficients and measurement probabilities of quantum states. E-Beads, which are obtained by removing compound correlation function components from the density operator (see section~\ref{sec:BEADSCorrel}) \cite{Mahler95}, offer the possibility to directly view the spatial orientation of entanglement-based connected correlation coefficients. 

Notably, this represents an important innovation as previous Ursell-function-based visualization approaches of correlation functions such as correlation matrix visuals (CMVs) \cite{Hazzard1, Hazzard2} achieve visualizations that merely approximate shapes which are proportional to $n$-qubit connected correlation coefficients along selected spatial directions and are less intuitive for correlation functions between more than two qubits. For the interested reader, we provide a detailed comparison and analysis of BEADS representations and CMVs of two-qubit connected correlations in supplementary section~\ref{sec:BEADSCMV}.

The applied separation of correlation function components, i.e., correlation coefficient types is currently restricted to visualizations of pure states. The more intricate nature of mixed states, which can have different types of correlation, e.g., classical correlation \cite{Werner}, entanglement \cite{Werner}, or other quantum correlations such as quantum discord \cite{Discord}, requires careful consideration and prevents a direct application of Ursell-functions such as in the case of pure states. Direct BEADS representations of mixed state entanglement-based correlation coefficients by Ursell-function-based approaches are an open problem which is currently under investigation and will be addressed in a future article.

The BEADS representation is symmetry adapted and thus reproduces spatial, e.g., rotational symmetries and permutation symmetries of correlations of a given state in a visual way. Since the BEADS representation is based on the decomposition of a density operator, it does not visualize the global phase. If the global phase is of interest, the BEADS representation can be supplemented by the state vector or state-vector-based alternative representations such as the dimensional circle notation (DCN) \cite{DCN}. 

We showed that a clear link between classical bits and qubits exists in the BEADS representation. In addition, due to the ability to visualize and observe rotations by single-qubit gates, the more intricate transformations induced by multiqubit gates, and the effects of projective measurements, together with the possibility to simultaneously represent all possible measurement outcomes, the BEADS representation is useful for didactic and scientific purposes with target audiences ranging from the general public to researchers in quantum information processing. Indeed, the visualization is particularly useful for understanding and designing quantum gates, predicting measurement outcomes and understanding and analyzing quantum algorithms. However, since BEADS are expectation-value-based, they do not provide a direct visualization of probability amplitudes which are central in the design of quantum algorithms. Hence, alternative representations such as DCN \cite{DCN} may be more suitable for such applications.

Besides offering a dynamical simulation software (\textit{QuBeads}, see appendix~\ref{app:BEADSQuBeads}), which allows to use the BEADS representation in a highly interactive manner, we find it beneficial to provide tangible three-dimensional BEADS models as a hands-on approach for educational purposes, thus allowing to literally grasp the otherwise abstract mathematical concepts of quantum mechanics and especially quantum computing.  Note that here, we do use the term model to denote a physical tactile object (alternatively referred to as a token). As was briefly mentioned in the introduction, the BEADS representation is not a mechanistic model of quantum computing and is not capable of performing computations on its own. Instead, it is a representation of the abstract quantum states of a system of qubits which is independent of the experimental platform which qubits are implemented on. For further reading we provide a detailed hierarchical overview on models and representations in context of the BEADS representation in supplementary section~\ref{ModelsTokens}.

In addition, we summarize the aspects mentioned in this discussion in tabular form in supplementary section~\ref{sec:BEADSComp} where we also provide a comparative assessment of existing alternative representations some of which were briefly addressed throughout this article.

\section{Conclusion}
In this article, we introduced the BEADS representation for visualizations of pure quantum states. We were able to show that it is possible to rigorously formulate a graphical representation that provides mathematically complete pictures of quantum states. The BEADS representation simultaneously captures important information of a quantum state such as correlation functions and reduced single-qubit density operators which are visualized in an intuitive way. By applying a separation of correlation function components which was initially proposed in \cite{Mahler95}, we achieve representations that allow to analyze the correlation properties of a given pure quantum state on a visual level. This renders the BEADS representation especially well suited for quantum information with applications ranging from education to science.

The BEADS representation provides a new way to see quantum information. In particular, the effect of extending the standard circuit representation to the BEADS-augmented circuit representation is akin to switching on the light to reveal what is happening during a quantum algorithm.

\section{Methods}\label{Methods}

\subsection{The LISA tensor operator basis}\label{LISA}
In order to map an operator $A$, e.g., the density operator describing a quantum state, to a corresponding BEADS representation, $A$ is decomposed in a suitable spherical tensor basis. In \cite{DROPS}, we introduced the so-called LISA spherical tensor basis, which is especially suited for applications in which the qubits are distinguishable. In particular, this is the case for quantum computing, for which the DiVincenzo criteria \cite{DiVincenzo} include single-qubit gates (as part of a universal set of quantum gates) and the capability to perform qubit-specific measurements. 

The LISA tensor operators $T_{j,m}^{\left(\ell\right)}$ with the rank $j$ and order $m\in{-j,...,j}$ form a complete basis and can be organized in sets $\ell$ that are hierarchically defined based on their \textit{\textbf{li}nearity} (i.e., the number of qubits on which the basis operator acts), \textit{\textbf{s}ubsystem} (the involved qubits on which the basis operator acts), and \textit{\textbf{a}uxiliary} criteria, such as permutation symmetry, parentage and (for systems consisting of more than five qubits) additional sublabels \cite{DROPS, LeinerDROPS}. Explicit expressions of the standard LISA operators for systems consisting of up to six qubits can be found in \cite{DROPS} and \cite{LeinerDROPS}.

\subsection{General visualization approach}\label{GenMap}

As shown in \cite{DROPS, LeinerDROPS}, any operator $A$ can be uniquely decomposed in the form
\begin{equation}
A=\sum_{\ell\in L}{A^{\left(\ell\right)}},
\end{equation}
where each of the components $A^{\left(\ell\right)}$ correspond to a specific set $\ell$, such that $A^{\left(\ell\right)}$ can be expressed as a linear combination of LISA basis operators $T_{j,m}^{\left(\ell\right)}$:
\begin{equation}\label{eq:DROPSdecomp}
A^{\left(\ell\right)}=\sum_{j\in J\left(\ell\right)}^{\ }\sum_{m=-j}^{j}{c_{jm}^{\left(\ell\right)}T_{j,m}^{\left(\ell\right)}\ },
\end{equation}
where the set of ranks $J\left(\ell\right)$ contains each value $j$ not more than once \cite{DROPS}.

For example, in the case of a system consisting of two qubits, $A^{\left(\ell\right)}$ can be one of the linear operators $A^{\{1\}}$ or $A^{\{2\}}$ acting only on the first or second qubit, respectively. Furthermore, A$^{\left(\ell\right)}$ can be the bilinear operator $A^{\{1,2\}}$ acting on both qubits, and the operator $A^{\{\emptyset\}}$, which is proportional to the identity operator. 

Similarly, in the case of a system consisting of three qubits, there can be linear operator components ($A^{\{1\}}$, $A^{\{2\}}$, $A^{\{3\}}$), bilinear operator components ($A^{\{1,2\}}$, $A^{\{1,3\}}$, $A^{\{2,3\}}$), and the identity term  $A^{\{\emptyset\}}$. In addition, there can be trilinear operator components acting on all three qubits. However, unlike for the bilinear components  $A^{\{k,l\}}$,  it is not possible to combine all trilinear components in a single set $\ell$ because in this case the condition would be violated that the set of ranks $J\left(\ell\right)$ must not contain any value $j$ more than once, as otherwise a bijective mapping between spherical tensor operators and simple spherical functions would not be possible (for details, see Table~1 of \cite{DROPS}). In the LISA basis, this problem is solved by adding an additional label $\tau_k$ based on the symmetry of operator components with respect to permutation of qubits. This results in the four trilinear operator components ($A^{\{k,l,m\ \tau_1\}}$, $A^{\{k,l,m\ \tau_2\}}$, $A^{\{k,l,m\ \tau_3\}}$, $A^{\{k,l,m\ \tau_4\}}$), where $\tau_1$ corresponds to completely symmetric and $\tau_4$ to completely antisymmetric permutation symmetry, whereas $\tau_2$ and $\tau_3$ correspond to symmetric and antisymmetric permutation symmetry with respect to a permutation of only the first and second qubit \cite{DROPS}).

In the DROPS representation \cite{DROPS}, each of the operator components $A^{\left(\ell\right)}$ is directly mapped to a unique spherical function $f_A^{(\ell)}$ (called a ?droplet function?) by simply replacing the spherical tensor operators in Eq.~\ref{eq:DROPSdecomp} by spherical harmonics:
\begin{equation}\label{eq:DROPSMapping}
A^{\left(\ell\right)}=\sum_{j\in J\left(\ell\right)}^{\ }\sum_{m=-j}^{j}{c_{j,m}^{\left(\ell\right)}T_{j,m}^{\left(\ell\right)}\ }\Longleftrightarrow\ f_A^{\left(\ell\right)}=\sum_{j\in J\left(\ell\right)}^{\ }\sum_{m=-j}^{j}{c_{j,m}^{\left(\ell\right)}Y_{j,m}}.
\end{equation}
Hence, any operator $A$ can be bijectively mapped on a discrete set of droplet functions $f_A^{(\ell)}$ \cite{DROPS}:
\begin{equation}
A=\sum_{\ell\in L}{A^{\left(\ell\right)}\ \Longleftrightarrow\ \bigcup_{\ell\in L}\ f_A^{\left(\ell\right)}}.
\end{equation}
The obtained set of spherical functions satisfies generalized Stratonovich conditions \cite{Brif98, Brif99, Stratonovich} for which the achieved mapping can be viewed as a generalized \textit{Wigner}-type representation \cite{DROPS}. 

\subsection{BEADS mapping}\label{BEADSMap}

Based on the mapping of Eq.~\ref{eq:DROPSMapping}, the following extensions are introduced in the BEADS representation:

\subsubsection*{Spherical function symmetry-adapted sets $\boldsymbol{\ell^\prime}$}

We introduce new sets $\ell^\prime$ with respect to the spherical function symmetries. In particular, we subdivide the original sets $\ell$ \cite{DROPS} are subdivided according to spherical function point symmetry, that is, the behavior under space inversion where odd point symmetry of spherical functions corresponds to a sign flip whereas even point symmetry implies invariance under space inversion. This is equivalent to subdividing spherical functions in $\ell$ according to even or odd ranks of the underlying LISA tensor operators such that modified mapping is
\begin{equation}
\sum_{\ell^\prime\in L}{A^{\left(\ell^\prime\right)}\ \Longleftrightarrow\ \bigcup_{\ell^\prime\in L}\ b_A^{\left(\ell^\prime\right)}}.
\end{equation}
As the point symmetries of spherical functions $b_A^{(\ell^\prime)}$ are well-defined, seeing the spherical functions from a single perspective provides complete information as the spherical function values on the hidden hemisphere can be directly inferred. Thus, we use this as a standard approach for the representation.
In the BEADS representation, spherical functions $b_A^{(\ell^\prime\ )}$ are called "beads". We use a general labelling scheme where involved qubits and additional labels are written in curly brackets and the spherical function point symmetry is provided as a subscript, e.g., $\{1,2,3\ \tau_1\}_{\textit{odd}}$. The sets $\ell^\prime$ and corresponding spherical function point symmetries and permutation symmetries for systems consisting of up to three qubits are summarized in Table~\ref{tab:TableSymmetries}.

\begin{table}[h!]
\caption{\label{tab:TableSymmetries}Overview of BEADS sets $\ell^\prime$ and corresponding spherical function and permutation symmetries in systems consisting of up to three qubits. All symmetry properties of beads corresponding to linearity $g > 2$ apply to T-, E-, and C-Beads. Fully permutation symmetric components are given by beads $\{k,l\}_{\textit{even}}$ and $\{k,l,m\ \tau_1\}_{\textit{odd}}$.}
\begin{tabular}{m{0.07\textwidth}<{\centering} m{0.15\textwidth}<{\centering} m{0.09\textwidth}<{\centering} m{0.15\textwidth}<{\centering} m{0.21\textwidth}<{\centering} m{0.12\textwidth}<{\centering}}
\hline\hline\\[-8pt]
Linearity $g$ & Bead $\ell^\prime$ & Ranks $j$ & Spherical function point symmetry & Permutation symmetry & Full permutation symmetry\\[14pt] \hline\hline \\[-8pt]
0 & $\{\emptyset\}$ & 0 & even & -- & -- \\[4pt] \cline{1-6} \\[-10pt]

1 & $\{k\}$ & 1 & odd & -- & -- \\[2pt]  \cline{1-6} \\[-12pt] 

\multirow{5}{*}{2} & \multirow{2}{*}{$\{k,l\}_{\textit{even}}$} & 0 & \multirow{2}{*}{even} & \multirow{2}{*}{{symmetric $\left(k,l\right)$}} & \multirow{2}{*}{\cmark} \\[2pt]
 & & 2 &  &  & \\[0pt]  \cline{2-6} \\[-12pt] 
 & $\{k,l\}_{\textit{odd}}$ & 1 & odd & {antisymmetric $\left(k,l\right)$} & \xmark \\[8pt]  \cline{1-6} \\[-12pt] 
\multirow{12}{*}{3} & \multirow{2}{*}{$\{k,l,m\ \tau_1\}_{\textit{odd}}$} & 1 & \multirow{2}{*}{odd} & \multirow{2}{*}{{symmetric $\left(k,l,m\right)$}} & \multirow{2}{*}{\cmark} \\[2pt]
 & & 3 &  &  & \\[0pt]  \cline{2-6} \\[-12pt]  
  & $\{k,l,m\ \tau_2\}_{\textit{odd}}$ & 1 & odd & \multirow{2}{*}{symmetric $\left(k,l\right)$} & \multirow{2}{*}{\xmark} \\[2pt]  \cline{2-4}\\[-10pt] 
  & $\{k,l,m\ \tau_2\}_{\textit{even}}$ & 2 & even &  &  \\[2pt] \cline{2-6}\\[-10pt] 
  & $\{k,l,m\ \tau_3\}_{\textit{odd}}$ & 1 & odd & \multirow{2}{*}{antisymmetric $\left(k,l\right)$} & \multirow{2}{*}{\xmark} \\[2pt]  \cline{2-4}\\[-10pt] 
  & $\{k,l,m\ \tau_3\}_{\textit{even}}$ & 2 & even &  &  \\[2pt]  \cline{2-6} \\[-10pt]
  & $\{k,l,m\ \tau_4\}_{\textit{even}}$ & 0 & even & {antisymmetric $\left(k,l,m\right)$}  & \xmark \\[4pt]  \hline\hline
\end{tabular}
\end{table}

\noindent
Further explanations on spherical function point symmetries and permutation symmetries are given in appendix~\ref{sec:BEADSSymmetries}.

\subsubsection*{Introduction of novel bead scaling factors $\boldsymbol{s_j^{\left(\ell^\prime\right)}\left(N,g\right)}$}

Compared to the Wigner-type DROPS representation, we introduce bead $\ell^\prime$ and rank $j$ specific scaling factors $s_j^{\left(\ell^\prime\right)}\left(N,g\right)$ which depend on the total number $N$ of qubits in the system and the linearity $g$.
By this, we scale the spherical harmonics of different ranks $j$, which belong to the same bead $\ell^\prime$, independently such that the resulting bead function values directly correspond to expectation values of interest.
The original mapping introduced in Eq.~\ref{eq:DROPSMapping} thus generalizes to
\begin{align}\label{eq:BEADSMapping}
A^{\left(\ell^\prime\right)}=\sum_{j\in J\left(\ell^\prime\right)}^{\ }\sum_{m=-j}^{j}{c_{j,m}^{\left(\ell^\prime\right)}T_{j,m}^{\left(\ell^\prime\right)}\ }\Longleftrightarrow\ b_A^{(\ell^\prime)}&=\sum_{j\in J\left(\ell^\prime\right)}^{\ }\sum_{m=-j}^{j}{{c^\prime}_{j,m}^{\left(\ell^\prime\right)}Y_{j,m}} \nonumber\\
&=\sum_{j\in J\left(\ell^\prime\right)}^{\ }\sum_{m=-j}^{j}{s_j^{\left(\ell^\prime\right)}\left(N,g\right) c_{j,m}^{\left(\ell^\prime\right)}Y_{j,m}}.
\end{align}
The scaling factors $s_j^{\left(\ell^\prime\right)}\left(N,g\right)$ are formed by a product of the three individual scaling factors:
\begin{equation}\label{eq:BEADSScalingFactor}
s_j^{\left(\ell^\prime\right)}\left(N,g\right)=\zeta\left(N\right)\cdot\xi_j^{\left(\ell^\prime\right)}\left(g\right)\cdot\eta_j.
\vspace{6pt}
\end{equation}
Here, $\zeta\left(N\right)$ depends on the total number $N$ of qubits in a system, $\eta_j$ is a rank-dependent scaling factor, and $\xi_j^{\left(\ell^\prime\right)}\left(g\right)$ denotes additional rank- and linearity-dependent factors which are specific for each bead. The latter are determined by one of two approaches:\newline

\begin{enumerate}[label=\textbf{\roman*)}]
\item Scaling factors of fully permutationally symmetric beads are calculated such that the resulting spherical function values directly correspond to experimentally measurable expectation values of Pauli product operators. We give a detailed derivation in appendix~\ref{sec:BEADSCanonicalScaling}. 
\item Scaling factors $\xi_j^{\left(\ell^\prime\right)}\left(g\right)$ of spherical functions corresponding to beads which do not represent fully permutation symmetric components are instead determined based on the global unitary bound of the underlying spherical tensor operator which is explained in appendix~\ref{sec:BEADSGUB}.\newline
\end{enumerate}

\noindent
In both cases, scaling factors are determined with respect to axial LISA tensor operators, i.e., LISA operators of order $m = 0$. Operator components of order $m \neq 0$ are scaled with the scaling factors determined for the corresponding axial operators. The scaling factors $\xi_j^{\left(\ell^\prime\right)}\left(g\right)$ for systems of up to three qubits are summarized in Table~\ref{tab:TableSF}.

Note that unlike in \cite{DROPS, LeinerDROPS} where non-Hermitian LISA tensor operators are mapped onto complex spherical harmonics, here, we use corresponding Hermitian LISA operators (see supplementary section~\ref{sec:BEADSLISASupp} for definitions of \textit{Hermitian LISA operators} for systems of up to three qubits) in derivations and descriptions throughout this chapter. These operators are mapped onto \textit{real spherical harmonics}. This is motivated by the fact that the BEADS representation is focused on visualizing density operators which are Hermitian by definition (yet, both approaches are entirely equivalent and yield the exact same results).  The introduced scaling factors give a set of spherical functions which do not satisfy all Stratonovich conditions. In particular, the scaling approach prevents satisfying the norm condition $\sum_{\ell^\prime\in L^\prime}\int_{S^2}^{\ }{b_A^{(\ell^\prime)}\left(\theta,\phi\right)b_{Id}^{(\ell^\prime)}\left(\theta,\phi\right)}d\mu=\text{Tr}\left(A\right)$ and hence, the BEADS representation cannot be interpreted as a generalized Wigner function. Instead, it can be viewed as generalization of the Husimi function (see section~\ref{sec:BEADSHusimi}).
Analogous to what was shown by Leiner et al. \cite{LeinerWQST}, it is then possible to derive corresponding scaled axial tensor operators which can be used to experimentally reconstruct the BEADS representation in tomography experiments. We give an overview of the reconstruction method and the required operators in appendix~\ref{sec:BEADSTomography}.

Importantly, the generalized mapping (Eq.~\ref{eq:BEADSMapping}) is still bijective since the expansion coefficients ${c^\prime}_{j,m}^{\left(\ell^\prime\right)}$ can always be mapped to the tensor operator coefficients

\begin{equation}
c_{j,m}^{\left(\ell^\prime\right)}=\frac{{c^\prime}_{j,m}^{\left(\ell^\prime\right)}}{s_j^{\left(\ell^\prime\right)}\left(n,g\right)}.
\vspace{2cm}
\end{equation}
\begin{table}[h!]
\caption{\label{tab:TableSF}Overview of scaling factors $\xi_j^{\left(\ell^\prime\right)}\left(g\right)$ for up to trilinear LISA tensor operators. The scaling method specifies whether the displayed scaling factor was calculated by using the canonical scaling approach (cf. section ~\ref{sec:BEADSCanonicalScaling}) or based on the global unitary bound (GUB, explained in section~\ref{sec:BEADSGUB}).}
\begin{tabular}{p{0.15\textwidth}<{\centering} p{0.15\textwidth}<{\centering} m{0.15\textwidth}<{\centering} m{0.15\textwidth}<{\centering}}
\hline\hline\\[-5pt]
Bead $\ell^\prime$ & Rank $j$ & Scaling factor $\xi_j^{\left(\ell^\prime\right)}\left(g\right)$ & Scaling method\\[12pt] \hline\hline \\[-8pt]
$\{\emptyset\}$ & 0 & -- & -- \\[6pt]  \hline \\[-8pt]
$\{k\}$	& 1 & -- & -- \\[6pt] \hline \\[-8pt]
\multirow[c]{2}{*}{$\{k,l\}_{\textit{even}}$} & 0 & $\sqrt{\frac{1}{3}}$ & canonical\\[6pt]
	& 2 & $\sqrt{\frac{2}{3}}$ & canonical\\[6pt]  \hline \\[-8pt]
$\{k,l\}_{\textit{odd}}$	& 1	& $\frac{1}{\sqrt2}$ &	GUB\\[6pt]  \hline \\[-8pt]
\multirow[c]{2}{*}{$\{k,l,m\ \tau_1\}_{\textit{odd}}$} & 1 & $\sqrt{\frac{3}{5}}$ &	canonical\\[6pt]
	& 3 & $\sqrt{\frac{2}{5}}$ &	canonical\\[6pt]  \hline \\[-8pt]
$\{k,l,m\ \tau_2\}_{\textit{odd}}$ &	1 & $\frac{3}{3+\sqrt3}$ &	GUB\\[6pt]  \hline \\[-8pt]
$\{k,l,m\ \tau_2\}_{\textit{even}}$ &	2 & $\frac{1}{\sqrt2}$ &	GUB\\[6pt]  \hline \\[-8pt]
$\{k,l,m\ \tau_3\}_{\textit{odd}}$ &	1 & $\frac{1}{\sqrt2}$ &	GUB\\[6pt]  \hline \\[-8pt]
$\{k,l,m\ \tau_3\}_{\textit{even}}$ &	2 & $\frac{1}{\sqrt2}$ &	GUB\\[6pt]  \hline \\[-8pt]
$\{k,l,m\ \tau_4\}_{\textit{even}}$ &	0 & $\frac{1}{\sqrt2}$ &	GUB\\[6pt]  \hline\hline
\end{tabular}
\end{table}  

\subsection{Correlation functions and separation of correlation function components}\label{sec:BEADSCorrel}

A common method to analyze correlations of a quantum state is given by $n$-partite correlation functions which correspond to expectation values of products of measurement outcomes \cite{Weinfurter}. For any $n$-qubit subsystem $\mathcal{G}$ of an $N$-qubit system, the correlation function is defined as
\begin{equation}\label{eq:CorrFunc}
T_{\mathcal{G}} = \text{Tr}\left(\rho\:O_{\mathcal{G}}\right) = \left\langle O_{\mathcal{G}} \right\rangle,
\end{equation}
where the observable $O_{\mathcal{G}}$ is given by
\begin{align}
O_{\mathcal{G}} = 
\bigotimes_{k = 1}^N \begin{cases}
\vec{r}_k \cdot \vec{\sigma} \qquad &\text{if } k \in \mathcal{G} \\
\mathds{1} \qquad  &\text{otherwise} \\
\end{cases}.
\end{align}
Here, $\vec{r}_k$ is the unit vector specifying the measurement direction of the $k$-th qubit and $\vec{\sigma} = (\sigma_x, \sigma_y, \sigma_z)$ is the vector of single-qubit Pauli matrices.
The correlation function value for a particular observable is also called \textit{(total) correlation coefficient} and adopts values between $-1$ and $1$ \cite{BrandtStatistics}.
Note that total correlation coefficients are sometimes referred to as disconnected correlation coefficients \cite{Tran17}. Importantly, it is not directly possible to draw any conclusions on the entanglement properties of a state using total correlation coefficients. 

As was shown by Schlienz et al. \cite{Mahler95, Mahler96}, in a qubit system described by a pure state $\ket{\psi}$ and the corresponding density operator $\rho=\ket{\psi}\bra{\psi}$, different contributions to the correlation function $T$ can be distinguished:\newline

\begin{enumerate}[label=\textbf{\roman*)}]
\item \textit{Connected} correlation functions $E$ which arise exclusively if a pure state is entangled \cite{Mahler95, Mahler96, Mahler97, Tran17}, that is, they emerge in presence of quantum correlation.
\item Correlation functions which factorize to products of single-qubit expectation values and/or lower-order connected correlation coefficients. Such contributions represent redundant information. We combine these coefficients as \textit{compound} correlation coefficients $C$.\newline
\end{enumerate}

\noindent
From a quantum information theoretic perspective \cite{Schilling2}, $n$-partite compound correlation functions do in general not emerge from genuine $n$-partite correlations of a pure state. Hence, such contributions  unnecessarily complicate and sometimes are obstructive for the study of correlation and entanglement properties. A detailed distinction between correlation functions or coefficients and the notion of correlation of a state in quantum information theory is provided in appendix~\ref{CorrQIT}.

In the BEADS representation, we use this separation of correlation coefficient types to obtain a modified operator $\widetilde{\rho}$ from the density operator that only has non-zero connected correlation functions and single-qubit components, which can be visualized by E-Beads and Q-Beads, respectively. This is achieved by expanding the density operator in an arbitrary basis, e.g., the Pauli basis and subtracting each basis operator weighted by the associated compound correlation coefficients. Moreover, we can use the separation procedure to quantify the ratio of connected to compound correlation coefficients in a specialized color scheme (see Figure~\ref{Figure:Fig6} and appendix~\ref{sec:BEADSColors}),

When further removing all single-qubit components and the identity from $\widetilde{\rho}$, the norm of the resulting operator is referred to as the \textit{entanglement norm} of a pure state and can serve as a measure to quantify the entanglement in a system \cite{Mahler95}. Using a decomposition of $\widetilde{\rho}$ into LISA tensor operators (see supplementary section~\ref{app:BEADSLISACorrel} for further details), it is possible to determine the entanglement norms corresponding to individual symmetry components.

$n$-partite connected correlation functions $E$ are obtained by applying the concept of $n$-partite \textit{Ursell} correlation functions \cite{Mahler95, Mahler96, Mahler97, Tran17, Kumar09, Ursell} which generalize the notion of covariance known from classical information theory \cite{Mahler97}. By that, all $n$-partite compound correlation coefficients are eliminated from the total correlation function $T$. 

Following this approach, $n$-qubit connected correlation functions $E$ can be defined 
recursively based on $N$-qubit total correlation functions $T$ which have the following expressions for two-qubit (Eq.~\ref{eq:Connected2Q}) and three-qubit systems:
\begin{align}
\label{eq:Connected2Q}
T_{12} &= C_{12} + E_{12}= \left\langle\sigma_{1\vec{r}_1}\right\rangle\left\langle\left.\sigma_{2\vec{r}_2}\right\rangle\right.+E_{12} \\
\nonumber \nonumber\\[-8pt]
T_{123} &= C_{123} + E_{123} \nonumber \\
& = \left\langle\sigma_{1\vec{r}_1}\right\rangle\left\langle\left.\sigma_{2\vec{r}_2}\right\rangle\left\langle\left.\sigma_{3\vec{r}_3}\right\rangle\right.\right.\nonumber\\
&+{\left\langle\left.\sigma_{1\vec{r}_1}\right\rangle\right.}E_{23}+\left\langle\left.\sigma_{2\vec{r}_2}\right\rangle E_{13}\right.+\left\langle\left.\sigma_{3\vec{r}_3}\right\rangle E_{12}\right.\nonumber\\
\label{eq:Connected3Q}&+E_{123}.
\end{align}
\noindent
Two-qubit connected correlation coefficients are thus given by \cite{Mahler95, Tran17, Kumar09}
\begin{equation}\label{eq:Connected2Q}
E_{12}=T_{12}-C_{12}=\left\langle\sigma_{1\vec{r}_1}\sigma_{2\vec{r}_2}\right\rangle-\left\langle\sigma_{1\vec{r}_1}\right\rangle\left\langle\left.\sigma_{2\vec{r}_2}\right\rangle\right.
\end{equation}
\noindent
and we obtain the modified two-qubit density operator as
\begin{equation}\label{eq:CoreelCoeff2Q}
\widetilde{\rho}=\rho-\frac{1}{4}\sum_{\alpha,\beta=x,y,z}{\left\langle\sigma_{1\alpha}\right\rangle\left\langle\sigma_{2\beta}\right\rangle\sigma_{1\alpha}\sigma_{2\beta}}.
\end{equation}
Three-qubit connected correlation coefficients are defined as \cite{Mahler95, Kumar09}
\begin{align}
E_{123}&=T_{123}-C_{123}\nonumber\\
&=\left\langle\sigma_{1\vec{r}_1}\sigma_{2\vec{r}_2}\sigma_{3\vec{r}_3}\right\rangle\nonumber\\
&-\left\langle\sigma_{1\vec{r}_1}\right\rangle E_{23}-\left\langle\sigma_{2\vec{r}_2}\right\rangle E_{13}-\left\langle\sigma_{3\vec{r}_3}\right\rangle E_{12}\nonumber\\
&-\left\langle\sigma_{1\vec{r}_1}\right\rangle\left\langle\sigma_{2\vec{r}_2}\right\rangle\left\langle\sigma_{3\vec{r}_3}\right\rangle.\label{eq:Connected3Qv}
\end{align}
Note that as discussed previously, $C_{123}$ in Eq.~\ref{eq:Connected3Q} and~\ref{eq:Connected3Qv} is a sum of multiple correlation coefficients which are formed by products of local single-qubit expectation values and/or bipartite lower-order connected correlation coefficients.

Indeed, it is possible to express these trilinear correlation coefficients as a linear combination of products of single-qubit components and $n$-qubit total correlations by applying the relations from Eq.~\ref{eq:Connected3Q} which gives
\begin{align}
E_{123}&=\left\langle\sigma_{1\vec{r}_1}\sigma_{2\vec{r}_2}\sigma_{3\vec{r}_3}\right\rangle-\left\langle\sigma_{1\vec{r}_1}
\right\rangle\left(\left\langle\sigma_{2\vec{r}_2}\sigma_{3\vec{r}_3}\right\rangle-\left\langle\sigma_{2\vec{r}_2}\right\rangle\left\langle\sigma_{3\vec{r}_3}\right\rangle\right)\nonumber\\
&-\left\langle\sigma_{2\vec{r}_2}\right\rangle\left(\left\langle\sigma_{1\vec{r}_1}\sigma_{3\vec{r}_3}\right\rangle-\left\langle\sigma_{1\vec{r}_1}\right\rangle\left\langle\sigma_{3\vec{r}_3}\right\rangle\right)-\left\langle\sigma_{3\vec{r}_3}\right\rangle\left(\left\langle\sigma_{1\vec{r}_1}\sigma_{2\vec{r}_2}\right\rangle-\left\langle\sigma_{1\vec{r}_1}\right\rangle\left\langle\sigma_{2\vec{r}_2}\right\rangle\right)\nonumber\\
&-\left\langle\sigma_{1\vec{r}_1}\right\rangle\left\langle\sigma_{2\vec{r}_2}\right\rangle\left\langle\sigma_{3\vec{r}_3}\right\rangle\nonumber\\
&=\left\langle\sigma_{1\vec{r}_1}\sigma_{2\vec{r}_2}\sigma_{3\vec{r}_3}\right\rangle-\left\langle\sigma_{1\vec{r}_1}\right\rangle\left\langle\sigma_{2\vec{r}_2}\sigma_{3\vec{r}_3}\right\rangle-\left\langle\sigma_{2\vec{r}_2}\right\rangle\left\langle\sigma_{1\vec{r}_1}\sigma_{3\vec{r}_3}\right\rangle-\left\langle\sigma_{3\vec{r}_3}\right\rangle\left\langle\sigma_{1\vec{r}_1}\sigma_{2\vec{r}_2}\right\rangle\nonumber\\
&+2\left\langle\sigma_{1\vec{r}_1}\right\rangle\left\langle\sigma_{2\vec{r}_2}\right\rangle\left\langle\sigma_{3\vec{r}_3}\right\rangle.
\end{align}
We arrive at the modified operator $\widetilde{\rho}$ for a three-qubit system:
\begin{align}
\widetilde{\rho}=\rho+\frac{1}{4}&\ \sum_{\alpha,\beta,\gamma=x,y,z}{\left\langle\sigma_{1\alpha}\right\rangle\left\langle\sigma_{2\beta}\right\rangle\left\langle\sigma_{3\gamma}\right\rangle\sigma_{1\alpha}\sigma_{2\beta}\sigma_{3\gamma}}\nonumber\\
-\frac{1}{8}&\left[\sum_{\alpha,\beta=x,y,z}{\left\langle\sigma_{1\alpha}\sigma_{2\beta}\right\rangle\sigma_{1\alpha}\sigma_{2\beta}}\sum_{\gamma=x,y,z}{\left\langle\sigma_{3\gamma}\right\rangle\sigma_{3\gamma}}\right.\nonumber\\
&+\sum_{\alpha,\gamma=x,y,z}{\left\langle\sigma_{1\alpha}\sigma_{3\gamma}\right\rangle\sigma_{1\alpha}\sigma_{3\gamma}}\sum_{\beta=x,y,z}{\left\langle\sigma_{2\beta}\right\rangle\sigma_{2\beta}}\nonumber\\
&+\left.\sum_{\beta,\gamma=x,y,z}{\left\langle\sigma_{2\beta}\sigma_{3\gamma}\right\rangle\sigma_{2\beta}\sigma_{3\gamma}}\sum_{\alpha=x,y,z}{\left\langle\sigma_{1\alpha}\right\rangle\sigma_{1\alpha}}\right]\nonumber\\
-\frac{1}{8}&\left[\sum_{\alpha,\beta=x,y,z}{\left\langle\sigma_{1\alpha}\right\rangle\left\langle\sigma_{2\beta}\right\rangle\sigma_{1\alpha}\sigma_{2\beta}}\right.\nonumber\\
&+\sum_{\alpha,\gamma=x,y,z}{\left\langle\sigma_{1\alpha}\right\rangle\left\langle\sigma_{3\gamma}\right\rangle\sigma_{1\alpha}\sigma_{3\gamma}}\nonumber\\
&+\left.\sum_{\beta,\gamma=x,y,z}{\left\langle\sigma_{2\beta}\right\rangle\left\langle\sigma_{3\gamma}\right\rangle\sigma_{2\beta}\sigma_{3\gamma}}\right].
\end{align}
\noindent
It should be pointed out, that due the large number of involved matrix operations, the calculation of $\widetilde{\rho}$ can become computationally expensive. The required computational cost increases exponentially with an increasing number of qubits in the observed system, yet it remains applicable for real-time simulations and visualizations in relatively small qubit systems.
\color{black}

\subsection{Representation of spherical functions}\label{Rep}
In the BEADS representation, we use three-dimensional polar plots to visualize spherical functions, that is, we plot the function on a spherical surface of constant size and represent the value at any vertex by a color defined by a color scheme. This is indeed different to other finite-dimensional phase-space representations such as DROPS \cite{DROPS} which are commonly plotted such that the magnitude of a spherical function value is represented by the distance from the origin to the spherical function surface in the corresponding direction. The choice of plotting methods is in general arbitrary. However, we deem our approach to be more suitable to provide a predictive visualization from all viewing perspectives. 

Moreover, the standard BEADS color scheme is specifically designed to visualize quantum states described by density operators (or of Hamiltonians) which are Hermitian by definition. For such state visualizations, the phase component is restricted to real values $\pm1$ and can be represented by two colors. This allows to optionally use a different connected correlation function color scheme to provide a clear distinction from compound correlations or single-qubit components.

We apply a red-green color scheme to represent any component which does not relate to entanglement in pure states. Entanglement-based connected correlation functions are instead visualized by applying a yellow-blue color scale. The choice of color schemes is arbitrary, and we present and discuss a selection of alternative schemes in appendix~\ref{sec:BEADSColors}. In particular, we propose alternative schemes which can be used in case of color blindness. For the visualization of non-Hermitian operators such as propagators (gates), adapted color schemes can be used \cite{DROPS}.

\subsection{Representation variants and simplifications}\label{sec:BEADSVariants}
\begin{figure}[H]
\centering
\includegraphics[width=.85\textwidth]{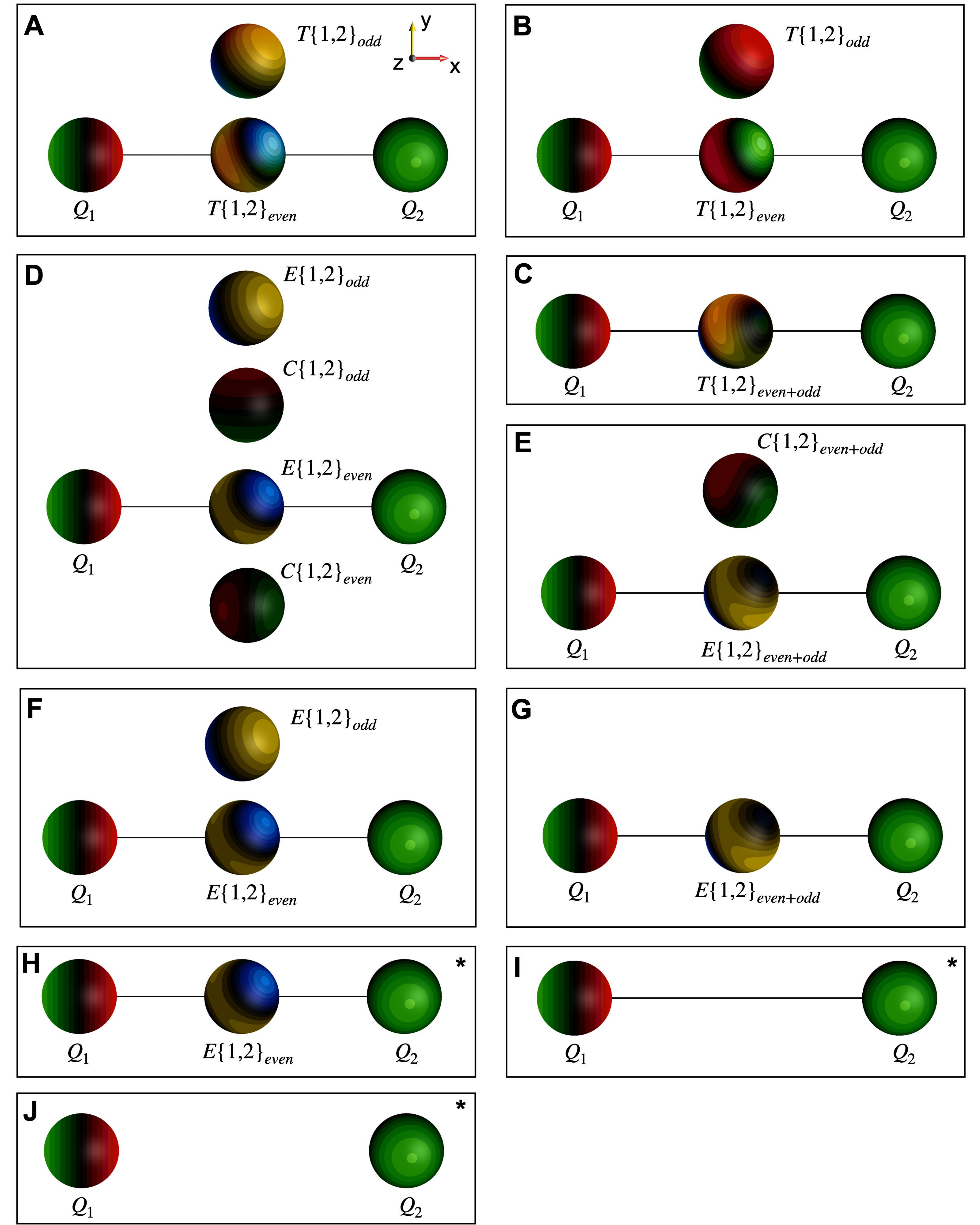}
\caption{\label{Figure:Fig10}Overview of BEADS variants. Possible variants of BEADS representations of quantum states (here, $\ket{\psi}=(-0.099893+0.207263i,\ 0.664132-0.077408i,-0.015899-0.323055i,\ 0.616169-0.125371i)^T$) include (\textbf{A}) total correlation functions with symmetry separation, (\textbf{B}) total correlation functions with symmetry separation using a single color scheme, (\textbf{C}) total correlation functions with combined symmetries, (\textbf{D})  separate correlation functions with full symmetry separation or (\textbf{E}) combined symmetries, (\textbf{F}) connected correlation functions with full symmetry separation or (\textbf{G}) combined beads, (\textbf{H}) fully permutation symmetric connected correlation functions, (\textbf{I}) no correlation function beads with or (\textbf{J}) without entanglement arcs. Variants \textbf{A}-\textbf{G} all carry the full information required to reconstruct the density operator if the colors of all surface points can be accessed. If the beads are shown only from one perspective, e.g., in the displayed top view, the variants with symmetry separation (\textbf{A}, \textbf{B}, \textbf{D}, and \textbf{F}) still make it possible to reconstruct the density operator because the information on the hidden hemisphere can be recovered due to the defined spherical function point symmetries of the beads (see section~\ref{sec:BEADSSymmetries}). The display variants \textbf{H}-\textbf{J} are marked with an asterisk to indicate that they only provide incomplete information about the density operator.}
\end{figure}

The BEADS representation can be tailored such that specific aspects of interest can be faithfully visualized in a simplified fashion. We now discuss different variants and simplifications of the BEADS representation, point out potential use-cases and address implications for the completeness of the representation. The display types \textbf{A} to \textbf{J} are summarized in table~\ref{tab:Variants} and shown in Fig.~\ref{Figure:Fig10} for the case of a two-qubit system. The most important display types \textbf{A}, \textbf{B}, \textbf{F}, and \textbf{H} are indicated by bold face letters in table~\ref{tab:Variants}.

In Fig.~\ref{Figure:Fig10} different BEADS visualizations of an arbitrary two-qubit quantum state which is highly asymmetric and partially entangled are shown. The most general complete BEADS variant is given in terms of (\textbf{A}) Q- and T-Beads including all symmetry components. This visualization is suitable to predict measurement outcomes of quantum states, yet connected correlation functions are merely indicated by the specialized coloring (cf. Fig.~\ref{Figure:Fig6} and appendix~\ref{sec:BEADSColors}). 

Instead of using the introduced total correlation color wheel (see Fig.~\ref{fig:FigColorTotCorr}) to quantify the ratio of connected and compound correlation coefficients in every spatial direction (\textbf{A}), it is always possible to just use a single color scheme, e.g., the red-green scheme to represent Q-Beads and total correlation coefficients (\textbf{B}). Clearly, this still allows to predict measurement outcomes without limitations, however, there is no visual indication on entanglement-based correlation coefficients.

\noindent
As is the standard approach used in the DROPS representation \cite{DROPS} beads of different point symmetries corresponding to the same subsystem (and permutation symmetry $\tau$-sublabel) can always be combined to create a more compact display as is shown for the display types \textbf{C}, \textbf{E}, and \textbf{G}. This reduces the number of beads while the displayed information remains complete. However, in this case it is not sufficient to view the representation from a single perspective, as, for spherical functions of undefined point symmetry, it is not possible to make assertions about the function values on the hidden hemisphere. These display types (\textbf{C}, \textbf{E}, \textbf{G}) are also disadvantageous in that they do generally not allow a uniform interpretation of the introduced BEADS color schemes. This is because the individual beads are designed such that the spherical function values are bounded (in terms of the underlying correlation coefficients) $-1\leq b^{(\ell^\prime)}(\theta,\phi)\leq1$ and by summing up multiple beads of different point symmetries, these bounds may well be exceeded, thus requiring extended color scales. In addition, components of different permutation symmetries are combined. For instance, the permutation-symmetric and antisymmetric bilinear components combine which does impede the direct outcome prediction of two-qubit symmetric measurements.

A natural extension (variants \textbf{D} and \textbf{E}) involves the previously outlined separation of correlation coefficient types into connected and compound correlation coefficients. While these variants provide direct insights into the intrinsic structures, that is, the topologies and symmetries of the different correlation function components, they are less intuitive to be used to predict measurement outcomes unless the visualized states are either fully separable or maximally entangled in which case the total correlation is entirely represented by one correlation coefficient type.

If one is instead only interested in characterizing the entanglement in the system, it is also possible (variants \textbf{F} and \textbf{G}) to omit all compound correlation functions. As was discussed earlier, this still yields a complete representation because compound correlation coefficients represent redundant information. 

One may further (\textbf{H}) omit beads that do not correspond to fully symmetric correlations if one is only interested in predicting measurement outcomes of symmetric measurements, which are encountered frequently in quantum information processing where measurements are commonly conducted along the z-axis, that is, in the computational basis. Indeed, this simplification is chosen in the sections~\ref{Grover} and~\ref{QuTel} as the examined circuits only involve multiqubit z-measurements. Visualizations of this type are not complete but very clear and sufficient to cover many examples of quantum algorithms and protocols.

One may even choose to omit all correlation beads as shown for the display types \textbf{I} and \textbf{J}. This may serve as an entry point when first introducing quantum mechanics which then allows to, e.g., focus on the effects of local gates in multiqubit separable states or the implications of entangling operations on the reduced states corresponding to individual qubits. It is further possible (\textbf{I}) to omit any correlation bead but to include the entanglement arc to indicate the magnitude (norm) of entanglement-related correlation components. It is clear, that this approach does by no means provide complete information, however, it is sufficient to indicate whether a pure state is entangled while providing a detailed representation of single-qubit components.

A minor simplification which is applied throughout this article is to omit the identity bead $\{\emptyset\}$. In quantum states, the expectation value of the identity operator $\mathbb{I}$ is always 1. The corresponding bead is given by a spherical harmonic of rank $j = 0$ that is, a spherical function with value 1 in all spatial directions which does not provide any further implications other than the direct representation of the state norm for which it can be omitted in the visualization.

Most importantly, combinations of the suggested simplifications are possible. For instance, it is always possible to visualize only fully permutation symmetric beads while viewing total correlation coefficients or separate correlation function types. 

\begin{table}[h!]
\vspace{-.5cm}
\centering
\caption{\label{tab:Variants}Overview and classification of BEADS display variants. The display types in bold face \textbf{A}, \textbf{B}, \textbf{D}, \textbf{F}, and \textbf{H} are most commonly used. All variants are classified with respect to the displayed types of beads, auxiliary elements (entanglement arcs), colors, symmetries, and the completeness of the representation.}
\vspace{.3cm}
\begin{tabular}{p{0.11\textwidth}<{\centering} p{0.03\textwidth}<{\centering}p{0.03\textwidth}<{\centering}p{0.03\textwidth}<{\centering}p{0.03\textwidth}<{\centering}p{0.07\textwidth}<{\centering}p{0.08\textwidth}<{\centering}p{0.1\textwidth}<{\centering}p{0.1\textwidth}<{\centering}p{0.10\textwidth}<{\centering}}
\hline\hline\\[-5pt]
\multirow[t]{2}{*}{\parbox[t]{1.5cm}{\centering Display \ type}} & \multicolumn{4}{c}{Beads} & \multirow[t]{2}{*}{\parbox[t]{1.cm}{\centering Ent. arc.}} & \multirow[t]{2}{*}{Colors} & \multirow[t]{2}{*}{\parbox[t]{1.5cm}{\centering Point \ sym.}} & \multirow[t]{2}{*}{\parbox[t]{1.5cm}{\centering Perm.\ sym.}} & \multirow[t]{2}{*}{\parbox[t]{1.7cm}{\centering Complete}} \\
& Q & T & E & C &\\[6pt] \hline\hline
&\\[-8pt] 
 \textbf{A} & \checkmark & \checkmark & -- & -- & \checkmark & \hatchcirc & $e\leftrightarrow o$ & $s\leftrightarrow a$, $\tau$ & \checkmark \\[6pt] \hline \\[-8pt]
 \textbf{B} & \checkmark & \checkmark & -- & -- & \checkmark & \emptycirc & $e\leftrightarrow o$ & $s\leftrightarrow a$, $\tau$ &\checkmark \\[6pt] \hline \\[-8pt]
 C & \checkmark & \checkmark & -- & -- & \checkmark & \hatchcirc & $e+o$ & $s+ a$, $\tau$ & \checkmark \\[6pt] \hline \\[-8pt]
 \textbf{D} & \checkmark & -- & \checkmark & \checkmark & \checkmark & \halfcirc & $e\leftrightarrow o$ &$s\leftrightarrow a$, $\tau$ & \checkmark \\[6pt] \hline \\[-8pt]
 E & \checkmark & -- & \checkmark & \checkmark & \checkmark & \halfcirc & $e+o$ & $s +a$, $\tau$ & \checkmark \\[6pt] \hline \\[-8pt]
 \textbf{F} & \checkmark & -- & \checkmark & -- & \checkmark & \halfcirc & $e\leftrightarrow o$ & $s\leftrightarrow a$, $\tau$  &\checkmark \\[6pt] \hline \\[-8pt]
 G & \checkmark & -- & \checkmark & -- & \checkmark & \halfcirc & $e+o$ & $s+ a$, $\tau$ & \checkmark\\[6pt] \hline \\[-8pt]
 \textbf{H} & \checkmark & -- & \checkmark & -- & \checkmark & \halfcirc  & $e\parallel o$ & $S$ & -- \\[6pt] \hline \\[-8pt]
 I & \checkmark & -- & -- & -- & \checkmark & \emptycirc  & $o$ & -- &--  \\[6pt] \hline \\[-8pt]
 J & \checkmark & -- & -- & -- & -- & \emptycirc & $o$ & -- & --  \\[6pt] \hline\hline
\end{tabular}
\end{table}%
\small{
\vspace{-.5cm}
\underline{Legend:}\newline
\begin{enumerate}[align=parleft, labelsep=.5cm, leftmargin = 1.6cm]
\setlength\itemsep{.0em}
\vspace{-12pt}
\item[\hatchcirc] Full total correlation color scheme (see Fig.~\ref{Figure:Fig6})
\item[\halfcirc] Separate color schemes for Q-Beads and compound correlations (red-green), and connected correlations (yellow-blue)
\item[\emptycirc] Simple red-green color scheme \\[-4pt]
\item[\mbox{$e\leftrightarrow o$}] Separate display of even and odd spherical function point reflection symmetry
\item[$e + o$] Combined display of even and odd spherical function point reflection symmetry
\item[\mbox{$e \parallel o$}] Either even or odd point reflection symmetry display in dependence of linearity
\item[$o$] Odd point reflection symmetry only \\[-4pt]
\item[\mbox{$s\leftrightarrow a$}] Separate display of bilinear permutationally symmetric and antisymmetric components
\item[\mbox{$s + a$}] Combined display of bilinear permutationally symmetric and antisymmetric components
\item[$\tau$] Separate display of $\tau$-permutation symmetries
\item[$S$] Totally permutation symmetric components only (e.g., corresponding to bilinear beads of even or trilinear $\tau_1$-beads of odd point reflection symmetry)
\end{enumerate} 
}

\normalsize

\subsection{Software}\label{Software}
We created a software package, \textit{QuBeads}, with which it is possible to simulate quantum circuits, i.e., the evolution of systems consisting of up to three qubits dynamically and which offers an easy-to-use interactive user interface. All figures in this paper were created from simulations in QuBeads. We provide a short overview of the software in appendix~\ref{app:BEADSQuBeads}.

\backmatter

\bmhead{Supplementary information}

We provide a supplementary information accompanying this article which has the following contents:
\begin{itemize}
\item Supplementary texts
\begin{itemize}
\item Theoretical details, sections S1.1--S1.4
\item Technical details, concepts and design, sections S2.1--S2.2
\item Comparison and classification, sections S3.1--S3.3
\item Application examples, section S4.1--S4.11
\end{itemize}
\end{itemize}

\bmhead{Acknowledgements}

The authors acknowledge helpful discussions with Christian Schilling, Barbara Kraus, and Thomas Schulte-Herbr\"uggen.

\section*{Declarations}

\begin{itemize}
\item Funding
	\begin{itemize}
	\item The authors acknowledge funding from the Digital Europe Programme for the project 
	DigiQ: Digitally Enhanced Quantum Technology Master (DigiQ) under grant agreement
ID 101084035
	\item Kekul\'e stipend of the Verband der chemischen Industrie e.V. (VCI)
	\item TUM EDU grant 3170573
	\item Munich Center of Quantum Science and Technology (MCQST)
	\item Munich Quantum Valley (MQV)
	\item Bavarian NMR Center (BNMRZ)
\end{itemize}
\item Authors declare that they have no competing interests.
\item Authors consent to publication.
\item All data are available in the main text, supplementary materials, or via the link to the software repository provided in appendix~\ref{app:BEADSQuBeads}.
\item Material availability: Cf. data availability
\item Code availability: Not applicable
\item Author contributions
\begin{itemize}
	\item Examples: DH, SG
	\item Conceptualization: DH, SG
	\item Methodology: DH, SG
	\item Investigation: DH, SG
	\item Visualization: DH, SG
	\item Supervision: SG
	\item Writing -- original draft: DH, SG
	\item Writing -- review \& editing: DH, SG

\end{itemize}
\end{itemize}

\noindent 

\begin{appendices}

\section{Spherical function point symmetries, permutation symmetries, and bit parities}\label{sec:BEADSSymmetries}
\begin{figure}[H]
\centering
\includegraphics[width=.95\textwidth]{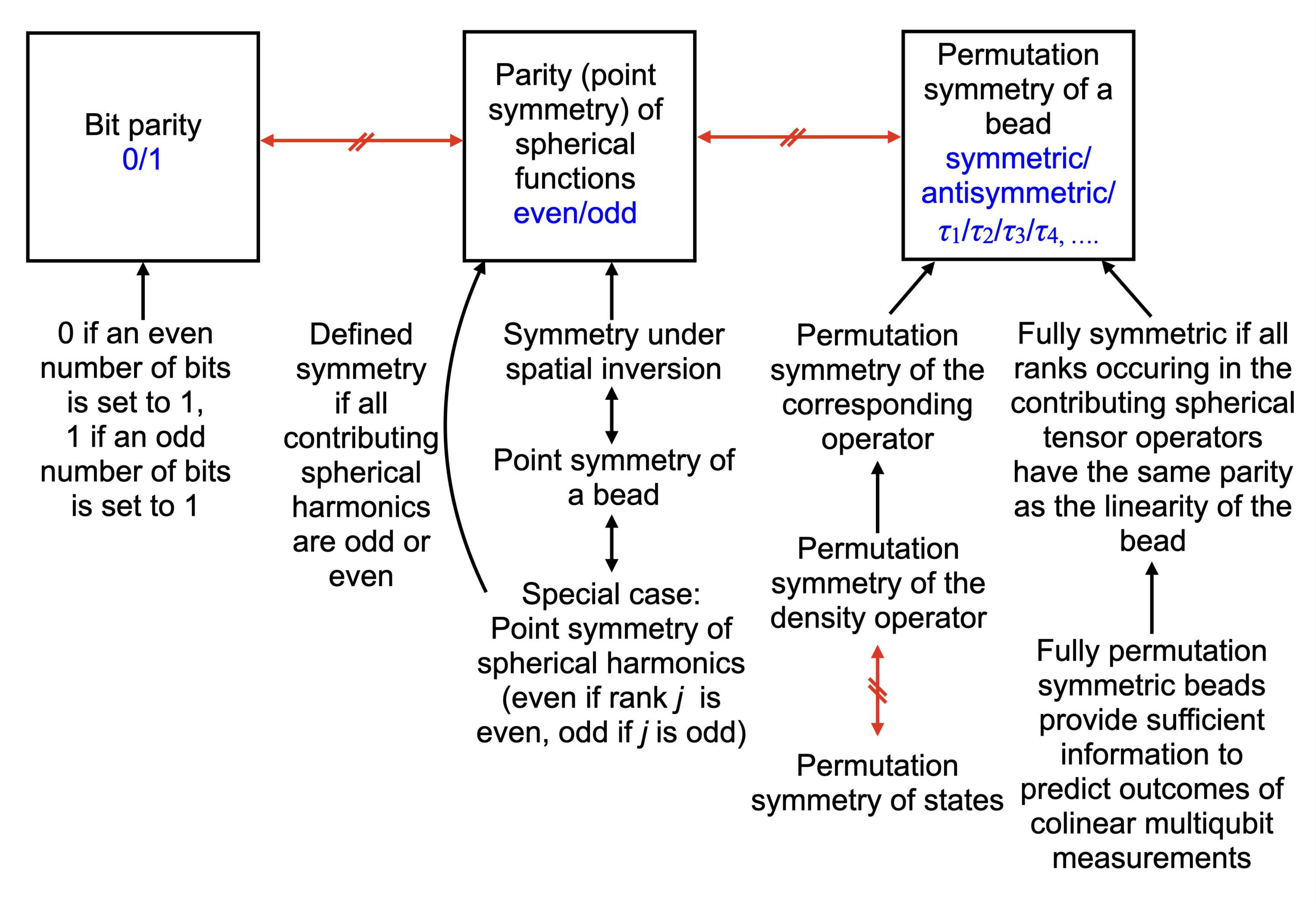}
\caption{\label{fig:FigSymmetryOverview}Overview of parities and symmetries in the BEADS representation. Red crossed out double-headed arrows indicate that two items are inequivalent. Labels which are used in the BEADS representation to describe specific properties are written in blue color.}
\end{figure}

The BEADS representation is highly symmetry-adapted. Thus, several different types of symmetries are visualized. In addition, as was discussed in section~\ref{2Q}, an interpretation in terms of bit parity probabilities is possible for some beads. In Fig.~\ref{fig:FigSymmetryOverview}, we provide an overview of properties captured by the BEADS representation which must not be confused with each other. 

\noindent
In the BEADS representation, bit parity, which was originally introduced to define the oddness of an integer, is used in a generalized sense to describe whether after a measurement the number of qubits in the down state of the respective measurement bases is even (parity~0) or odd (parity~1). 
\newpage
\vspace{2cm}

\begin{figure}[H]
\centering
\vspace{2cm}
\includegraphics[width=.95\textwidth]{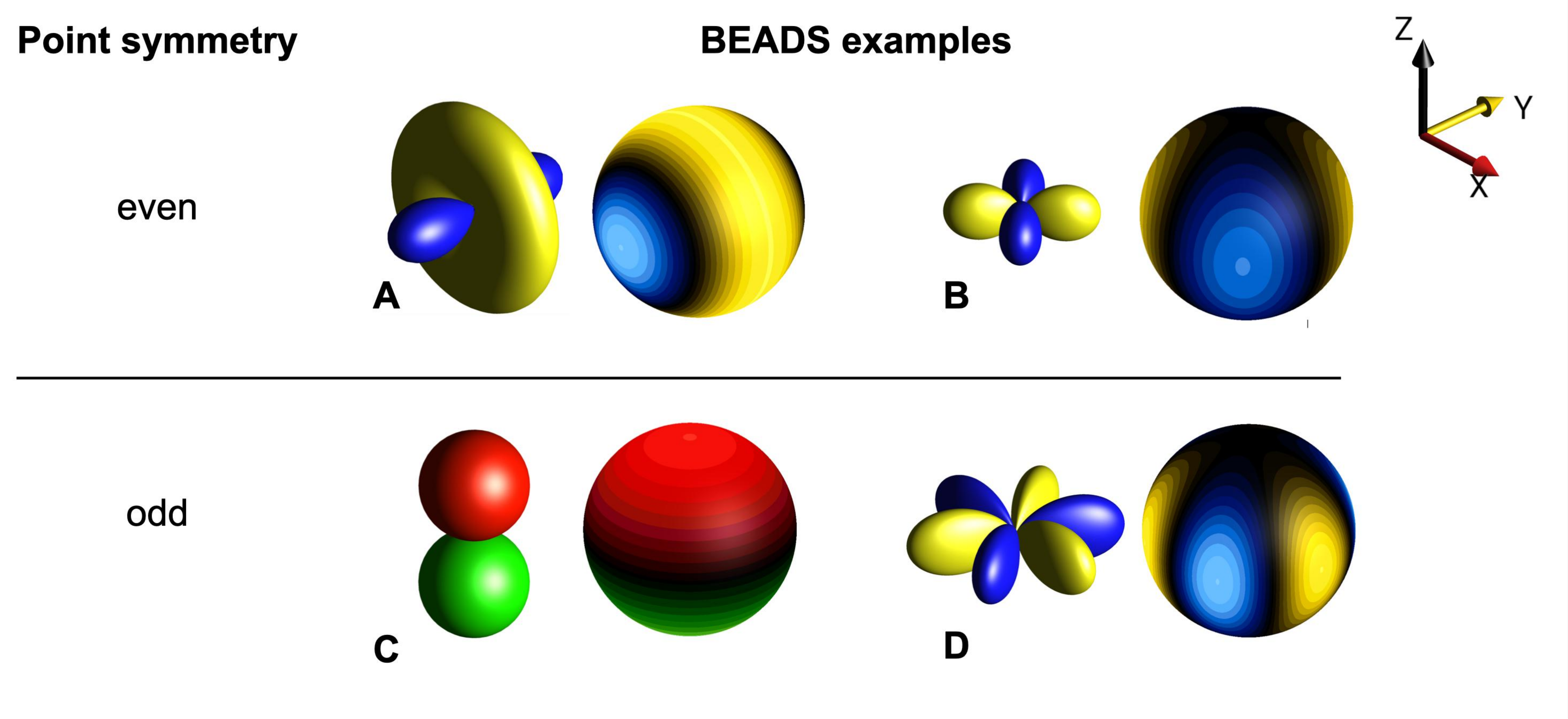}
\caption{\label{fig:FigPointSymmetry}Spherical function point symmetry in the BEADS representation. We plot exemplary beads of various states by using the magnitude of the spherical function values as distance from the origin and by standard spherical color plots. Beads have even (even ranks $j$) or odd (odd ranks $j$) spherical function point symmetry. Beads of even point symmetry such as the $E\{1,2\}_{\textit{even}}$ E-Bead visualized exemplarily (\textbf{A}) for the Bell state $\ket{\Psi^+}=1/\sqrt2\left(\ket{00}+\ket{11}\right)$, or (\textbf{B}) $E\{1,2,3\ \tau_2\}_{\textit{even}}$ which is here shown for the state $\ket{\psi}=1/2\left(\ket{000}-\ket{011}-\ket{101}+\ket{110}\right)$ are invariant under space inversion, i.e., for any spherical direction, have the same value in the opposite direction. Odd point symmetry such as given by (\textbf{C}) Q-Beads, e.g., $Q_1$ exemplarily visualized in the state $\ket{0}$ or (\textbf{D}) $E\{1,2,3\ \tau_1\}_{\textit{odd}}$ shown for the GHZ state $\ket{\text{GHZ}}=1/\sqrt2\left(\ket{000}+\ket{111}\right)$ undergo sign inversion under spatial inversion, that is, for any arbitrary spherical direction, these beads have oppositely signed values in opposite directions.}
\vspace{.5cm}
\end{figure}

\noindent
Bit parity must not be confused with the parity of a spherical function. The latter is equivalent to the point reflection symmetry of a spherical function, that is, the symmetry under space inversion. As pointed out in section~\ref{BEADSMap}, we use labels \textit{even} (no sign inversion) and \textit{odd} (sign inversion) to indicate the point symmetry of beads. Illustrative examples of different spherical function point symmetries are provided in Fig.~\ref{fig:FigPointSymmetry}. 

\newpage

\noindent
The permutation symmetry associated with a bead corresponds to the permutation symmetry of a visualized operator component. As outlined in section~\ref{LISA}, we use the LISA tensor operator basis to decompose operators, specifically the density operator describing quantum states. The symmetry of the \textit{density operator} under permutation of particles (that is, particle exchange) is not identical to the symmetry of a given \textit{state} under exchange of qubit labels because a change of sign of the \textit{state} corresponds simply to a different global phase and hence corresponds to the identical density operator. For example, for a system of two qubits, the singlet state function $\ket{\Psi^-}=1/\sqrt2\left(\ket{01}-\ket{10}\right)$ is \textit{antisymmetric} under exchange of qubit labels, but the corresponding density operator
$\rho=1/2\left(\ket{01}-\ket{10}\right)\left(\bra{01}-\bra{10}\right)=\left(\mathbb{I}-X_1X_2-Y_1Y_2-Z_1Z_2\right)/4$, using the short forms $X$, $Y$, and $Z$ to denote corresponding Pauli matrices, is fully permutation \textit{symmetric}. 

Contributing spherical tensor operators of beads corresponding to full permutation symmetry have ranks $j$ that are of the same parity as the linearity $g$ of the bead. For instance, the fully permutationally symmetric bilinear ($g = 2$) bead $E\{1,2\}_{\textit{even}}$ corresponds to spherical tensor operators with ranks $j \in\left\{0,2\right\}$, i.e., both, the linearity and all possible ranks are even.

\begin{figure}[H]
\centering
\vspace{1.5cm}
\includegraphics[width=1.\textwidth]{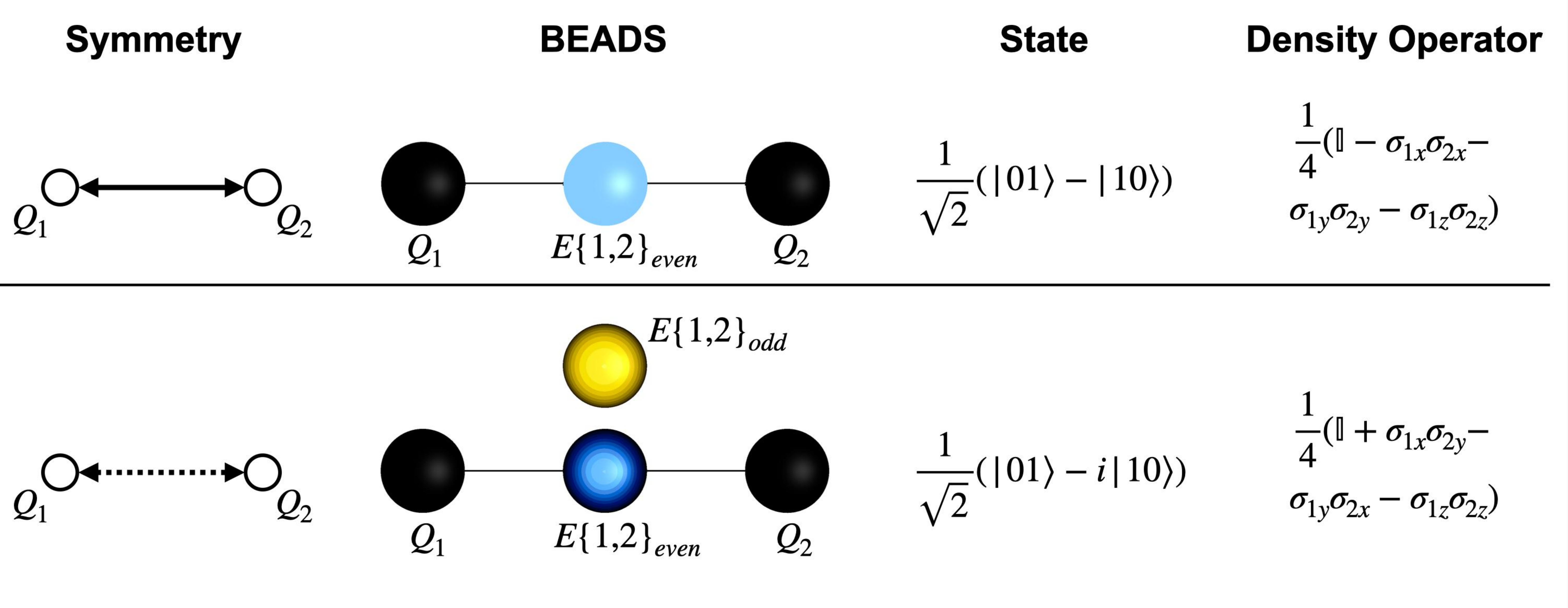}
\caption{\label{fig:FigPermSym2Q}Two-qubit permutation (particle exchange) symmetries in the BEADS representation. Two-qubit correlation components such as given in the displayed maximally entangled states can be symmetric (solid double-headed arrow) or antisymmetric (dashed double-headed arrow). In the BEADS representation of quantum states, these symmetries refer to the density operator and not to the state. For example, we find that the Bell state $\ket{\Psi^-}=1/\sqrt2\left(\ket{01}-\ket{10}\right)$ only has fully permutation symmetric density operator components, i.e., is only represented by the permutation and point symmetric component $E\{1,2\}_{\textit{even}}$ whereas the state $\ket{\psi}=1/\sqrt2\left(\ket{01}-i\ket{10}\right)$ has a maximum antisymmetric component $E\{1,2\}_{\textit{odd}}$ (corresponding to $\sigma_{1x}\sigma_{1y}-\sigma_{1y}\sigma_{2x}$). Note that the bottom state still has an additional symmetric component $E\{1,2\}_{\textit{even}}$ (corresponding to $-\sigma_{1z}\sigma_{2z}$).}
\end{figure}

\begin{figure}[H]
\centering
\includegraphics[width=1\textwidth]{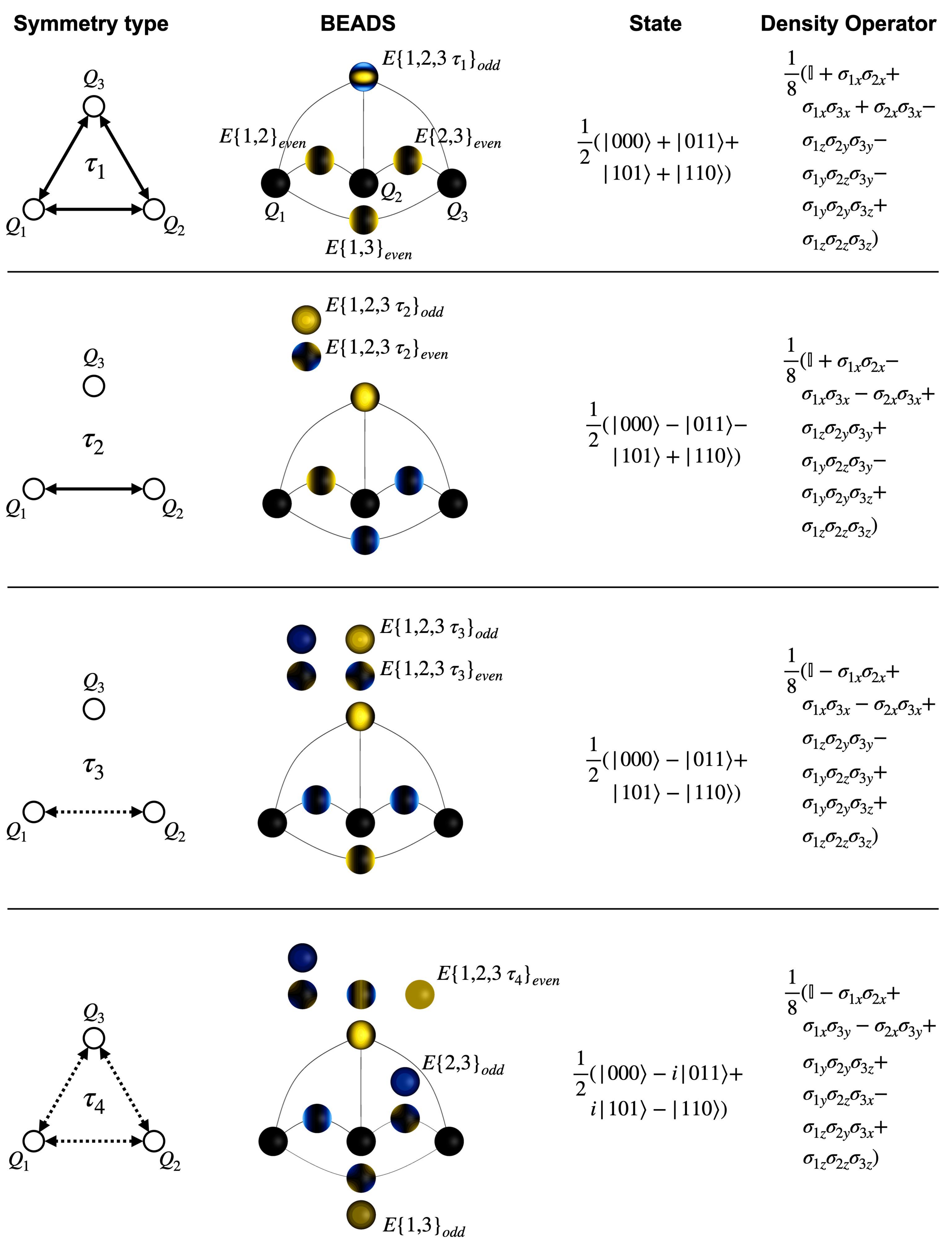}
\caption{\label{fig:FigPermSym3Q}Three-qubit permutation symmetries in the BEADS representation. By construction \cite{DROPS} (see section~\ref{GenMap}), three-qubit correlation components of the density operator, such as given for the displayed maximally entangled states, can be totally symmetric ($\tau_1$), symmetric with respect to permutations of qubits 1 and 2 ($\tau_2$), antisymmetric with respect to permutations of qubits 1 and 2 ($\tau_3$) or totally antisymmetric ($\tau_4$) but do not correspond to the symmetries of a state under particle exchange.}
\end{figure}

Permutation symmetries of two- and three-qubit states are visualized by examples in Fig.~\ref{fig:FigPermSym2Q} and Fig.~\ref{fig:FigPermSym3Q}, respectively. 

\newpage

\section{BEADS scaling factors of permutation symmetric density operator components}\label{sec:BEADSCanonicalScaling}

Compared to the Wigner-type DROPS representation of operators \cite{DROPS}, we introduce additional scaling factors $s_j^{\left(\ell^\prime\right)}\left(N,g\right)$  for the spherical functions in the BEADS representation (see Eq.~\ref{eq:BEADSMapping} and~\ref{eq:BEADSScalingFactor} in section~\ref{BEADSMap}). In a first approach, we choose these scaling factors such that the spherical function values directly correspond to experimentally measurable expectation values of Pauli product operators. Although the approach is valid for arbitrary Pauli product operators, for concreteness and simplicity, we focus the following discussion on expectation values of (products of) Pauli-Z operators, using the compact notation $Z_k =\sigma_{kz}$, such as the expectation values $\left\langle Z_k \right\rangle$, $\left\langle Z_kZ_l \right\rangle$, or $\left\langle Z_k Z_l Z_m \right\rangle$, etc., of linear operators $Z_k$, bilinear operators $Z_kZ_l$, trilinear operators $Z_k Z_l Z_m$,  etc., respectively. These are the standard expectation values which are directly measurable in most experimental implementations of quantum information processing tasks and quantum computing algorithms.
In the LISA basis, Pauli-Z (product) operators only contribute to tensor operators $T_{j,0}^{\left(\ell^\prime\right)}$ (i.e., operators $T_{j,m}^{\left(\ell^\prime\right)}$ with $m=0$). For example, the linear LISA operator $T_{1,0}^{\left\{k\right\}}$ (which acts exclusively on the set ${\ell^\prime=\left\{k\right\}}$ consisting only of the $k$-th qubit) is proportional to $Z_k$ \cite{DROPS}:
\begin{align}
T_{1,0}^{\left\{k\right\}}&=\frac{1}{\sqrt{2^N}}\cdot Z_k, \\
\label{eq:CanScalingLin}Z_k &=\sqrt{2^N}\cdot T_{1,0}^{\left\{k\right\}}.
\end{align}
The scaling factor $\frac{1}{\sqrt{2^N}}$ normalizes the LISA basis operator $T_{10}^{\left\{k\right\}}$ for a quantum system consisting of $N$ qubits. Conversely, $Z_k =\sqrt{2^N} \cdot T_{10}^{\left\{k\right\}}$ and hence
\begin{equation}
\left\langle Z_k \right\rangle =\zeta\left(N\right)\cdot\left\langle T_{1,0}^{\left\{k\right\}} \right\rangle,
\end{equation}
with the $N$-dependent scaling factor
\begin{equation}\label{eq:Zeta}
\zeta\left(N\right)=\sqrt{2^N}.
\vspace{6pt}
\end{equation}
As shown in \cite{LeinerWQST}, the expectation value $T_{j,0}^{\{k\}}$ of an axial LISA basis operator is proportional to the value of the corresponding droplet function $f_j^{\left\{k\right\}}(\theta,\phi)$ in the z-direction (i.e., at the north pole, where $\theta=0$ and $\phi$ is arbitrary, e.g., $\phi=0$):
\begin{equation}
\left\langle T_{j,0}^{\left(\ell^\prime\right)} \right\rangle = \eta_j \cdot f_j^{\left\{k\right\}}\left(0,0\right),
\end{equation}
with
\begin{equation}\label{eq:Eta}
\eta_j=\sqrt{4\pi/\left(2j+1\right)}.
\vspace{6pt}
\end{equation}
Hence, with the help of the two scaling factors $\zeta(N)$ (Eq.~\ref{eq:Zeta}) and $\eta_1$ (Eq.~\ref{eq:Eta}), we can express an expectation value $\left\langle Z_k \right\rangle$ in terms of the value of the droplet function $f_1^{\left\{k\right\}}(0,0)$ in the z-direction:
\begin{equation}
\left\langle Z_k \right\rangle = \zeta\left(N\right)\cdot \eta_1\cdot {f_1^{\left\{k\right\}}\left(0,0\right)}.
\vspace{6pt}
\end{equation}
Similarly, the expectation value $\left\langle Z_k Z_l \right\rangle$  of a bilinear Pauli-Z product operator can be expressed in terms of spherical functions $f_j^{\left\{k,l\right\}_{\textit{even}}}(0,0)$ in the z-direction. However, this case is more complicated because the operator $Z_kZ_l$ contributes not only to one but to several bilinear LISA basis operators $T_{j,0}^{\left\{k,l\right\}_{\textit{even}}}$. Although bilinear LISA basis operators $T_{j,0}^{\left\{k,l\right\}_{even/odd}}$ can have ranks $j\in\left\{0,1,2\right\}$, the operator $Z_k Z_l$  only contributes to the two LISA basis operators $T_{0,0}^{\left\{k,l\right\}_{\textit{even}}}$ and $T_{2,0}^{\left\{k,l\right\}_{\textit{even}}}$ \cite{DROPS} which in the BEADS representation are comprised in the fully symmetric subset ${\ell^\prime=\left\{k,l\right\}_{\textit{even}}}$:
\begin{align}
T_{0,0}^{\left\{k,l\right\}_{\textit{even}}} &= \frac{1}{\sqrt{2^N}}\frac{Z_k Z_l + \left(X_k X_l +Y_k Y_l \right)}{\sqrt3}, \\
T_{2,0}^{\left\{k,l\right\}_{\textit{even}}}&= \frac{1}{\sqrt{2^N}}{\frac{2Z_k Z_l -\left(X_k X_l +Y_k Y_l\right)}{\sqrt6}}.
\end{align}
By multiplying the second equation by $\sqrt{2}$ and adding this to the first equation, the term $\left(X_k X_l +Y_k Y_l\right)$ is eliminated and we obtain
\begin{align}
T_{0,0}^{\left\{k,l\right\}_{\textit{even}}}+\sqrt{2}\:T_{2,0}^{\left\{k,l\right\}_{\textit{even}}}&=\frac{1}{\sqrt{2^N}}\left(\frac{Z_k Z_l}{\sqrt3}+\frac{2\:Z_k Z_l}{\sqrt3}\right) \nonumber\\
&=\frac{1}{\sqrt{2^N}}\frac{3\:Z_k Z_l}{\sqrt3}=\frac{1}{\sqrt{2^N}}{{\sqrt3\:Z}_k Z_l}. 
\end{align}
Hence, we can express the bilinear Pauli product operator $Z_k Z_l$ as the following linear combination of $T_{0,0}^{\left\{k,l\right\}_{\textit{even}}}$ and $T_{2,0}^{\left\{k,l\right\}_{\textit{even}}}$:
\begin{align}\label{eq:CanScalingBil}
Z_kZ_l&=\sqrt{2^N}\left(\frac{1}{\sqrt3}\cdot T_{0,0}^{\left\{k,l\right\}_{\textit{even}}}+\sqrt{\frac{2}{3}}\cdot T_{2,0}^{\left\{k,l\right\}_{\textit{even}}}\right) \nonumber\\
&= \zeta\left(N\right)\left(\xi_0^{\left\{k,l\right\}_{\textit{even}}}\left(2\right)\cdot T_{0,0}^{\left\{k,l\right\}_{\textit{even}}}+\xi_2^{\left\{k,l\right\}_{\textit{even}}}\left(2\right)\cdot T_{2,0}^{\left\{k,l\right\}_{\textit{even}}}\right),
\end{align}
where we used the scaling factor $\zeta(N)=\sqrt{2^N}$ (cf. Eq.~\ref{eq:Zeta}) and the scaling factors $\xi_j(g)$ for 
$j\in\left\{0,2\right\}$ and linearity $g=2$ (corresponding to the bilinearity of the LISA basis operators $T_{0,0}^{\left\{k,l\right\}_{\textit{even}}}$ and $T_{2,0}^{\left\{k,l\right\}_{\textit{even}}})$ with $\xi_0^{\left\{k,l\right\}_{\textit{even}}}\left(2\right)=\sqrt{1/3}$ and $\xi_2^{\left\{k,l\right\}_{\textit{even}}}\left(2\right)=\sqrt{2/3}$.
Based on Eq.~\ref{eq:CanScalingBil}, the expectation value $\left\langle Z_kZ_l \right\rangle$  can finally be expressed in the form
\begin{align}
\left\langle Z_kZ_l \right\rangle &=\zeta(N)\cdot\left(\xi_0^{\left\{k,l\right\}_{\textit{even}}}\left(2\right) \cdot \left\langle T_{0,0}^{\left\{k,l\right\}_{\textit{even}}} \right\rangle +\xi_2^{\left\{k,l\right\}_{\textit{even}}}\left(2\right)\cdot \left\langle T_{2,0}^{\left\{k,l\right\}_{\textit{even}}} \right\rangle\right)\nonumber\\
&=\zeta(N)\cdot\left(\xi_0^{\left\{k,l\right\}_{\textit{even}}}\left(2\right)\cdot\eta_0\cdot f_0^{\left\{k,l\right\}_{\textit{even}}}+\xi_2^{\left\{k,l\right\}_{\textit{even}}}\left(2\right)\cdot\eta_2 \cdot f_2^{\left\{k,l\right\}_{\textit{even}}}\right)\nonumber\\
&=\zeta(N)\cdot\xi_0^{\left\{k,l\right\}_{\textit{even}}}\left(2\right)\cdot\eta_0 \cdot f_0^{\left\{k,l\right\}_{\textit{even}}}+\zeta(N)\cdot\xi_2^{\left\{k,l\right\}_{\textit{even}}}\left(2\right)\cdot\eta_2\cdot f_2^{\left\{k,l\right\}_{\textit{even}}} \nonumber\\
&={s}_0^{\left\{k,l\right\}_{\textit{even}}}\left(N,2\right)\cdot f_0^{\left\{k,l\right\}_{\textit{even}}}+s_2^{\left\{k,l\right\}_{\textit{even}}}\left(N,2\right)\cdot f_2^{\left\{k,l\right\}_{\textit{even}}},
\end{align}
with the overall scaling factors
\begin{align}
s_j^{\left(\ell^\prime\right)}\left(N,g\right)&=\zeta(N)\cdot\xi_j^{\left(\ell^\prime\right)}\left(g\right)\cdot\eta_j \nonumber\\
&=\sqrt{2^N}\cdot\xi_j^{\left(\ell^\prime\right)}\left(g\right)\cdot\sqrt{4\pi/\left(2j+1\right)}.
\vspace{1cm}
\end{align}
For example, for a system consisting of three qubits ($N=3$) we find
\begin{align}
s_0^{\left\{k,l\right\}_{\textit{even}}}\left(3,2\right)&=\sqrt{2^3}\cdot\xi_0^{\left\{k,l\right\}_{\textit{even}}}\left(2\right)\cdot\sqrt{4\pi/1} \nonumber\\
&=2\sqrt{2}\cdot\sqrt{1/3}\cdot 2\sqrt\pi \nonumber\\
&=4\sqrt{2\pi/3} \nonumber\\
&\approx 5.78881
\end{align}
and
\begin{align}
s_2^{\left\{k,l\right\}_{\textit{even}}}\left(3,2\right)&=\sqrt{2^3}\cdot\xi_2^{\left\{k,l\right\}_{\textit{even}}}\left(2\right)\cdot\sqrt{4\pi/5} \nonumber\\
&=2\sqrt{2}\cdot\sqrt{2/3}\ \cdot 2\sqrt{\pi/5} \nonumber\\
&=8\sqrt{\pi/15}\nonumber\\
&\approx 3.66116.
\end{align}
\noindent
Similarly, the trilinear ($g=3$) Pauli product operator $Z_k Z_l Z_m$ can be expressed as the following linear combination of LISA basis operators:
\begin{equation}
Z_k Z_l Z_m =\zeta\left(N\right)\left(\xi_1^{\left\{k,l,m\ \tau_1\right\}_{\textit{odd}}}\left(3\right)\cdot T_{1,0}^{\left\{k,l,m\ \tau_1\right\}_{\textit{odd}}}+\xi_3^{\left\{k,l,m\ \tau_1\right\}_{\textit{odd}}}\left(3\right)\cdot T_{3,0}^{\left\{k,l,m\ \tau_1\right\}_{\textit{odd}}}\right),
\end{equation}
\noindent
with $\xi_1(3)=\sqrt{3/5}$ and $\xi_3(3)=\sqrt{2/5}$ and
\begin{equation}\label{eq:CanScalingTril}
\left\langle Z_k Z_l Z_m \right\rangle =s_1^{\left\{k,l,m\ \tau_1\right\}_{\textit{odd}}}\left(n,2\right){\cdot f}_1^{\left\{k,l,m\ \tau_1\right\}_{\textit{odd}}}+s_3^{\left\{k,l,m\ \tau_1\right\}_{\textit{odd}}}\left(n,2\right){\cdot f}_3^{\left\{k,l,m\ \tau_1\right\}_{\textit{odd}}}.
\end{equation}
\noindent
In general, a $g$-linear Pauli-Z product operator can be expressed as a linear combination
\begin{equation}
\zeta\left(N\right)\sum_{j}{\xi_j^{\left(\ell^\prime\right)}\left(g\right)}T_{j,0}^{\left(\ell^\prime\right)},
\end{equation}
\noindent
where $j\in\left\{0,2,\ 4,\ ...\ ,\ g\right\}$ if $g$ is even and $j\in\left\{1,3,\ 5,\ ...\ ,\ g\right\}$ if $g$ is odd and the set $\ell^\prime$ includes the $g$ qubits on which the $g$-linear Pauli product operator acts. For $g\leq6$ the scaling factors $\xi_j^{\left(\ell^\prime\right)}\left(g\right)$  are summarized in table~\ref{tab:CanSF}. Note that the linear combination $\sum_{j}{\xi_j^{\left(\ell^\prime\right)}\left(g\right)\ }T_{j,0}^{\left(\ell^\prime\right)}$ represents a normalized operator as the unscaled LISA operators are normalized and the scaling factors $\xi_j^{\left(\ell^\prime\right)}\left(g\right)$ satisfy the relation $\sum_{j}\left(\xi_j^{\left(\ell^\prime\right)}\left(g\right)\right)^2=1$.

It is important to point out that axial LISA tensor operators that contribute to Pauli-Z operators (cf. Eq.~\ref{eq:CanScalingLin}, \ref{eq:CanScalingBil}, and \ref{eq:CanScalingTril}) naturally correspond to fully symmetric density operator components with respect to permutations of qubits. We scale all non-axial operators $T_{j,(m\ \neq\ 0)}^{\left(\ell^\prime\right)}$ of the same rank $j$ and set $\ell^\prime$ with the same scaling factors as the corresponding axial operators~$T_{j,0}^{\left(\ell^\prime\right)}$.

Axial tensor operators $T_{j,0}^{\left(\ell^\prime\right)}$ and non-axial tensor operators $T_{j,m\ \neq\ 0}^{\left(\ell^\prime\right)}$ corresponding to components which are not fully symmetric with respect to permutations of qubits can in principle be scaled in an analogue fashion. However, the resulting bead spherical function values may be larger than 1 which prevents a uniform interpretation of the color schemes introduced in appendix~\ref{sec:BEADSColors}. Thus, to ensure a uniform interpretation of color scales, such components are scaled by using alternative scaling factors which are based on the global unitary bounds of the underlying tensor operators (see appendix~\ref{sec:BEADSGUB}). 
\begin{table}[h!]
\centering
\caption{\label{tab:CanSF}Scaling factors $\xi_j^{\left(\ell^\prime\right)}\left(g\right)$ of fully permutation symmetric axial tensor operators $T_{j,0}^{\left(\ell^\prime\right)}$ of rank $j$ and linearity $g$ that contribute to Pauli-Z operators.}
\vspace{.3cm}
\begin{tabular}{p{0.17\textwidth}<{\centering} p{0.088\textwidth}<{\centering} p{0.088\textwidth}<{\centering} p{0.088\textwidth}<{\centering} p{0.088\textwidth}<{\centering} p{0.088\textwidth}<{\centering} p{0.088\textwidth}<{\centering} p{0.088\textwidth}<{\centering}}
\hline\hline\\[-5pt]
$\xi_j^{\left(\ell^\prime\right)}\left(g\right)$ &	$j=0$ &	$j=1$ &	$j=2$ &	$j=3$ &	$j=4$ &	$j=5$ &	$j=6$\\[12pt] \hline\hline \\[-8pt]
$g=0$	& 1 & - & - & - & - & - & - \\[6pt]  \hline \\[-8pt]
$g=1$	&-&	1&	-&	-&	-&	-&	-\\[6pt]  \hline \\[-8pt]
$g=2$ &	$\sqrt{1/3}$ &	- &	$\sqrt{2/3}$ &	- &	- &	- &	-\\[6pt]  \hline \\[-8pt]
$g=3$, $\ell^\prime=\tau_1[3]$ & - &	$\sqrt{3/5}$ &	- &	$\sqrt{2/5}$ &	- &	- &	-\\[6pt]  \hline \\[-8pt]
$g=4$, $\ell^\prime=\tau_1[4]$ &$\sqrt{7/35}$	&-	& $\sqrt{20/35}$	&-&	$\sqrt{8/35}$& - &	-\\[6pt]  \hline \\[-8pt]
$g=5$, $\ell^\prime=\tau_1[5]$ & -&	$\sqrt{27/63}$ &	- & $\sqrt{28/63}$ &	- &	$\sqrt{8/63}$	& -\\[6pt]  \hline \\[-8pt]
$g=6$, $\ell^\prime=\tau_1[6]$ & $\sqrt{33/231}$ &	- &	$\sqrt{110/231}$ &-&	$\sqrt{72/231}$	&-&	$\sqrt{16/231}$\\[6pt]  \hline\hline
\end{tabular}
\end{table}  

\section{Global unitary bound scaling factors of density operator components which are not fully permutation symmetric} \label{sec:BEADSGUB}
As we briefly discussed in section~\ref{BEADSMap}, in the BEADS representation, we can scale the spherical functions of fully permutation symmetric density operator components by factors $s_j^{\left(\ell^\prime\right)}\left(N,g\right)$ such that the resulting spherical function values match expectation values of Pauli-Z operators. While the introduced approach (see section~\ref{sec:BEADSCanonicalScaling}) would in principle be applicable to any bead, the resulting scaling factors may cause the absolute bead function values to become larger than~1. This is because non-fully permutation symmetric tensor operator components correspond to linear combinations of correlation expectation values of joint measurements for which physical scenarios exist where the individual expectation values constructively add up. 

For instance, $T_{1,0}^{\left\{1,2\right\}_{\textit{odd}}}=\left(\sigma_{1x}\sigma_{1y}-\sigma_{1y}\sigma_{1x}\right)/(2\sqrt2)$. For the state $\ket{\psi} = \left(\ket{01}-i\ket{10}\right)$, ${\left\langle \sigma_{1x}\sigma_{1y} \right\rangle=1}$ and $\left\langle \sigma_{1y}\sigma_{1x} \right\rangle =-1$ such that we would find that by using scaling factors as introduced in appendix~\ref{sec:BEADSCanonicalScaling}, the bead spherical function value along the z-axis would be $b^{\left\{1,2\right\}_{\textit{odd}}}\left(0,0\right)=2$. In fact, this would prevent a uniform interpretation of the color schemes introduced in section~\ref{sec:BEADSColors}. Specifically, this is the case for beads that represent bilinear antisymmetric components or those corresponding to permutational symmetries $\tau_k$, with $k>1$. 

Here, we suggest a different scaling approach for density operator components that are not fully permutation symmetric by defining additional scaling factors $\xi_j^{\left(\ell^\prime\right)}\left(g\right)$ which ensure the absolute spherical function values to be upper-bounded by 1. These scaling factors can be found by considering the \textit{global unitary bounds} between arbitrary pure state density operators and LISA tensor operator components $T_{j,m}^{\left(\ell^\prime\right)}$. Note that it is sufficient to consider pure states as the maximum beads spherical function values $\max|b^{\left(\ell^\prime\right)}\left(\theta,\phi\right)|$ over all pure states are equal or larger than the values obtained for mixed states.

The global unitary bound $u=\left\langle {UAU^\dag}\middle| B\right\rangle$ corresponds to the maximum overlap between a unitarily transformed initial operator $A$ and a target operator $B$, where $A$ and $B$ are normalized. If both operators are Hermitian, the global unitary bound can be readily calculated using eigen decompositions of $A$ and $B$ \cite{GUB1}.

To determine the scaling factors, we are interested in the global unitary bound $u_{j,0}^{\left(\ell^\prime\right)}$ between an arbitrary pure state and a bead of interest. Without loss of generality we chose $A$ as the state $\rho_0=\ket{0}^{\otimes N}\bra{0}^{\otimes N}$ and $B$ as the axial LISA operator $T_{j,0}^{\left(\ell^\prime\right)}$ associated to the bead of interest. The global unitary bound is then given by \cite{GUB1, GUB2, GUB3}
\begin{equation}
u_{j,0}^{\left(\ell^\prime\right)}=\max_U \left\langle T_{j,m}^{\left(\ell^\prime\right)} \right\rangle = \max_U \left\langle U\rho_0U^\dag \middle|  T_{j,m}^{\left(\ell^\prime\right)} \right\rangle = \max_U \left\langle U\ket{0}^{\otimes N}\bra{0}^{\otimes N}U^\dag \middle|  T_{j,m}^{\left(\ell^\prime\right)} \right\rangle.
\vspace{6pt}
\end{equation}
\noindent
and the maximally overlapping state is the normalized eigenvector $\ket{V}$ (where $\ket{V}=U\ket{0}^{\otimes N}$) corresponding to the largest absolute eigenvalue of $T_{j,m}^{\left(\ell^\prime\right)}$.

The scaling factors $\xi_j^{\left(\ell^\prime\right)}\left(g\right)$  can then be determined by considering the global unitary bound $u_{j,0}^{\left(\ell^\prime\right)}$ and the scaling factor $\zeta\left(N\right)$ (see section~\ref{sec:BEADSCanonicalScaling}):
\begin{equation}\label{eq:GUBScaling}
\xi_j^{\left(\ell^\prime\right)}\left(g\right)=\left(\zeta\left(N\right)\cdot u_{j,0}^{\left(\ell^\prime\right)}\right)^{-1}.
\vspace{6pt}
\end{equation}

\noindent
A list of global unitary bounds of two- and three-qubit LISA tensor operators and corresponding scaling factors $\xi_j^{\left(\ell^\prime\right)}\left(g\right)$ that are applied in the BEADS representation is given in Table~\ref{tab:GUBSF}.
Note that the global unitary bound approach causes beads to be scaled such that resulting numerical ranges of beads values are bounded by $\pm1$ which adds quantitative information on how closely the individual symmetry components of a visualized density operator approach the corresponding global unitary bounds, that is, how large the symmetry components in a quantum state are compared to the maximally reachable components.

State expressions and BEADS representations of three-qubit global unitary bound states which incorporate the calculated scaling factors are shown in appendix~\ref{app:BEADSGUBStates}. 

\begin{table}[h!]
\centering
\caption{\label{tab:GUBSF}Additional scaling factors $\xi_j^{\left(\ell^\prime\right)}\left(g\right)$ for bilinear and trilinear beads which do not correspond to fully permutation symmetric components of the density operator calculated based on the global unitary bounds of the corresponding LISA tensor operator.}
\vspace{.3cm}
\begin{tabular}{m{0.15\textwidth}<{\centering} m{0.15\textwidth}<{\centering} m{0.15\textwidth}<{\centering} m{0.15\textwidth}<{\centering}}
\hline\hline\\[-10pt]

LISA tensor operator & \multicolumn{2}{c}{Global unitary bound } & Scaling factor $\xi_j^{\left(\ell^\prime\right)}\left(g\right)$\\[6pt]
 & $N=2$ & $N=3$ & \\[5pt] \hline\hline \\[-8pt]

$T_{1,0}^{\{k,l\}_{\textit{odd}}}$ & $\frac{1}{\sqrt2}$ & $\frac{1}{2}$ & $\frac{1}{\sqrt2}$\\[6pt]  \hline \\[-8pt]
$T_{1,0}^{\{k,l,m\ \tau_2\}_{\textit{odd}}}$ & -- & $\sqrt{\frac{\sqrt3+2}{12}}$ & $\frac{3}{3+\sqrt3}$\\[6pt]  \hline \\[-8pt]
$T_{2,0}^{\{k,l,m\ \tau_2\}_{\textit{even}}}$ & -- & $\frac{1}{2}$ & $\frac{1}{\sqrt2}$\\[6pt]  \hline \\[-8pt]
$T_{1,0}^{\{k,l,m\ \tau_3\}_{\textit{odd}}}$ & -- & $\frac{1}{2}$ & $\frac{1}{\sqrt2}$\\[6pt]  \hline \\[-8pt]
$T_{2,0}^{\{k,l,m\ \tau_3\}_{\textit{even}}}$ & -- & $\frac{1}{2}$ & $\frac{1}{\sqrt2}$\\[6pt]  \hline \\[-8pt]
$T_{0,0}^{\{k,l,m\ \tau_4\}_{\textit{even}}}$ & -- & $\frac{1}{2}$ & $\frac{1}{\sqrt2}$\\[6pt]  \hline\hline
\end{tabular}
\vspace{-.5cm}
\end{table}  

\noindent
Applying scaling factors as described in Eq.~\ref{eq:GUBScaling} can lead to beads appearing very bright even if the underlying unscaled expectation values are relatively small. This is specifically the case for permutational symmetries $\tau_k$, with $k>1$. Thus, an alternative scaling, which we call \textit{natural expectation value scaling}, is given by omitting $u_{j,0}^{\left(\ell^\prime\right)}$ in Eq.~\ref{eq:GUBScaling} such that the beads spherical function values directly correspond to the expectation values of the underlying LISA tensor operators.

\section{Axial tensor operators and tomography}\label{sec:BEADSTomography}

As was previously shown by Leiner et al. \cite{LeinerWQST,LeinerWQPT}, the DROPS representation can be reconstructed experimentally by using tomography techniques. Indeed, this can be achieved as the droplet function $f_j^{\left(\ell\right)}(\beta,\alpha)$ corresponding to a specific set $\ell$ is proportional to the expectation value of the associated rotated axial LISA basis operator $R_{\alpha\beta}T_{j,0}^{\left(\ell\right)}$ for which it can be directly mapped to the corresponding axial spherical harmonics function $Y_{j,0}$ which has a value of
\begin{equation}\label{eq:DROPSTomo}
f_j^{\left(\ell\right)}\left(\beta,\alpha\right)=\sqrt{\frac{2j+1}{4\pi}}\cdot\left\langle R_{\alpha\beta}\ T_{j,0}^{\left(\ell\right)} \right\rangle
\vspace{6pt}
\end{equation}

\noindent
along angular direction $\left(\beta,\alpha\right)$ and where $R_{\alpha\beta}$ denotes a rotation matrix \cite{LeinerWQST}. In particular, for an (unrotated) axial LISA basis operator $T_{j,0}^{\left(\ell\right)}$, the value of the droplet function along the z-axis is given by
\begin{equation}
f_j^{\left(\ell\right)}\left(0,0\right)=\sqrt{\frac{2j+1}{4\pi}}\cdot\left\langle T_{j,0}^{\left(\ell\right)}\right\rangle.
\vspace{6pt}
\end{equation}

\noindent
Since the BEADS representation remains bijective, despite introducing additional scaling factors $s_j^{\left(\ell^\prime\right)}\left(N,g\right)$, it can also be reconstructed experimentally by directly applying the scaled non-normalized rotated axial tensor operators ${\widetilde{T}}_{j,0}^{\left(\ell^\prime\right)} = \zeta(N)\xi_j^{\left(\ell^\prime\right)}\left(g\right)T_{j,0}^{\left(\ell^\prime\right)}$ in Eq.~\ref{eq:DROPSTomo}. The spherical function bead value along angular direction $\left(\beta,\alpha\right)$ is then given by
\begin{equation}
b_j^{\left(\ell^\prime\right)}\left(\beta,\alpha\right)=\left\langle R_{\alpha\beta}{\widetilde{T}}_{j,0}^{\left(\ell^\prime\right)} \right\rangle = \left\langle R_{\alpha\beta}\ \sqrt{2^N}\ {\xi_j^{\left(\ell^\prime\right)}\left(g\right)}\ T_{j,0}^{\left(\ell^\prime\right)} \right\rangle,
\vspace{6pt}
\end{equation}
\noindent
where $\zeta(N)=\sqrt{2^N}$ and $\xi_j^{\left(\ell^\prime\right)}\left(g\right)$ are the scaling factors introduced in sections~\ref{sec:BEADSCanonicalScaling} and~\ref{sec:BEADSGUB} and $N$ is the number of qubits in the system. Scaled axial LISA tensor operators up to linearity $g=3$ are provided in section~\ref{app:BEADSTomog} of the supplementary material. 

\section{Correlation functions and correlation in quantum information theory}\label{CorrQIT}

As discussed in section~\ref{sec:BEADSCorrel}, the BEADS representation is able to visualize different types of correlation functions of a pure state. Such correlation functions must not be confused with the more general notion of correlation of a quantum state in quantum information theory. We now provide a brief overview of both concepts and explain how they are related.

Correlation functions are useful for inferring important aspects of the correlation of a given quantum state. For instance, in a bipartite system described by a density operator $\rho_{AB}$, the connected correlation function (also cf. Eq.~\ref{eq:Connected2Q})
\begin{equation}\label{eq:Covariance}
E_{AB} = \left\langle O_A \otimes O_B \right\rangle_{\rho_{AB}} - \left\langle O_A \right\rangle_{\rho_{A}} \left\langle O_B \right\rangle_{\rho_{B}}
\end{equation}
using observables $O_A$ and $O_B$, which act on the corresponding subsystems and the reduced density matrices $\rho_{A} = \text{Tr}_{B}(\rho)$ and $\rho_{B} = \text{Tr}_{A}(\rho)$, directly reveals $\rho_{AB}$ to be uncorrelated if for all possible combinations of observables, the correlation function is zero \cite{Horodecki, Schilling}. Here, a state being uncorrelated is equivalent to the state being a product state, i.e., $\rho_{AB} = \rho_A \otimes \rho_B$. Importantly, this applies independent of whether the state is pure or mixed. For general $N$-partite systems, the same analysis can be done by using $N$-partite connected correlation functions as introduced in section~\ref{sec:BEADSCorrel}.

Note that Eq.~\ref{eq:Covariance} is essentially the covariance as known from classical information theory \cite{BrandtStatistics} applied to quantum mechanics.

In case of pure states $\rho = \ket{\psi}\bra{\psi}$, the connected correlation function quantifies manifestations of quantum correlations \cite{Horodecki} which for pure states exclusively correspond to entanglement \cite{Mahler95, Mahler96, Mahler97, Tran17}. In other words, if the connected correlation function is non-zero for some set of observables, $\rho$ is entangled and the effect on measurement outcomes is quantified by the corresponding connected correlation coefficient.

In contrast, non-zero connected correlation functions of mixed states ${\rho = \sum_k p_k \ket{\psi_k}\bra{\psi_k}}$ do not specifically signify quantum correlations let alone entanglement but are an indicator of correlation in general. If calculated as shown in section~\ref{sec:BEADSCorrel}, it is impossible to distinguish between different types of correlation of a general mixed state, that is, classical correlation \cite{Werner}, quantum discord \cite{Discord}, or entanglement \cite{Horodecki} based on connected correlation functions. This aspect is important in that it clearly differentiates quantum information theoretic studies which apply a separation of correlation types of mixed states as, e.g., proposed in \cite{Schilling, Schilling2} from the separation of correlation function contributions of pure states applied in the BEADS representation.

The total correlation functions $T$ as defined in Eq.~\ref{eq:CorrFunc} does \textit{not} directly quantify "total correlations" of quantum states in a quantum information theoretic sense. Total correlation is commonly described by the quantum mutual information \cite{Lindblad73, Henderson01}
\begin{equation}\label{eq:MutInf}
I(\rho)=S(\rho_A) + S(\rho_B) - S(\rho),
\end{equation}
where $S(\rho)=-\text{Tr}(\rho \ln(\rho))$ is the von Neumann entropy. It quantifies the information in a quantum state -- independent of the type of the underlying correlation -- which is not encoded in the reduced states $\rho_A$ and $\rho_B$ \cite{Schilling}. For systems of distinguishable particles, $I(\rho)$ can be thought of as the geometric distance of a state $\rho$ to the closest product state $\zeta$ \cite{Schilling2}. 

The total correlation function $T = E + C$ (see section~\ref{sec:BEADSCorrel}), on the other hand, is the sum of the connected correlation function $E$ and all compound correlation function contributions $C$, i.e., all components of $T$ which can be factorized to expectation values of local observables and/or lower-order connected correlation functions. For pure states, where entanglement is the only source of correlation, it thus follows from the previous findings that in fact the \textit{connected} correlation function quantifies manifestations of total correlations. This can also be deduced from the quantum mutual information of a pure state $\rho$ which is given by \cite{Horodecki}
\begin{equation}
I(\rho)=S(\rho_A) + S(\rho_B) = 2S_E(\rho)
\end{equation}
and where $S_E(\rho) = S(\rho_A) = S(\rho_B)$ is the entanglement entropy of the state \cite{Schilling}. Note that $S(\rho)$ (cf. Eq.~\ref{eq:MutInf}) of a pure state is 0 due to all pure state density operators being rank 1 projectors, that is, there is no uncertainty with respect to the statistical description of the state as there is only one non-zero eigenvalue $\lambda=1$.

Compound correlation functions, which we have previously described as representing redundant information, do not relate to the correlation of a pure state. The contributing terms correspond to trivial correlations of measurement outcomes. As a simple example, consider the uncorrelated product state $\ket{00}$ for which computational basis measurement outcomes are perfectly correlated. In this case, the bipartite total zz-correlation coefficient is given by $T^{zz}_{12} = C^{zz}_{12}=\left\langle \sigma_{1z} \right\rangle\left\langle \sigma_{2z} \right\rangle = 1$. Obviously, the entire information is already encoded in the reduced states and any local operations on one qubit will not affect the measurement outcomes of the other qubit. In case of the given state, the correlation of measurement results that one observes and which are captured by $C_{12}^{zz}$ merely arise from the fact that both qubits are in the same (pure) reduced state.

As a consequence, whereas connected correlation functions emerge from quantum correlations of a pure state, compound correlation functions do not relate to the correlation of a pure state at all. We currently investigate possibilities to isolate correlation function contributions which relate to the mentioned types of correlation of mixed states.

\section{BEADS color schemes}\label{sec:BEADSColors}
The BEADS representation uses two color schemes to represent different types of information that are contained in a quantum state. In this article, we use a \textit{red-green discontinuous} color scale which features shaded red and green colors to represent expectation values (and bit parity probabilities in case of fully symmetric contributions) corresponding to local single-qubit components and compound correlation function contributions. On the other hand, connected correlation coefficients (E-Beads) are represented by a \textit{yellow-blue discontinuous} color scheme. Both scales are shown in Fig.~\ref{Figure:Fig1},~\ref{Figure:Fig3} and~\ref{Figure:Fig5}. Note that here, we use diverging color schemes which are black for the special case where the corresponding expectation value is $\left\langle O \right\rangle =0$ (or where the related bit parity probability $p$ is 0.5) and which diverge to a bright red or yellow color for the case $\left\langle O \right\rangle =1$  ($p = 0$) and a bright green or blue color for the case $\left\langle O \right\rangle =-1$ ($p = 1$).
 
Indeed, it would be possible to utilize a single uniform color scheme for all correlation components. However, different colorings increase the distinguishability of the visualized information and, in particular, highlight the importance of connected correlation functions which occur in pure entangled states in quantum computing.

The introduced color scales can be chosen to be continuous or discontinuous. By default, we use color scales which are shaded discontinuously despite, at a first glance, using a continuous color scheme theoretically appears to be the best option to faithfully represent the information encoded in a quantum state as expectation values $\left\langle O \right\rangle$ can vary continuously between $-1$ and~1. However, the human eye is poorly adapted to recognizing subtle differences in color. Hence, to improve the accessibility of the BEADS representation, we intentionally accept the loss of a negligible portion of information that comes with the proposed discontinuity to enable a direct and fast quantitative estimation of relevant numerical information by counting the bands, e.g., between the black band and a color of interest. This is a well-known approach which is commonly applied in the context of contour lines in surface plots or geographic maps. 

\begin{figure}[H]
\centering
\includegraphics[width=.9\textwidth]{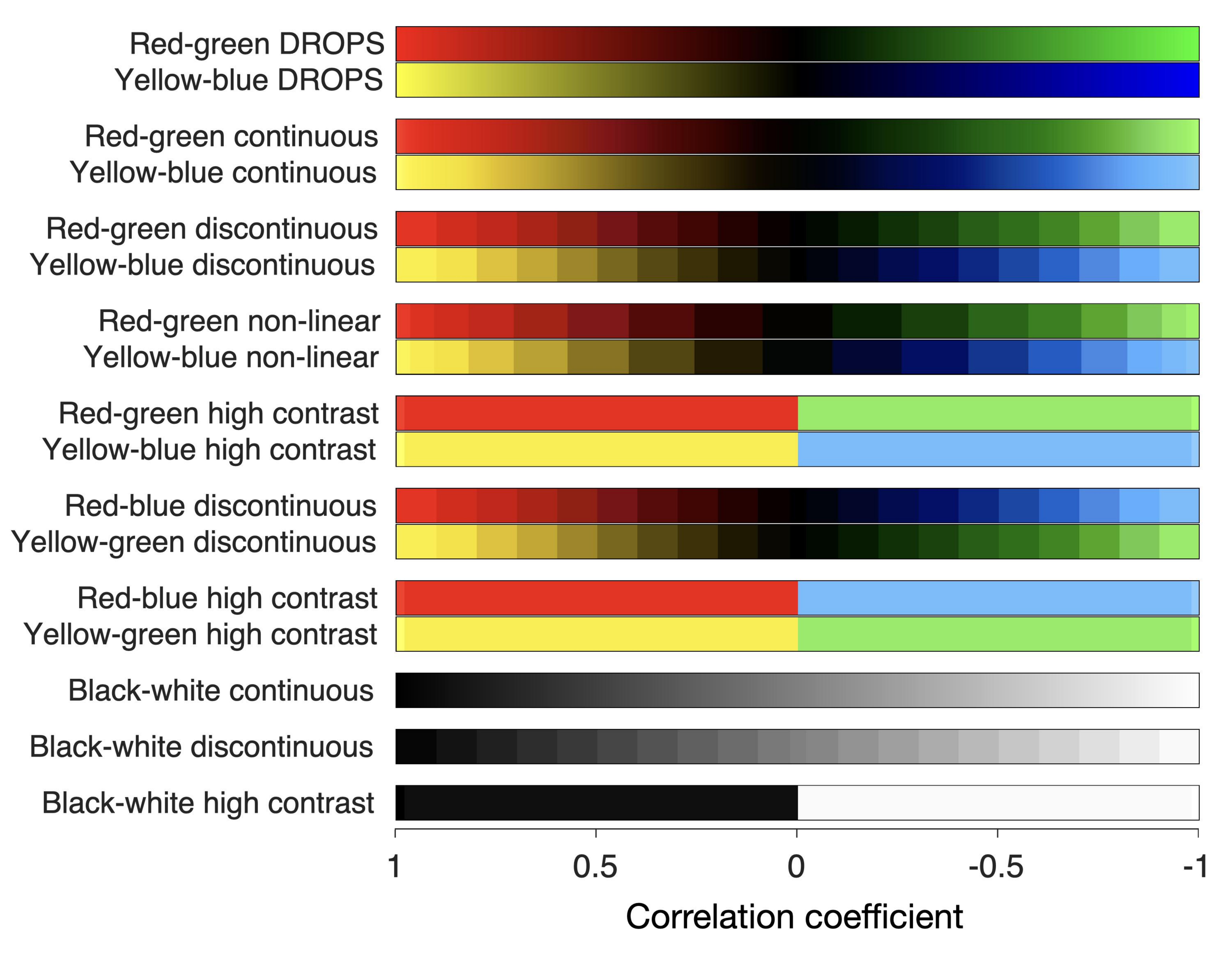}
\caption{\label{fig:FigColorScales}BEADS color scheme variants. Color scales which are intended to be used jointly are shown in pairs. The DROPS scales refer to the color scheme used in SpinDrops \cite{SpinDrops}. The scales are given in terms of bit parity probabilities and expectation values. For discontinuous color schemes, each color increment covers a range of 0.1. The non-linear color scheme, results in equiangular color increments in case of Q-Beads. For each pair of color scales, the upper scheme corresponds to single-qubit and compound correlations, whereas the lower color scheme encodes for entanglement-associated connected correlations. All color scales which are not continuous use special slightly brighter colors for expectation values $\left\langle M \right\rangle \approx \pm1$. In addition, high contrast color scales are grey (black-white high contrast) or black (all remaining high contrast scales) for values$\left\langle M \right\rangle \approx 0$.}
\end{figure}

Alternative color schemes can nonetheless be applied to the BEADS representation without limitations to represent different levels of information and to account for special requirements. We present additional exemplary continuous and discontinuous color scale variants in Fig.~\ref{fig:FigColorScales}. The DROPS color schemes which are adopted from the SpinDrops software \cite{SpinDrops} are constructed by a simple linear interpolation of red to black to green. Compared to the red-green and yellow-blue discontinuous color scales, color shadings are hard to distinguish for roughly similar expectation values. The same applies to the remaining continuous variants despite using optimized colors to enhance detail perceptibility.
 
The BEADS representation is not restricted to vivid colors, i.e., all information can also be visualized by using greyscale schemes. Yet, since different nuances of grey are in general harder to distinguish, we consider non-greyscale color schemes as being practically more suitable. However, a very simplistic greyscale color scheme which we denote \textit{black-white high contrast} (and also the full color high contrast analogues \textit{red-green}, \textit{yellow-blue}, \textit{red-blue} and \textit{yellow-green high contrast}) can be used as a possible convenient entry point to the BEADS representation as it allows to visualize a reduced set of information qualitatively which is suited to explain basic concepts and properties with respect to projective measurements, correlations, and symmetries in a simplified manner (see supplementary Figure~\ref{app:BEADSHighContrast}). High contrast BEADS representations can also be easily sketched by hand.

We are well-aware that combining red and green color in one color scale (and to a lesser degree yellow and blue) may pose a severe problem for people with color blindness. For such cases, one can apply shuffled versions of the standard scales which combine red with blue as well as yellow with green, thus, avoiding conflicting colors.

\subsubsection*{The total correlation function color scheme}
Using two colormaps to distinguish between correlation coefficient types and a correlation function separation approach as introduced in section~\ref{sec:BEADSCorrel} further allows us to add another layer of information with respect to representing total correlation coefficients: As we have shown, e.g., in Fig.~\ref{Figure:Fig5} in section~\ref{2Q}, any T-Bead (total correlation function bead) can be represented such that its color along an arbitrary direction corresponds to the absolute total correlation coefficient by its brightness and the ratio between connected to compound correlations by its hue.

\begin{figure}[H]
\vspace{-.7cm}
\centering
\begin{minipage}{.5\textwidth}
\includegraphics[width=.9\textwidth]{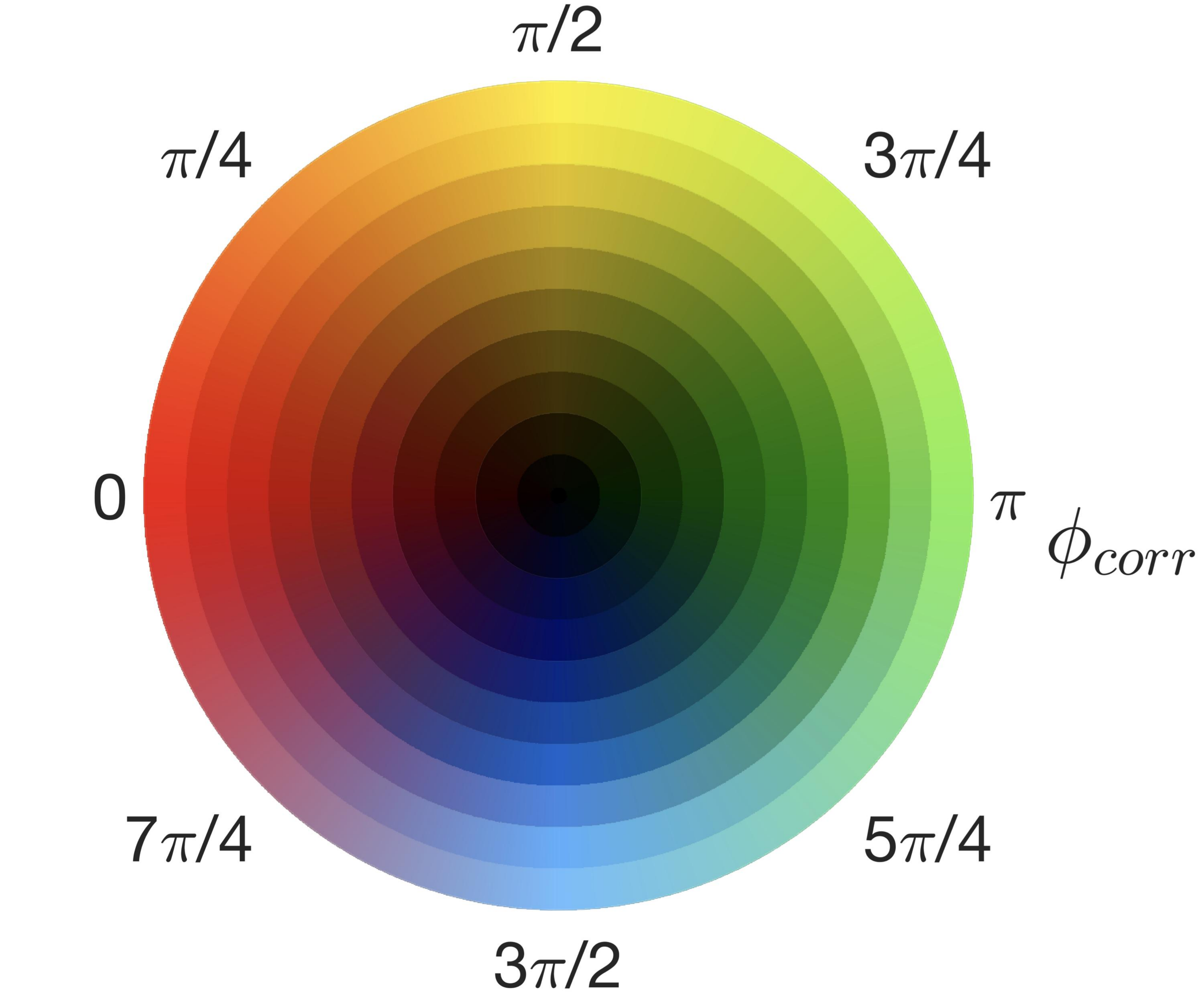}
\end{minipage}%
\begin{minipage}{.5\textwidth}
\includegraphics[width=.9\textwidth]{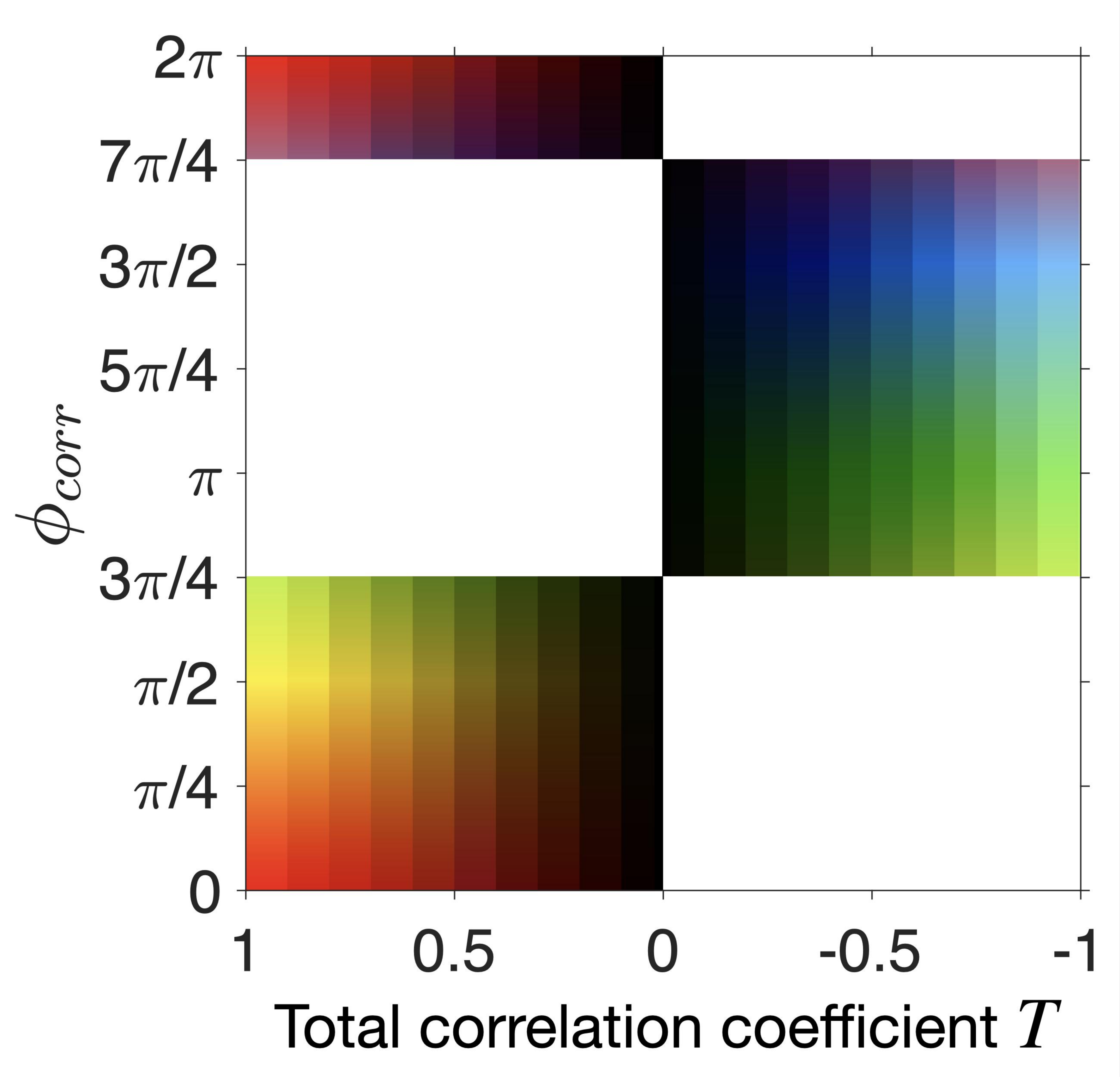}
\end{minipage}
\caption{\label{fig:FigColorTotCorr}Total correlation BEADS color wheel (left) and $\phi_{corr}$ vs. $T$ 2D plot (right). In the color wheel, the magnitude of the total correlation coefficient corresponds to the radius from the center and the ratio of the individual correlation function components determines the angular direction $\phi_{corr}$ in which to read off the color. The connected and compound correlation function color scales are given by the vertical and horizontal diameters of the color wheel, respectively. Plotting $\phi_{corr}$ against $T$ illustrates two special cases at $3\pi/4 = 135^\circ$ and $7\pi/4 = 315^\circ$ where the total correlation coefficient is always zero due to the components $E$ and $C$ having the same magnitudes but opposite signs.}
\end{figure}

\begin{figure}[H]
\centering
\includegraphics[width=1.\textwidth]{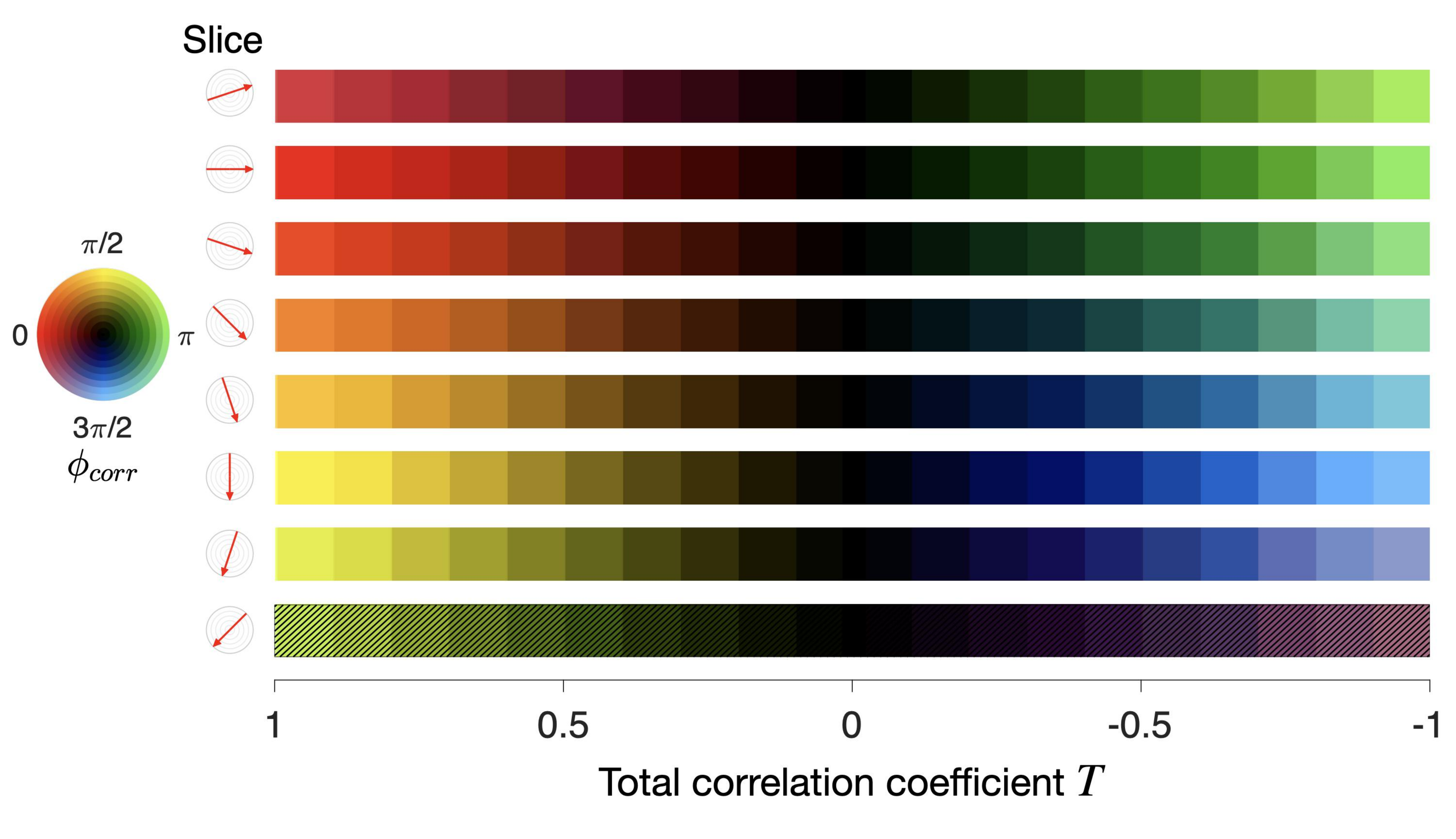}
\caption{\label{fig:FigColorSlices}Total correlation function color scales for different ratios of correlation function contributions. All color scales correspond to diametric slices of the color wheel (left) at specific angular directions indicated by arrows, respectively. The displayed cases, from top to bottom, comprise correlation function component ratios $E:C$ of (-1):3, 0:1, 1:3, 1:1, 3:1, 1:0, 3:(-1), and 1:(-1). Note that for the last scale, all colors except black are physically unreachable ($T=0$) which is indicated by a hatching.}
\end{figure}

\begin{figure}[H]
\centering
\includegraphics[width=1.\textwidth]{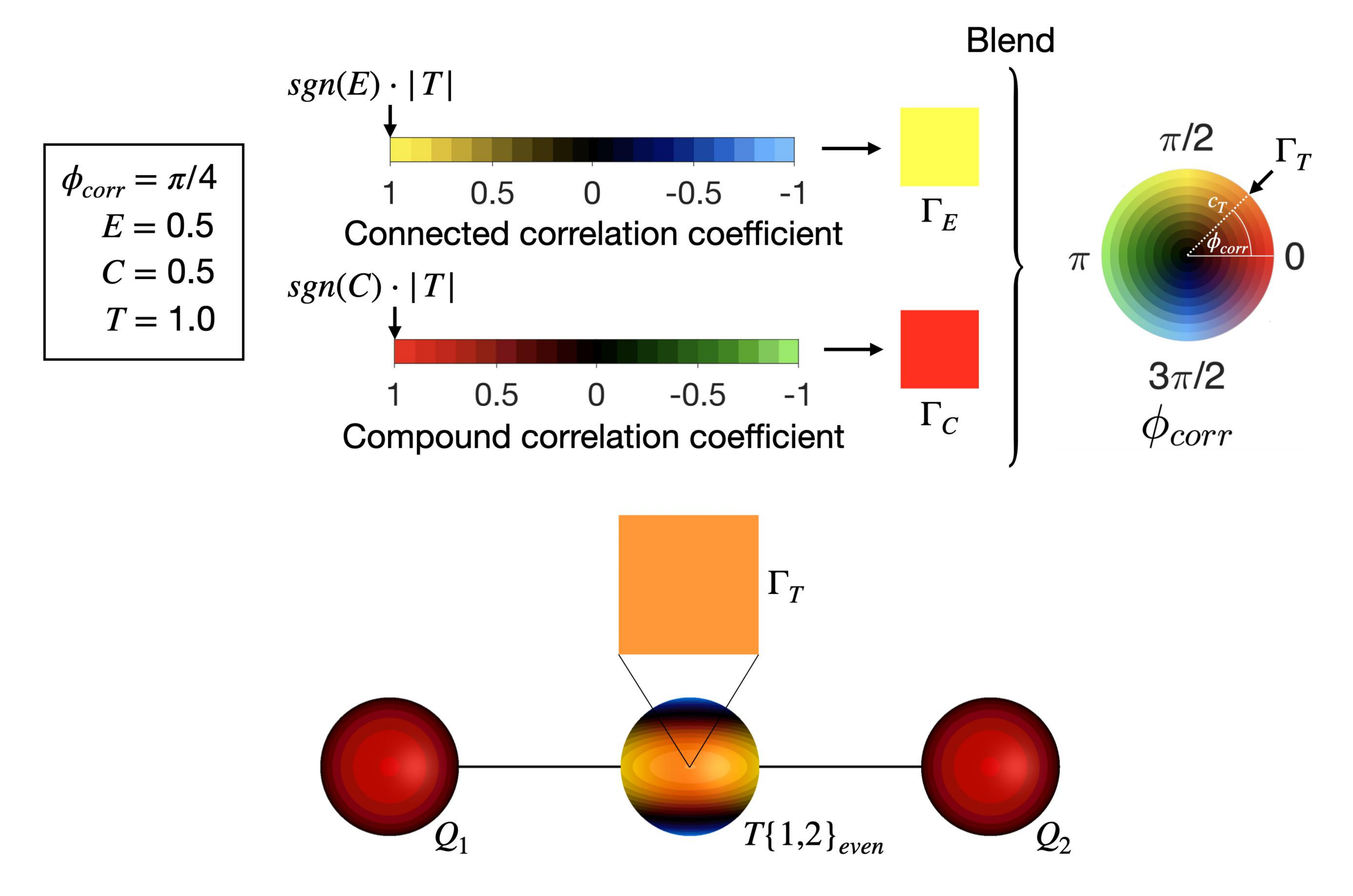}
\caption{\label{fig:FigColorSchmidt}Total correlation color generation in the BEADS representation. In the BEADS representation of the Schmidt form state $\ket{\psi}=\cos{(\pi/8)}\ket{00}+\sin{(\pi/8)}\ket{11}$, the bilinear total correlation color $\Gamma_T$ along the z-axis is determined by picking colors $\Gamma_E$ and $\Gamma_C$ from the basic color scales at $\text{sgn}(c_k)\cdot|T|$, where $\text{sgn}$ is the sign function and $c_k\in\left\{E,C\right\}$, which are then blended as described by Eq.~\ref{eq:ColorBlend}.}
\end{figure}

In the following, we use the short notations $E$ (entanglement-based connected correlation coefficient), $C$ (compound correlation coefficients) and $T =E+C$ (total correlation coefficient) to denote \textit{correlation function components} which are obtained as explained in section~\ref{sec:BEADSCorrel}. To calculate the T-Bead color along an arbitrary direction, we first determine the correlation function component colors $\mathrm{\Gamma}_E$ and $\mathrm{\Gamma}_C$ by reading off the colors from the corresponding color scales at the absolute total correlation coefficient multiplied with the signs of $E$ or $C$, i.e., at  $\text{sgn}(c_k)\cdot\left|T\right|$ where $c_k\in\left\{E,C\right\}$. We then blend these two colors, expressed as RGB vectors, based on a blending angle $\vartheta_{blend}$ which can be calculated as 
\begin{equation}
\vartheta_{blend}=\text{atan}{\frac{\left|E\right|}{\left|C\right|}}.
\end{equation}
The T-Bead vertex color is then given by
\begin{equation}\label{eq:ColorBlend}
\mathrm{\Gamma}_T=\mathrm{\Gamma}_C+\frac{{2\vartheta}_{blend}}{\pi}\left(\mathrm{\Gamma}_E-\mathrm{\Gamma}_C\right)=\left(\begin{matrix}R_C\\G_C\\B_C\\\end{matrix}\right)+\frac{2\vartheta_{blend}}{\pi}\left[\left(\begin{matrix}R_E\\G_E\\B_E\\\end{matrix}\right)-\left(\begin{matrix}R_C\\G_C\\B_C\\\end{matrix}\right)\right].
\end{equation}
Following this approach and by applying the default discontinuous color schemes (see Fig.~\ref{fig:FigColorScales}), we thus obtain the color wheel shown in Fig.~\ref{fig:FigColorTotCorr} that specifies a unique color for any possible combination of correlation components and where the radius from the center corresponds to the absolute total correlation coefficient. Moreover, the horizontal and vertical diameters correspond to the original unblended color scales, respectively. Note that here, we use the correlation angle $\phi_{corr}=\text{atan}{\frac{E}{C}}$ to describe the ratio between the correlation function components. Plotting $\phi_{corr}$ against the total correlation coefficient $T$ further illustrates that the sign of $T$ changes at $\phi_{corr}=3\pi/4=135^\circ$ and $\phi_{corr}=7\pi/4=315^\circ$. For these values, $E=-C$, and thus, $T=0$ which implies that such combinations of correlation coefficients will always be represented by black color independent of the magnitudes of $E$ and $C$. Based on these findings one may further infer that not all depicted colors and the underlying combinations of correlation coefficients are physically reachable. However, we show the full color wheel to clarify the construction.
\noindent
Knowing the ratios between compound to connected correlation coefficients, we can also extract corresponding characteristic color scales as diameters of the color wheel (see Fig.~\ref{fig:FigColorTotCorr}) in the angular direction specified by $\phi_{corr}$. Fig.~\ref{fig:FigColorSlices} shows some color scales for different ratios of correlation function components $(E:C)$.
\vspace{1cm}
To further clarify the method, an exemplary visualization of a Schmidt form state (corresponding to a Schmidt angle $\theta=\pi/4$) is provided in Fig.~\ref{fig:FigColorSchmidt} that illustrates the construction in detail. Here, $C=E=0.5$, and thus, $T=1.0$. First, we pick the compound correlation coefficient color $\mathrm{\Gamma}_C$ at
$\text{sgn}(C)\cdot|T|$ from the red-green discontinuous color scale and the connected correlation coefficient color $\mathrm{\Gamma}_E$ is obtained at $\text{sgn}(E)\cdot|T|$ from the yellow-blue discontinuous color scale. Both colors are then blended according to Eq.~\ref{eq:ColorBlend} by using the mixing angle $\vartheta_{blend}=\text{atan}(1)=\pi/4=45^\circ$ which gives the total correlation coefficient color $\mathrm{\Gamma}_T$.

\section{Comparison with other representations}
\subsection{The BEADS representation as a generalization of the Husimi function and relations to the Majorana stellar representation}\label{sec:BEADSHusimi}
As was briefly mentioned in section~\ref{BEADSMap}, the BEADS representation can be viewed as a generalized Husimi function~\cite{Husimi}. A quantum state $\ket{\psi}$ (or its corresponding density operator $\rho$) can be represented by the Husimi function $\mathcal{H}$ (avoiding the generally used symbol $Q$ to prevent confusion with Q-Beads in the following) which is defined for arbitrary spatial directions $\left(\theta,\phi\right)$ as \cite{HusimiMintert, HusimiSugita}
\begin{equation}
\mathcal{H}\left(\theta,\phi\right)=\left|\left\langle s\left(\theta,\phi\right)\middle|\psi\right\rangle\right|^2=\left\langle s\left(\theta,\phi\right)\middle|\rho\middle| s\left(\theta,\phi\right)\right\rangle.
\end{equation}

\noindent
Note that for simplicity of calculation, this definition omits the originally introduced normalization by an additional factor $1/\pi$ \cite{Husimi} which is required to achieve normalization with respect to integration. Hence, for distinguishable particles such as qubits, $\mathcal{H}$ is the expectation value with respect to separable coherent states $\ket{s\left(\theta,\phi\right)}$. In an $N$-qubit system, $\ket{s\left(\theta,\phi\right)}$ represent the subset of coherent product states for which all Bloch vectors co-align in spatial directions $\left(\theta,\phi\right)$, that is,
\begin{equation}\label{eq:Husimi}
\ket{s\left(\theta,\phi\right)}=\left[\cos{\frac{\theta}{2}}\ket{0}+\sin{\frac{\theta}{2}}e^{i\phi}\ket{1}\right]^{\otimes N}.
\end{equation}

\noindent
By definition, $\mathcal{H}$ is non-negative and upper-bounded for quantum states, i.e., $0\le\mathcal{H}\left(\theta,\phi\right)\le1$. It is possible to show that the BEADS representation can be interpreted as a generalized Husimi function by a simple relation between $\mathcal{H}$ and the beads spherical functions $b^{(\ell^\prime)}$. 

\noindent
The Husimi function $\mathcal{H}$ is indeed obtained as the scaled sum of the identity operator bead $b^{\left\{\emptyset\right\}}$, all Q-Beads $b^{\left\{k\right\}}$ and all totally permutation symmetric total correlation function beads $b_T^{\left({\ell^\prime}_{sym}\right)}$, where ${\ell^\prime}_{sym}$ denotes sets which correspond to full permutation symmetry (cf. section~\ref{sec:BEADSSymmetries}). For instance, for systems consisting of $N = 2$ and $N = 3$ qubits, Husimi functions $\mathcal{H}_N$ are given by:
\begin{align}
\label{eq:Husimi2Q}\mathcal{H}_2&=\frac{1}{4}\left(b^{\left\{\emptyset\right\}}+b^{\left\{1\right\}}+b^{\left\{2\right\}}+b^{\left\{1,2\right\}_{\textit{even}}}\right)=\frac{1}{4}\left(\emptyset+Q_1+Q_2+T\left\{1,2\right\}_{\textit{even}}\right),\\
\mathcal{H}_3&=\frac{1}{8}\left(b^{\left\{\emptyset\right\}}+b^{\left\{1\right\}}+b^{\left\{2\right\}}+b^{\left\{3\right\}}+b_T^{\left\{1,2\right\}_{\textit{even}}}+b_T^{\left\{1,3\right\}_{\textit{even}}}+b_T^{\left\{2,3\right\}_{\textit{even}}}+b_T^{\left\{1,2,3\ \tau_1\right\}_{\textit{odd}}}\right)\nonumber\\
\label{eq:Husimi3Q}&=\frac{1}{8}\left(\emptyset+Q_1+Q_2+Q_3+T\left\{1,2\right\}_{\textit{even}}+T\left\{1,3\right\}_{\textit{even}}+T\left\{2,3\right\}_{\textit{even}}\right.\nonumber\\
&+\left.T\left\{1,2,3\ \tau_1\right\}_{\textit{odd}}\right).
\end{align}

\noindent
In the above relations, spherical function values of beads are interpreted as expectation values $(-1\le b^{(\ell^\prime)}\left(\theta,\phi\right)\le1)$. 
It is immediately clear, that the Husimi $\mathcal{H}$ representation of a quantum state hence only captures the identity operator, single-qubit, and the fully permutationally symmetric components of a density operator. In contrast, the BEADS representation provides a separation of these components and further includes components corresponding to other permutation symmetries thus generalizing the function $\mathcal{H}$.

The Majorana stellar representation \cite{Majorana32} of symmetric quantum states is closely related to the Husimi function and thus also to the BEADS representation. Symmetric $N$-qubit pure states $\ket{\psi_{\textit{sym}}}$ can be expressed in terms of the $N+1$ symmetric Dicke states \cite{Devi12}
\begin{equation}
D_{N,l}=\left|\frac{N}{2}, l-\frac{N}{2}\right\rangle = \frac{1}{\sqrt{B^N_l}} \bigg(\ket{\underbrace{00\dots}_{l}\underbrace{11\dots}_{N-l}} + \text{all permutations} \bigg),
\end{equation}
where $B^N_l = \frac{N!}{l!(N-l)!}$ is the binomial coefficient, such that $\ket{\psi_{\textit{sym}}}=\sum_{l=0}^N c_l D_{N,l}$. We then obtain the stellar representation of $\ket{\psi_{\textit{sym}}}$ as the stereographic projection of the $N$ roots of the Majorana polynomial \cite{Devi12}
\begin{equation}
P(z) = \sum_{l=0}^N(-1)^l \sqrt{B^N_l}\:c_l\:z^l
\end{equation}
on the Riemann sphere using the south pole as the projection point. If $n<N$ roots are obtained, additional roots are added at $z = \infty$. The obtained projection points are called \textit{stars}. Majorana stars are always antipodal to roots of the corresponding Husimi function. Note that whereas Majorana representations are limited to symmetric states, they can also be obtained for mixed states \cite{Braun20}.

\begin{figure}[H]
\centering
\includegraphics[width=1.\textwidth]{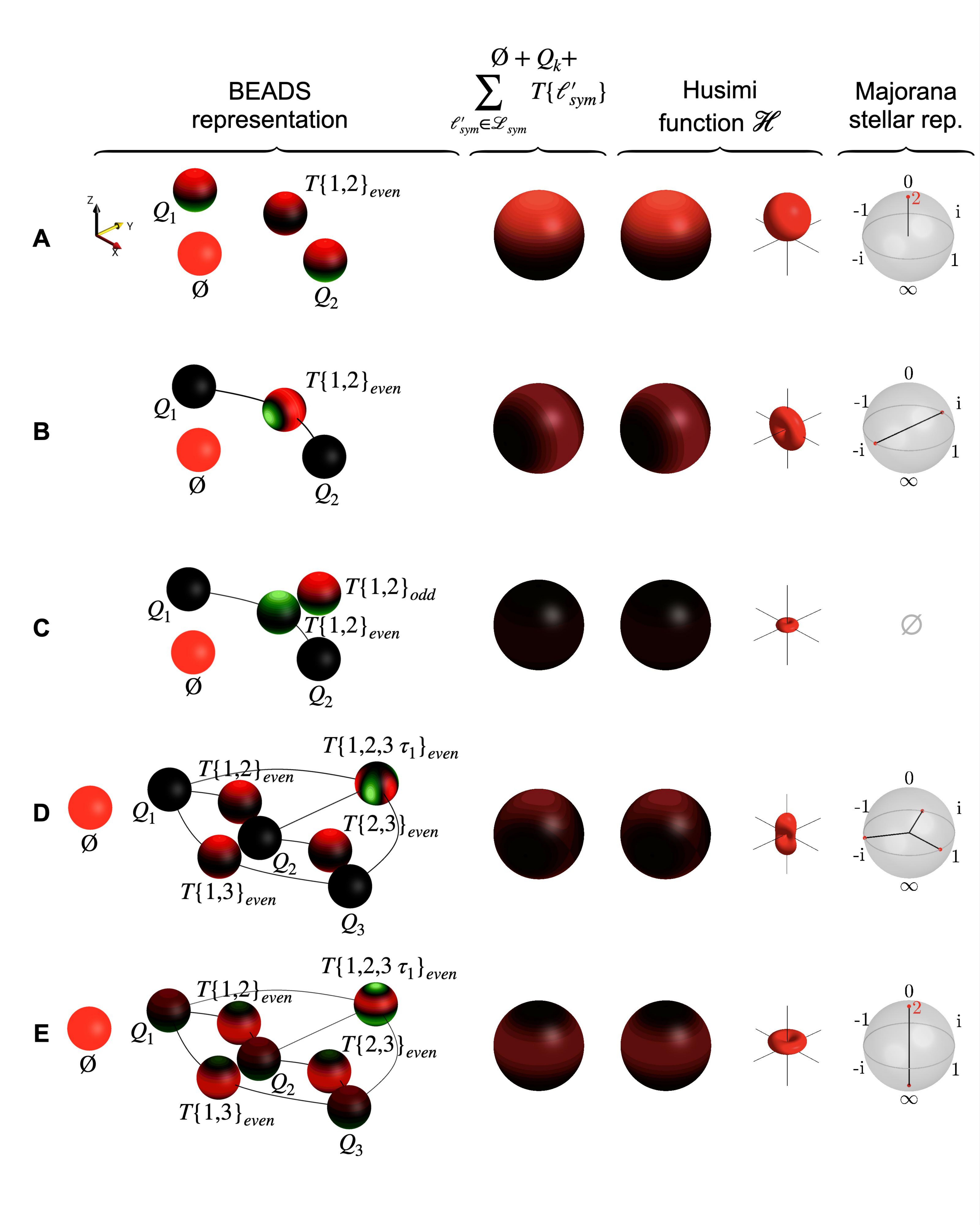}
\vspace{-1cm}
\caption{\label{fig:FigHusimi}The Husimi function corresponds to a linear combination of beads. Combining the identity bead, all Q-Beads, and all fully permutation symmetric T-Beads of a state gives the Husimi representation (by suitable rescaling, see Eq.~\ref{eq:Husimi2Q} and~\ref{eq:Husimi3Q}). Using a single colorscheme (red-green discontinuous), the BEADS and Husimi representations (calculated based on Eq.~\ref{eq:Husimi},~\ref{eq:Husimi2Q} and~\ref{eq:Husimi3Q}) of (\textbf{A}) $\ket{00}$, (\textbf{B}) $\ket{\Phi^+}=1/\sqrt2\left(\ket{00}+\ket{11}\right)$, (\textbf{C}) $\ket{\psi}=1/\sqrt2\left(\ket{01}-i\ket{10}\right)$, (\textbf{D}) $\ket{\text{GHZ}}=1/\sqrt2\left(\ket{000}+\ket{111}\right)$, and (\textbf{E}) $\ket{\text{W}}=1/\sqrt3\left(\ket{001}+\ket{010}+\ket{100}\right)$ are visualized for comparison. The absolute function value is plotted as distance from the origin in the rightmost column to improve the visibility of small function values such as in the highly antisymmetric case (\textbf{C}). For comparison, the corresponding Majorana stellar representations are provided on the right for symmetric states. Each star is represented by a red point which is connected to the Riemann sphere center by a line for better visibility. Multiplicities $m > 1$ of stars are indicated by a red number. Each star is antipodal to a root of the corresponding Husimi function.}
\end{figure}

In Fig.~\ref{fig:FigHusimi}, we provide a comparison of various states visualized by the standard BEADS, the Husimi $\mathcal{H}$, and -- if possible -- the Majorana stellar representation which illustrates the equivalence of $\mathcal{H}$ and the linear combinations of beads introduced in Eq.~\ref{eq:Husimi2Q} and~\ref{eq:Husimi3Q}, and highlights connections to the Majorana representation. It is obvious, that asymmetric density operators result in Husimi representations with small numeric values as non-permutation-symmetric components are neglected.

\subsection{Comparison of the BEADS and Bloch vector representations}\label{sec:BEADSBloch}
In Fig.~\ref{fig:FigDiscussionBloch}, we illustrate the equivalence of Q-Beads and Bloch vectors. Here, it is obvious that the Bloch vector strictly points to the red pole of the Q-Bead. However, Q-Beads prove to be easily interpretable in terms of expectation values and measurement probabilities independent of the viewing perspective. Indeed, this is not the case for Bloch vectors which are virtually indistinguishable when the view direction is parallel or anti-parallel to the Bloch vector. More importantly, in order to correctly determine the components of a Bloch vector in a straightforward fashion, which is required for determining single-qubit expectation values, the view direction must be orthogonal to the Bloch vector components of interest. In the BEADS representation, this can be achieved independent of the viewing perspective.

\begin{figure}[H]
\centering
\includegraphics[width=1.\textwidth]{FigDiscussionBloch.pdf}
\caption{\label{fig:FigDiscussionBloch}Comparison of Q-Beads and the Bloch vector. Q-Beads and Bloch vectors \cite{Bloch, FeynmanBloch} are visualized for different single-qubit states in oblique and top views. The Bloch vector is equivalent to the Q-Bead and exactly points to the red pole of the corresponding Q-Bead which can be readily seen by overlapping both representations. However, the direction of a Bloch vector is not unambiguously identifiable when the viewing direction co-aligns with the Bloch vector. This is shown for the states $\ket{0}$ and $\ket{1}$ where the Bloch vectors are parallel or anti-parallel to the z-axis, respectively. In top view, i.e., when looking in the negative z-direction, the Bloch vectors corresponding to both states are virtually indistinguishable and in particular, in case of mixed states it is impossible to determine the z-component of the Bloch vector whereas the Q-Bead colors ensure an unambiguous assignment. Moreover, Bloch vector components can only be determined if the view direction is orthogonal to the components of interest. For instance, predicting the expectation values $\left\langle\sigma_z\right\rangle$ requires to correctly determine the z-component of the Bloch vector which is difficult, while in the BEADS representation $\left\langle\sigma_z\right\rangle$ is represented by the Q-Bead colors at the north pole which can be interpreted with relative ease.}
\end{figure}

\subsection{Comparison of the BEADS and DROPS representations}\label{sec:BEADSDROPS}
As the BEADS representation is a derivative of the DROPS representation \cite{DROPS} and both visualizations may appear the same at first glance, we provide a detailed comparison in Fig.~\ref{fig:FigDiscussionDROPS} which demonstrates important differences between the representations. 
\begin{figure}[H]
\centering
\vspace{-1cm}
\includegraphics[width=1.\textwidth]{FigDiscussionDROPS.pdf}
\caption{\label{fig:FigDiscussionDROPS}Comparison between BEADS and DROPS. For a simple and direct visual comparison of shapes and surface patterns, axially symmetric spherical BEADS and DROPS \cite{DROPS} representations of $g$-linear scaled Pauli-Z operators are shown which reach a maximum function value of 1. The spherical functions are plotted with absolute function values as distance from the origin (first row) and standard (spherical, second to fourth row) polar plots. Note the differences in bilinear and trilinear shapes highlighted by zoomed views. In case of BEADS, spherical function values along the z-axis directly correspond to expectation values $\left\langle\sigma_{kz}\right\rangle$, $\left\langle\sigma_{kz}\sigma_{lz}\right\rangle$ (where $k,l \in \{1,2,3\}$), and $\left\langle\sigma_{1z}\sigma_{2z}\sigma_{3z}\right\rangle$ obtained for the state $\ket{000}$. However, the DROPS representation of $\ket{000}$ corresponds to the extremely small droplets plotted in the first row and the remaining DROPS spherical functions are scaled for a direct comparison with the BEADS representation. The bottom plots display the underlying (unscaled) spherical functions of Pauli-Z operators which reduce to simple powers of cosine functions in the BEADS picture. The beads function values directly correspond to Pauli-Z expectation values which are all zero for $\theta = \pi/2$. In the DROPS picture the spherical functions of Pauli-Z operators are represented by more complicated linear combinations of cosine functions.}
\vspace{-1cm}
\end{figure}
\noindent
DROPS, which is defined such that the integral over the spherical functions is normalized, can be described as a generalized Wigner function. While the DROPS representation can reveal symmetry properties of a visualized operator or reveal basic rotations, compared to BEADS, it does not provide direct quantitative information on entanglement-based correlation functions, or in general, any correlation function and hence it is significantly less useful for understanding gates or algorithms in quantum information.

\section{QuBeads -- a dynamical simulation software using the BEADS representation}\label{app:BEADSQuBeads}
We developed a powerful interactive software named \textit{QuBeads application}, \textit{QuBeads app}, or simply \textit{QuBeads} which uses the BEADS representation to simulate and visualize the dynamics of qubit systems in quantum circuits. QuBeads is a modified descendant of the \textit{SpinDrops} app \cite{SpinDrops} which was developed to offer DROPS visualizations in the context of nuclear magnetic resonance. Currently, the QuBeads software is able to support visualizations of system of up to three qubits and will be extended to feature larger qubit systems in the near future. The software is highly responsive, offers all variants of the BEADS representation discussed in this chapter, and features a palette of the most common quantum gates and supports the setup of custom quantum circuits following a user-friendly intuitive modular approach that does not require any coding. Indeed, all BEADS visualizations in this chapter were generated using the QuBeads software. 
QuBeads can be downloaded and installed on all common operating systems. The software (beta version) and detailed installation guides are available via the following link:
\newline

\noindent
\href{https://github.com/denhub97/QuBeads}{https://github.com/denhub97/QuBeads}\newline

\noindent
In QuBeads, we offer two main visualization modes: the standard BEADS mode shown in Fig.~\ref{fig:FigQuBeads1} which provides a visualization of the current system state at any point during the simulation and the so-called BEADS-augmented circuit mode displayed in Fig.~\ref{fig:FigQuBeads2} which provides a visualization of the chosen quantum circuit and the BEADS representations of the system states after each operation (i.e., gates or measurements).

\begin{figure}[H]
\centering
\includegraphics[width=1.0\textwidth]{FigQuBeads1.pdf}
\caption{\label{fig:FigQuBeads1}Standard BEADS mode in the \textit{QuBeads} app. A custom quantum circuit can be simulated dynamically. The system state is monitored by a corresponding BEADS representation at runtime. In QuBeads, quantum circuits can be configured by an intuitive modular setup which uses a graphical interface (see smaller window) and does not require any coding.}
\end{figure}

\begin{figure}[H]
\centering
\includegraphics[width=1.0\textwidth]{FigQuBeads2.pdf}
\caption{\label{fig:FigQuBeads2}BEADS-augmented circuit mode in the \textit{QuBeads} app. A custom quantum circuit and beads corresponding to the system states after each individual operation are visualized. Possible measurement outcomes are simulated and optionally shown simultaneously. }
\end{figure}




\end{appendices}

\newpage
\bibliography{BEADS}

\clearpage
\setcounter{page}{0}
\begin{titlepage}
\begin{center}
\Large BEADS: A canonical visualization of quantum states for applications in quantum information processing \\[.5cm] \large Supplementary information \\[1cm]

Dennis Huber$^{1,2}$ and Steffen Glaser$^{1,2^*}$\\[.5cm]

$^1$TUM School of Natural Sciences, Technical University of Munich, Lichtenbergstrasse 4, Garching, 85748, Bavaria, Germany.\\
$^2$Munich Center for Quantum Science and Technology (MCQST), M\"unchen, 80799, Bavaria, Munich.\\[1cm]
$^*$Corresponding author(s) E-mail(s): \textcolor{blue}{glaser@tum.de}\\
Contributing author(s): \textcolor{blue}{dennis.huber@tum.de}
\end{center}
\end{titlepage}
\normalsize

\section*{Table of Contents}
\startcontents
\printcontents{ }{1}{}
  
\beginsupplement
    
\newpage
\section{Theory}\label{Supp:Theory}
\subsection{Hermitian LISA tensor operators}\label{sec:BEADSLISASupp}
Hermitian LISA operators for (sub)systems consisting of up to $n = 3$ qubits are defined as follows. Operators corresponding to different sets $\ell^\prime$ are separated by double lines whereas operators which correspond to the same set $\ell^\prime$ but have different rank $j$ are separated by single lines.

\begin{flalign}
\hline\hline\nonumber\\[-8pt]
&T_{0,0}^{\{\emptyset\}}=\frac{1}{\sqrt{2^N}}\mathbb{I} &&\\[8pt]
\hline\hline\nonumber
\end{flalign}
\begin{subequations}
\vspace{-1.2cm}
\begin{flalign}
&T_{1,-1}^{\left\{k\right\}}=\frac{1}{\sqrt{2^N}}\sigma_{ky}&&\\
&T_{1,0}^{\left\{k\right\}}=\frac{1}{\sqrt{2^N}}\sigma_{kz}&&\\
&T_{1,1}^{\left\{k\right\}}=\frac{1}{\sqrt{2^N}}\sigma_{kx}&&\\[10pt]
\hline\hline\nonumber
\end{flalign}
\end{subequations}
\vspace{-1.2cm}
\begin{flalign}
&T_{0,0}^{\left\{k,l\right\}_{\textit{even}}}=\frac{1}{\sqrt{2^N}}\frac{1}{\sqrt3}\left(\sigma_{kx}\sigma_{lx}+\sigma_{ky}\sigma_{ly}+\sigma_{kz}\sigma_{lz}\right)&&\\[10pt]
\hline\nonumber
\end{flalign}
\begin{subequations}
\vspace{-1.2cm}
\begin{flalign}
&T_{2,-2}^{\left\{k,l\right\}_{\textit{even}}}=\frac{1}{\sqrt{2^N}}\frac{1}{\sqrt2}\left(\sigma_{kx}\sigma_{ly}+\sigma_{ky}\sigma_{lx}\right)&&\\
&T_{2,-1}^{\left\{k,l\right\}_{\textit{even}}}=\frac{1}{\sqrt{2^N}}\frac{1}{\sqrt2}\left(\sigma_{ky}\sigma_{lz}+\sigma_{kz}\sigma_{ly}\right)&&\\
&T_{2,0}^{\left\{k,l\right\}_{\textit{even}}}=\frac{1}{\sqrt{2^N}}\frac{1}{\sqrt6}\left(2\sigma_{kz}\sigma_{lz}-\sigma_{kx}\sigma_{lx}-\sigma_{ky}\sigma_{ly}\right)&&\\
&T_{2,1}^{\left\{k,l\right\}_{\textit{even}}}=\frac{1}{\sqrt{2^N}}\frac{1}{\sqrt2}\left(\sigma_{kz}\sigma_{lx}+\sigma_{kx}\sigma_{lz}\right)&&\\
&T_{2,2}^{\left\{k,l\right\}_{\textit{even}}}=\frac{1}{\sqrt{2^N}}\frac{1}{\sqrt2}\left(\sigma_{kx}\sigma_{lx}-\sigma_{ky}\sigma_{ly}\right)\\[10pt]
\hline\hline\nonumber
\end{flalign}
\end{subequations}
\begin{subequations}
\vspace{-1.4cm}
\begin{flalign}
&T_{1,-1}^{\left\{k,l\right\}_{\textit{odd}}}=\frac{1}{\sqrt{2^N}}\frac{1}{\sqrt2}\left(\sigma_{kz}\sigma_{lx}-\sigma_{kx}\sigma_{lz}\right)&&\\
&T_{1,0}^{\left\{k,l\right\}_{\textit{odd}}}=\frac{1}{\sqrt{2^N}}\frac{1}{\sqrt2}\left(\sigma_{kx}\sigma_{ly}-\sigma_{ky}\sigma_{lx}\right)&&\\
&T_{1,1}^{\left\{k,l\right\}_{\textit{odd}}}=\frac{1}{\sqrt{2^N}}\frac{1}{\sqrt2}\left(\sigma_{ky}\sigma_{lz}-\sigma_{kz}\sigma_{ly}\right)&&\\[10pt]
\hline\hline\nonumber
\end{flalign}
\end{subequations}

\vspace{-.5cm}
\begin{subequations}
\begin{flalign}
\hline\hline\nonumber\\[-8pt]
T_{1,-1}^{\left\{k,l,m\ \tau_1\right\}_{\textit{odd}}}=\frac{1}{\sqrt{2^N}}\frac{1}{\sqrt{15}}\bigl[&3\sigma_{ky}\sigma_{ly}\sigma_{my}+\left(\sigma_{ky}\sigma_{lx}\sigma_{mx}+\sigma_{kx}\sigma_{ly}\sigma_{mx}+\sigma_{kx}\sigma_{lx}\sigma_{my}\right)\nonumber&&\\
+&\left(\sigma_{ky}\sigma_{lz}\sigma_{mz}+\sigma_{kz}\sigma_{ly}\sigma_{mz}+\sigma_{kz}\sigma_{lz}\sigma_{my}\right)\bigr]&&\\
T_{1,0}^{\left\{k,l,m\ \tau_1\right\}_{\textit{odd}}}=\frac{1}{\sqrt{2^N}}\frac{1}{\sqrt{15}}\bigl[&3\sigma_{kz}\sigma_{lz}\sigma_{mz}+\left(\sigma_{kz}\sigma_{lx}\sigma_{mx}+\sigma_{kx}\sigma_{lz}\sigma_{mx}+\sigma_{kx}\sigma_{lx}\sigma_{mz}\right)\nonumber&&\\
+&\left(\sigma_{kz}\sigma_{ly}\sigma_{my}+\sigma_{ky}\sigma_{lz}\sigma_{my}+\sigma_{ky}\sigma_{ly}\sigma_{mz}\right)\bigr]&&\\
T_{1,1}^{\left\{k,l,m\ \tau_1\right\}_{\textit{odd}}}=\frac{1}{\sqrt{2^N}}\frac{1}{\sqrt{15}}\bigl[&3\sigma_{kx}\sigma_{lx}\sigma_{mx}+\left(\sigma_{kx}\sigma_{ly}\sigma_{my}+\sigma_{ky}\sigma_{lx}\sigma_{my}+\sigma_{ky}\sigma_{ly}\sigma_{mx}\right)\nonumber&&\\
+&\left(\sigma_{kx}\sigma_{lz}\sigma_{mz}+\sigma_{kz}\sigma_{lx}\sigma_{mz}+\sigma_{kz}\sigma_{lz}\sigma_{mx}\right)\bigr]&&\\[10pt]
\hline\nonumber
\end{flalign}
\end{subequations}

\begin{subequations}
\vspace{-1.2cm}
\begin{flalign}
&T_{3,-3}^{\left\{k,l,m\ \tau_1\right\}_{\textit{odd}}}=\frac{1}{\sqrt{2^N}}\frac{1}{2}\left[-\sigma_{ky}\sigma_{ly}\sigma_{my}+\left(\sigma_{ky}\sigma_{lx}\sigma_{mx}+\sigma_{kx}\sigma_{ly}\sigma_{mx}+\sigma_{kx}\sigma_{lx}\sigma_{my}\right)\right]&&
\end{flalign}
\vspace{-.8cm}
\begin{flalign}
T_{3,-2}^{\left\{k,l,m\ \tau_1\right\}_{\textit{odd}}}=\frac{1}{\sqrt{2^N}}\frac{1}{\sqrt6}\bigl[&\left(\sigma_{kx}\sigma_{ly}\sigma_{mz}+\sigma_{ky}\sigma_{lz}\sigma_{mx}+\sigma_{kz}\sigma_{lx}\sigma_{my}\right)\nonumber&&\\
+&\left(\sigma_{kx}\sigma_{lz}\sigma_{my}+\sigma_{kz}\sigma_{ly}\sigma_{mx}+\sigma_{ky}\sigma_{lx}\sigma_{mz}\right)\bigr]&&
\end{flalign}
\vspace{-.8cm}
\begin{flalign}
T_{3,-1}^{\left\{k,l,m\ \tau_1\right\}_{\textit{odd}}}=\frac{1}{\sqrt{2^N}}\frac{1}{2\sqrt{15}}&\bigl[-3\sigma_{ky}\sigma_{ly}\sigma_{my}-\left(\sigma_{ky}\sigma_{lx}\sigma_{mx}+\sigma_{kx}\sigma_{ly}\sigma_{mx}+\sigma_{kx}\sigma_{lx}\sigma_{my}\right)\nonumber&&\\
&+4\left(\sigma_{ky}\sigma_{lz}\sigma_{mz}+\sigma_{kz}\sigma_{ly}\sigma_{mz}+\sigma_{kz}\sigma_{lz}\sigma_{my}\right)\bigr]&&
\end{flalign}
\vspace{-.8cm}
\begin{flalign}
T_{3,0}^{\left\{k,l,m\ \tau_1\right\}_{\textit{odd}}}=\frac{1}{\sqrt{2^N}}\frac{1}{\sqrt{10}}\bigl[&2\sigma_{kz}\sigma_{lz}\sigma_{mz}-\left(\sigma_{kz}\sigma_{lx}\sigma_{mx}+\sigma_{kx}\sigma_{lz}\sigma_{mx}+\sigma_{kx}\sigma_{lx}\sigma_{mz}\right)\nonumber&&\\
+&\left(\sigma_{kz}\sigma_{ly}\sigma_{my}+\sigma_{ky}\sigma_{lz}\sigma_{my}+\sigma_{ky}\sigma_{ly}\sigma_{mz}\right)\bigr]&&
\end{flalign}
\vspace{-.8cm}
\begin{flalign}
T_{3,1}^{\left\{k,l,m\ \tau_1\right\}_{\textit{odd}}}=\frac{1}{\sqrt{2^N}}\frac{1}{2\sqrt{15}}\bigl[&{-3\sigma_{kx}\sigma_{lx}\sigma_{mx}}-\left(\sigma_{kx}\sigma_{ly}\sigma_{my}+\sigma_{ky}\sigma_{lx}\sigma_{my}+\sigma_{ky}\sigma_{ly}\sigma_{mx}\right)\nonumber&&\\
&+4\left(\sigma_{kx}\sigma_{lz}\sigma_{mz}+\sigma_{kz}\sigma_{lx}\sigma_{mz}+\sigma_{kz}\sigma_{lz}\sigma_{mx}\right)\bigr]&&
\end{flalign}
\vspace{-.8cm}
\begin{flalign}
T_{3,2}^{\left\{k,l,m\ \tau_1\right\}_{\textit{odd}}}=\frac{1}{\sqrt{2^N}}\frac{1}{\sqrt6}\bigr[&\left(\sigma_{kz}\sigma_{lx}\sigma_{mx}+\sigma_{kx}\sigma_{lz}\sigma_{mx}+\sigma_{kx}\sigma_{lx}\sigma_{mz}\right)\nonumber&&\\
-&\left(\sigma_{kz}\sigma_{ly}\sigma_{my}+\sigma_{ky}\sigma_{lz}\sigma_{my}+\sigma_{ky}\sigma_{ly}\sigma_{mz}\right)\bigr]&&
\end{flalign}
\vspace{-.8cm}
\begin{flalign}
&T_{3,3}^{\left\{k,l,m\ \tau_1\right\}_{\textit{odd}}}=\frac{1}{\sqrt{2^N}}\frac{1}{2}\left[\sigma_{kx}\sigma_{lx}\sigma_{mx}-\left(\sigma_{kx}\sigma_{ly}\sigma_{my}+\sigma_{ky}\sigma_{lx}\sigma_{my}+\sigma_{ky}\sigma_{ly}\sigma_{mx}\right)\right]&&\\[10pt]
\hline\hline\nonumber
\end{flalign}
\end{subequations}

\begin{subequations}
\vspace{-1.2cm}
\begin{flalign}
T_{1,-1}^{\left\{k,l,m\ \tau_2\right\}_{\textit{odd}}}=\frac{1}{\sqrt{2^N}}\frac{1}{\sqrt{12}}\bigl[&\left(-2\sigma_{kx}\sigma_{lx}\sigma_{my}+\sigma_{ky}\sigma_{lx}\sigma_{mx}+\sigma_{kx}\sigma_{ly}\sigma_{mx}\right)\nonumber&&\\
+&\left(-2\sigma_{kz}\sigma_{lz}\sigma_{my}+\sigma_{ky}\sigma_{lz}\sigma_{mz}+\sigma_{kz}\sigma_{ly}\sigma_{mz}\right)\bigr]&&\\
T_{1,0}^{\left\{k,l,m\ \tau_2\right\}_{\textit{odd}}}=\frac{1}{\sqrt{2^N}}\frac{1}{\sqrt{12}}\bigl[&\left(-2\sigma_{kx}\sigma_{lx}\sigma_{mz}+\sigma_{kz}\sigma_{lx}\sigma_{mx}+\sigma_{kx}\sigma_{lz}\sigma_{mx}\right)\nonumber&&\\
+&\left({-2\sigma}_{ky}\sigma_{ly}\sigma_{mz}+\sigma_{kz}\sigma_{ly}\sigma_{my}+\sigma_{ky}\sigma_{lz}\sigma_{my}\right)\bigr]&&\\
T_{1,1}^{\left\{k,l,m\ \tau_2\right\}_{\textit{odd}}}=\frac{1}{\sqrt{2^N}}\frac{1}{\sqrt{12}}\bigl[&\left(-2\sigma_{ky}\sigma_{ly}\sigma_{mx}+\sigma_{kx}\sigma_{ly}\sigma_{my}+\sigma_{ky}\sigma_{lx}\sigma_{my}\right)\nonumber&&\\
+&\left({-2\sigma}_{kz}\sigma_{lz}\sigma_{mx}+\sigma_{kx}\sigma_{lz}\sigma_{mz}+\sigma_{kz}\sigma_{lx}\sigma_{mz}\right)\bigr]&&
\end{flalign}
\end{subequations}

\begin{subequations}
\vspace{-.6cm}
\begin{flalign}
\hline\hline\nonumber\\[-8pt]
T_{2,-2}^{\left\{k,l,m\ \tau_2\right\}_{\textit{even}}}=\frac{1}{\sqrt{2^N}}\frac{1}{\sqrt{12}}\bigl[&\left(2\sigma_{kx}\sigma_{lx}\sigma_{mz}-\sigma_{kz}\sigma_{lx}\sigma_{mx}-\sigma_{kx}\sigma_{lz}\sigma_{mx}\right)\nonumber&&\\
+&\left({-2\sigma}_{ky}\sigma_{ly}\sigma_{mz}+\sigma_{kz}\sigma_{ly}\sigma_{my}+\sigma_{ky}\sigma_{lz}\sigma_{my}\right)\bigr]&&\\
T_{2,-1}^{\left\{k,l,m\ \tau_2\right\}_{\textit{even}}}=\frac{1}{\sqrt{2^N}}\frac{1}{\sqrt{12}}\bigl[&\left(2\sigma_{ky}\sigma_{ly}\sigma_{mx}-\sigma_{kx}\sigma_{ly}\sigma_{my}-\sigma_{ky}\sigma_{lx}\sigma_{my}\right)\nonumber&&\\
+&\left({-2\sigma}_{kz}\sigma_{lz}\sigma_{mx}+\sigma_{kx}\sigma_{lz}\sigma_{mz}+\sigma_{kz}\sigma_{lx}\sigma_{mz}\right)\bigr]&&
\end{flalign}
\vspace{-.9cm}
\begin{flalign}
&T_{2,0}^{\left\{k,l,m\ \tau_2\right\}_{\textit{even}}}=\frac{1}{\sqrt{2^N}}\frac{1}{2}\bigl[-\sigma_{kx}\sigma_{lz}\sigma_{my}-\sigma_{kz}\sigma_{lx}\sigma_{my}+\sigma_{ky}\sigma_{lz}\sigma_{mx}+\sigma_{kz}\sigma_{ly}\sigma_{mx}\bigr]&&
\end{flalign}
\vspace{-.8cm}
\begin{flalign}
T_{2,1}^{\left\{k,l,m\ \tau_2\right\}_{\textit{even}}}=\frac{1}{\sqrt{2^N}}\frac{1}{\sqrt{12}}\bigl[&\left(-2\sigma_{kx}\sigma_{lx}\sigma_{my}+\sigma_{ky}\sigma_{lx}\sigma_{mx}+\sigma_{kx}\sigma_{ly}\sigma_{mx}\right)\nonumber&&\\
+&\left(2\sigma_{kz}\sigma_{lz}\sigma_{my}-\sigma_{ky}\sigma_{lz}\sigma_{mz}-\sigma_{kz}\sigma_{ly}\sigma_{mz}\right)\bigr]&&\\
T_{2,2}^{\left\{k,l,m\ \tau_2\right\}_{\textit{even}}}=\frac{1}{\sqrt{2^N}}\frac{1}{\sqrt{12}}\bigl[&\left({-2\sigma}_{kx}\sigma_{ly}\sigma_{mz}+\sigma_{ky}\sigma_{lz}\sigma_{mx}+\sigma_{kz}\sigma_{lx}\sigma_{my}\right)\nonumber&&\\
+&\left(-{2\sigma}_{ky}\sigma_{lx}\sigma_{mz}+\sigma_{kx}\sigma_{lz}\sigma_{my}+\sigma_{kz}\sigma_{ly}\sigma_{mx}\right)\bigr]&&\\[10pt]
\hline\hline\nonumber
\end{flalign}
\end{subequations}

\begin{subequations}
\vspace{-1.2cm}
\begin{flalign}
T_{1,-1}^{\left\{k,l,m\ \tau_3\right\}_{\textit{odd}}}=\frac{1}{\sqrt{2^N}}\frac{1}{2}\left[\left(\sigma_{ky}\sigma_{lx}\sigma_{mx}-\sigma_{kx}\sigma_{ly}\sigma_{mx}\right)+\left(\sigma_{ky}\sigma_{lz}\sigma_{mz}-\sigma_{kz}\sigma_{ly}\sigma_{mz}\right)\right]&&\\
T_{1,0}^{\left\{k,l,m\ \tau_3\right\}_{\textit{odd}}}=\frac{1}{\sqrt{2^N}}\frac{1}{2}\left[\left(\sigma_{kz}\sigma_{lx}\sigma_{mx}-\sigma_{kx}\sigma_{lz}\sigma_{mx}\right)+\left(\sigma_{kz}\sigma_{ly}\sigma_{my}-\sigma_{ky}\sigma_{lz}\sigma_{my}\right)\right]&&\\
T_{1,1}^{\left\{k,l,m\ \tau_3\right\}_{\textit{odd}}}=\frac{1}{\sqrt{2^N}}\frac{1}{2}\left[\left(\sigma_{kx}\sigma_{ly}\sigma_{my}-\sigma_{ky}\sigma_{lx}\sigma_{my}\right)+\left(\sigma_{kx}\sigma_{lz}\sigma_{mz}-\sigma_{kz}\sigma_{lx}\sigma_{mz}\right)\right]&&\\[10pt]
\hline\hline\nonumber
\end{flalign}
\end{subequations}

\begin{subequations}
\vspace{-1.2cm}
\begin{flalign}
&T_{2,-2}^{\left\{k,l,m\ \tau_3\right\}_{\textit{even}}}=\frac{1}{\sqrt{2^N}}\frac{1}{2}\left[\left(-\sigma_{kz}\sigma_{lx}\sigma_{mx}+\sigma_{kx}\sigma_{lz}\sigma_{mx}\right)+\left(\sigma_{kz}\sigma_{ly}\sigma_{my}-\sigma_{ky}\sigma_{lz}\sigma_{my}\right)\right]&&\\
&T_{2,-1}^{\left\{k,l,m\ \tau_3\right\}_{\textit{even}}}=\frac{1}{\sqrt{2^N}}\frac{1}{2}\left[\left({-\sigma}_{kx}\sigma_{ly}\sigma_{my}+\sigma_{ky}\sigma_{lx}\sigma_{my}\right)+\left(\sigma_{kx}\sigma_{lz}\sigma_{mz}-\sigma_{kz}\sigma_{lx}\sigma_{mz}\right)\right]&&
\end{flalign}
\vspace{-.8cm}
\begin{flalign}
T_{2,0}^{\left\{k,l,m\ \tau_3\right\}_{\textit{even}}}=\frac{1}{\sqrt{2^N}}\frac{1}{\sqrt{12}}\bigl[&\left({-2\sigma}_{kx}\sigma_{ly}\sigma_{mz}+\sigma_{ky}\sigma_{lz}\sigma_{mx}+\sigma_{kz}\sigma_{lx}\sigma_{my}\right)\nonumber&&\\
+&\left({2\sigma}_{ky}\sigma_{lx}\sigma_{mz}-\sigma_{kx}\sigma_{lz}\sigma_{my}-\sigma_{kz}\sigma_{ly}\sigma_{mx}\right)\bigr]&&
\end{flalign}
\vspace{-.9cm}
\begin{flalign}
&T_{2,1}^{\left\{k,l,m\ \tau_3\right\}_{\textit{even}}}=\frac{1}{\sqrt{2^N}}\frac{1}{2}\left[\left(-\sigma_{ky}\sigma_{lz}\sigma_{mz}+\sigma_{kz}\sigma_{ly}\sigma_{mz}\right)+\left(\sigma_{ky}\sigma_{lx}\sigma_{mx}-\sigma_{kx}\sigma_{ly}\sigma_{mx}\right)\right]&&\\
&T_{2,2}^{\left\{k,l,m\ \tau_3\right\}_{\textit{even}}}=\frac{1}{\sqrt{2^N}}\frac{1}{2}\left[\left(-\sigma_{kx}\sigma_{lz}\sigma_{my}+\sigma_{kz}\sigma_{lx}\sigma_{my}\right)+\left(\sigma_{kz}\sigma_{ly}\sigma_{mx}-\sigma_{ky}\sigma_{lz}\sigma_{mx}\right)\right]&&\\[10pt]
\hline\hline\nonumber
\end{flalign}
\end{subequations}
\vspace{-1.2cm}
\begin{flalign}
T_{0,0}^{\left\{k,l,m\ \tau_4\right\}_{\textit{even}}}=\frac{1}{\sqrt{2^N}}\frac{1}{\sqrt6}\bigl[&\left(\sigma_{kx}\sigma_{ly}\sigma_{mz}+\sigma_{ky}\sigma_{lz}\sigma_{mx}+\sigma_{kz}\sigma_{lx}\sigma_{my}\right)\nonumber&&\\
-&\left(\sigma_{kx}\sigma_{lz}\sigma_{my}+\sigma_{ky}\sigma_{lx}\sigma_{mz}+\sigma_{kz}\sigma_{ly}\sigma_{mx}\right)\bigr]&&\\[10pt]
\hline\hline\nonumber
\end{flalign}

\subsection{Global unitary bound states of LISA tensor operators}\label{app:BEADSGUBStates}
\begin{figure}[H]
\centering
\includegraphics[width=1\textwidth]{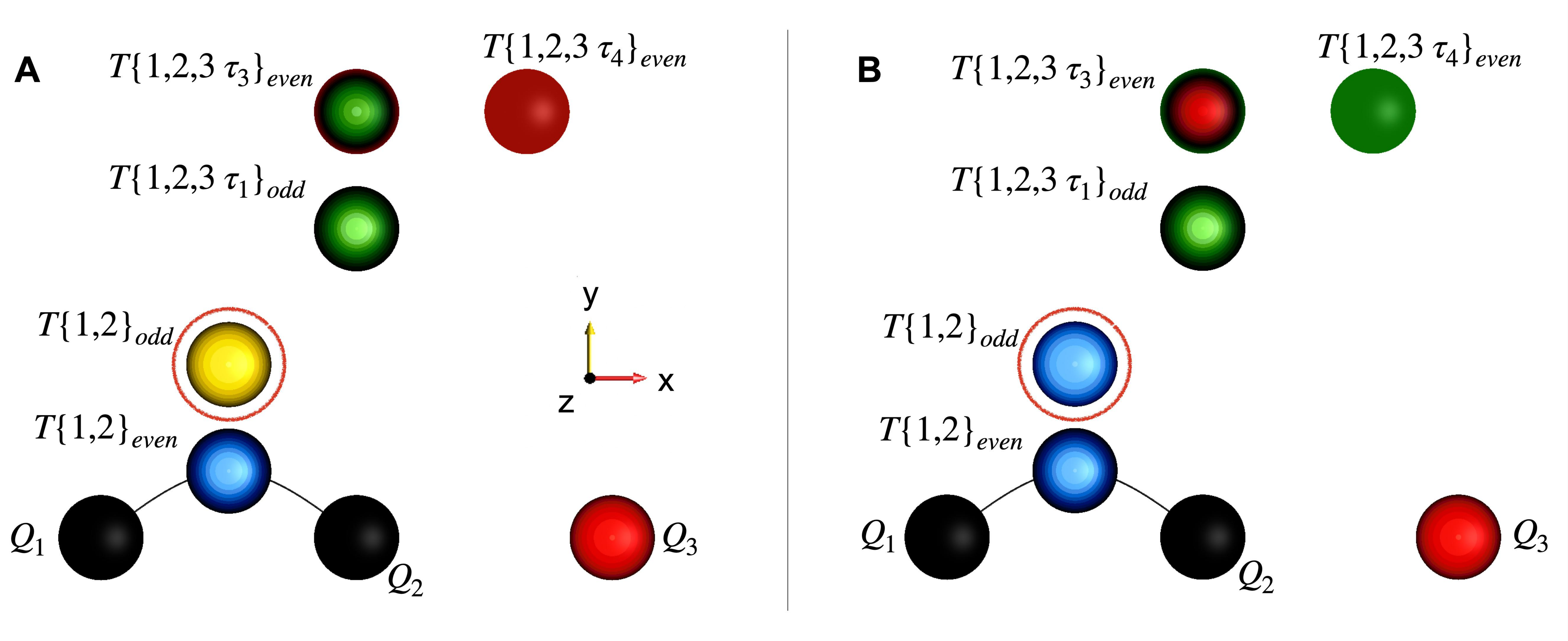}
\caption{\label{fig:Fig12odd}BEADS representation of (\textbf{A}) $\ket{\psi_+^{\left\{1,2\right\}_{\textit{odd}}}}$ and (\textbf{B}) $\ket{\psi_-^{\left\{1,2\right\}_{\textit{odd}}}}$ for $\theta=0$. The visualized states reach maximum values $\pm1$ for the bead $T\{1,2\}_{\textit{odd}}$ marked by red circles.}
\end{figure}

In the following, expressions and BEADS representations of three-qubit pure states corresponding to the global unitary bounds of the tensor operators listed in Table~\ref{tab:GUBSF} of section~\ref{sec:BEADSGUB} are presented. This includes states $\ket{\psi_+^{\left(\ell^\prime\right)}}$ which achieve maximally positive expectation values and states $\ket{\psi_-^{\left(\ell^\prime\right)}}$ with maximally negative expectation values along the z-axis with respect to the symmetry component represented by the bead $b^{\left(\ell^\prime\right)}$. These states are in general not maximally entangled for which the total correlation function BEADS representation is shown, respectively. Note that in cases where the global unitary bounds correspond to degenerate eigenvalues, any normalized vector that is a combination, expressed by an additional angular parameter $\theta$, of associated orthogonal eigenvectors, is also an eigenvector and a state representing the global unitary bound of the corresponding tensor operator. The three-qubit global unitary bound states corresponding to the LISA operators listed in table~\ref{tab:GUBSF} are:\newline

\noindent
\begin{subequations}\label{eq:GUBStart}
\vspace{-.3cm}
\begin{flalign}
\hline\hline &&\nonumber \\[-8pt]
&\ell^\prime=\left\{1,2\right\}_{\textit{odd}}: &&\nonumber\\
&\ket{\psi_+^{\left\{1,2\right\}_{\textit{odd}}}}=\cos{\theta}\left[\frac{1}{\sqrt2}\left(\ket{010}-i\ket{100}\right)\right]+\sin{\theta}\left[\frac{1}{\sqrt2}\left(\ket{011}-i\ket{101}\right)\right] &&\\
&\ket{\psi_-^{\left\{1,2\right\}_{\textit{odd}}}}=\cos{\theta}\left[\frac{1}{\sqrt2}\left(\ket{010}+i\ket{100}\right)\right]+\sin{\theta}\left[\frac{1}{\sqrt2}\left(\ket{011}+i\ket{101}\right)\right]&&\\[6pt] \hline\nonumber
\end{flalign}
\end{subequations}
\begin{subequations}
\vspace{-1.4cm}
\begin{flalign}
&\ell^\prime=\left\{1,3\right\}_{\textit{odd}}: &&\nonumber\\
&\ket{\psi_+^{\left\{1,3\right\}_{\textit{odd}}}}=\cos{\theta}\left[\frac{1}{\sqrt2}\left(\ket{001}-i\ket{100}\right)\right]+\sin{\theta}\left[\frac{1}{\sqrt2}\left(\ket{011}-i\ket{110}\right)\right] &&\\
&\ket{\psi_-^{\left\{1,3\right\}_{\textit{odd}}}}=\cos{\theta}\left[\frac{1}{\sqrt2}\left(\ket{001}+i\ket{100}\right)\right]+\sin{\theta}\left[\frac{1}{\sqrt2}\left(\ket{011}+i\ket{110}\right)\right]&&\\[6pt] \hline\nonumber
\end{flalign}
\end{subequations}
\begin{subequations}
\vspace{-1.4cm}
\begin{flalign}
&\ell^\prime=\left\{2,3\right\}_{\textit{odd}}: &&\nonumber\\
&\ket{\psi_+^{\left\{2,3\right\}_{\textit{odd}}}}=\cos{\theta}\left[\frac{1}{\sqrt2}\left(\ket{001}-i\ket{010}\right)\right]+\sin{\theta}\left[\frac{1}{\sqrt2}\left(\ket{101}-i\ket{110}\right)\right] &&\\
&\ket{\psi_-^{\left\{2,3\right\}_{\textit{odd}}}}=\cos{\theta}\left[\frac{1}{\sqrt2}\left(\ket{001}+i\ket{010}\right)\right]+\sin{\theta}\left[\frac{1}{\sqrt2}\left(\ket{101}+i\ket{110}\right)\right]&&\\[6pt] \hline\nonumber
\end{flalign}
\end{subequations}
\begin{subequations}
\vspace{-1.4cm}
\begin{flalign}
&\ell^\prime=\left\{1,2,3\ \tau_2\right\}_{\textit{odd}}: &&\nonumber\\
&\ket{\psi_+^{\left\{1,2,3\ \tau_2\right\}_{\textit{odd}}}}=\frac{1}{2}\sqrt{1+\frac{1}{\sqrt3}}\left(\ket{011}+\ket{101}\right)-\frac{1}{\sqrt2}\sqrt{1-\frac{1}{\sqrt3}}\ket{110}&&\\
&\ket{\psi_-^{\left\{1,2,3\ \tau_2\right\}_{\textit{odd}}}}=\frac{1}{2}\sqrt{1+\frac{1}{\sqrt3}}\left(\ket{010}+\ket{100}\right)-\frac{1}{\sqrt2}\sqrt{1-\frac{1}{\sqrt3}}\ket{001}&&\\[6pt] \hline\nonumber
\end{flalign}
\end{subequations}
\vspace{-1.4cm}
\begin{flalign}
\ell^\prime=\left\{1,2,3\ \tau_2\right\}_{\textit{even}}: &&\nonumber
\end{flalign}
\begin{subequations}
\vspace{-.8cm}
\begin{flalign}
\ket{\psi_+^{\left\{1,2,3\ \tau_2\right\}_{\textit{even}}}}=\cos{\theta}&\left[\frac{1}{2}\left(\ket{100}+\ket{010}\right)-\frac{i}{\sqrt2}\ket{001}\right]\nonumber&&\\
+\sin{\theta}&\left[\frac{1}{2}\left(\ket{011}+\ket{101}\right)-\frac{i}{\sqrt2}\ket{110}\right]&&\\
\ket{\psi_-^{\left\{1,2,3\ \tau_2\right\}_{\textit{even}}}}=\cos{\theta}&\left[\frac{1}{2}\left(\ket{100}+\ket{010}\right)+\frac{i}{\sqrt2}\ket{001}\right]\nonumber&&\\
+\sin{\theta}&\left[\frac{1}{2}\left(\ket{011}+\ket{101}\right)+\frac{i}{\sqrt2}\ket{110}\right]&&\\[6pt] \hline\nonumber
\end{flalign}
\end{subequations}
\vspace{-1.2cm}
\begin{flalign}
\ell^\prime=\left\{1,2,3\ \tau_3\right\}_{\textit{odd}}: &&\nonumber
\end{flalign}
\begin{subequations}
\vspace{-.8cm}
\begin{flalign}
\ket{\psi_+^{\left\{1,2,3\ \tau_3\right\}_{\textit{odd}}}}=\cos{\theta}&\left[\frac{1}{2}\left(\ket{010}-\ket{100}\right)+\frac{1}{\sqrt2}\ket{001}\right]\nonumber&&\\
+\sin{\theta}&\left[\frac{1}{2}\left(\ket{011}-\ket{101}\right)+\frac{1}{\sqrt2}\ket{110}\right]&&\\
\ket{\psi_-^{\left\{1,2,3\ \tau_3\right\}_{\textit{odd}}}}=\cos{\theta}&\left[\frac{1}{2}\left(\ket{100}-\ket{010}\right)+\frac{1}{\sqrt2}\ket{001}\right]\nonumber&&\\
+\sin{\theta}&\left[\frac{1}{2}\left(\ket{101}-\ket{011}\right)+\frac{1}{\sqrt2}\ket{110}\right]&&\\[6pt] \hline\nonumber
\end{flalign}
\end{subequations}
\vspace{-1.2cm}
\begin{flalign}
\ell^\prime=\left\{1,2,3\ \tau_3\right\}_{\textit{even}}: &&\nonumber
\end{flalign}
\begin{subequations}
\vspace{-.8cm}
\begin{flalign}
\ket{\psi_+^{\left\{1,2,3\ \tau_3\right\}_{\textit{even}}}}=\cos{\theta}&\left[\frac{1}{\sqrt6}\ket{001}+i\frac{1-\sqrt6}{2}\ket{010}+i\frac{1+\sqrt6}{2}\ket{100}\right]\nonumber&&\\
+\sin{\theta}&\left[\frac{1}{\sqrt6}\ket{110}+i\frac{1-\sqrt6}{2}\ket{101}+i\frac{1+\sqrt6}{2}\ket{011}\right]&&\\
\ket{\psi_-^{\left\{1,2,3\ \tau_3\right\}_{\textit{even}}}}=\cos{\theta}&\left[\frac{1}{\sqrt6}\ket{001}+i\frac{1+\sqrt6}{2}\ket{010}+i\frac{1-\sqrt6}{2}\ket{100}\right]\nonumber&&\\
+\sin{\theta}&\left[\frac{1}{\sqrt6}\ket{110}+i\frac{1+\sqrt6}{2}\ket{101}+i\frac{1-\sqrt6}{2}\ket{011}\right]&&\\[6pt] \hline\nonumber
\end{flalign}
\end{subequations}
\vspace{-1.4cm}
\begin{flalign}
\ell^\prime=\left\{1,2,3\ \tau_4\right\}_{\textit{even}}: &&\nonumber
\end{flalign}
\begin{subequations}\label{eq:GUBEnd}
\vspace{-1.1cm}
\begin{flalign}
\ket{\psi_+^{\left\{1,2,3\ \tau_4\right\}_{\textit{even}}}}=\cos{\theta}&\left[\frac{1}{\sqrt3}\left(\ket{001}-\frac{1+\sqrt3i}{2}\ket{010}-\frac{1-\sqrt3i}{2}\ket{100}\right)\right]\nonumber&&\\
+\sin{\theta}&\left[\frac{1}{\sqrt3}\left(\ket{110}-\frac{1+\sqrt3i}{2}\ket{101}-\frac{1-\sqrt3i}{2}\ket{011}\right)\right]&&\\
\ket{\psi_-^{\left\{1,2,3\ \tau_4\right\}_{\textit{even}}}}=\cos{\theta}&\left[\frac{1}{\sqrt3}\left(\ket{001}-\frac{1-\sqrt3i}{2}\ket{010}-\frac{1+\sqrt3i}{2}\ket{100}\right)\right]\nonumber&&\\
+\sin{\theta}&\left[\frac{1}{\sqrt3}\left(\ket{110}-\frac{1-\sqrt3i}{2}\ket{101}-\frac{1+\sqrt3i}{2}\ket{011}\right)\right]&&\\[6pt] \hline\hline\nonumber
\end{flalign}
\end{subequations}\\[-1.1cm]
For the simple case where $\theta = 0$, the density operators $\rho_\pm^{\left(\ell^\prime\right)}$ corresponding to the states $\ket{\psi_\pm^{\left(\ell^\prime\right)}}$ defined in Eq.~\ref{eq:GUBStart} to~\ref{eq:GUBEnd} read as follows. Here, the components which contribute to the bead of interest $b^{(\ell^\prime)}$ are marked in \textcolor{blue}{blue color}. The corresponding BEADS representations are shown in Fig.~\ref{fig:Fig12odd} to~\ref{fig:FigTau4}.
\begin{flalign}
\hline\hline\nonumber\\[-8pt]
\ell^\prime=\left\{1,2 \right\}_{\textit{odd}}: &&\nonumber\\[-30pt]\nonumber
\end{flalign}
\vspace{-.4cm}
\begin{flalign}
\rho^{\left\{1,2 \right\}_{\textit{odd}}}_{\pm,\,\theta=0} = \frac{1}{8} &\left( \mathbb{I} - \sigma_{3z}  \textcolor{blue}{\pm \sigma_{1x}\sigma_{2y} \mp  \sigma_{1y}\sigma_{2x}} - \sigma_{1z}\sigma_{2z} \mp \sigma_{1x}\sigma_{2y}\sigma_{3z}\right. &&\nonumber\\
&\pm \left.\sigma_{1y}\sigma_{2x}\sigma_{3z} + \sigma_{1z}\sigma_{2z}\sigma_{3z} \right) &&\\[6pt] \hline\nonumber\\[-30pt]\nonumber
\vspace{-.9cm}
\end{flalign}%
$\ell^\prime=\left\{1,3 \right\}_{\textit{odd}}:$
\begin{flalign}
\vspace{-.7cm}
\rho^{\left\{1,3 \right\}_{\textit{odd}}}_{\pm,\,\theta=0} =\frac{1}{8} &\left( \mathbb{I} - \sigma_{2z}  \textcolor{blue}{\pm \sigma_{1x}\sigma_{3y} \mp  \sigma_{1y}\sigma_{3x}} - \sigma_{1z}\sigma_{3z} \mp \sigma_{1x}\sigma_{2z}\sigma_{3y} \right.\nonumber\\
&\pm \left. \sigma_{1y}\sigma_{2z}\sigma_{3x} + \sigma_{1z}\sigma_{2z}\sigma_{3z} \right) &&\\[6pt] \hline\nonumber\\[-30pt]\nonumber
\vspace{-.8cm}
\end{flalign}
$\ell^\prime=\left\{2,3 \right\}_{\textit{odd}}:$
\begin{flalign}
\vspace{-.6cm}
\rho^{\left\{2,3 \right\}_{\textit{odd}}}_{\pm,\,\theta=0} =\frac{1}{8} &\left( \mathbb{I} - \sigma_{1z}  \textcolor{blue}{\pm \sigma_{2x}\sigma_{3y} \mp  \sigma_{2y}\sigma_{3x}} - \sigma_{2z}\sigma_{3z} \mp \sigma_{1z}\sigma_{2x}\sigma_{3y}\right.\nonumber\\
&\pm \left.\sigma_{1z}\sigma_{2y}\sigma_{3x} + \sigma_{1z}\sigma_{2z}\sigma_{3z} \right) &&\\[6pt] \hline\nonumber\\[-30pt]\nonumber
\end{flalign}
$\ell^\prime=\left\{1,2,3\ \tau_2\right\}_{\textit{odd}}:$
\begin{flalign}
\vspace{-2cm}
\rho^{\left\{1,2,3\ \tau_2\right\}_{\textit{odd}}}_{\pm,\,\theta=0} =\frac{1}{8} &\bigg[ \left(\mathbb{I} \pm \sigma_{1z}\sigma_{2z}\sigma_{3z}\right) \nonumber&&\\
&+ \frac{1}{3 + \sqrt3} \left(\mp \sigma_{1z} \mp \sigma_{2z} - \sigma_{1z}\sigma_{3z} - \sigma_{2z}\sigma_{3z} \right) \nonumber&&\\
&+\frac{1}{\sqrt3} \big(\mp \sigma_{3z} - \sigma_{1z}\sigma_{2z} - \sigma_{1x}\sigma_{3x} -\sigma_{1y}\sigma_{2y} - \sigma_{2x}\sigma_{3x} - \sigma_{2y}\sigma_{3y}\nonumber&&\\
&\phantom{\quad\ \ \frac{1}{\sqrt3}} \textcolor{blue}{\pm \sigma_{1x}\sigma_{2z}\sigma_{3x} \pm \sigma_{1y}\sigma_{2z}\sigma_{3y} \pm \sigma_{1z}\sigma_{2x}\sigma_{3x} \pm \sigma_{1z}\sigma_{2y}\sigma_{3y}}\big) \nonumber&&\\
&+\frac{3+\sqrt3}{6} \big(\sigma_{1x}\sigma_{2x} + \sigma_{1y}\sigma_{2y}  \textcolor{blue}{\mp \sigma_{1x}\sigma_{2x}\sigma_{3z} \pm \sigma_{1y}\sigma_{2y}\sigma_{3z}} \big)\bigg] &&\\[6pt] \hline\nonumber\\[-30pt]\nonumber
\vspace{-.8cm}
\end{flalign}
$\ell^\prime=\left\{1,2,3\ \tau_2\right\}_{\textit{even}}:$
\begin{flalign}
\vspace{-.6cm}
\rho^{\left\{1,2,3\ \tau_2\right\}_{\textit{even}}}_{\pm,\,\theta=0} =\frac{1}{8} &\bigg[ \left(\mathbb{I} - \sigma_{1z}\sigma_{2z}\sigma_{3z}\right)
 \nonumber&&\\
&+\frac{1}{2} \big( \sigma_{1z} + \sigma_{2z} + \sigma_{1x}\sigma_{2x}+\sigma_{1y}\sigma_{2y} - \sigma_{1z}\sigma_{3z} - \sigma_{2z}\sigma_{3z}\nonumber&&\\
&\phantom{+\frac{1}{2}} +  \sigma_{1x}\sigma_{2x}\sigma_{3z} +  \sigma_{1y}\sigma_{2y}\sigma_{3z}\big) \nonumber&&\\
&+\frac{1}{\sqrt2} \big( \mp \sigma_{1x}\sigma_{3y} \pm \sigma_{1y}\sigma_{3x} \mp \sigma_{2x}\sigma_{3y} \pm \sigma_{2y}\sigma_{3x}  \nonumber&&\\
&\phantom{+\frac{1}{\sqrt2}\big(\ }  \textcolor{blue}{ \mp  \sigma_{1x}\sigma_{2z}\sigma_{3y}\pm  \sigma_{1y}\sigma_{2z}\sigma_{3x} \mp  \sigma_{1z}\sigma_{2x}\sigma_{3y} \pm  \sigma_{1z}\sigma_{2y}\sigma_{3x}} \big)\bigg]&&\\[6pt] \hline\nonumber
\end{flalign}\\[-.7cm]
$\ell^\prime=\left\{1,2,3\ \tau_3\right\}_{\textit{odd}}:$
\begin{flalign}
\vspace{-.8cm}
\rho^{\left\{1,2,3\ \tau_3\right\}_{\textit{odd}}}_{\pm,\,\theta=0} =\frac{1}{8} &\bigg[ \left(\mathbb{I} - \sigma_{1z}\sigma_{2z}\sigma_{3z}\right) \nonumber&&\\
& +\frac{1}{2} \big( \sigma_{1z} + \sigma_{2z} - \sigma_{1x}\sigma_{2x} - \sigma_{1y}\sigma_{2y} - \sigma_{1z}\sigma_{3z} - \sigma_{2z}\sigma_{3z} \nonumber&&\\
& \phantom{+\frac{1}{2} }-\sigma_{1x}\sigma_{2x}\sigma_{3z} - \sigma_{1y}\sigma_{2y}\sigma_{3z}\big) \nonumber&&\\
& +\frac{1}{\sqrt2} \big( \mp \sigma_{1x}\sigma_{3x} \mp \sigma_{1y}\sigma_{3y} \pm \sigma_{2x}\sigma_{3x} \pm \sigma_{2y}\sigma_{3y} \nonumber&&\\
&\phantom{+\sqrt2 \big(\ \ \:}\textcolor{blue}{\mp  \sigma_{1x}\sigma_{2z}\sigma_{3x} \mp  \sigma_{1y}\sigma_{2z}\sigma_{3y} \pm  \sigma_{1z}\sigma_{2x}\sigma_{3x} \pm  \sigma_{1z}\sigma_{2y}\sigma_{3y}} \big)\bigg]&&\\[6pt] \hline\nonumber \\[-30pt]\nonumber
\end{flalign}
$\ell^\prime=\left\{1,2,3\ \tau_3\right\}_{\textit{even}}:$
\begin{flalign}
\vspace{-.6cm}
\rho^{\left\{1,2,3\ \tau_3\right\}_{\textit{even}}}_{\pm,\,\theta=0} =\frac{1}{8} &\bigg[ 
\left(\mathbb{I} - \sigma_{1z}\sigma_{2z}\sigma_{3z}\right)\nonumber&&\\
&+\frac{1}{6} \big(
\sigma_{1z} 
+\sigma_{2z} 
- \sigma_{1x}\sigma_{2x} 
- \sigma_{1y}\sigma_{2y} 
- \sigma_{1z}\sigma_{3z} 
- \sigma_{2z}\sigma_{3z} \nonumber&&\\
&\phantom{+1}- \sigma_{1x}\sigma_{2x}\sigma_{3z}
- \sigma_{1y}\sigma_{2y}\sigma_{3z}
\big)\nonumber&&\\
&+\frac{1}{3} \big(
\sigma_{1x} \sigma_{3x}
+\sigma_{1y} \sigma_{3y}
+ \sigma_{2x} \sigma_{3x}
+ \sigma_{2y} \sigma_{3y}\nonumber&&\\
&\phantom{+1}+ \sigma_{1x}\sigma_{2z}\sigma_{3x}
+ \sigma_{1y}\sigma_{2z}\sigma_{3y}
+ \sigma_{1z}\sigma_{2x}\sigma_{3x}
+ \sigma_{1z}\sigma_{2y}\sigma_{3y}
\big)\nonumber&&\\
&+\frac{2}{3} \big(
\sigma_{3z} 
-\sigma_{1z} \sigma_{2z}
\big)\nonumber&&\\
&+\frac{1}{\sqrt6} \big(
\sigma_{1x} \sigma_{3y}
-\sigma_{1y} \sigma_{3x}
- \sigma_{2x} \sigma_{3y}
+\sigma_{1y} \sigma_{3x}\nonumber&&\\
&\phantom{+\sqrt6\ }\textcolor{blue}{+\sigma_{1x}\sigma_{2z}\sigma_{3y}
- \sigma_{1y}\sigma_{2z}\sigma_{3x}
- \sigma_{1z}\sigma_{2x}\sigma_{3y}
+ \sigma_{1z}\sigma_{2y}\sigma_{3x}}
\big)\nonumber&&\\
&+\frac{2}{\sqrt6} \big(
\sigma_{1x} \sigma_{2y}
-\sigma_{1y} \sigma_{2x}
\textcolor{blue}{+\sigma_{1x}\sigma_{2y}\sigma_{3z}
-\sigma_{1y}\sigma_{2x}\sigma_{3z}}
\big)\bigg]&&\\[6pt] \hline\nonumber \\[-30pt]\nonumber
\end{flalign}
$\ell^\prime=\left\{1,2,3\ \tau_4\right\}_{\textit{even}}:$
\begin{flalign}
\vspace{-.6cm}
\rho^{\left\{1,2,3\ \tau_4\right\}_{\textit{even}}}_{\pm,\,\theta=0} =\frac{1}{8} &\bigg[ 
\left(\mathbb{I} - \sigma_{1z}\sigma_{2z}\sigma_{3z}\right)\nonumber&&\\
&+\frac{1}{3} \big( 
\sigma_{1z} 
+\sigma_{2z} 
+\sigma_{3z}
- \sigma_{1x}\sigma_{2x} 
- \sigma_{1y}\sigma_{2y} 
- \sigma_{1z}\sigma_{2z}\nonumber&&\\
&\quad\:-\sigma_{1x} \sigma_{3x}
-\sigma_{1y} \sigma_{3y}
-\sigma_{1z} \sigma_{3z}
-\sigma_{2x} \sigma_{3x}
-\sigma_{2y} \sigma_{3y}
-\sigma_{2z} \sigma_{3z}\nonumber&&\\
&\quad\:-\sigma_{1x} \sigma_{2x} \sigma_{3z}
-\sigma_{1x} \sigma_{2z} \sigma_{3x}
-\sigma_{1y} \sigma_{2y} \sigma_{3z}\nonumber&&\\
&\quad\:-\sigma_{1y} \sigma_{2z} \sigma_{3y}
-\sigma_{1z} \sigma_{2x} \sigma_{3x}
-\sigma_{1z} \sigma_{2y} \sigma_{3y}
\big)\nonumber&&\\
&+\frac{1}{\sqrt3} \big(
\pm \sigma_{1x} \sigma_{2y}
\mp \sigma_{1y} \sigma_{2x}
\mp \sigma_{1x} \sigma_{3y}
\pm \sigma_{1y} \sigma_{3x}
\pm \sigma_{2x} \sigma_{3y}
\mp \sigma_{2y} \sigma_{3x}\nonumber&&\\
&\qquad\ \ \ \ \textcolor{blue}{\pm \sigma_{1x} \sigma_{2y} \sigma_{3z}
\mp \sigma_{1x} \sigma_{2z} \sigma_{3y}
\mp \sigma_{1y} \sigma_{2x} \sigma_{3z}}\nonumber&&\\
&\qquad\ \ \ \ \textcolor{blue}{\pm \sigma_{1y} \sigma_{2z} \sigma_{3x}
\pm \sigma_{1z} \sigma_{2x} \sigma_{3y}
\mp \sigma_{1z} \sigma_{2y} \sigma_{3x}}
\big)
\bigg]&&\\[6pt] \hline\hline\nonumber
\end{flalign}
\newpage
\begin{figure}[H]
\centering
\vspace{1.5cm}
\includegraphics[width=1\textwidth]{Fig13odd.pdf}
\caption{\label{fig:Fig13odd}BEADS representation of (\textbf{A}) $\ket{\psi_+^{\left\{1,3\right\}_{\textit{odd}}}}$ and (\textbf{B}) $\ket{\psi_-^{\left\{1,3\right\}_{\textit{odd}}}}$ for $\theta=0$. The visualized states reach maximum values $\pm1$ for the bead $T\{1,3\}_{\textit{odd}}$ marked by red circles.}
\end{figure}

\begin{figure}[H]
\centering
\includegraphics[width=1\textwidth]{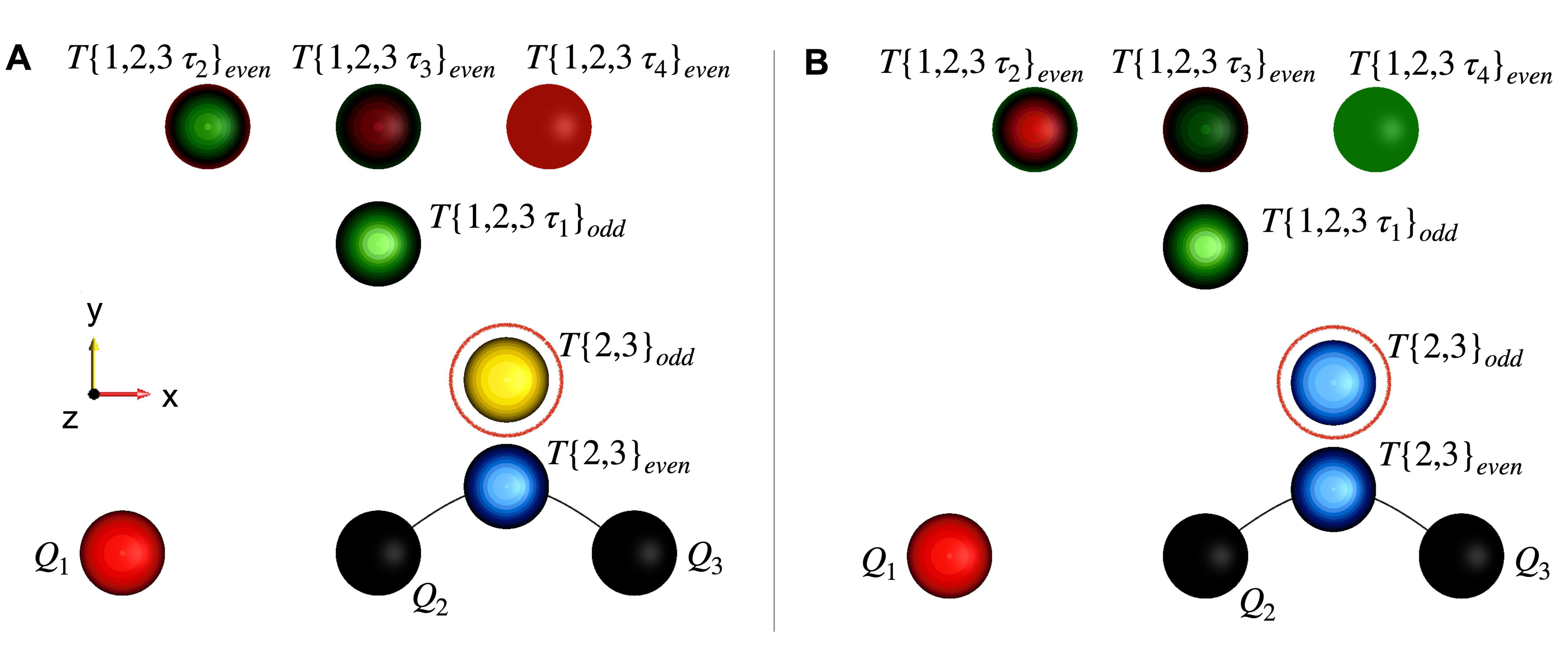}
\caption{\label{fig:Fig23odd}BEADS representation of (\textbf{A}) $\ket{\psi_+^{\left\{2,3\right\}_{\textit{odd}}}}$ and (\textbf{B}) $\ket{\psi_-^{\left\{2,3\right\}_{\textit{odd}}}}$ for $\theta=0$. The visualized states reach maximum values $\pm1$ for the bead $T\{2,3\}_{\textit{odd}}$ marked by red circles.}
\end{figure}

\begin{figure}[H]
\centering
\vspace{1.5cm}
\includegraphics[width=1\textwidth]{FigTau2odd.pdf}
\caption{\label{fig:FigTau2odd}BEADS representation of (\textbf{A}) $\ket{\psi_+^{\left\{1,2,3\ \tau_2\right\}_{\textit{odd}}}}$ and (\textbf{B}) $\ket{\psi_-^{\left\{1,2,3\ \tau_2\right\}_{\textit{odd}}}}$. The visualized states reach maximum values $\pm1$ for the bead $T\{1,2,3\ \tau_2\}_{\textit{odd}}$ marked by red circles.}
\end{figure}

\begin{figure}[H]
\centering
\includegraphics[width=1\textwidth]{FigTau2even.pdf}
\caption{\label{fig:FigTau2even}BEADS representation of (\textbf{A}) $\ket{\psi_+^{\left\{1,2,3\ \tau_2\right\}_{\textit{even}}}}$ and (\textbf{B}) $\ket{\psi_-^{\left\{1,2,3\ \tau_2\right\}_{\textit{even}}}}$ for $\theta=0$. The visualized states reach maximum values $\pm1$ for the bead $T\{1,2,3\ \tau_2\}_{\textit{even}}$ marked by red circles.}
\end{figure}

\begin{figure}[H]
\centering
\vspace{.3cm}
\includegraphics[width=1\textwidth]{FigTau3odd.pdf}
\caption{\label{fig:FigTau3odd}BEADS representation of (\textbf{A}) $\ket{\psi_+^{\left\{1,2,3\ \tau_3\right\}_{\textit{odd}}}}$ and (\textbf{B}) $\ket{\psi_-^{\left\{1,2,3\ \tau_3\right\}_{\textit{odd}}}}$ for $\theta=0$. The visualized states reach maximum values $\pm1$ for the bead $T\{1,2,3\ \tau_3\}_{\textit{odd}}$ marked by red circles.}
\end{figure}

\begin{figure}[H]
\centering
\includegraphics[width=1\textwidth]{FigTau3even.pdf}
\caption{\label{fig:FigTau3even}BEADS representation of (\textbf{A}) $\ket{\psi_+^{\left\{1,2,3\ \tau_3\right\}_{\textit{even}}}}$ and (\textbf{B}) $\ket{\psi_-^{\left\{1,2,3\ \tau_3\right\}_{\textit{even}}}}$ for $\theta=0$. The visualized states reach maximum values $\pm1$ for the bead $T\{1,2,3\ \tau_3\}_{\textit{even}}$ marked by red circles.}
\end{figure}

\begin{figure}[H]
\centering
\vspace{.8cm}
\includegraphics[width=1\textwidth]{FigTau4.pdf}
\caption{\label{fig:FigTau4}BEADS representation of (\textbf{A}) $\ket{\psi_+^{\left\{1,2,3\ \tau_4\right\}_{\textit{even}}}}$ and (\textbf{B}) $\ket{\psi_-^{\left\{1,2,3\ \tau_4\right\}_{\textit{even}}}}$ for $\theta=0$. The visualized states reach maximum values $\pm1$ for the bead $T\{1,2,3\ \tau_4\}_{\textit{even}}$ marked by red circles.}
\end{figure}

\subsection{Axial tensor operators}\label{app:BEADSTomog}
Using Pauli matrices $\sigma_{k\epsilon}$ corresponding to the $k$-th qubit, where $\epsilon\in\{x,y,z\}$, and the identity operator $\mathds{1}$, the scaled axial LISA tensor operators ${\widetilde{T}}_{j,0}^{\left(\ell^\prime\right)}=\zeta(N)\xi_j^{\left(\ell^\prime\right)}\left(g\right)T_{j,0}^{\left(\ell^\prime\right)}$ for systems consisting of up to $N=3$ qubits are:

\begin{flalign}
\hline\hline\nonumber&&\\[-8pt]
&\ell^\prime=\{\emptyset\}:&&\nonumber\\
&{\widetilde{T}}_{0,0}^{\{\emptyset\}}=\mathds{1}&&\\[8pt]
\hline\nonumber&&\\[-8pt]
&\ell^\prime=\{k\}:&&\nonumber\\
&{\widetilde{T}}_{1,0}^{\left\{k\right\}}=\sigma_{kz}&&\\[8pt]
\hline\nonumber&&\\[-8pt]
&\ell^\prime=\left\{k,l\right\}_{\textit{even}}:&&\nonumber\\
&{\widetilde{T}}_{0,0}^{\left\{k,l\right\}_{\textit{even}}}=\frac{1}{3}\left(\sigma_{kx}\sigma_{lx}+{\sigma_{ky}\sigma}_{ly}+\sigma_{kz}\sigma_{lz}\right)&&\\
&{\widetilde{T}}_{2,0}^{\left\{k,l\right\}_{\textit{even}}}=\frac{1}{3}\left({-\sigma}_{kx}\sigma_{lx}-{\sigma_{ky}\sigma}_{ly}+2\sigma_{kz}\sigma_{lz}\right)&&\\[8pt]
\hline\nonumber&&\\[-8pt]
&\ell^\prime=\left\{k,l\ \right\}_{\textit{odd}}:&&\nonumber\\
&{\widetilde{T}}_{1,0}^{\left\{k,l\right\}_{\textit{odd}}}=\frac{1}{2}\left({\sigma_{kx}\sigma}_{ly}-{\sigma_{ky}\sigma}_{lx}\right)&&\\[8pt]
\hline\nonumber&&
\end{flalign}
\vspace{-1.4cm}
\begin{flalign}
\ell^\prime=\left\{k,l,m\ \tau_1\right\}_{\textit{odd}}: &&\nonumber
\end{flalign}
\begin{subequations}
\vspace{-.8cm}
\begin{flalign}
{\widetilde{T}}_{1,0}^{\left\{k,l,m\ \tau_1\right\}_{\textit{odd}}}=\frac{2}{5}\bigl[&\left(\sigma_{kx}\sigma_{lx}\sigma_{mz}+\sigma_{kx}\sigma_{lz}\sigma_{mx}+\sigma_{kz}\sigma_{lx}\sigma_{mx}\right)\nonumber&&\\
+&\left(\sigma_{ky}\sigma_{ly}\sigma_{mz}+\sigma_{ky}\sigma_{lz}\sigma_{my}+\sigma_{kz}\sigma_{ly}\sigma_{my}\right)+{{3\sigma}_{kz}\sigma_{lz}\sigma}_{mz}\bigr]&&\\
{\widetilde{T}}_{3,0}^{\left\{k,l,m\ \tau_1\right\}_{\textit{odd}}}=\frac{2}{5}\bigl[&-\left(\sigma_{kx}\sigma_{lx}\sigma_{mz}+{{\sigma_{kx}\sigma}_{lz}\sigma}_{mx}+\sigma_{kz}\sigma_{lx}\sigma_{mx}\right)\nonumber&&\\
&-\left(\sigma_{ky}\sigma_{ly}\sigma_{mz}+\sigma_{ky}\sigma_{lz}\sigma_{my}+\sigma_{kz}\sigma_{ly}\sigma_{my}\right)+2\sigma_{kz}\sigma_{lz}\sigma_{mz}\bigr]&&\\[6pt] \hline\nonumber
\end{flalign}
\end{subequations}
\vspace{-1.4cm}
\begin{flalign}
\ell^\prime=\left\{k,l,m\ \tau_2\right\}_{\textit{odd}}: &&\nonumber
\end{flalign}
\begin{subequations}
\vspace{-.8cm}
\begin{flalign}
{\widetilde{T}}_{1,0}^{\left\{k,l,m\ \tau_2\right\}_{\textit{odd}}}=\frac{\sqrt3}{3+\sqrt3}\bigl[&\left(\sigma_{kx}\sigma_{lz}\sigma_{mx}+\sigma_{kz}\sigma_{lx}\sigma_{mx}\right)+\left(\sigma_{ky}\sigma_{lz}\sigma_{my}+\sigma_{kz}\sigma_{ly}\sigma_{my}\right)\nonumber&&\\
-2&\left(\sigma_{kx}\sigma_{lx}\sigma_{mz}+\sigma_{ky}\sigma_{ly}\sigma_{mz}\right)\bigr]&&\\[6pt] \hline\nonumber
\end{flalign}
\end{subequations}
\vspace{-1.4cm}
\begin{flalign}
&\ell^\prime=\left\{k,l,m\ \tau_2\right\}_{\textit{even}}:&&\nonumber\\
&{\widetilde{T}}_{2,0}^{\left\{k,l,m\ \tau_2\right\}_{\textit{even}}}=\frac{1}{\sqrt2}\left[\left(\sigma_{ky}\sigma_{lz}\sigma_{mx}+\sigma_{kz}\sigma_{ly}\sigma_{mx}\right)-\left(\sigma_{kx}\sigma_{lz}\sigma_{my}+\sigma_{kz}\sigma_{lx}\sigma_{my}\right)\right]&&\\[8pt]
\hline\nonumber&&\\[-8pt]
&\ell^\prime=\left\{k,l,m\ \tau_3\right\}_{\textit{odd}}:&&\nonumber\\
&{\widetilde{T}}_{1,0}^{\left\{k,l,m\ \tau_3\right\}_{\textit{odd}}}=\frac{1}{\sqrt2}\left[\left(\sigma_{kz}\sigma_{lx}\sigma_{mx}-\sigma_{kx}\sigma_{lz}\sigma_{mx}\right)+\left(\sigma_{kz}\sigma_{ly}\sigma_{my}-\sigma_{ky}\sigma_{lz}\sigma_{my}\right)\right]&&\\[-16pt]\nonumber
\end{flalign}
\begin{flalign}
\hline\nonumber\\[-8pt]
&\ell^\prime=\left\{k,l,m\ \tau_3\right\}_{\textit{even}}: &&\nonumber
\end{flalign}
\begin{subequations}
\vspace{-1.cm}
\begin{flalign}
{\widetilde{T}}_{2,0}^{\left\{k,l,m\ \tau_3\right\}_{\textit{even}}}=\frac{1}{\sqrt6}\bigl[-&\left(2\sigma_{kx}\sigma_{ly}\sigma_{mz}+\sigma_{kx}\sigma_{lz}\sigma_{my}-\sigma_{kz}\sigma_{lx}\sigma_{my}\right)\nonumber&&\\
+&\left(2\sigma_{ky}\sigma_{lx}\sigma_{mz}+\sigma_{ky}\sigma_{lz}\sigma_{mx}-{\ \sigma}_{kz}\sigma_{ly}\sigma_{mx}\right)\bigr]&&\\[6pt] \hline\nonumber
\end{flalign}
\end{subequations}
\vspace{-1.4cm}
\begin{flalign}
&\ell^\prime=\left\{k,l,m\ \tau_4\right\}_{\textit{even}}: &&\nonumber
\end{flalign}
\begin{subequations}
\vspace{-1.cm}
\begin{flalign}
{\widetilde{T}}_{0,0}^{\left\{k,l,m\ \tau_4\right\}_{\textit{even}}}=\frac{1}{\sqrt3}\bigl[&\sigma_{kx}\sigma_{ly}\sigma_{mz}+\sigma_{ky}\sigma_{lz}\sigma_{mx}+\sigma_{kz}\sigma_{lx}\sigma_{my}\nonumber&&\\
-&\sigma_{kx}\sigma_{lz}\sigma_{my}-\sigma_{ky}\sigma_{lx}\sigma_{mz}-\sigma_{kz}\sigma_{ly}\sigma_{mx}\bigr]&&\\[6pt] \hline\hline\nonumber
\end{flalign}
\end{subequations}

\subsection{Separation of correlation function components using LISA tensor operators}\label{app:BEADSLISACorrel}
For pure states, a modified operator $\widetilde{\rho}$ that, besides single qubit components, only has non-zero connected correlation coefficients can be obtained using Hermitian LISA operators $T_{j,m}^{\left(\ell^\prime\right)}$ instead of Pauli operators which were applied in section~\ref{sec:BEADSCorrel}. In general, this connected correlated operator can hence be written as
\begin{equation}
\widetilde{\rho}=\rho\ -\sum_{j\in J\left(\ell^\prime\right)}\sum_{p=-j}^{j}C_{j,m}^{\left(\ell^\prime\right)}T_{j,m}^{\left(\ell^\prime\right)},
\end{equation}
\noindent
where $C_{j,m}^{\left(\ell^\prime\right)}$ is the sum of all \textit{compound correlation} coefficients corresponding to the tensor operator $T_{j,m}^{\left(\ell^\prime\right)}$. Bipartite compound correlation coefficients which factorize into linear combinations of expectation values of local observables (single-qubit operators) can be calculated as follows:
\begin{flalign}
\hline\hline\nonumber&&\\[-8pt]
&\text{Rank } j = 0 \nonumber&&\\
&C_{0,0}^{\left\{\kappa,\lambda\right\}_{\textit{even}}}=\frac{1}{\sqrt{2^N}}\frac{1}{\sqrt3}\left(\left\langle\sigma_{\kappa x}\right\rangle\left\langle\sigma_{\lambda x}\right\rangle+\left\langle\sigma_{\kappa y}\right\rangle\left\langle\sigma_{\lambda y}\right\rangle+\left\langle\sigma_{\kappa z}\right\rangle\left\langle\sigma_{\lambda z}\right\rangle\right)&&\\[6pt] \hline\nonumber \\[-8pt]
&\text{Rank } j = 1 \nonumber&&
\end{flalign}
\begin{subequations}
\vspace{-1.cm}
\begin{flalign}
&C_{1,-1}^{\left\{\kappa,\lambda\right\}_{\textit{odd}}}=\frac{1}{\sqrt{2^N}}\frac{1}{\sqrt2}\left(\left\langle\sigma_{\kappa z}\right\rangle\left\langle\sigma_{\lambda x}\right\rangle-\left\langle\sigma_{\kappa x}\right\rangle\left\langle\sigma_{\lambda z}\right\rangle\right)&&\\
&C_{1,0}^{\left\{\kappa,\lambda\right\}_{\textit{odd}}}=\frac{1}{\sqrt{2^N}}\frac{1}{\sqrt2}\left(\left\langle\sigma_{\kappa x}\right\rangle\left\langle\sigma_{\lambda y}\right\rangle-\left\langle\sigma_{\kappa y}\right\rangle\left\langle\sigma_{\lambda x}\right\rangle\right)&&\\
&C_{1,1}^{\left\{\kappa,\lambda\right\}_{\textit{odd}}}=\frac{1}{\sqrt{2^N}}\frac{1}{\sqrt2}\left(\left\langle\sigma_{\kappa y}\right\rangle\left\langle\sigma_{\lambda z}\right\rangle-\left\langle\sigma_{\kappa z}\right\rangle\left\langle\sigma_{\lambda y}\right\rangle\right)&&\\[6pt] \hline\nonumber \\[-1.4cm]\nonumber
\end{flalign}
\end{subequations}
\vspace{-.5cm}
\begin{subequations}
\begin{flalign}
&\text{Rank } j = 2 \nonumber &&\\
&C_{2,-2}^{\left\{\kappa,\lambda\right\}_{\textit{even}}}=\frac{1}{\sqrt{2^N}}\frac{1}{\sqrt2}\left(\left\langle\sigma_{\kappa y}\right\rangle\left\langle\sigma_{\lambda x}\right\rangle+\left\langle\sigma_{\kappa x}\right\rangle\left\langle\sigma_{\lambda y}\right\rangle\right)&&\\
&C_{2,-1}^{\left\{\kappa,\lambda\right\}_{\textit{even}}}=\frac{1}{\sqrt{2^N}}\frac{1}{\sqrt2}\left(\left\langle\sigma_{\kappa z}\right\rangle\left\langle\sigma_{\lambda y}\right\rangle+\left\langle\sigma_{\kappa y}\right\rangle\left\langle\sigma_{\lambda z}\right\rangle\right)&&\\
&C_{2,0}^{\left\{\kappa,\lambda\right\}_{\textit{even}}}=\frac{1}{\sqrt{2^N}}\frac{1}{\sqrt6}\left(-\left\langle\sigma_{\kappa x}\right\rangle\left\langle\sigma_{\lambda x}\right\rangle-\left\langle\sigma_{\kappa y}\right\rangle\left\langle\sigma_{\lambda y}\right\rangle+2\left\langle\sigma_{\kappa z}\right\rangle\left\langle\sigma_{\lambda z}\right\rangle\right)&&\\
&C_{2,1}^{\left\{\kappa,\lambda\right\}_{\textit{even}}}=\frac{1}{\sqrt{2^N}}\frac{1}{\sqrt2}\left(\left\langle\sigma_{\kappa x}\right\rangle\left\langle\sigma_{\lambda z}\right\rangle+\left\langle\sigma_{\kappa z}\right\rangle\left\langle\sigma_{\lambda x}\right\rangle\right)\\
&C_{2,2}^{\left\{\kappa,\lambda\right\}_{\textit{even}}}=\frac{1}{\sqrt{2^N}}\frac{1}{\sqrt2}\left(\left\langle\sigma_{\kappa x}\right\rangle\left\langle\sigma_{\lambda x}\right\rangle-\left\langle\sigma_{\kappa y}\right\rangle\left\langle\sigma_{\lambda y}\right\rangle\right)
\\[6pt] \hline\hline\nonumber
\end{flalign}
\end{subequations}\\[-1.cm]
Here, $\left\{k,\lambda\right\}$ denotes an arbitrary bilinear subsystem of qubits and $N$ is the total number of qubits in the system. Trilinear compound correlation coefficients expressed in terms of LISA tensor operator read as follows. Note that as discussed in section~\ref{sec:BEADSCorrel}, every trilinear tensor operator has multiple contributing compound correlation coefficients which include a component that factorizes to expectation values of local observables only and three contributions which are formed as products of the expectation value of a local observable and a lower-order connected correlation coefficient, respectively.
\begin{subequations}
\begin{flalign}
\hline\hline\nonumber&&\\[-8pt]
C_{1,-1,\text{\textit{local}}}^{\left\{\kappa,\lambda,\mu\ \tau_1\right\}_{\textit{odd}}}&=\frac{1}{\sqrt{2^N}}\frac{1}{\sqrt{15}}\left(\left\langle\sigma_{\kappa x}\right\rangle\left\langle\sigma_{\lambda x}\right\rangle\left\langle\sigma_{\mu y}\right\rangle+\left\langle\sigma_{\kappa x}\right\rangle\left\langle\sigma_{\lambda y}\right\rangle\left\langle\sigma_{\mu x}\right\rangle+\left\langle\sigma_{\kappa y}\right\rangle\left\langle\sigma_{\lambda x}\right\rangle\left\langle\sigma_{\mu x}\right\rangle\right.\nonumber&&\\
&+\left.3\left\langle\sigma_{\kappa y}\right\rangle\left\langle\sigma_{\lambda y}\right\rangle\left\langle\sigma_{\mu y}\right\rangle+\left\langle\sigma_{\kappa y}\right\rangle\left\langle\sigma_{\lambda z}\right\rangle\left\langle\sigma_{\mu z}\right\rangle+\left\langle\sigma_{\kappa z}\right\rangle\left\langle\sigma_{\lambda y}\right\rangle\left\langle\sigma_{\mu z}\right\rangle\right.\nonumber&&\\
&+\left.\left\langle\sigma_{\kappa z}\right\rangle\left\langle\sigma_{\lambda z}\right\rangle\left\langle\sigma_{\mu y}\right\rangle\right)&&
\end{flalign}
\vspace{-1.cm}
\begin{flalign}
C_{1,-1,\text{\textit{low-ord.}}}^{\left\{\kappa,\lambda,\mu\ \tau_1\right\}_{\textit{odd}},(\kappa\lambda)\mu}&=\frac{1}{\sqrt{2^N}}\frac{1}{\sqrt{15}}\left(\left\langle\sigma_{\kappa x}\sigma_{\lambda x}\right\rangle\left\langle\sigma_{\mu y}\right\rangle+\left\langle\sigma_{\kappa x}\sigma_{\lambda y}\right\rangle\left\langle\sigma_{\mu x}\right\rangle+\left\langle\sigma_{\kappa y}\sigma_{\lambda x}\right\rangle\left\langle\sigma_{\mu x}\right\rangle\right.\nonumber&&\\
&+\left.3\left\langle\sigma_{\kappa y}\sigma_{\lambda y}\right\rangle\left\langle\sigma_{\mu y}\right\rangle+\left\langle\sigma_{\kappa y}\sigma_{\lambda z}\right\rangle\left\langle\sigma_{\mu z}\right\rangle+\left\langle\sigma_{\kappa z}\sigma_{\lambda y}\right\rangle\left\langle\sigma_{\mu z}\right\rangle\right.\nonumber&&\\
&+\left.\left\langle\sigma_{\kappa z}\sigma_{\lambda z}\right\rangle\left\langle\sigma_{\mu y}\right\rangle\right)&&
\end{flalign}

\begin{flalign}
C_{1,0,\text{\textit{local}}}^{\left\{\kappa,\lambda,\mu\ \tau_1\right\}_{\textit{odd}}}&=\frac{1}{\sqrt{2^N}}\frac{1}{\sqrt{15}}\left(\left\langle\sigma_{\kappa x}\right\rangle\left\langle\sigma_{\lambda x}\right\rangle\left\langle\sigma_{\mu z}\right\rangle+\left\langle\sigma_{\kappa x}\right\rangle\left\langle\sigma_{\lambda z}\right\rangle\left\langle\sigma_{\mu x}\right\rangle+\left\langle\sigma_{\kappa y}\right\rangle\left\langle\sigma_{\lambda y}\right\rangle\left\langle\sigma_{\mu z}\right\rangle\right.\nonumber&&\\
&+\left\langle\sigma_{\kappa y}\right\rangle\left\langle\sigma_{\lambda z}\right\rangle\left\langle\sigma_{\mu y}\right\rangle+\left\langle\sigma_{\kappa z}\right\rangle\left\langle\sigma_{\lambda x}\right\rangle\left\langle\sigma_{\mu x}\right\rangle+\left\langle\sigma_{\kappa z}\right\rangle\left\langle\sigma_{\lambda y}\right\rangle\left\langle\sigma_{\mu y}\right\rangle\nonumber&&\\
&+\left.3\left\langle\sigma_{\kappa z}\right\rangle\left\langle\sigma_{\lambda z}\right\rangle\left\langle\sigma_{\mu z}\right\rangle\right)&&
\end{flalign}
\vspace{-.8cm}
\begin{flalign}
C_{1,0,\text{\textit{low-ord.}}}^{\left\{\kappa,\lambda,\mu\ \tau_1\right\}_{\textit{odd}},(\kappa\lambda)\mu}&=\frac{1}{\sqrt{2^N}}\frac{1}{\sqrt{15}}\left(\left\langle\sigma_{\kappa x}\sigma_{\lambda x}\right\rangle\left\langle\sigma_{\mu z}\right\rangle+\left\langle\sigma_{\kappa x}\sigma_{\lambda z}\right\rangle\left\langle\sigma_{\mu x}\right\rangle+\left\langle\sigma_{\kappa y}\sigma_{\lambda y}\right\rangle\left\langle\sigma_{\mu z}\right\rangle\right.\nonumber&&\\
&+\left\langle\sigma_{\kappa y}\sigma_{\lambda z}\right\rangle\left\langle\sigma_{\mu y}\right\rangle+\left\langle\sigma_{\kappa z}\sigma_{\lambda x}\right\rangle\left\langle\sigma_{\mu x}\right\rangle+\left\langle\sigma_{\kappa z}\sigma_{\lambda y}\right\rangle\left\langle\sigma_{\mu y}\right\rangle\nonumber&&\\
&+\left.3\left\langle\sigma_{\kappa z}\sigma_{\lambda z}\right\rangle\left\langle\sigma_{\mu z}\right\rangle\right)&&
\end{flalign}

\begin{flalign}
C_{1,1,\text{\textit{local}}}^{\left\{\kappa,\lambda,\mu\ \tau_1\right\}_{\textit{odd}}}&=\frac{1}{\sqrt{2^N}}\frac{1}{\sqrt{15}}\left(3\left\langle\sigma_{\kappa x}\right\rangle\left\langle\sigma_{\lambda x}\right\rangle\left\langle\sigma_{\mu x}\right\rangle+\left\langle\sigma_{\kappa x}\right\rangle\left\langle\sigma_{\lambda y}\right\rangle\left\langle\sigma_{\mu y}\right\rangle+\left\langle\sigma_{\kappa x}\right\rangle\left\langle\sigma_{\lambda z}\right\rangle\left\langle\sigma_{\mu z}\right\rangle\right.\nonumber&&\\
&+\left\langle\sigma_{\kappa y}\right\rangle\left\langle\sigma_{\lambda x}\right\rangle\left\langle\sigma_{\mu y}\right\rangle+\left\langle\sigma_{\kappa y}\right\rangle\left\langle\sigma_{\lambda y}\right\rangle\left\langle\sigma_{\mu x}\right\rangle+\left\langle\sigma_{\kappa z}\right\rangle\left\langle\sigma_{\lambda x}\right\rangle\left\langle\sigma_{\mu z}\right\rangle\nonumber&&\\
&+\left.\left\langle\sigma_{\kappa z}\right\rangle\left\langle\sigma_{\lambda z}\right\rangle\left\langle\sigma_{\mu x}\right\rangle\right)&&
\end{flalign}
\vspace{-.8cm}
\begin{flalign}
C_{1,1,\text{\textit{low-ord.}}}^{\left\{\kappa,\lambda,\mu\ \tau_1\right\}_{\textit{odd}},(\kappa\lambda)\mu}&=\frac{1}{\sqrt{2^N}}\frac{1}{\sqrt{15}}\left(3\left\langle\sigma_{\kappa x}\sigma_{\lambda x}\right\rangle\left\langle\sigma_{\mu x}\right\rangle+\left\langle\sigma_{\kappa x}\sigma_{\lambda y}\right\rangle\left\langle\sigma_{\mu y}\right\rangle+\left\langle\sigma_{\kappa x}\sigma_{\lambda z}\right\rangle\left\langle\sigma_{\mu z}\right\rangle\right.\nonumber&&\\
&+\left\langle\sigma_{\kappa y}\sigma_{\lambda x}\right\rangle\left\langle\sigma_{\mu y}\right\rangle+\left\langle\sigma_{\kappa y}\sigma_{\lambda y}\right\rangle\left\langle\sigma_{\mu x}\right\rangle+\left\langle\sigma_{\kappa z}\sigma_{\lambda x}\right\rangle\left\langle\sigma_{\mu z}\right\rangle\nonumber&&\\
&+\left.\left\langle\sigma_{\kappa z}\sigma_{\lambda z}\right\rangle\left\langle\sigma_{\mu x}\right\rangle\right)&& \\[6pt]\hline\nonumber\\[-1.4cm]\nonumber
\end{flalign}
\end{subequations}
\vspace{-.5cm}
\begin{subequations}
\begin{flalign}
C_{3,-3,\text{\textit{local}}}^{\left\{\kappa,\lambda,\mu\ \tau_1\right\}_{\textit{odd}}}&=\frac{1}{\sqrt{2^N}}\frac{1}{2}\left(\left\langle\sigma_{\kappa x}\right\rangle\left\langle\sigma_{\lambda x}\right\rangle\left\langle\sigma_{\mu y}\right\rangle+\left\langle\sigma_{\kappa x}\right\rangle\left\langle\sigma_{\lambda y}\right\rangle\left\langle\sigma_{\mu x}\right\rangle+\left\langle\sigma_{\kappa y}\right\rangle\left\langle\sigma_{\lambda x}\right\rangle\left\langle\sigma_{\mu x}\right\rangle\right.\nonumber&&\\
&-\left.\left\langle\sigma_{\kappa y}\right\rangle\left\langle\sigma_{\lambda y}\right\rangle\left\langle\sigma_{\mu y}\right\rangle\right)&&
\end{flalign}
\vspace{-.6cm}
\begin{flalign}
C_{3,-3,\text{\textit{low-ord.}}}^{\left\{\kappa,\lambda,\mu\ \tau_1\right\}_{\textit{odd}},(\kappa\lambda)\mu}&=\frac{1}{\sqrt{2^N}}\frac{1}{2}\left(\left\langle\sigma_{\kappa x}\sigma_{\lambda x}\right\rangle\left\langle\sigma_{\mu y}\right\rangle+\left\langle\sigma_{\kappa x}\sigma_{\lambda y}\right\rangle\left\langle\sigma_{\mu x}\right\rangle+\left\langle\sigma_{\kappa y}\sigma_{\lambda x}\right\rangle\left\langle\sigma_{\mu x}\right\rangle\right.\nonumber&&\\
&-\left.\left\langle\sigma_{\kappa y}\sigma_{\lambda y}\right\rangle\left\langle\sigma_{\mu y}\right\rangle\right)&&
\end{flalign}
\vspace{-.2cm}
\begin{flalign}
C_{3,-2,\text{\textit{local}}}^{\left\{\kappa,\lambda,\mu\ \tau_1\right\}_{\textit{odd}}}&=\frac{1}{\sqrt{2^N}}\ \frac{1}{\sqrt6}\left(\left\langle\sigma_{\kappa x}\right\rangle\left\langle\sigma_{\lambda y}\right\rangle\left\langle\sigma_{\mu z}\right\rangle+\left\langle\sigma_{\kappa x}\right\rangle\left\langle\sigma_{\lambda z}\right\rangle\left\langle\sigma_{\mu y}\right\rangle+\left\langle\sigma_{\kappa y}\right\rangle\left\langle\sigma_{\lambda x}\right\rangle\left\langle\sigma_{\mu z}\right\rangle\right.\nonumber&&\\
&+\left\langle\sigma_{\kappa y}\right\rangle\left\langle\sigma_{\lambda z}\right\rangle\left\langle\sigma_{\mu x}\right\rangle+\left\langle\sigma_{\kappa z}\right\rangle\left\langle\sigma_{\lambda x}\right\rangle\left\langle\sigma_{\mu y}\right\rangle+\left.\left\langle\sigma_{\kappa z}\right\rangle\left\langle\sigma_{\lambda y}\right\rangle\left\langle\sigma_{\mu x}\right\rangle\right)&&
\end{flalign}
\vspace{-.6cm}
\begin{flalign}
C_{3,-2,\text{\textit{low-ord.}}}^{\left\{\kappa,\lambda,\mu\ \tau_1\right\}_{\textit{odd}},(\kappa\lambda)\mu}&=\frac{1}{\sqrt{2^N}}\ \frac{1}{\sqrt6}\left(\left\langle\sigma_{\kappa x}\sigma_{\lambda y}\right\rangle\left\langle\sigma_{\mu z}\right\rangle+\left\langle\sigma_{\kappa x}\sigma_{\lambda z}\right\rangle\left\langle\sigma_{\mu y}\right\rangle+\left\langle\sigma_{\kappa y}\sigma_{\lambda x}\right\rangle\left\langle\sigma_{\mu z}\right\rangle\right.\nonumber&&\\
&+\left\langle\sigma_{\kappa y}\sigma_{\lambda z}\right\rangle\left\langle\sigma_{\mu x}\right\rangle+\left\langle\sigma_{\kappa z}\sigma_{\lambda x}\right\rangle\left\langle\sigma_{\mu y}\right\rangle+\left.\left\langle\sigma_{\kappa z}\sigma_{\lambda y}\right\rangle\left\langle\sigma_{\mu x}\right\rangle\right)&&
\end{flalign}
\vspace{-.2cm}
\begin{flalign}
C_{3,-1,\text{\textit{local}}}^{\left\{\kappa,\lambda,\mu\ \tau_1\right\}_{\textit{odd}}}&=\frac{1}{\sqrt{2^N}}\ \frac{1}{2\sqrt{15}\ }\left(-\left(\left\langle\sigma_{\kappa x}\right\rangle\left\langle\sigma_{\lambda x}\right\rangle\left\langle\sigma_{\mu y}\right\rangle+\left\langle\sigma_{\kappa x}\right\rangle\left\langle\sigma_{\lambda y}\right\rangle\left\langle\sigma_{\mu x}\right\rangle\right.\right.\nonumber&&\\
&+\left.\left\langle\sigma_{\kappa y}\right\rangle\left\langle\sigma_{\lambda x}\right\rangle\left\langle\sigma_{\mu x}\right\rangle\right)+4\left(\left\langle\sigma_{\kappa y}\right\rangle\left\langle\sigma_{\lambda z}\right\rangle\left\langle\sigma_{\mu z}\right\rangle+\left\langle\sigma_{\kappa z}\right\rangle\left\langle\sigma_{\lambda y}\right\rangle\left\langle\sigma_{\mu z}\right\rangle\right.\nonumber&&\\
&+\left.\left\langle\sigma_{\kappa z}\right\rangle\left\langle\sigma_{\lambda z}\right\rangle\left\langle\sigma_{\mu y}\right\rangle\right)-3\left.\left\langle\sigma_{\kappa y}\right\rangle\left\langle\sigma_{\lambda y}\right\rangle\left\langle\sigma_{\mu y}\right\rangle\right)&&
\end{flalign}
\vspace{-.2cm}
\begin{flalign}
C_{3,-1,\text{\textit{low-ord.}}}^{\left\{\kappa,\lambda,\mu\ \tau_1\right\}_{\textit{odd}},(\kappa\lambda)\mu}&=\frac{1}{\sqrt{2^N}}\ \frac{1}{2\sqrt{15}\ }\left(-\left(\left\langle\sigma_{\kappa x}\sigma_{\lambda x}\right\rangle\left\langle\sigma_{\mu y}\right\rangle+\left\langle\sigma_{\kappa x}\sigma_{\lambda y}\right\rangle\left\langle\sigma_{\mu x}\right\rangle\right.\right.\nonumber&&\\
&+\left.\left\langle\sigma_{\kappa y}\sigma_{\lambda x}\right\rangle\left\langle\sigma_{\mu x}\right\rangle\right)+\left.4\left(\left\langle\sigma_{\kappa y}\sigma_{\lambda z}\right\rangle\left\langle\sigma_{\mu z}\right\rangle+\left\langle\sigma_{\kappa z}\sigma_{\lambda y}\right\rangle\left\langle\sigma_{\mu z}\right\rangle\right.\right.\nonumber&&\\
&+\left.\left.\left\langle\sigma_{\kappa z}\sigma_{\lambda z}\right\rangle\left\langle\sigma_{\mu y}\right\rangle\right)-3\left\langle\sigma_{\kappa y}\sigma_{\lambda y}\right\rangle\left\langle\sigma_{\mu y}\right\rangle\right)&&
\end{flalign}
\vspace{-.2cm}
\begin{flalign}
C_{3,0,\text{\textit{local}}}^{\left\{\kappa,\lambda,\mu\ \tau_1\right\}_{\textit{odd}}}&=\frac{1}{\sqrt{2^N}}\ \frac{1}{\sqrt{10}\ }\left(-\left\langle\sigma_{\kappa x}\right\rangle\left\langle\sigma_{\lambda x}\right\rangle\left\langle\sigma_{\mu z}\right\rangle-\left\langle\sigma_{\kappa x}\right\rangle\left\langle\sigma_{\lambda z}\right\rangle\left\langle\sigma_{\mu x}\right\rangle\right.\nonumber&&\\
&-\left\langle\sigma_{\kappa y}\right\rangle\left\langle\sigma_{\lambda y}\right\rangle\left\langle\sigma_{\mu z}\right\rangle-\left\langle\sigma_{\kappa y}\right\rangle\left\langle\sigma_{\lambda z}\right\rangle\left\langle\sigma_{\mu y}\right\rangle-\ \left\langle\sigma_{\kappa z}\right\rangle\left\langle\sigma_{\lambda x}\right\rangle\left\langle\sigma_{\mu x}\right\rangle\nonumber&&\\
&-\left.\left\langle\sigma_{\kappa z}\right\rangle\left\langle\sigma_{\lambda y}\right\rangle\left\langle\sigma_{\mu y}\right\rangle+2\left\langle\sigma_{\kappa z}\right\rangle\left\langle\sigma_{\lambda z}\right\rangle\left\langle\sigma_{\mu z}\right\rangle\right)&&
\end{flalign}
\vspace{-.8cm}
\begin{flalign}
C_{3,0,\text{\textit{low-ord.}}}^{\left\{\kappa,\lambda,\mu\ \tau_1\right\}_{\textit{odd}},(\kappa\lambda)\mu}&=\frac{1}{\sqrt{2^N}}\ \frac{1}{\sqrt{10}\ }\left(-\left\langle\sigma_{\kappa x}\sigma_{\lambda x}\right\rangle\left\langle\sigma_{\mu z}\right\rangle-\left\langle\sigma_{\kappa x}\sigma_{\lambda z}\right\rangle\left\langle\sigma_{\mu x}\right\rangle\right.\nonumber&&\\
&-\left\langle\sigma_{\kappa y}\sigma_{\lambda y}\right\rangle\left\langle\sigma_{\mu z}\right\rangle-\left\langle\sigma_{\kappa y}\sigma_{\lambda z}\right\rangle\left\langle\sigma_{\mu y}\right\rangle-\left\langle\sigma_{\kappa z}\sigma_{\lambda x}\right\rangle\left\langle\sigma_{\mu x}\right\rangle\nonumber&&\\
&-\left.\left\langle\sigma_{\kappa z}\sigma_{\lambda y}\right\rangle\left\langle\sigma_{\mu y}\right\rangle+2\left\langle\sigma_{\kappa z}\sigma_{\lambda z}\right\rangle\left\langle\sigma_{\mu z}\right\rangle\right)&& 
\end{flalign}
\vspace{-.2cm}
\begin{flalign}
C_{3,1,\text{\textit{local}}}^{\left\{\kappa,\lambda,\mu\ \tau_1\right\}_{\textit{odd}}}&=\frac{1}{\sqrt{2^N}}\ \frac{1}{2\sqrt{15}\ }\left(-\left(\left\langle\sigma_{\kappa y}\right\rangle\left\langle\sigma_{\lambda x}\right\rangle\left\langle\sigma_{\mu y}\right\rangle+\left\langle\sigma_{\kappa x}\right\rangle\left\langle\sigma_{\lambda y}\right\rangle\left\langle\sigma_{\mu y}\right\rangle\right.\right.\nonumber&&\\
&+\left.\left\langle\sigma_{\kappa y}\right\rangle\left\langle\sigma_{\lambda y}\right\rangle\left\langle\sigma_{\mu x}\right\rangle\right)+4\left(\left\langle\sigma_{\kappa x}\right\rangle\left\langle\sigma_{\lambda z}\right\rangle\left\langle\sigma_{\mu z}\right\rangle+\left\langle\sigma_{\kappa z}\right\rangle\left\langle\sigma_{\lambda x}\right\rangle\left\langle\sigma_{\mu z}\right\rangle\right.\nonumber&&\\
&+\left.\left\langle\sigma_{\kappa z}\right\rangle\left\langle\sigma_{\lambda z}\right\rangle\left\langle\sigma_{\mu x}\right\rangle\right)\left.-3\left\langle\sigma_{\kappa x}\right\rangle\left\langle\sigma_{\lambda x}\right\rangle\left\langle\sigma_{\mu x}\right\rangle\right)&&
\end{flalign}
\vspace{-.8cm}
\begin{flalign}
C_{3,1,\text{\textit{low-ord.}}}^{\left\{\kappa,\lambda,\mu\ \tau_1\right\}_{\textit{odd}},(\kappa\lambda)\mu}&=\frac{1}{\sqrt{2^N}}\ \frac{1}{2\sqrt{15}\ }\left(-\left(\left\langle\sigma_{\kappa x}\sigma_{\lambda y}\right\rangle\left\langle\sigma_{\mu y}\right\rangle+\left\langle\sigma_{\kappa y}\sigma_{\lambda x}\right\rangle\left\langle\sigma_{\mu y}\right\rangle\right.\right.\nonumber&&\\
&+\left.\left\langle\sigma_{\kappa y}\sigma_{\lambda y}\right\rangle\left\langle\sigma_{\mu x}\right\rangle\right)+4\left(\left\langle\sigma_{\kappa z}\sigma_{\lambda x}\right\rangle\left\langle\sigma_{\mu z}\right\rangle+\left\langle\sigma_{\kappa x}\sigma_{\lambda z}\right\rangle\left\langle\sigma_{\mu z}\right\rangle\right.\nonumber&&\\
&+\left.\left.\left\langle\sigma_{\kappa z}\sigma_{\lambda z}\right\rangle\left\langle\sigma_{\mu x}\right\rangle\right)-3\left\langle\sigma_{\kappa x}\sigma_{\lambda x}\right\rangle\left\langle\sigma_{\mu x}\right\rangle\right)&&
\end{flalign}
\begin{flalign}
C_{3,2,\text{\textit{local}}}^{\left\{\kappa,\lambda,\mu\ \tau_1\right\}_{\textit{odd}}}&=\frac{1}{\sqrt{2^N}}\ \frac{1}{\sqrt6}\left(\left\langle\sigma_{\kappa x}\right\rangle\left\langle\sigma_{\lambda x}\right\rangle\left\langle\sigma_{\mu z}\right\rangle+\left\langle\sigma_{\kappa x}\right\rangle\left\langle\sigma_{\lambda z}\right\rangle\left\langle\sigma_{\mu x}\right\rangle-\left\langle\sigma_{\kappa y}\right\rangle\left\langle\sigma_{\lambda y}\right\rangle\left\langle\sigma_{\mu z}\right\rangle\right.\nonumber&&\\
&-\left.\left\langle\sigma_{\kappa y}\right\rangle\left\langle\sigma_{\lambda z}\right\rangle\left\langle\sigma_{\mu y}\right\rangle+\left\langle\sigma_{\kappa z}\right\rangle\left\langle\sigma_{\lambda x}\right\rangle\left\langle\sigma_{\mu x}\right\rangle-\left\langle\sigma_{\kappa z}\right\rangle\left\langle\sigma_{\lambda y}\right\rangle\left\langle\sigma_{\mu y}\right\rangle\right)&&
\end{flalign}
\vspace{-.6cm}
\begin{flalign}
C_{3,2,\text{\textit{low-ord.}}}^{\left\{\kappa,\lambda,\mu\ \tau_1\right\}_{\textit{odd}},(\kappa\lambda)\mu}&=\frac{1}{\sqrt{2^N}}\ \frac{1}{\sqrt6}\left(\left\langle\sigma_{\kappa x}\sigma_{\lambda x}\right\rangle\left\langle\sigma_{\mu z}\right\rangle+\left\langle\sigma_{\kappa x}\sigma_{\lambda z}\right\rangle\left\langle\sigma_{\mu x}\right\rangle-\left\langle\sigma_{\kappa y}\sigma_{\lambda y}\right\rangle\left\langle\sigma_{\mu z}\right\rangle\right.\nonumber&&\\
&-\left.\left\langle\sigma_{\kappa y}\sigma_{\lambda z}\right\rangle\left\langle\sigma_{\mu y}\right\rangle+\left\langle\sigma_{\kappa z}\sigma_{\lambda x}\right\rangle\left\langle\sigma_{\mu x}\right\rangle-\left\langle\sigma_{\kappa z}\sigma_{\lambda y}\right\rangle\left\langle\sigma_{\mu y}\right\rangle\right)&&
\end{flalign}
\vspace{.4cm}
\begin{flalign}
C_{3,3,\text{\textit{local}}}^{\left\{\kappa,\lambda,\mu\ \tau_1\right\}_{\textit{odd}}}&=\frac{1}{\sqrt{2^N}}\frac{1}{2}\left(\left\langle\sigma_{\kappa x}\right\rangle\left\langle\sigma_{\lambda x}\right\rangle\left\langle\sigma_{\mu x}\right\rangle-\left\langle\sigma_{\kappa x}\right\rangle\left\langle\sigma_{\lambda y}\right\rangle\left\langle\sigma_{\mu y}\right\rangle-\left\langle\sigma_{\kappa y}\right\rangle\left\langle\sigma_{\lambda x}\right\rangle\left\langle\sigma_{\mu y}\right\rangle\right.\nonumber&&\\
&-\left.\left\langle\sigma_{\kappa y}\right\rangle\left\langle\sigma_{\lambda y}\right\rangle\left\langle\sigma_{\mu x}\right\rangle\right)&&
\end{flalign}
\vspace{-.6cm}
\begin{flalign}
C_{3,3,\text{\textit{low-ord.}}}^{\left\{\kappa,\lambda,\mu\ \tau_1\right\}_{\textit{odd}},(\kappa\lambda)\mu}&=\frac{1}{\sqrt{2^N}}\frac{1}{2}\left(\left\langle\sigma_{\kappa x}\sigma_{\lambda x}\right\rangle\left\langle\sigma_{\mu x}\right\rangle-\left\langle\sigma_{\kappa x}\sigma_{\lambda y}\right\rangle\left\langle\sigma_{\mu y}\right\rangle-\left\langle\sigma_{\kappa y}\sigma_{\lambda x}\right\rangle\left\langle\sigma_{\mu y}\right\rangle\right.\nonumber&&\\
&-\left.\left\langle\sigma_{\kappa y}\sigma_{\lambda y}\right\rangle\left\langle\sigma_{\mu x}\right\rangle\right)&&\\[6pt] \hline\hline\nonumber&&
\end{flalign}
\end{subequations}
\vspace{-1.3cm}
\begin{subequations}
\begin{flalign}
C_{1,-1,\text{\textit{local}}}^{\left\{\kappa,\lambda,\mu\ \tau_2\right\}_{\textit{odd}}}&=\frac{1}{\sqrt{2^N}}\frac{1}{\sqrt{12}}\left(-2\left(\left\langle\sigma_{\kappa x}\right\rangle\left\langle\sigma_{\lambda x}\right\rangle\left\langle\sigma_{\mu y}\right\rangle+\left\langle\sigma_{\kappa z}\right\rangle\left\langle\sigma_{\lambda z}\right\rangle\left\langle\sigma_{\mu y}\right\rangle\right)\right.\nonumber&&\\
&+\left\langle\sigma_{\kappa x}\right\rangle\left\langle\sigma_{\lambda y}\right\rangle\left\langle\sigma_{\mu x}\right\rangle
+\left\langle\sigma_{\kappa y}\right\rangle\left\langle\sigma_{\lambda x}\right\rangle\left\langle\sigma_{\mu x}\right\rangle+\left\langle\sigma_{\kappa y}\right\rangle\left\langle\sigma_{\lambda z}\right\rangle\left\langle\sigma_{\mu z}\right\rangle\nonumber&&\\
&+\left.\left\langle\sigma_{\kappa z}\right\rangle\left\langle\sigma_{\lambda y}\right\rangle\left\langle\sigma_{\mu z}\right\rangle\right)&&
\end{flalign}
\vspace{-.8cm}
\begin{flalign}
C_{1,-1,\text{\textit{low-ord.}}}^{\left\{\kappa,\lambda,\mu\ \tau_2\right\}_{\textit{odd}},(\kappa\lambda)\mu}&=\frac{1}{\sqrt{2^N}}\frac{1}{\sqrt{12}}\left(-2\left(\left\langle\sigma_{\kappa x}\sigma_{\lambda x}\right\rangle\left\langle\sigma_{\mu y}\right\rangle+\left\langle\sigma_{\kappa z}\sigma_{\lambda z}\right\rangle\left\langle\sigma_{\mu y}\right\rangle\right)\right.\nonumber&&\\
&+\left\langle\sigma_{\kappa x}\sigma_{\lambda y}\right\rangle\left\langle\sigma_{\mu x}\right\rangle+\left\langle\sigma_{\kappa y}\sigma_{\lambda x}\right\rangle\left\langle\sigma_{\mu x}\right\rangle+\left\langle\sigma_{\kappa y}\sigma_{\lambda z}\right\rangle\left\langle\sigma_{\mu z}\right\rangle\nonumber&&\\
&+\left.\left\langle\sigma_{\kappa z}\sigma_{\lambda y}\right\rangle\left\langle\sigma_{\mu z}\right\rangle\right)&&
\end{flalign}
\vspace{-.2cm}
\begin{flalign}
C_{1,0,\text{\textit{local}}}^{\left\{\kappa,\lambda,\mu\ \tau_2\right\}_{\textit{odd}}}&=\frac{1}{\sqrt{2^N}}\frac{1}{\sqrt{12}}\left(-2\left(\left\langle\sigma_{\kappa x}\right\rangle\left\langle\sigma_{\lambda x}\right\rangle\left\langle\sigma_{\mu z}\right\rangle+\left\langle\sigma_{\kappa y}\right\rangle\left\langle\sigma_{\lambda y}\right\rangle\left\langle\sigma_{\mu z}\right\rangle\right)\right.\nonumber&&\\
&+\left\langle\sigma_{\kappa x}\right\rangle\left\langle\sigma_{\lambda z}\right\rangle\left\langle\sigma_{\mu x}\right\rangle+\left\langle\sigma_{\kappa y}\right\rangle\left\langle\sigma_{\lambda z}\right\rangle\left\langle\sigma_{\mu y}\right\rangle+\left\langle\sigma_{\kappa z}\right\rangle\left\langle\sigma_{\lambda x}\right\rangle\left\langle\sigma_{\mu x}\right\rangle\nonumber&&\\
&+\left.\left\langle\sigma_{\kappa z}\right\rangle\left\langle\sigma_{\lambda y}\right\rangle\left\langle\sigma_{\mu y}\right\rangle\right)&&
\end{flalign}
\vspace{-.8cm}
\begin{flalign}
C_{1,0,\text{\textit{low-ord.}}}^{\left\{\kappa,\lambda,\mu\ \tau_2\right\}_{\textit{odd}},(\kappa\lambda)\mu}&=\frac{1}{\sqrt{2^N}}\frac{1}{\sqrt{12}}\left(-2\left(\left\langle\sigma_{\kappa x}\sigma_{\lambda x}\right\rangle\left\langle\sigma_{\mu z}\right\rangle+\left\langle\sigma_{\kappa y}\sigma_{\lambda y}\right\rangle\left\langle\sigma_{\mu z}\right\rangle\right)\right.\nonumber&&\\
&+\left\langle\sigma_{\kappa x}\sigma_{\lambda z}\right\rangle\left\langle\sigma_{\mu x}\right\rangle+\left\langle\sigma_{\kappa y}\sigma_{\lambda z}\right\rangle\left\langle\sigma_{\mu y}\right\rangle+\left\langle\sigma_{\kappa z}\sigma_{\lambda x}\right\rangle\left\langle\sigma_{\mu x}\right\rangle\nonumber&&\\
&+\left.\left\langle\sigma_{\kappa z}\sigma_{\lambda y}\right\rangle\left\langle\sigma_{\mu y}\right\rangle\right)&&
\end{flalign}
\vspace{-.2cm}
\begin{flalign}
C_{1,1,\text{\textit{local}}}^{\left\{\kappa,\lambda,\mu\ \tau_2\right\}_{\textit{odd}}}&=\frac{1}{\sqrt{2^N}}\frac{1}{\sqrt{12}}\left(-2\left(\left\langle\sigma_{\kappa y}\right\rangle\left\langle\sigma_{\lambda y}\right\rangle\left\langle\sigma_{\mu x}\right\rangle+\left\langle\sigma_{\kappa z}\right\rangle\left\langle\sigma_{\lambda z}\right\rangle\left\langle\sigma_{\mu x}\right\rangle\right)\right.\nonumber&&\\
&+\left\langle\sigma_{\kappa x}\right\rangle\left\langle\sigma_{\lambda y}\right\rangle\left\langle\sigma_{\mu y}\right\rangle+\left\langle\sigma_{\kappa x}\right\rangle\left\langle\sigma_{\lambda z}\right\rangle\left\langle\sigma_{\mu z}\right\rangle+\left\langle\sigma_{\kappa y}\right\rangle\left\langle\sigma_{\lambda x}\right\rangle\left\langle\sigma_{\mu y}\right\rangle\nonumber&&\\
&+\left.\left\langle\sigma_{\kappa z}\right\rangle\left\langle\sigma_{\lambda x}\right\rangle\left\langle\sigma_{\mu z}\right\rangle\right)&&
\end{flalign}
\vspace{-.8cm}
\begin{flalign}
C_{1,1,\text{\textit{low-ord.}}}^{\left\{\kappa,\lambda,\mu\ \tau_2\right\}_{\textit{odd}},(\kappa\lambda)\mu}&=\frac{1}{\sqrt{2^N}}\frac{1}{\sqrt{12}}\left(-2\left(\left\langle\sigma_{\kappa y}\sigma_{\lambda y}\right\rangle\left\langle\sigma_{\mu x}\right\rangle+\left\langle\sigma_{\kappa z}\sigma_{\lambda z}\right\rangle\left\langle\sigma_{\mu x}\right\rangle\right)\right.\nonumber&&\\
&+\left\langle\sigma_{\kappa x}\sigma_{\lambda y}\right\rangle\left\langle\sigma_{\mu y}\right\rangle+\left\langle\sigma_{\kappa x}\sigma_{\lambda z}\right\rangle\left\langle\sigma_{\mu z}\right\rangle+\left\langle\sigma_{\kappa y}\sigma_{\lambda x}\right\rangle\left\langle\sigma_{\mu y}\right\rangle\nonumber&&\\
&+\left.\left\langle\sigma_{\kappa z}\sigma_{\lambda x}\right\rangle\left\langle\sigma_{\mu z}\right\rangle\right)&&\\[8pt]
\hline\hline\nonumber
\end{flalign}
\end{subequations}
\vspace{-1.cm}
\begin{subequations}
\begin{flalign}
C_{2,-2,\ class.}^{\left\{\kappa,\lambda,\mu\ \tau_2\right\}_{\textit{even}}}&=\frac{1}{\sqrt{2^N}}\frac{1}{\sqrt{12}}\left(2\left(\left\langle\sigma_{\kappa x}\right\rangle\left\langle\sigma_{\lambda x}\right\rangle\left\langle\sigma_{\mu z}\right\rangle-\left\langle\sigma_{\kappa y}\right\rangle\left\langle\sigma_{\lambda y}\right\rangle\left\langle\sigma_{\mu z}\right\rangle\right)\right.\nonumber&&\\
&-\left\langle\sigma_{\kappa x}\right\rangle\left\langle\sigma_{\lambda z}\right\rangle\left\langle\sigma_{\mu x}\right\rangle+\left\langle\sigma_{\kappa y}\right\rangle\left\langle\sigma_{\lambda z}\right\rangle\left\langle\sigma_{\mu y}\right\rangle-\left\langle\sigma_{\kappa z}\right\rangle\left\langle\sigma_{\lambda x}\right\rangle\left\langle\sigma_{\mu x}\right\rangle\nonumber&&\\
&+\left.\left\langle\sigma_{\kappa z}\right\rangle\left\langle\sigma_{\lambda y}\right\rangle\left\langle\sigma_{\mu y}\right\rangle\right)&&
\end{flalign}
\vspace{-.6cm}
\begin{flalign}
C_{2,-2,\text{\textit{low-ord.}}}^{\left\{\kappa,\lambda,\mu\ \tau_2\right\}_{\textit{even}},(\kappa\lambda)\mu}&=\frac{1}{\sqrt{2^N}}\frac{1}{\sqrt{12}}\left(2\left(\left\langle\sigma_{\kappa x}\sigma_{\lambda x}\right\rangle\left\langle\sigma_{\mu z}\right\rangle-\left\langle\sigma_{\kappa y}\sigma_{\lambda y}\right\rangle\left\langle\sigma_{\mu z}\right\rangle\right)\right.\nonumber&&\\
&-\left\langle\sigma_{\kappa x}\sigma_{\lambda z}\right\rangle\left\langle\sigma_{\mu x}\right\rangle+\left\langle\sigma_{\kappa y}\sigma_{\lambda z}\right\rangle\left\langle\sigma_{\mu y}\right\rangle-\left\langle\sigma_{\kappa z}\sigma_{\lambda x}\right\rangle\left\langle\sigma_{\mu x}\right\rangle\nonumber&&\\
&+\left.\left\langle\sigma_{\kappa z}\sigma_{\lambda y}\right\rangle\left\langle\sigma_{\mu y}\right\rangle\right)&&
\end{flalign}
\begin{flalign}
C_{2,-1,class}^{\left\{\kappa,\lambda,\mu\ \tau_2\right\}_{\textit{even}}}&=\frac{1}{\sqrt{2^N}}\frac{1}{\sqrt{12}}\left(2\left(\left\langle\sigma_{\kappa y}\right\rangle\left\langle\sigma_{\lambda y}\right\rangle\left\langle\sigma_{\mu x}\right\rangle-\left\langle\sigma_{\kappa z}\right\rangle\left\langle\sigma_{\lambda z}\right\rangle\left\langle\sigma_{\mu x}\right\rangle\right)\right.\nonumber&&\\
&-\left\langle\sigma_{\kappa x}\right\rangle\left\langle\sigma_{\lambda y}\right\rangle\left\langle\sigma_{\mu y}\right\rangle+\left\langle\sigma_{\kappa x}\right\rangle\left\langle\sigma_{\lambda z}\right\rangle\left\langle\sigma_{\mu z}\right\rangle-\left\langle\sigma_{\kappa y}\right\rangle\left\langle\sigma_{\lambda x}\right\rangle\left\langle\sigma_{\mu y}\right\rangle\nonumber&&\\
&+\left.\left\langle\sigma_{\kappa z}\right\rangle\left\langle\sigma_{\lambda x}\right\rangle\left\langle\sigma_{\mu z}\right\rangle\right)&&
\end{flalign}
\vspace{-.8cm}
\begin{flalign}
C_{2,-1,\text{\textit{low-ord.}}}^{\left\{\kappa,\lambda,\mu\ \tau_2\right\}_{\textit{even}},(\kappa\lambda)\mu}&=\frac{1}{\sqrt{2^N}}\frac{1}{\sqrt{12}}\left(2\left(\left\langle\sigma_{\kappa y}\sigma_{\lambda y}\right\rangle\left\langle\sigma_{\mu x}\right\rangle-\left\langle\sigma_{\kappa z}\sigma_{\lambda z}\right\rangle\left\langle\sigma_{\mu x}\right\rangle\right)\right.\nonumber&&\\
&-\left\langle\sigma_{\kappa x}\sigma_{\lambda y}\right\rangle\left\langle\sigma_{\mu y}\right\rangle+\left\langle\sigma_{\kappa x}\sigma_{\lambda z}\right\rangle\left\langle\sigma_{\mu z}\right\rangle-\left\langle\sigma_{\kappa y}\sigma_{\lambda x}\right\rangle\left\langle\sigma_{\mu y}\right\rangle\nonumber&&\\
&+\left.\left\langle\sigma_{\kappa z}\sigma_{\lambda x}\right\rangle\left\langle\sigma_{\mu z}\right\rangle\right)&&
\end{flalign}
\begin{flalign}
C_{2,0,\text{\textit{local}}}^{\left\{\kappa,\lambda,\mu\ \tau_2\right\}_{\textit{even}}}&=\frac{1}{\sqrt{2^N}}\frac{1}{2}\left(-\left\langle\sigma_{\kappa x}\right\rangle\left\langle\sigma_{\lambda z}\right\rangle\left\langle\sigma_{\mu y}\right\rangle+\left\langle\sigma_{\kappa y}\right\rangle\left\langle\sigma_{\lambda z}\right\rangle\left\langle\sigma_{\mu x}\right\rangle-\left\langle\sigma_{\kappa z}\right\rangle\left\langle\sigma_{\lambda x}\right\rangle\left\langle\sigma_{\mu y}\right\rangle\right.\nonumber&&\\
&+\left.\left\langle\sigma_{\kappa z}\right\rangle\left\langle\sigma_{\lambda y}\right\rangle\left\langle\sigma_{\mu x}\right\rangle\right)&&
\end{flalign}
\vspace{-.8cm}
\begin{flalign}
C_{2,0,\text{\textit{low-ord.}}}^{\left\{\kappa,\lambda,\mu\ \tau_2\right\}_{\textit{even}},(\kappa\lambda)\mu}&=\frac{1}{\sqrt{2^N}}\frac{1}{2}\left(-\left\langle\sigma_{\kappa x}\sigma_{\lambda z}\right\rangle\left\langle\sigma_{\mu y}\right\rangle+\left\langle\sigma_{\kappa y}\sigma_{\lambda z}\right\rangle\left\langle\sigma_{\mu x}\right\rangle-\left\langle\sigma_{\kappa z}\sigma_{\lambda x}\right\rangle\left\langle\sigma_{\mu y}\right\rangle\right.\nonumber&&\\
&+\left.\left\langle\sigma_{\kappa z}\sigma_{\lambda y}\right\rangle\left\langle\sigma_{\mu x}\right\rangle\right)&&
\end{flalign}
\begin{flalign}
C_{2,1,\text{\textit{local}}}^{\left\{\kappa,\lambda,\mu\ \tau_2\right\}_{\textit{even}}}&=\frac{1}{\sqrt{2^N}}\frac{1}{\sqrt{12}}\left(2\left(\left\langle\sigma_{\kappa z}\right\rangle\left\langle\sigma_{\lambda z}\right\rangle\left\langle\sigma_{\mu y}\right\rangle-\left\langle\sigma_{\kappa x}\right\rangle\left\langle\sigma_{\lambda x}\right\rangle\left\langle\sigma_{\mu y}\right\rangle\right)\right.\nonumber&&\\
&+\left\langle\sigma_{\kappa x}\right\rangle\left\langle\sigma_{\lambda y}\right\rangle\left\langle\sigma_{\mu x}\right\rangle+\left\langle\sigma_{\kappa y}\right\rangle\left\langle\sigma_{\lambda x}\right\rangle\left\langle\sigma_{\mu x}\right\rangle-\left\langle\sigma_{\kappa y}\right\rangle\left\langle\sigma_{\lambda z}\right\rangle\left\langle\sigma_{\mu z}\right\rangle\nonumber&&\\
&-\left.\left\langle\sigma_{\kappa z}\right\rangle\left\langle\sigma_{\lambda y}\right\rangle\left\langle\sigma_{\mu z}\right\rangle\right)&&
\end{flalign}
\vspace{-.8cm}
\begin{flalign}
C_{2,1,\text{\textit{low-ord.}}}^{\left\{\kappa,\lambda,\mu\ \tau_2\right\}_{\textit{even}},(\kappa\lambda)\mu}&=\frac{1}{\sqrt{2^N}}\frac{1}{\sqrt{12}}\left(2\left(\left\langle\sigma_{\kappa z}\sigma_{\lambda z}\right\rangle\left\langle\sigma_{\mu y}\right\rangle-\left\langle\sigma_{\kappa x}\sigma_{\lambda x}\right\rangle\left\langle\sigma_{\mu y}\right\rangle\right)\right.\nonumber&&\\
&+\left\langle\sigma_{\kappa x}\sigma_{\lambda y}\right\rangle\left\langle\sigma_{\mu x}\right\rangle+\left\langle\sigma_{\kappa y}\sigma_{\lambda x}\right\rangle\left\langle\sigma_{\mu x}\right\rangle-\left\langle\sigma_{\kappa y}\sigma_{\lambda z}\right\rangle\left\langle\sigma_{\mu z}\right\rangle\nonumber&&\\
&-\left.\left\langle\sigma_{\kappa z}\sigma_{\lambda y}\right\rangle\left\langle\sigma_{\mu z}\right\rangle\right)&&
\end{flalign}
\begin{flalign}
C_{2,2,\text{\textit{local}}}^{\left\{\kappa,\lambda,\mu\ \tau_2\right\}_{\textit{even}}}&=\frac{1}{\sqrt{2^N}}\frac{1}{\sqrt{12}}\left(-2\left(\left\langle\sigma_{\kappa x}\right\rangle\left\langle\sigma_{\lambda y}\right\rangle\left\langle\sigma_{\mu z}\right\rangle+\left\langle\sigma_{\kappa y}\right\rangle\left\langle\sigma_{\lambda x}\right\rangle\left\langle\sigma_{\mu z}\right\rangle\right)\right.\nonumber&&\\
&+\left\langle\sigma_{\kappa x}\right\rangle\left\langle\sigma_{\lambda z}\right\rangle\left\langle\sigma_{\mu y}\right\rangle+\left\langle\sigma_{\kappa y}\right\rangle\left\langle\sigma_{\lambda z}\right\rangle\left\langle\sigma_{\mu x}\right\rangle+\left\langle\sigma_{\kappa z}\right\rangle\left\langle\sigma_{\lambda x}\right\rangle\left\langle\sigma_{\mu y}\right\rangle\nonumber&&\\
&+\left.\left\langle\sigma_{\kappa z}\right\rangle\left\langle\sigma_{\lambda y}\right\rangle\left\langle\sigma_{\mu x}\right\rangle\right)&&
\end{flalign}
\vspace{-.6cm}
\begin{flalign}
C_{2,2,\text{\textit{low-ord.}}}^{\left\{\kappa,\lambda,\mu\ \tau_2\right\}_{\textit{even}},(\kappa\lambda)\mu}&=\frac{1}{\sqrt{2^N}}\frac{1}{\sqrt{12}}\left(-2\left(\left\langle\sigma_{\kappa x}\sigma_{\lambda y}\right\rangle\left\langle\sigma_{\mu z}\right\rangle+\left\langle\sigma_{\kappa y}\sigma_{\lambda x}\right\rangle\left\langle\sigma_{\mu z}\right\rangle\right)\right.\nonumber&&\\
&+\left\langle\sigma_{\kappa x}\sigma_{\lambda z}\right\rangle\left\langle\sigma_{\mu y}\right\rangle+\left\langle\sigma_{\kappa y}\sigma_{\lambda z}\right\rangle\left\langle\sigma_{\mu x}\right\rangle+\left\langle\sigma_{\kappa z}\sigma_{\lambda x}\right\rangle\left\langle\sigma_{\mu y}\right\rangle\nonumber&&\\
&+\left.\left\langle\sigma_{\kappa z}\sigma_{\lambda y}\right\rangle\left\langle\sigma_{\mu x}\right\rangle\right)&&\\[8pt] \hline\hline\nonumber
\end{flalign}
\end{subequations}
\vspace{-1.4cm}
\begin{subequations}
\begin{flalign}
C_{1,-1,\text{\textit{local}}}^{\left\{\kappa,\lambda,\mu\ \tau_3\right\}_{\textit{odd}}}&=\frac{1}{\sqrt{2^N}}\frac{1}{2}\left(-\left\langle\sigma_{\kappa x}\right\rangle\left\langle\sigma_{\lambda y}\right\rangle\left\langle\sigma_{\mu x}\right\rangle+\left\langle\sigma_{\kappa y}\right\rangle\left\langle\sigma_{\lambda x}\right\rangle\left\langle\sigma_{\mu x}\right\rangle+\left\langle\sigma_{\kappa y}\right\rangle\left\langle\sigma_{\lambda z}\right\rangle\left\langle\sigma_{\mu z}\right\rangle\right.\nonumber&&\\
&-\left.\left\langle\sigma_{\kappa z}\right\rangle\left\langle\sigma_{\lambda y}\right\rangle\left\langle\sigma_{\mu z}\right\rangle\right)&&
\end{flalign}
\vspace{-.8cm}
\begin{flalign}
C_{1,-1,\text{\textit{low-ord.}}}^{\left\{\kappa,\lambda,\mu\ \tau_3\right\}_{\textit{odd}},(\kappa\lambda)\mu}&=\frac{1}{\sqrt{2^N}}\frac{1}{2}\left(-\left\langle\sigma_{\kappa x}\sigma_{\lambda y}\right\rangle\left\langle\sigma_{\mu x}\right\rangle+\left\langle\sigma_{\kappa y}\sigma_{\lambda x}\right\rangle\left\langle\sigma_{\mu x}\right\rangle+\left\langle\sigma_{\kappa y}\sigma_{\lambda z}\right\rangle\left\langle\sigma_{\mu z}\right\rangle\right.\nonumber&&\\
&-\left.\left\langle\sigma_{\kappa z}\sigma_{\lambda y}\right\rangle\left\langle\sigma_{\mu z}\right\rangle\right)&&
\end{flalign}
\begin{flalign}
C_{1,0,\text{\textit{local}}}^{\left\{\kappa,\lambda,\mu\ \tau_3\right\}_{\textit{odd}}}&=\frac{1}{\sqrt{2^N}}\frac{1}{2}\left(-\left\langle\sigma_{\kappa x}\right\rangle\left\langle\sigma_{\lambda z}\right\rangle\left\langle\sigma_{\mu x}\right\rangle-\left\langle\sigma_{\kappa y}\right\rangle\left\langle\sigma_{\lambda z}\right\rangle\left\langle\sigma_{\mu y}\right\rangle+\left\langle\sigma_{\kappa z}\right\rangle\left\langle\sigma_{\lambda x}\right\rangle\left\langle\sigma_{\mu x}\right\rangle\right.\nonumber&&\\
&+\left.\left\langle\sigma_{\kappa z}\right\rangle\left\langle\sigma_{\lambda y}\right\rangle\left\langle\sigma_{\mu y}\right\rangle\right)&&
\end{flalign}
\vspace{-.8cm}
\begin{flalign}
C_{1,0,\text{\textit{low-ord.}}}^{\left\{\kappa,\lambda,\mu\ \tau_3\right\}_{\textit{odd}},(\kappa\lambda)\mu}&=\frac{1}{\sqrt{2^N}}\frac{1}{2}\left(-\left\langle\sigma_{\kappa x}\sigma_{\lambda z}\right\rangle\left\langle\sigma_{\mu x}\right\rangle-\left\langle\sigma_{\kappa y}\sigma_{\lambda z}\right\rangle\left\langle\sigma_{\mu y}\right\rangle+\left\langle\sigma_{\kappa z}\sigma_{\lambda x}\right\rangle\left\langle\sigma_{\mu x}\right\rangle\right.\nonumber&&\\
&+\left.\left\langle\sigma_{\kappa z}\sigma_{\lambda y}\right\rangle\left\langle\sigma_{\mu y}\right\rangle\right)&&
\end{flalign}
\begin{flalign}
C_{1,1,\text{\textit{local}}}^{\left\{\kappa,\lambda,\mu\ \tau_3\right\}_{\textit{odd}}}&=\frac{1}{\sqrt{2^N}}\frac{1}{2}\left(\left\langle\sigma_{\kappa x}\right\rangle\left\langle\sigma_{\lambda y}\right\rangle\left\langle\sigma_{\mu y}\right\rangle+\left\langle\sigma_{\kappa x}\right\rangle\left\langle\sigma_{\lambda z}\right\rangle\left\langle\sigma_{\mu z}\right\rangle-\left\langle\sigma_{\kappa y}\right\rangle\left\langle\sigma_{\lambda x}\right\rangle\left\langle\sigma_{\mu y}\right\rangle\right.\nonumber&&\\
&-\left.\ \left\langle\sigma_{\kappa z}\right\rangle\left\langle\sigma_{\lambda x}\right\rangle\left\langle\sigma_{\mu z}\right\rangle\right)&&
\end{flalign}
\vspace{-.8cm}
\begin{flalign}
C_{1,1,\text{\textit{low-ord.}}}^{\left\{\kappa,\lambda,\mu\ \tau_3\right\}_{\textit{odd}},(\kappa\lambda)\mu}&=\frac{1}{\sqrt{2^N}}\frac{1}{2}\left(\left\langle\sigma_{\kappa x}\sigma_{\lambda y}\right\rangle\left\langle\sigma_{\mu y}\right\rangle+\left\langle\sigma_{\kappa x}\sigma_{\lambda z}\right\rangle\left\langle\sigma_{\mu z}\right\rangle-\left\langle\sigma_{\kappa y}\sigma_{\lambda x}\right\rangle\left\langle\sigma_{\mu y}\right\rangle\right.\nonumber&&\\
&-\left.\left\langle\sigma_{\kappa z}\sigma_{\lambda x}\right\rangle\left\langle\sigma_{\mu z}\right\rangle\right)&&\\[11pt]
\hline\hline\nonumber
\end{flalign}
\end{subequations}
\vspace{-1.4cm}
\begin{subequations}
\begin{flalign}
C_{2,-2,\text{\textit{local}}}^{\left\{\kappa,\lambda,\mu\ \tau_3\right\}_{\textit{even}}}&=\frac{1}{\sqrt{2^N}}\frac{1}{2}\left(\left\langle\sigma_{\kappa x}\right\rangle\left\langle\sigma_{\lambda z}\right\rangle\left\langle\sigma_{\mu x}\right\rangle-\left\langle\sigma_{\kappa y}\right\rangle\left\langle\sigma_{\lambda z}\right\rangle\left\langle\sigma_{\mu y}\right\rangle-\left\langle\sigma_{\kappa z}\right\rangle\left\langle\sigma_{\lambda x}\right\rangle\left\langle\sigma_{\mu x}\right\rangle\right.\nonumber&&\\
&+\left.\left\langle\sigma_{\kappa z}\right\rangle\left\langle\sigma_{\lambda y}\right\rangle\left\langle\sigma_{\mu y}\right\rangle\right)&&
\end{flalign}
\vspace{-.8cm}
\begin{flalign}
C_{2,-2,\text{\textit{low-ord.}}}^{\left\{\kappa,\lambda,\mu\ \tau_3\right\}_{\textit{even}},(\kappa\lambda)\mu}&=\frac{1}{\sqrt{2^N}}\frac{1}{2}\left(\left\langle\sigma_{\kappa x}\sigma_{\lambda z}\right\rangle\left\langle\sigma_{\mu x}\right\rangle-\left\langle\sigma_{\kappa y}\sigma_{\lambda z}\right\rangle\left\langle\sigma_{\mu y}\right\rangle-\left\langle\sigma_{\kappa z}\sigma_{\lambda x}\right\rangle\left\langle\sigma_{\mu x}\right\rangle\right.\nonumber&&\\
&+\left.\left\langle\sigma_{\kappa z}\sigma_{\lambda y}\right\rangle\left\langle\sigma_{\mu y}\right\rangle\right)&&
\end{flalign}
\begin{flalign}
C_{2,-1,\text{\textit{local}}}^{\left\{\kappa,\lambda,\mu\ \tau_3\right\}_{\textit{even}}}&=\frac{1}{\sqrt{2^N}}\frac{1}{2}\left(-\left\langle\sigma_{\kappa x}\right\rangle\left\langle\sigma_{\lambda y}\right\rangle\left\langle\sigma_{\mu y}\right\rangle+\left\langle\sigma_{\kappa x}\right\rangle\left\langle\sigma_{\lambda z}\right\rangle\left\langle\sigma_{\mu z}\right\rangle+\left\langle\sigma_{\kappa y}\right\rangle\left\langle\sigma_{\lambda x}\right\rangle\left\langle\sigma_{\mu y}\right\rangle\right.\nonumber&&\\
&-\left.\left\langle\sigma_{\kappa z}\right\rangle\left\langle\sigma_{\lambda x}\right\rangle\left\langle\sigma_{\mu z}\right\rangle\right)&&
\end{flalign}
\vspace{-.8cm}
\begin{flalign}
C_{2,-1,\text{\textit{low-ord.}}}^{\left\{\kappa,\lambda,\mu\ \tau_3\right\}_{\textit{even}},(\kappa\lambda)\mu}&=\frac{1}{\sqrt{2^N}}\frac{1}{2}\left(\left\langle\sigma_{\kappa x}\sigma_{\lambda y}\right\rangle\left\langle\sigma_{\mu y}\right\rangle+\left\langle\sigma_{\kappa x}\sigma_{\lambda z}\right\rangle\left\langle\sigma_{\mu z}\right\rangle+\left\langle\sigma_{\kappa y}\sigma_{\lambda x}\right\rangle\left\langle\sigma_{\mu y}\right\rangle\right.\nonumber&&\\
&-\left.\left\langle\sigma_{\kappa z}\sigma_{\lambda x}\right\rangle\left\langle\sigma_{\mu z}\right\rangle\right)&&
\end{flalign}
\begin{flalign}
C_{2,0,\text{\textit{local}}}^{\left\{\kappa,\lambda,\mu\ \tau_3\right\}_{\textit{even}}}&=\frac{1}{\sqrt{2^N}}\frac{1}{\sqrt{12}}\left(2\left(\left\langle\sigma_{\kappa y}\right\rangle\left\langle\sigma_{\lambda x}\right\rangle\left\langle\sigma_{\mu z}\right\rangle-\left\langle\sigma_{\kappa x}\right\rangle\left\langle\sigma_{\lambda y}\right\rangle\left\langle\sigma_{\mu z}\right\rangle\right)\right.\nonumber&&\\
&-\left\langle\sigma_{\kappa x}\right\rangle\left\langle\sigma_{\lambda z}\right\rangle\left\langle\sigma_{\mu y}\right\rangle+\left\langle\sigma_{\kappa y}\right\rangle\left\langle\sigma_{\lambda z}\right\rangle\left\langle\sigma_{\mu x}\right\rangle+\left\langle\sigma_{\kappa z}\right\rangle\left\langle\sigma_{\lambda x}\right\rangle\left\langle\sigma_{\mu y}\right\rangle\nonumber&&\\
&-\left.\left\langle\sigma_{\kappa z}\right\rangle\left\langle\sigma_{\lambda y}\right\rangle\left\langle\sigma_{\mu x}\right\rangle\right)&&
\end{flalign}
\vspace{-.8cm}
\begin{flalign}
C_{2,0,\text{\textit{low-ord.}}}^{\left\{\kappa,\lambda,\mu\ \tau_3\right\}_{\textit{even}},(\kappa\lambda)\mu}&=\frac{1}{\sqrt{2^N}}\frac{1}{\sqrt{12}}\left(2\left(\left\langle\sigma_{\kappa y}\sigma_{\lambda x}\right\rangle\left\langle\sigma_{\mu z}\right\rangle-\left\langle\sigma_{\kappa x}\sigma_{\lambda y}\right\rangle\left\langle\sigma_{\mu z}\right\rangle\right)\right.\nonumber&&\\
&-\left\langle\sigma_{\kappa x}\sigma_{\lambda z}\right\rangle\left\langle\sigma_{\mu y}\right\rangle+\left\langle\sigma_{\kappa y}\sigma_{\lambda z}\right\rangle\left\langle\sigma_{\mu x}\right\rangle+\left\langle\sigma_{\kappa z}\sigma_{\lambda x}\right\rangle\left\langle\sigma_{\mu y}\right\rangle\nonumber&&\\
&-\left.\left\langle\sigma_{\kappa z}\sigma_{\lambda y}\right\rangle\left\langle\sigma_{\mu x}\right\rangle\right)&&
\end{flalign}
\begin{flalign}
C_{2,1,\text{\textit{local}}}^{\left\{\kappa,\lambda,\mu\ \tau_3\right\}_{\textit{even}}}&=\frac{1}{\sqrt{2^N}}\frac{1}{2}\left(-\left\langle\sigma_{\kappa x}\right\rangle\left\langle\sigma_{\lambda y}\right\rangle\left\langle\sigma_{\mu x}\right\rangle+\left\langle\sigma_{\kappa y}\right\rangle\left\langle\sigma_{\lambda x}\right\rangle\left\langle\sigma_{\mu x}\right\rangle-\left\langle\sigma_{\kappa y}\right\rangle\left\langle\sigma_{\lambda z}\right\rangle\left\langle\sigma_{\mu z}\right\rangle\right.\nonumber&&\\
&+\left.\left\langle\sigma_{\kappa z}\right\rangle\left\langle\sigma_{\lambda y}\right\rangle\left\langle\sigma_{\mu z}\right\rangle\right)&&
\end{flalign}
\vspace{-.8cm}
\begin{flalign}
C_{2,1,\text{\textit{low-ord.}}}^{\left\{\kappa,\lambda,\mu\ \tau_3\right\}_{\textit{even}},(\kappa\lambda)\mu}&=\frac{1}{\sqrt{2^N}}\frac{1}{2}\left(-\left\langle\sigma_{\kappa x}\sigma_{\lambda y}\right\rangle\left\langle\sigma_{\mu x}\right\rangle+\left\langle\sigma_{\kappa y}\sigma_{\lambda x}\right\rangle\left\langle\sigma_{\mu x}\right\rangle-\left\langle\sigma_{\kappa y}\sigma_{\lambda z}\right\rangle\left\langle\sigma_{\mu z}\right\rangle\right.\nonumber&&\\
&+\left.\left\langle\sigma_{\kappa z}\sigma_{\lambda y}\right\rangle\left\langle\sigma_{\mu z}\right\rangle\right)&&
\end{flalign}
\begin{flalign}
C_{2,2,\text{\textit{local}}}^{\left\{\kappa,\lambda,\mu\ \tau_3\right\}_{\textit{even}}}&=\frac{1}{\sqrt{2^N}}\frac{1}{2}\left(-\left\langle\sigma_{\kappa x}\right\rangle\left\langle\sigma_{\lambda z}\right\rangle\left\langle\sigma_{\mu y}\right\rangle-\left\langle\sigma_{\kappa y}\right\rangle\left\langle\sigma_{\lambda z}\right\rangle\left\langle\sigma_{\mu x}\right\rangle+\left\langle\sigma_{\kappa z}\right\rangle\left\langle\sigma_{\lambda x}\right\rangle\left\langle\sigma_{\mu y}\right\rangle\right.\nonumber&&\\
&+\left.\left\langle\sigma_{\kappa z}\right\rangle\left\langle\sigma_{\lambda y}\right\rangle\left\langle\sigma_{\mu x}\right\rangle\right)&&
\end{flalign}
\vspace{-.8cm}
\begin{flalign}
C_{2,2,\text{\textit{low-ord.}}}^{\left\{\kappa,\lambda,\mu\ \tau_2\right\}_{\textit{even}},(\kappa\lambda)\mu}&=\frac{1}{\sqrt{2^N}}\frac{1}{2}\left(-\left\langle\sigma_{\kappa x}\sigma_{\lambda z}\right\rangle\left\langle\sigma_{\mu y}\right\rangle-\left\langle\sigma_{\kappa y}\sigma_{\lambda z}\right\rangle\left\langle\sigma_{\mu x}\right\rangle+\left\langle\sigma_{\kappa z}\sigma_{\lambda x}\right\rangle\left\langle\sigma_{\mu y}\right\rangle\right.\nonumber&&\\
&+\left.\left\langle\sigma_{\kappa z}\sigma_{\lambda y}\right\rangle\left\langle\sigma_{\mu x}\right\rangle\right)&&\\[11pt] \hline\hline\nonumber
\end{flalign}
\end{subequations}
\vspace{-1.4cm}
\begin{subequations}
\begin{flalign}
C_{0,0,\text{\textit{local}}}^{\left\{\kappa,\lambda,\mu\ \tau_4\right\}_{\textit{even}}}&=\frac{1}{\sqrt{2^N}}\frac{1}{\sqrt6}\left(\left\langle\sigma_{\kappa x}\right\rangle\left\langle\sigma_{\lambda y}\right\rangle\left\langle\sigma_{\mu z}\right\rangle-\left\langle\sigma_{\kappa x}\right\rangle\left\langle\sigma_{\lambda z}\right\rangle\left\langle\sigma_{\mu y}\right\rangle-\left\langle\sigma_{\kappa y}\right\rangle\left\langle\sigma_{\lambda x}\right\rangle\left\langle\sigma_{\mu z}\right\rangle\right.\nonumber&&\\
&+\left.\left\langle\sigma_{\kappa y}\right\rangle\left\langle\sigma_{\lambda z}\right\rangle\left\langle\sigma_{\mu x}\right\rangle+\left\langle\sigma_{\kappa z}\right\rangle\left\langle\sigma_{\lambda x}\right\rangle\left\langle\sigma_{\mu y}\right\rangle-\left\langle\sigma_{\kappa z}\right\rangle\left\langle\sigma_{\lambda y}\right\rangle\left\langle\sigma_{\mu x}\right\rangle\right)&&
\end{flalign}
\vspace{-.8cm}
\begin{flalign}
C_{0,0,\text{\textit{low-ord.}}}^{\left\{\kappa,\lambda,\mu\ \tau_4\right\}_{\textit{even}},(\kappa\lambda)\mu}&\frac{1}{\sqrt{2^N}}\frac{1}{\sqrt6}\left(\left\langle\sigma_{\kappa x}\sigma_{\lambda y}\right\rangle\left\langle\sigma_{\mu z}\right\rangle-\left\langle\sigma_{\kappa x}\sigma_{\lambda z}\right\rangle\left\langle\sigma_{\mu y}\right\rangle-\left\langle\sigma_{\kappa y}\sigma_{\lambda x}\right\rangle\left\langle\sigma_{\mu z}\right\rangle\right.\nonumber&&\\
&+\left.\left\langle\sigma_{\kappa y}\sigma_{\lambda z}\right\rangle\left\langle\sigma_{\mu x}\right\rangle+\left\langle\sigma_{\kappa z}\sigma_{\lambda x}\right\rangle\left\langle\sigma_{\mu y}\right\rangle-\left\langle\sigma_{\kappa z}\sigma_{\lambda y}\right\rangle\left\langle\sigma_{\mu x}\right\rangle\right)&&\\[11pt] \hline\hline\nonumber
\end{flalign}
\end{subequations}

\noindent
Here, $(\kappa\lambda)\mu$ denotes a \textit{bipartitioning} of the three qubits denoted by indices $\kappa$, $\lambda$, and $\mu$ into a bilinear (indicated by parentheses) and a linear contribution which is reflected in the applied expectation values of Pauli product operators in the preceding equations. To eliminate all trilinear compound correlation coefficients, it is required to consider any possible bipartitioning of the involved qubits, i.e., $(\kappa\lambda)\mu$, $(\kappa\mu)\lambda$ , and $(\lambda\mu)\kappa$.
The modified operator $\widetilde{\rho}$ of a three-qubit system then results in the following expression (Eq.~\ref{eq:RhoCorrelLISA}):
\begin{align}\label{eq:RhoCorrelLISA}
\widetilde{\rho}=\rho&-\sum_{{\ell^\prime}_{\textit{bil}}\in\mathcal{B}}\ \sum_{j\in J\left({\ell^\prime}_{\textit{bil}}\right)}\ \sum_{m=-j}^{j}C_{j,m}^{\left({\ell^\prime}_{\textit{bil}}\right)}T_{j,m}^{\left({\ell^\prime}_{\textit{bil}}\right)}\nonumber\\
&-\sum_{{\ell^\prime}_{\textit{tril}}\in\mathcal{T}}\ \sum_{j\in J\left({\ell^\prime}_{\textit{tril}}\right)}\ \sum_{m=-j}^{j}\Bigg(C_{j,m,\text{\textit{low-ord.}}}^{\left({\ell^\prime}_{\textit{tril}}\right),\ \left(12\right)3}+C_{j,m,\text{\textit{low-ord.}}}^{\left({\ell^\prime}_{\textit{tril}}\right),\left(13\right)2}\nonumber\\
&\phantom{-\sum_{{\ell^\prime}_{\textit{tril}}\in\mathcal{M}}\ \sum_{j\in J\left({\ell^\prime}_{\textit{tril}}\right)}\ \sum_{m=-j}^{j}\Bigg(}+C_{j,m,\text{\textit{low-ord.}}}^{\left({\ell^\prime}_{\textit{tril}}\right),\left(23\right)1}-2C_{j,m,\text{\textit{local}}}^{\left({\ell^\prime}_{\textit{tril}}\right)}\Bigg)\ T_{j,m}^{\left({\ell^\prime}_{\textit{tril}}\right)}.
\end{align}
Here, the sets $\mathcal{B}=\{\left\{1,2\right\}_{\textit{even}},\:\left\{1,2\right\}_{\textit{odd}},\:\left\{1,3\right\}_{\textit{even}},\:\left\{1,3\right\}_{\textit{odd}},$ $\left\{2,3\right\}_{\textit{even}},\:\left\{2,3\right\}_{\textit{odd}}\}$ and $\mathcal{T}=\{\left\{1,2,3\ \tau_1\right\}_{\textit{odd}},\:\left\{1,2,3\ \tau_2\right\}_{\textit{even}},\:\left\{1,2,3\ \tau_2\right\}_{\textit{odd}},\:\left\{1,2,3\ \tau_3\right\}_{\textit{even}},$ $\left\{1,2,3\ \tau_3\right\}_{\textit{odd}},\:\left\{1,2,3\ \tau_4\right\}_{\textit{even}}\}$ correspond to the bilinear and trilinear LISA tensor operator subsets and the corresponding beads.

\subsection{Separation of correlation function components -- graphical analysis}

\begin{figure}[H]
\centering
\includegraphics[width=1\textwidth]{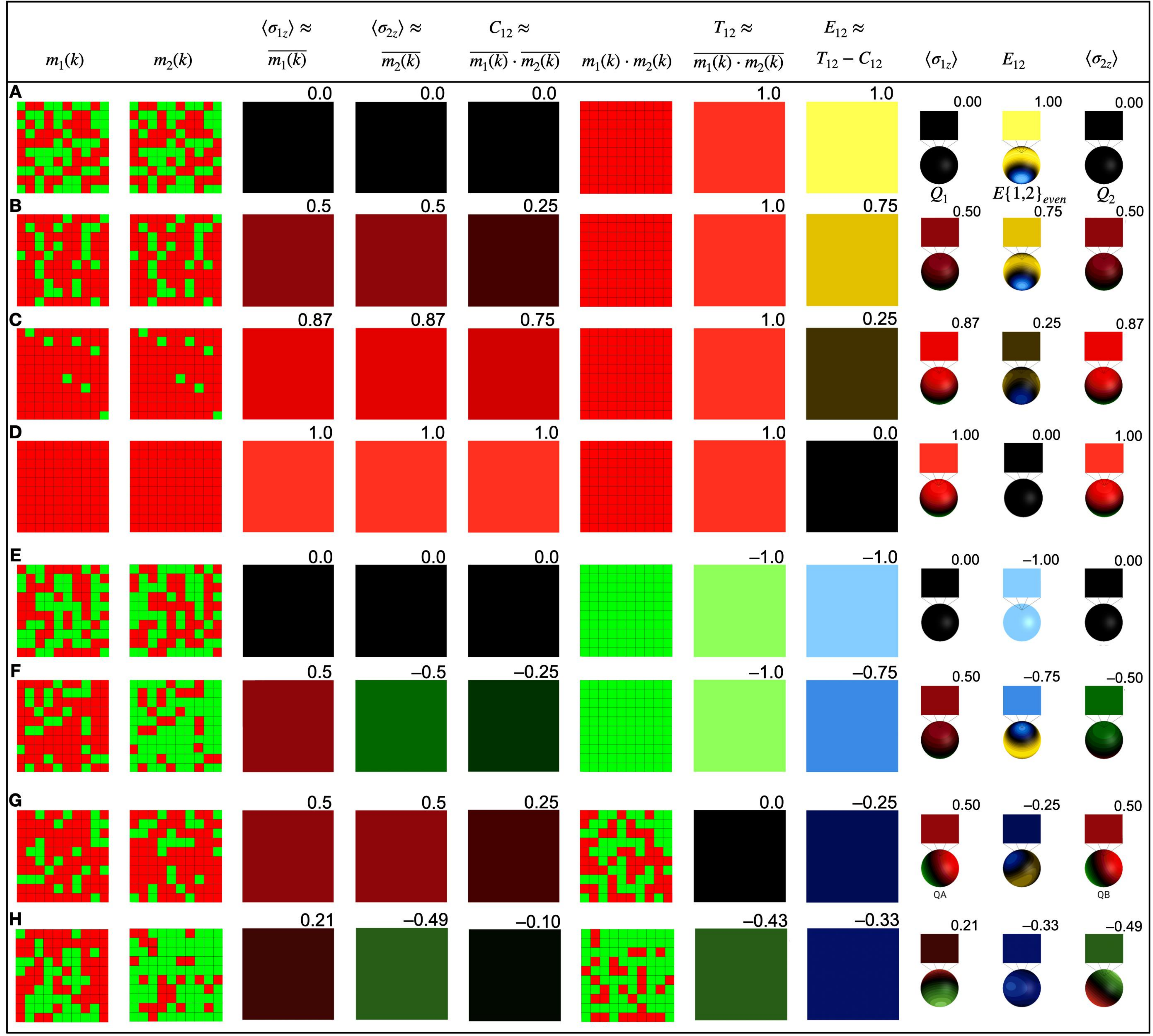}
\caption{\label{fig:FigUrsell}Separation of correlation function components in two-qubit pure states based on simulated measurement outcomes \screen. Total and compound correlation functions ($T_{12}$ and $C_{12}$) in a two-qubit system can be calculated based on measurement outcomes $m_1\left(k\right)$ and $m_2\left(k\right)$. Here, the total zz-correlation $T_{12}\approx\overline{m_1\left(k\right)\cdot m_2\left(k\right)}$ can be approximated from the individual computational basis measurement outcomes as introduced in Fig.~\ref{Figure:Fig2}. All values are given as correlation coefficients (expectation values), i.e., bright red color corresponds to a value of 1 and green to $-1$. The entanglement-based connected two-qubit zz-correlation coefficient $E_{12}$ is then obtained by subtracting $C_{12}$ from $T_{12}$. The visualized states are: The Schmidt form states (\textbf{A}) $\ket{\psi_A}=1/\sqrt2\left(\ket{00}+\ket{11}\right)$, (\textbf{B}) $\ket{\psi_B}=\cos{(\pi/6)}\ket{00}+\sin{(\pi/6)}\ket{11}$, (\textbf{C}) $\ket{\psi_C}=\cos{(\pi/12)}\ket{00}+\sin{(\pi/12)}\ket{11}$, and (\textbf{D}) $\ket{\psi_D}=\ket{00}$, the singlet state (\textbf{E}) $\ket{\psi_E}=1/\sqrt2\left(\ket{01}-\ket{10}\right)$, (\textbf{F}) the generalized Schmidt form state $\ket{\psi_F}=\cos{(\pi/6)}\ket{01}+\sin{(\pi/6)}\ket{10}$, (\textbf{G}) $\ket{\psi_G}=1/\sqrt2\ket{00}+1/2\left(\ket{01} +\ket{10}\right)$, and (\textbf{H}) $\ket{\psi_H}=\left(-0.25090-0.099137i,0.622241+0.377673i,-0.143036-0.404289i,-0.349897\right.$ $\left.+0.301740i\right)^T$.}
\end{figure}

\begin{figure}[H]
\centering
\includegraphics[width=.8\textwidth]{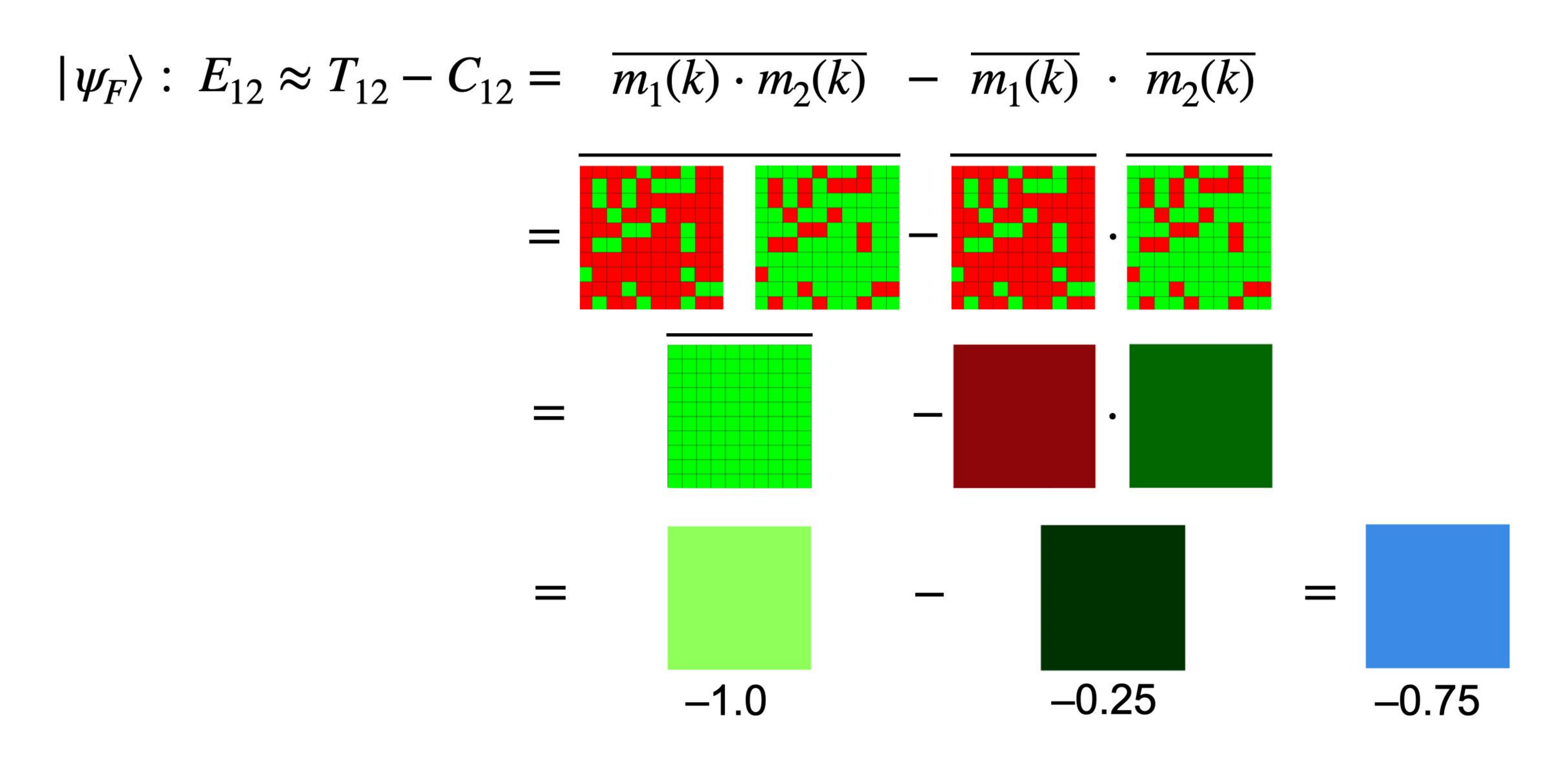}
\caption{\label{fig:FigUrsellCalc}Schematic graphical calculation of the connected zz-correlation \textcolor{red}{coefficient} $E_{12}$ for the state (\textbf{F})~$\ket{\psi_F}=\cos{(\pi/6)}\ket{01}+\sin{(\pi/6)}\ket{10}$ visualized in Fig.~\ref{fig:FigUrsell}. The connected correlation coefficient is obtained by subtracting the compound correlation coefficient which, for a finite number of measurements, is approximately given by the product of averages of single-qubit measurement outcomes $C_{12}\approx\overline{m_1(k)}\cdot\overline{m_2(k)}$, from the total correlation $T_{12}\approx\overline{m_1(k)\cdot m_2(k)}$. The corresponding colors of $T_{12}$ and $C_{12}$ can be approximated as was explained in Fig.~\ref{Figure:Fig2} of section~\ref{LightInt}.}
\end{figure}

To further clarify the concepts introduced in section~\ref{sec:BEADSCorrel}, Fig.~\ref{fig:FigUrsell} provides a graphical illustration how entanglement-based $n$-qubit connected correlations can be determined experimentally. Two-qubit states are exemplarily visualized for which computational basis measurement outcomes $m_A(k)$ and $m_B(k)$ were simulated based on which the total and the compound correlation coefficients $T_{12}$ and $C_{12}$ can be approximated. The connected correlation coefficient $E_{12}$ can then be calculated as the difference between the total and compound correlation coefficients according to Eq.~\ref{eq:CoreelCoeff2Q}.
Additionally, an exemplary schematic graphical calculation is shown in Fig.~\ref{fig:FigUrsellCalc}.

\subsection{Asymmetric correlations}\label{sec:BEADSAsym}
In quantum computing, measurements are commonly performed along the z-axis. Other measurement directions, if required by the experiment or algorithm, with Bell tests (see section~\ref{sec:BEADSBellTest}) being a prominent example, are not always directly achievable in some hardware architectures.
In general, the expectation value of a particular measurement is given by the expectation value of a measurement operator $O$:
\begin{equation}
\left\langle O \right\rangle = \left\langle \psi | O | \psi \right\rangle = \text{Tr}\left(\rho O\right).
\end{equation}

\noindent
As is shown in section~\ref{sec:BEADSBellTest}, it is always possible to transform the state of the qubits that are to be measured by local rotations such that the intended measurement can be implemented as a z- (or any other fully symmetric) measurement.

Nonetheless, as BEADS is a complete visualization, the question arises how, when being given arbitrary measurement directions for a set of qubits, to deduce $\left\langle O \right\rangle$, or the closely related bit parity probability, solely based on the BEADS representation of the qubit system state. 

Recall that in case of fully symmetric measurements, i.e., all qubits are measured along the same direction, it is sufficient to directly read off the correlation coefficient from the corresponding fully permutationally symmetric T-Bead (or E-Bead in case of maximally entangled states) along the measurement direction (see section~\ref{2Q}). In the following, we show how the BEADS representation can be used to obtain correlation expectation values of asymmetric measurements. 

\textbf{\textit{Note that we present this as a proof of principle rather than a recommended strategy to determine correlations of outcomes of asymmetric measurements, as alternative approaches such as applying local transformations to perform symmetric measurements are directly and intuitively visualizable in the BEADS representation (see section~\ref{sec:BEADSBellTest}).}}

As was pointed out in section~\ref{2Q}, T-Beads which do not correspond to fully permutation symmetric components cannot be interpreted directly. For such beads, the spherical function value along an arbitrary direction generally represents correlation coefficients of joint measurements, i.e., of a linear combination of observables. For example, the value of the antisymmetric bead $T\{1,2\}_{\textit{odd}}$ along the z-axis corresponds to $\left(\sigma_{1x}\sigma_{2y}-\sigma_{1y}\sigma_{2x}\right)/2$, and thus, includes the expectation values of two different products of local observables. This also implies that because of these multiple contributions, translating the expectation values obtained from such beads to bit parity probabilities (see section~\ref{2Q}) is mathematically feasible, yet the resulting values do only have indirect physical implications for distinct projective measurements. Instead, such T-Bead values can be applied in the calculation of correlation coefficients of asymmetric measurements, i.e., measurements for which the individual measurement directions do not co-align.

\subsubsection*{Two-qubit measurements}
For any symmetric or asymmetric combination of measurement directions, we can deduce the correlation coefficient by formulating the corresponding measurement operator. In case of a two-qubit measurement of an arbitrary bilinear subsystem $\{k,l\}$ with measurement directions ${\vec{r}}_k=\left(r_{kx},r_{ky},r_{kz}\right)^T$ corresponding to spherical coordinates $(\theta_k,\phi_k)$ and ${\vec{r}}_l=\left(r_{lx},r_{ly},r_{lz}\right)^T$ corresponding to $(\theta_l,\phi_l)$, it is possible to express the corresponding measurement operator
$O=\sigma_{{k}\vec{r}_k}\sigma_{{l}\vec{r}_l}$ in terms of Pauli operators as
\begin{align}
O&=\left(r_{kx}\sigma_{kx}+r_{ky}\sigma_{ky}+r_{kz}\sigma_{kz}\right)\left(r_{lx}\sigma_{lx}+r_{ly}\sigma_{ly}+r_{lz}\sigma_{lz}\right)\nonumber\\
&=r_{kz}r_{lz}\ \sigma_{kz}\sigma_{lz}+\ r_{kx}r_{lz}\ \sigma_{kx}\sigma_{lz}+r_{ky}r_{lz}\ \sigma_{ky}\sigma_{lz}\nonumber\\
&+r_{kz}r_{lx}\ \sigma_{kz}\sigma_{lx}+r_{kx}r_{lx}\ \sigma_{kx}\sigma_{lx}+r_{ky}r_{lx}\ \sigma_{ky}\sigma_{lx}\nonumber\\
&+r_{kz}r_{ly}\ \sigma_{kz}\sigma_{ly}+r_{kx}r_{ly}\ \sigma_{kx}\sigma_{ly}+r_{ky}r_{ly}\ \sigma_{ky}\sigma_{ly},
\end{align}
\noindent
which can be rewritten in terms of spherical coordinates and unscaled LISA tensor operators $T_{j,m}^{\left(\ell^\prime \right)}$:
\begin{align}
O&=\cos{\theta_1}\cos{\theta_2}\ \left(\frac{1}{\sqrt3}T_{0,0}^{\left\{k,l\right\}_{\textit{even}}}+\frac{2}{\sqrt6}T_{2,0}^{\left\{k,l\right\}_{\textit{even}}}\right)\nonumber\\
&+\sin{\theta_1}\cos{\phi_1}\cos{\theta_2}\ \left(\frac{1}{\sqrt2}T_{2,1}^{\left\{k,l\right\}_{\textit{even}}}-\frac{1}{\sqrt2}T_{1,-1}^{\left\{k,l\right\}_{\textit{odd}}}\right)\nonumber\\
&+\sin{\theta_1}\sin{\phi_1}\cos{\theta_2}\ \left(\frac{1}{\sqrt2}T_{2,-1}^{\left\{k,l\right\}_{\textit{even}}}+\frac{1}{\sqrt2}T_{1,1}^{\left\{k,l\right\}_{\textit{odd}}}\right)\nonumber\\
&+\cos{\theta_1\sin{\theta_2\cos{\phi_2}}}\left(\frac{1}{\sqrt2}T_{2,1}^{\left\{k,l\right\}_{\textit{even}})}+\frac{1}{\sqrt2}T_{1,-1}^{\left\{k,l\right\}_{\textit{odd}}}\right)\nonumber\\
&+\sin{\theta_1}{\cos{\phi_1}\sin{\theta_2\cos{\phi_2}}}\left(\frac{1}{\sqrt3}T_{0,0}^{\left\{k,l\right\}_{\textit{even}}}-\frac{1}{\sqrt6}T_{2,0}^{\left\{k,l\right\}_{\textit{even}}}+\frac{1}{\sqrt2}T_{2,2}^{\left\{k,l\right\}_{\textit{even}}}\right)\nonumber\\
&+\sin{\theta_1}{\sin{\phi_1}\sin{\theta_2\cos{\phi_2}}}\left(\frac{1}{\sqrt2}T_{2,-2}^{\left\{k,l\right\}_{\textit{even}}}-\frac{1}{\sqrt2}T_{1,0}^{\left\{k,l\right\}_{\textit{odd}}}\right)\nonumber\\
&+\cos{\theta_1}\sin{\theta_2}\sin{\phi_2}\left(\frac{1}{\sqrt2}T_{2,-1}^{\left\{k,l\right\}_{\textit{even}}}-\frac{1}{\sqrt2}T_{1,1}^{\left\{k,l\right\}_{\textit{odd}}}\right)\nonumber\\
&+ \sin{\theta_1} \cos{\phi_1} \sin{\theta_2} \sin{\phi_2} \left(\frac{1}{\sqrt2} 
T_{2,-2}^{\left\{1,2\right\}_{\textit{even}}} + \frac{1}{\sqrt2} T_{1,0}^{\left\{k,l\right\}_{\textit{odd}}} \right)\nonumber\\
&+ \sin{\theta_1} \sin{\phi_1} \sin{\theta_2} \sin{\phi_2} \left(\frac{1}{\sqrt3}T_{0,0}^{\left\{k,l\right\}_{\textit{even}}}-\frac{1}{\sqrt6}T_{2,0}^{\left\{k,l\right\}_{\textit{even}}}-\frac{1}{\sqrt2}T_{2,2}^{\left\{k,l\right\}_{\textit{even}}}\right).
\end{align}

\noindent
We further include the previously introduced BEADS scaling factors for each tensor operator (see section~\ref{BEADSMap}) and form the differences $\mathrm{\Delta}_\theta\ =\ \theta_k-\theta_l$ and $\mathrm{\Delta}_\phi=\phi_k-\phi_l$  as well as the sums
 $\mathrm{\Sigma}_\theta\ =\theta_k+\theta_l$ and $\mathrm{\Sigma}_\phi=\phi_k+\phi_l$ of polar and azimuthal angles, respectively. Hence, the expectation value of $O$ reads
 \begin{align}\label{eq:AsymMaster2Q}
\left\langle O \right\rangle &= \left\langle T_{0,0}^{\left\{k,l\right\}_{\textit{even}}\prime} \right\rangle\left[\cos{\Delta_\theta}\cos^2{\frac{\Delta_\phi}{2}}+\cos{\Sigma_\theta}\sin^2{\frac{\Delta_\phi}{2}}\right]\nonumber\\
 &+ \left\langle T_{1,-1}^{\left\{k,l\right\}_{\textit{odd}}\prime} \right\rangle \left[\sin{\Sigma_\theta}\sin{\frac{\Delta_\phi}{2}}\sin{\frac{\Sigma_\phi}{2}}-\sin{\Delta_\theta}\cos{\frac{\Delta_\phi}{2}}\cos{\frac{\Sigma_\phi}{2}}\right]\nonumber\\
 &+ \left\langle T_{1,0}^{\left\{k,l\right\}_{\textit{odd}}\prime} \right\rangle \left[\frac{\sin{\Delta_\phi}}{2}\left(\cos{\Sigma_\theta}-\cos{\Delta_\theta}\right)\right] \nonumber\\
 &+ \left\langle T_{1,1}^{\left\{k,l\right\}_{\textit{odd}}\prime} \right\rangle \left[\sin{\Delta_\theta}\sin{\frac{\Sigma_\phi}{2}}\cos{\frac{\Delta_\phi}{2}}+\sin{\Sigma_\theta}\sin{\frac{\Delta_\phi}{2}}\cos{\frac{\Sigma_\phi}{2}}\right] \nonumber\\
 &+ \left\langle T_{2,-2}^{\left\{k,l\right\}_{\textit{even}}\prime} \right\rangle \left[\frac{\sin{\Sigma_\phi}}{2}\left(\cos{\Delta_\theta}-\cos{\Sigma_\theta}\right)\right] \nonumber\\
 &+ \left\langle T_{2,-1}^{\left\{k,l\right\}_{\textit{even}}\prime} \right\rangle \left[\sin{\Delta_\theta}\sin{\frac{\Delta_\phi}{2}}\cos{\frac{\Sigma_\phi}{2}}+\sin{\Sigma_\theta}\sin{\frac{\Sigma_\phi}{2}}\cos{\frac{\Delta_\phi}{2}}\right] \nonumber\\
 &+ \left\langle T_{2,0}^{\left\{k,l\right\}_{\textit{even}}\prime} \right\rangle \left[\frac{\cos{\Delta_\theta}}{2}\left(1-\frac{\cos{\Delta_\phi}}{2}\right)+\frac{\cos{\Sigma_\theta}}{2}\left(1+\frac{\cos{\Delta_\phi}}{2}\right)\right] \nonumber\\
 &+ \left\langle T_{2,1}^{\left\{k,l\right\}_{\textit{even}}\prime} \right\rangle \left[-\sin{\Delta_\theta}\sin{\frac{\Delta_\phi}{2}}\sin{\frac{\Sigma_\phi}{2}}+\sin{\Sigma_\theta}\cos{\frac{\Delta_\phi}{2}}\cos{\frac{\Sigma_\phi}{2}}\right] \nonumber\\
 &+ \left\langle T_{2,2}^{\left\{k,l\right\}_{\textit{even}}\prime} \right\rangle \left[\frac{\cos{\Sigma_\phi}}{2}\left(\cos{\Delta_\theta}-\cos{\Sigma_\theta}\right)\right].
 \end{align}
 
\noindent
Here, $\left\langle T_{j,m}^{(\ell^\prime)\prime} \right\rangle$ denotes scaled expectation values of the LISA tensor operator $T_{j,m}^{\left(\ell^\prime \right)}$ that can be obtained from the corresponding bead. Note that these expectation values are, in general, not to be confused with those of the rotated scaled axial tensor operators $R_{\alpha\beta}\ {\widetilde{T}}_{j,0}^{\left(\ell^\prime\right)}$ introduced in section~\ref{sec:BEADSTomography} which are applied to reconstruct the entire BEADS representation experimentally. Here, we aim to isolate the individual tensor operator components that, when deduced from the corresponding spherical function, may include further scaling factors specific to the tensor operator order, respectively.

Expectation values of rank $j=1$ components can be directly obtained by reading off the spherical function values of the antisymmetric bead $T\{k,l\}_{\textit{odd}}$ along the following Cartesian coordinate axes. $\left\langle T_{1,-1}^{\left\{k,l\right\}_{\textit{odd}}\prime} \right\rangle$ is given by the value of $T\{k,l\}_{\textit{odd}}$ along the y-axis, $\left\langle T_{1,0}^{\left\{k,l\right\}_{\textit{odd}}\prime} \right\rangle$ is given by value of $T\{k,l\}_{\textit{odd}}$ along the z-axis, and $\left\langle T_{1,1}^{\left\{k,l\right\}_{\textit{odd}}\prime} \right\rangle$ is given by the value of $T\{k,l\}_{\textit{odd}}$ along the x-axis.

To determine the expectation values of all $j=\left\{0,\ 2\right\}$ components, we can take (linearly independent) values of the symmetric bead $T\{k,l\}_{\textit{even}}$ along six angular directions which involve the z-axis $R_z=(0^\circ,0^\circ)$, the x-axis $R_x=(90^\circ,0^\circ)$, the y-axis $R_y=(90^\circ,90^\circ)$, the xy-bisecting axis $R_{xy}=(90^\circ,45^\circ)$, the xz bisecting axis $R_{xz}=(45^\circ,0^\circ)$, and the yz-bisecting axis $R_{yz}=(45^\circ,90^\circ)$. The scaled tensor operator component expectation values can then be determined by solving the underlying system of equations which yields the solutions:
\begin{subequations}\label{eq:AsymTVals2Q}
\begin{align}
\left\langle T_{0,0}^{\left\{k,l\right\}_{\textit{even}}\prime} \right\rangle &= \frac{1}{3}\left(R_x+R_y+R_z\right), \\
\left\langle T_{2,-2}^{\left\{k,l\right\}_{\textit{even}}\prime} \right\rangle &=R_{xy}-\frac{1}{2}\left(R_x+R_y\right),  \\
\left\langle T_{2,-1}^{\left\{k,l\right\}_{\textit{even}}\prime} \right\rangle &= R_{yz}-\frac{1}{2}\left(R_y+R_z\right), \\
\left\langle T_{2,0}^{\left\{k,l\right\}_{\textit{even}}\prime} \right\rangle &= \frac{1}{3}\left(2R_z-R_x-R_y\right),\ \  \\
\left\langle T_{2,1}^{\left\{k,l\right\}_{\textit{even}}\prime} \right\rangle &= R_{xz}-\frac{1}{2}\left(R_x+R_z\right),  \\
\left\langle T_{2,2}^{\left\{k,l\right\}_{\textit{even}}\prime} \right\rangle &= \frac{1}{2}\left(R_x-R_y\right).
\end{align}
\end{subequations}
Based on Eq.~\ref{eq:AsymMaster2Q} we can then find expressions of special cases for which the calculation simplifies greatly.\newline

\noindent
\underline{Measurements in the transverse (xy) plane}\newline

\noindent
Measuring two qubits in the transverse plane implies that $\theta_1=\theta_2=90^\circ$, $\Delta_\theta=0^\circ$, and
$\Sigma_\theta=180^\circ$. Thus, Eq.~\ref{eq:AsymMaster2Q} reduces to 
\begin{align}
\left\langle O_{xy} \right\rangle =&\cos{\Delta_\phi}\ \left(\left\langle T_{0,0}^{\left\{k,l\right\}_{\textit{even}}\prime} \right\rangle-\left\langle T_{2,0}^{\left\{k,l\right\}_{\textit{even}}\prime} \right\rangle\right)-\sin{\Delta_\phi \left\langle T_{1,0}^{\left\{k,l\right\}_{\textit{odd}}\prime} \right\rangle}\nonumber\\
+&\sin{\Sigma_\phi}\left\langle T_{2,-2}^{\left\{k,l\right\}_{\textit{even}}\prime} \right\rangle+\cos{\Sigma_\phi}\left\langle T_{2,2}^{\left\{k,l\right\}_{\textit{even}}\prime} \right\rangle.
\end{align}\newpage

\noindent
\underline{Measurements in vertical half planes}\newline

\noindent
In case of two-qubit measurements with measurement directions that span a closed vertical half plane delimited by the z-axis, $\phi=\phi_1=\phi_2$, $\Delta_\phi=0$, and $\Sigma_\phi=2\phi$ for which Eq.~\ref{eq:AsymMaster2Q} becomes
\begin{align}\label{eq:VerticalPlane}
\left\langle O_{\textit{vert}} \right\rangle =&\cos{\Delta_\theta}\left(\left\langle T_{0,0}^{\left\{k,l\right\}_{\textit{even}}\prime} \right\rangle+\frac{1}{4}\left\langle T_{2,0}^{\left\{k,l\right\}_{\textit{even}}\prime} \right\rangle\right)+\frac{3\cos{\Sigma_\theta}}{4}\left\langle T_{2,0}^{\left\{k,l\right\}_{\textit{even}}\prime} \right\rangle\nonumber\\
+&\sin{\Delta_\theta}\left(\sin{\phi}\left\langle T_{1,1}^{\left\{k,l\right\}_{\textit{odd}}\prime} \right\rangle-\cos{\phi}\left\langle T_{1,-1}^{\left\{k,l\right\}_{\textit{odd}}\prime} \right\rangle\right)\nonumber\\
+&\sin{\Sigma_\theta}\left(\sin{\phi}\left\langle T_{2,-1}^{\left\{k,l\right\}_{\textit{even}}\prime} \right\rangle+\cos{\phi}\left\langle T_{2,1}^{\left\{k,l\right\}_{\textit{even}}\prime} \right\rangle\right)\nonumber\\
+&\frac{\cos{\Delta_\theta}-\cos{\Sigma_\theta}}{2}\left(\sin{2\phi}\left\langle T_{2,-2}^{\left\{k,l\right\}_{\textit{even}}\prime} \right\rangle+\cos{2\phi}\left\langle T_{2,2}^{\left\{k,l\right\}_{\textit{even}}\prime} \right\rangle\right).
\end{align}
In Bell tests (section~\ref{sec:BEADSBellTest}), measurements are commonly implemented such that one qubit is measured along the z-axis $(\theta_1=\ 0^\circ)$ or x-axis $(\theta_1=90^\circ)$  while the second qubit is measured along a tilted axis $(\theta_2\ =\ 45^\circ)$ or $\theta_2=135^\circ)$ and $\phi_1=\phi_2=0$ which all corresponds to measurement directions in the xz-half plane. Hence, for measurements in the ($\pm$x)z-half planes we can further simplify Eq.~\ref{eq:VerticalPlane} which gives 
\begin{align}\label{eq:MeasXZ}
\left\langle O_{xz} \right\rangle =&\cos{\Delta_\theta}\left(\left\langle T_{0,0}^{\left\{k,l\right\}_{\textit{even}}\prime} \right\rangle+\frac{\left\langle T_{2,0}^{\left\{k,l\right\}_{\textit{even}}\prime} \right\rangle}{4}+\frac{\left\langle T_{2,2}^{\left\{k,l\right\}_{\textit{even}}\prime} \right\rangle}{2}\right) \nonumber\\
+&\cos{\Sigma_\theta}\left(\frac{3\left\langle T_{2,0}^{\left\{k,l\right\}_{\textit{even}}\prime} \right\rangle}{4}-\frac{\left\langle T_{2,2}^{\left\{k,l\right\}_{\textit{even}}\prime} \right\rangle}{2}\right) \nonumber\\
\pm&\sin{\Sigma_\theta}\left\langle T_{2,1}^{\left\{k,l\right\}_{\textit{even}}\prime} \right\rangle\mp\sin{\Delta_\theta}\left\langle T_{1,-1}^{\left\{k,l\right\}_{\textit{odd}}\prime} \right\rangle.
\end{align}
\noindent
Similarly, for measurements in the ($\pm$y)z -half plane Eq.~\ref{eq:VerticalPlane} simplifies to
\begin{align}
\left\langle O_{yz} \right\rangle =&\cos{\Delta_\theta}\left(\left\langle T_{0,0}^{\left\{k,l\right\}_{\textit{even}}\prime} \right\rangle+\frac{\left\langle T_{2,0}^{\left\{k,l\right\}_{\textit{even}}\prime} \right\rangle}{4}-\frac{\left\langle T_{2,2}^{\left\{k,l\right\}_{\textit{even}}\prime} \right\rangle}{2}\right) \nonumber\\
+&\cos{\Sigma_\theta}\left(\frac{3\left\langle T_{2,0}^{\left\{k,l\right\}_{\textit{even}}\prime} \right\rangle}{4}-\frac{\left\langle T_{2,2}^{\left\{k,l\right\}_{\textit{even}}\prime} \right\rangle}{2}\right) \nonumber\\
\pm&\sin{\Sigma_\theta}\left\langle T_{2,-1}^{\left\{k,l\right\}_{\textit{even}}\prime} \right\rangle\pm\sin{\Delta_\theta}\left\langle T_{1,1}^{\left\{k,l\right\}_{\textit{odd}}\prime} \right\rangle.
\end{align}

\noindent
\subsubsection*{Three-qubit measurements}
Tripartite correlations of measurement outcomes can be determined in an analogue fashion. If one aims to measure three qubits $\{k,l,m\}$ in three independent arbitrary measurement directions $\vec{r}_k=\left(r_{kx},r_{ky},r_{kz}\right)^T$, $\vec{r}_l=\left(r_{lx},r_{ly},r_{lz}\right)^T$, and $\vec{r}_m=\left(r_{mx},r_{my},r_{mz}\right)^T$, the corresponding correlation coefficients can then, based on the BEADS representation of the system state prior to the measurement, be determined by the following equation. Here, using angular directions in terms of spherical coordinates does not provide any benefit with respect to simplifications. Hence, all expressions are given in terms of Cartesian components.
\begin{align}
\left\langle O \right\rangle = &\left\langle T_{1,-1}^{\left\{k,l,m\ \tau_1\right\}_{\textit{odd}}\prime} \right\rangle \bigg[\frac{1}{3}\left(r_{ky}r_{lz}r_{3m}+r_{kz}r_{ly}r_{mz}+r_{kz}r_{lz}r_{my}\right)\nonumber\\
&\phantom{\left\langle T_{1,0}^{\left\{k,l,m\ \tau_1\right\}_{\textit{odd}}\prime} \right\rangle}+\left(r_{ky}r_{lx}r_{mx}+r_{kx}r_{ly}r_{mx}+r_{kx}r_{lx}r_{my}\right)+\ r_{ky}r_{ly}r_{my}\bigg] \nonumber\\
+&\left\langle T_{1,0}^{\left\{k,l,m\ \tau_1\right\}_{\textit{odd}}\prime} \right\rangle \bigg[\frac{1}{3}\left(r_{kz}r_{lx}r_{mx}\ +\ r_{kx}r_{lz}r_{mx}\ +\ r_{kx}r_{lx}r_{mz}\right)\nonumber\\
&\phantom{\left\langle T_{1,0}^{\left\{k,l,m\ \tau_1\right\}_{\textit{odd}}\prime} \right\rangle}+\left(r_{kz}r_{ly}r_{my}+r_{ky}r_{lz}r_{my}+r_{ky}r_{ly}r_{mz}\right)+r_{kz}r_{lz}r_{mz}\bigg] \nonumber\\
+&\left\langle T_{1,1}^{\left\{k,l,m\ \tau_1\right\}_{\textit{odd}}\prime} \right\rangle \bigg[\frac{1}{3}\left(r_{kx}r_{lz}r_{mz}+r_{kz}r_{lx}r_{mz}+r_{kz}r_{lz}r_{mx}\right)\nonumber\\
&\phantom{\left\langle T_{1,0}^{\left\{k,l,m\ \tau_1\right\}_{\textit{odd}}\prime} \right\rangle}+\left(r_{kx}r_{ly}r_{my}+r_{ky}r_{ly}r_{mx}+r_{ky}r_{lx}r_{my}\right)+r_{kx}r_{lx}r_{mx}\bigg] \nonumber\\
+&\left\langle T_{3,-3}^{\left\{k,l,m\ \tau_1\right\}_{\textit{odd}}\prime} \right\rangle \bigg[\left(r_{ky}r_{lx}r_{mx}+r_{kx}r_{ly}r_{mx}+r_{kx}r_{lx}r_{my}\right)-r_{ky}r_{ly}r_{my}\bigg]\nonumber\\
+&\left\langle T_{3,-2}^{\left\{k,l,m\ \tau_1\right\}_{\textit{odd}}\prime} \right\rangle \frac{\sqrt3}{2}\bigg[r_{ky}r_{lx}r_{mz}+r_{kx}r_{ly}r_{mz}+r_{ky}r_{lz}r_{mx}\nonumber\\
&\phantom{\left\langle T_{1,0}^{\left\{k,l,m\ \tau_1\right\}_{\textit{odd}}\prime} \right\rangle\frac{\sqrt3}{2}}+r_{kz}r_{ly}r_{mx}+r_{kz}r_{lx}r_{my}+r_{kx}r_{lz}r_{my}\bigg]\nonumber\\
+&\left\langle T_{3,-1}^{\left\{k,l,m\ \tau_1\right\}_{\textit{odd}}\prime} \right\rangle \frac{1}{3}\bigg[-\left(r_{ky}r_{lx}r_{mx}+r_{kx}r_{ly}r_{mx}+r_{kx}r_{lx}r_{my}\right)-3r_{ky}r_{ly}r_{my}\nonumber\\
&\phantom{\left\langle T_{1,0}^{\left\{k,l,m\ \tau_1\right\}_{\textit{odd}}\prime} \right\rangle\frac{1}{3}}+4\left(r_{ky}r_{lz}r_{mz}+r_{kz}r_{ly}r_{mz}+r_{kz}r_{lz}r_{my}\right)\bigg]\nonumber\\
+&\left\langle T_{3,0}^{\left\{k,l,m\ \tau_1\right\}_{\textit{odd}}\prime} \right\rangle\frac{1}{2}\bigg[-\left(r_{kz}r_{lx}r_{mx}+r_{kx}r_{lx}r_{mz}+r_{kx}r_{lz}r_{mx}\right)\nonumber\\
&\phantom{\left\langle T_{1,0}^{\left\{k,l,m\ \tau_1\right\}_{\textit{odd}}\prime} \right\rangle\frac{1}{2}}-\left(r_{kz}r_{ly}r_{my}+r_{ky}r_{lz}r_{my}+r_{ky}r_{ly}r_{mz}\right)+2r_{kz}r_{lz}r_{mz}\bigg]\nonumber\\
+&\left\langle T_{3,1}^{\left\{k,l,m\ \tau_1\right\}_{\textit{odd}}\prime} \right\rangle \frac{1}{3}\bigg[-\left(r_{kx}r_{ly}r_{my}+r_{ky}r_{lx}r_{my}+r_{ky}r_{ly}r_{mx}\right)-3r_{kx}r_{lx}r_{mx}\nonumber\\
&\phantom{\left\langle T_{1,0}^{\left\{k,l,m\ \tau_1\right\}_{\textit{odd}}\prime} \right\rangle\frac{1}{3}}+4\left(r_{kx}r_{lz}r_{mz}+r_{kz}r_{lx}r_{mz}+r_{kz}r_{lz}r_{mx}\right)\bigg]\nonumber\\
+&\left\langle T_{3,2}^{\left\{k,l,m\ \tau_1\right\}_{\textit{odd}}\prime} \right\rangle \frac{\sqrt3}{2}\bigg[\left(r_{kz}r_{lx}r_{mx}+r_{kx}r_{lz}r_{mx}+r_{kx}r_{lx}r_{mz}\right)\nonumber\\
&\phantom{\left\langle T_{1,0}^{\left\{k,l,m\ \tau_1\right\}_{\textit{odd}}\prime} \right\rangle\frac{\sqrt3}{2}}-\left(r_{ky}r_{ly}r_{mz}+r_{ky}r_{lz}r_{my}+r_{kz}r_{ly}r_{my}\right)\bigg]\nonumber\\
+&\left\langle T_{3,3}^{\left\{k,l,m\ \tau_1\right\}_{\textit{odd}}\prime} \right\rangle \bigg[-r_{kx}r_{ly}r_{my}-r_{ky}r_{lx}r_{my}-r_{ky}r_{ly}r_{mx}+r_{kx}r_{lx}r_{mx}\bigg]\nonumber\\
+&\left\langle T_{1,-1}^{\left\{k,l,m\ \tau_2\right\}_{\textit{odd}}\prime} \right\rangle \frac{1+\sqrt3}{6}\bigg[-2\left(r_{kz}r_{lz}r_{my}+r_{kx}r_{lx}r_{my}\right)\ +\left({\ r}_{ky}r_{lz}r_{mz}+r_{kz}r_{ly}r_{mz}\right)\nonumber\\
&\phantom{\left\langle T_{1,0}^{\left\{k,l,m\ \tau_1\right\}_{\textit{odd}}\prime} \right\rangle}+\left(r_{ky}r_{lx}r_{mx}+r_{kx}r_{ly}r_{mx}\right)\bigg] \nonumber\\
+&\left\langle T_{1,0}^{\left\{k,l,m\ \tau_2\right\}_{\textit{odd}}\prime} \right\rangle \frac{1+\sqrt3}{6}\bigg[-2\left(r_{kx}r_{lx}r_{mz}+r_{ky}r_{ly}r_{mz}\right)+\left(r_{kz}r_{lx}r_{mx}+r_{kx}r_{lz}r_{mx}\right)\nonumber\\
&\phantom{\left\langle T_{1,0}^{\left\{k,l,m\ \tau_1\right\}_{\textit{odd}}\prime} \right\rangle}+\left(r_{kz}r_{ly}r_{my}+r_{ky}r_{lz}r_{my}\right)\bigg] \nonumber\\
+&\left\langle T_{1,1}^{\left\{k,l,m\ \tau_2\right\}_{\textit{odd}}\prime} \right\rangle \frac{1+\sqrt3}{6}\bigg[-2\left(r_{kz}r_{lz}r_{mx}+r_{ky}r_{ly}r_{mx}\right)+\left(r_{kx}r_{ly}r_{my}+r_{ky}r_{lx}r_{my}\right)\nonumber\\
&\phantom{\left\langle T_{1,0}^{\left\{k,l,m\ \tau_1\right\}_{\textit{odd}}\prime} \right\rangle}+\left(r_{kx}r_{lz}r_{mz}+r_{kz}r_{lx}r_{mz}\right)\bigg] \nonumber\\
+&\left\langle T_{2,-2}^{\left\{k,l,m\ \tau_2\right\}_{\textit{even}}\prime} \right\rangle \frac{\sqrt2}{3}\bigg[2\left(r_{kx}r_{lx}r_{mz}-r_{ky}r_{ly}r_{mz}\right)-\left(r_{kx}r_{lz}r_{mx}+r_{kz}r_{lx}r_{mx}\right)\nonumber\\
&\phantom{\left\langle T_{1,0}^{\left\{k,l,m\ \tau_1\right\}_{\textit{odd}}\prime} \right\rangle}+\left(r_{ky}r_{lz}r_{my}+r_{kz}r_{ly}r_{my}\right)\bigg]\nonumber\\
+&\left\langle T_{2,-1}^{\left\{k,l,m\ \tau_2\right\}_{\textit{even}}\prime} \right\rangle \frac{\sqrt2}{3}\bigg[2\left(r_{ky}r_{ly}r_{mx}-r_{kz}r_{lz}r_{mx}\right)-\left(r_{ky}r_{lx}r_{my}+r_{kx}r_{ly}r_{my}\right)\nonumber\\
&\phantom{\left\langle T_{1,0}^{\left\{k,l,m\ \tau_1\right\}_{\textit{odd}}\prime} \right\rangle}+\left({r_{kz}r_{lx}r_{mz}\ +r}_{kx}r_{lz}r_{mz}\right)\bigg]\nonumber\\
+&\left\langle T_{2,0}^{\left\{k,l,m\ \tau_2\right\}_{\textit{even}}\prime} \right\rangle \frac{1}{\sqrt2}\bigg[r_{ky}r_{lz}r_{mx}+r_{kz}r_{ly}r_{mx}-\left(r_{kx}r_{lz}r_{my}+r_{kz}r_{lx}r_{my}\right)\bigg]\nonumber\\
+&\left\langle T_{2,1}^{\left\{k,l,m\ \tau_2\right\}_{\textit{even}}\prime} \right\rangle \frac{\sqrt2}{3}\bigg[2\left(r_{kz}r_{lz}r_{my}-r_{kx}r_{lx}r_{my}\right)-\left(r_{kz}r_{ly}r_{mz}+r_{ky}r_{lz}r_{mz}\right)\nonumber\\
&\phantom{\left\langle T_{1,0}^{\left\{k,l,m\ \tau_1\right\}_{\textit{odd}}\prime} \right\rangle}+\left(r_{kx}r_{ly}r_{mx}+r_{ky}r_{lx}r_{mx}\right)\bigg]\nonumber\\
+&\left\langle T_{2,2}^{\left\{k,l,m\ \tau_2\right\}_{\textit{even}}\prime} \right\rangle \frac{\sqrt2}{3}\bigg[-2\left(r_{ky}r_{lx}r_{mz}+r_{kx}r_{ly}r_{mz}\right)+\left(r_{kz}r_{lx}r_{my}+r_{kx}r_{lz}r_{my}\right)\nonumber\\
&\phantom{\left\langle T_{1,0}^{\left\{k,l,m\ \tau_1\right\}_{\textit{odd}}\prime} \right\rangle}+\left(r_{kz}r_{ly}r_{mx}+r_{ky}r_{lz}r_{mx}\right)\bigg]\nonumber
\end{align}
\begin{align}
+&\left\langle T_{1,-1}^{\left\{k,l,m\ \tau_3\right\}_{\textit{odd}}\prime} \right\rangle \frac{1}{\sqrt2}\bigg[\left(r_{ky}r_{lz}r_{mz}-r_{kz}r_{ly}r_{mz}\right)+\left(r_{ky}r_{lx}r_{mx}-r_{kx}r_{ly}r_{mx}\right)\bigg]\nonumber\\
+&\left\langle T_{1,0}^{\left\{k,l,m\ \tau_3\right\}_{\textit{odd}}\prime} \right\rangle \frac{1}{\sqrt2}\bigg[\left(r_{kz}r_{lx}r_{mx}-r_{kx}r_{lz}r_{mx}\right)+\left(r_{kz}r_{ly}r_{my}-r_{ky}r_{lz}r_{my}\right)\bigg]\nonumber\\
+&\left\langle T_{1,1}^{\left\{k,l,m\ \tau_3\right\}_{\textit{odd}}\prime} \right\rangle \frac{1}{\sqrt2}\bigg[\left(r_{kx}r_{lz}r_{mz}-r_{kz}r_{lx}r_{mz}\right)+\left(r_{kx}r_{ly}r_{my}-r_{ky}r_{lx}r_{my}\right)\bigg]\nonumber\\
+&\left\langle T_{2,-2}^{\left\{k,l,m\ \tau_3\right\}_{\textit{even}}\prime} \right\rangle \sqrt{\frac{2}{3}}\bigg[\left(r_{kx}r_{lz}r_{mx}-r_{kz}r_{lx}r_{mx}\right)+\left(r_{kz}r_{ly}r_{my}-r_{ky}r_{lz}r_{my}\right)\bigg]\nonumber\\
+&\left\langle T_{2,-1}^{\left\{k,l,m\ \tau_3\right\}_{\textit{even}}\prime} \right\rangle \sqrt{\frac{2}{3}}\bigg[\left(r_{kx}r_{lz}r_{mz}-r_{kz}r_{lx}r_{mz}\right)+\left(r_{ky}r_{lx}r_{my}-r_{kx}r_{ly}r_{my}\right)\bigg]\nonumber\\
+&\left\langle T_{2,0}^{\left\{k,l,m\ \tau_3\right\}_{\textit{even}}\prime} \right\rangle \frac{1}{\sqrt6}\bigg[2\left(r_{ky}r_{lx}r_{mz}-r_{kx}r_{ly}r_{mz}\right)+\left(r_{ky}r_{lz}r_{mx}-r_{kz}r_{ly}r_{mx}\right)\nonumber\\
&\phantom{\left\langle T_{1,0}^{\left\{k,l,m\ \tau_1\right\}_{\textit{odd}}\prime} \right\rangle}-\left(r_{kx}r_{lz}r_{my}+r_{kz}r_{lx}r_{my}\right)\bigg]\nonumber\\
+&\left\langle T_{2,1}^{\left\{k,l,m\ \tau_3\right\}_{\textit{even}}\prime} \right\rangle \sqrt{\frac{2}{3}}\bigg[\left(r_{kz}r_{ly}r_{mz}-r_{ky}r_{lz}r_{mz}\right)+\left(r_{ky}r_{lx}r_{mx}-r_{kx}r_{ly}r_{mx}\right)\bigg]\nonumber\\
+&\left\langle T_{2,2}^{\left\{k,l,m\ \tau_3\right\}_{\textit{even}}\prime} \right\rangle \sqrt{\frac{2}{3}}\bigg[\left(r_{kz}r_{ly}r_{mx}-r_{ky}r_{lz}r_{mx}\right)+\left(r_{kz}r_{lx}r_{my}-r_{kx}r_{lz}r_{my}\right)\bigg]\nonumber\\
+&\left\langle T_{0,0}^{\left\{k,l,m\ \tau_4\right\}_{\textit{even}}\prime} \right\rangle \frac{1}{\sqrt3}\bigg[\left(r_{kx}r_{ly}r_{mz}-r_{ky}r_{lx}r_{mz}\right)+\left(r_{ky}r_{lz}r_{mx}-r_{kz}r_{ly}r_{mx}\right)\nonumber\\
\label{eq:AsymMaster3Q}&\phantom{\left\langle T_{1,0}^{\left\{k,l,m\ \tau_1\right\}_{\textit{odd}}\prime} \right\rangle}+\left(r_{kz}r_{lx}r_{my}-r_{kx}r_{lz}r_{my}\right)\bigg]
\end{align}

\noindent
Individual scaled expectation values $\left\langle T_{j,m}^{\left(\ell^\prime\right)\prime} \right\rangle$ can again be deduced based on values of the involved beads along specific directions. The $T\{k,l,m\ \tau_1\}_{\textit{odd}}$ T-bead is composed of ten different spherical tensor operator components. Hence, we require ten different (linearly independent) spherical function values of $T\{k,l,m\ \tau_1\}_{\textit{odd}}$ which can be obtained at angular directions $R_z=(0^\circ,0^\circ)$, $R_x=(90^\circ,0^\circ)$, $R_{-y}=(90^\circ,270^\circ)$, $R_{xy}=(90^\circ,45^\circ)$, $R_{\left(-x\right)y}=(90^\circ,135^\circ)$, $R_1=\left(\cos^{-1}{\sqrt{\frac{3}{5}}},0^\circ\right)$, $R_2=\left(\cos^{-1}{\left(-\sqrt{\frac{3}{5}}\right)},0^\circ\right)$, $R_3=\left(\cos^{-1}{\sqrt{\frac{3}{5}}},45^\circ\right)$, $R_4=\left(\cos^{-1}{\left(-\sqrt{\frac{3}{5}}\right)},45^\circ\right)$, and 
$R_5=(\cos^{-1}{\sqrt{\frac{3}{5}}},135^\circ)$ where $\cos^{-1}{\sqrt{\frac{3}{5}}}\approx39.23^\circ$ and $\cos^{-1}{\left(-\sqrt{\frac{3}{5}}\right)}\approx140.77^\circ$. In principle, the choice of angular directions is arbitrary as long the underlying equations specifying the spherical function values along the chosen directions are linearly independent. Here, we chose directions which correspond to roots of the involved spherical harmonics such that the obtained expressions become as simple as possible. The desired expectation values can then be calculated as:\newline

\begin{subequations}\label{eq:AsymTVals3Q}
\vspace{-.5cm}
\begin{flalign}
\left\langle T_{1,-1}^{\left\{k,l,m\ \tau_1\right\}_{\textit{odd}}\prime} \right\rangle &= \frac{2}{15}R_x-\frac{2}{5}R_y+\frac{\sqrt2}{15}R_{xy}+\frac{\sqrt2}{5}R_{\left(-x\right)y}+\frac{\sqrt5}{6}\left(R_3+R_4\right)&&\nonumber\\
&-\frac{\sqrt{10}}{12}\left(R_1+R_2\right),&&\\
\left\langle T_{1,0}^{\left\{k,l,m\ \tau_1\right\}_{\textit{odd}}\prime} \right\rangle &= -\frac{2\sqrt3}{15}R_x+\frac{\sqrt{15}}{6}\left(R_5-R_4\right)+\frac{\sqrt{30}}{12}\left(R_1+R_2\right)&&\nonumber\\
&+\frac{\sqrt6}{15}\left(R_{xy}-R_{\left(-x\right)y}\right),&&\\
\left\langle T_{1,1}^{\left\{k,l,m\ \tau_1\right\}_{\textit{odd}}\prime} \right\rangle &=\frac{4}{15}R_x+\frac{\sqrt2}{5}\left(R_{xy}-R_{\left(-x\right)y}\right)+\frac{\sqrt{10}}{12}\left(R_1+R_2\right),&&\\[11pt]
\left\langle T_{3,-3}^{\left\{k,l,m\ \tau_1\right\}_{\textit{odd}}\prime} \right\rangle &= \frac{1}{2}R_{-y}+\frac{\sqrt2}{4}\left(R_{xy}+R_{\left(-x\right)y}\right),&&\\[11pt]
\left\langle T_{3,-2}^{\left\{k,l,m\ \tau_1\right\}_{\textit{odd}}\prime} \right\rangle &= \frac{2}{9}R_x+\frac{\sqrt2}{9}\left(-R_{xy}+R_{\left(-x\right)y}\right)+\frac{5\sqrt5}{18}\left(R_3-R_5\right)&&\nonumber\\
&-\frac{5\sqrt{10}}{36}\left(R_1+R_2\right),&&\\
\left\langle T_{3,-1}^{\left\{k,l,m\ \tau_1\right\}_{\textit{odd}}\prime} \right\rangle &= \frac{2}{15}R_x+\frac{1}{10}R_y-\frac{11\sqrt2}{60}R_{xy}-\ \frac{\sqrt2}{20}R_{\left(-x\right)y}+\frac{\sqrt5}{6}\left(R_3+R_4\right)&&\nonumber\\
-\frac{\sqrt{10}}{12}\left(R_1+R_2\right),&&\\
\left\langle T_{3,0}^{\left\{k,l,m\ \tau_1\right\}_{\textit{odd}}\prime} \right\rangle &= R_z+\ \frac{2\sqrt3}{15}R_x+\frac{\sqrt{15}}{6}\left(R_4-R_5\right)-\frac{\sqrt{30}}{12}\left(R_1+R_2\right)&&\nonumber\\
&+\frac{\sqrt6}{15}\left({-R}_{xy}+R_{\left(-x\right)y}\right),&&\\
\left\langle T_{3,1}^{\left\{k,l,m\ \tau_1\right\}_{\textit{odd}}\prime} \right\rangle &= -\frac{7}{30}R_x+\frac{\sqrt2}{20}\left(-R_{xy}+R_{\left(-x\right)y}\right)+\frac{\sqrt{10}}{12}\left(R_1+R_2\right),&&\\[11pt]
\left\langle T_{3,2}^{\left\{k,l,m\ \tau_1\right\}_{\textit{odd}}\prime} \right\rangle &= \frac{2}{9}R_x-\frac{5\sqrt5}{18}\left(R_2-R_1-R_4+R_5\right)-\frac{5\sqrt{10}}{36}\left(R_1+R_2\right)&&\nonumber\\
&+\frac{\sqrt2}{9}\left(-R_{xy}+R_{\left(-x\right)y}\right),&&\\
\left\langle T_{3,3}^{\left\{k,l,m\ \tau_1\right\}_{\textit{odd}}\prime} \right\rangle &= \frac{1}{2}R_x+\frac{1}{2\sqrt2}\left({-R}_{xy}+R_{\left(-x\right)y}\right).
\end{flalign}
\end{subequations}
\noindent
The expectation values of rank $j=1$ components of $T\{k,l,m\ \tau_2\}_{\textit{odd}}$  or $T\{k,l,m\ \tau_3\}_{\textit{odd}}$ can be deduced by directly reading off values along the Cartesian coordinate axes as we previously explained in terms of antisymmetric two-qubit measurements.
Rank $j=2$ component expectation values of $T\{k,l,m\ \tau_2\}_{\textit{even}}$ or $T\{k,l,m\ \tau_3\}_{\textit{even}}$ can be calculated from the spherical function values at $R_z=(0^\circ,0^\circ)$, $R_x=(90^\circ,0^\circ)$, $R_y=(90^\circ,90^\circ)$,
$R_{xy}=(90^\circ,45^\circ)$, and $R_{xz}=(45^\circ,0^\circ)$ by the following relations:
\vspace{-.2cm}
\begin{subequations}
\begin{align}
\left\langle T_{2,-2}^{\left\{k,l,m\ \tau_k\right\}_{\textit{even}}\prime} \right\rangle &= R_{xy}+\frac{1}{2}R_z,\\
\left\langle T_{2,-1}^{\left\{k,l,m\ \tau_k\right\}_{\textit{even}}\prime} \right\rangle &= R_{yz}+\frac{1}{2}R_x,\\
\left\langle T_{2,0}^{\left\{k,l,m\ \tau_k\right\}_{\textit{even}}\prime} \right\rangle &= R_z,\\
\left\langle T_{2,1}^{\left\{k,l,m\ \tau_k\right\}_{\textit{even}}\prime} \right\rangle &= R_{xz}-\frac{1}{2}\left(R_x+R_z\right),\\
\left\langle T_{2,2}^{\left\{k,l,m\ \tau_k\right\}_{\textit{even}}\prime} \right\rangle &= R_x+\frac{1}{2}R_z.
\end{align}
\end{subequations}
where $k\in\left\{2,3\right\}$. $T_{0,0}^{\{k,l,m\ \tau_4\}_{\textit{even}}}$ can be directly determined by reading off the value of the corresponding bead along any arbitrary direction.

\newpage
\section{Technical details, concepts, and design}\label{sec:TechDetails}
\subsection{Plotting variants}\label{sec:BEADSPlotting}
In the BEADS representation, three-dimensional colored sphere polar plots are used to visualize spherical functions, that is, the beads functions are plotted on a spherical surface of constant size and the value at any vertex is represented by a color defined by a color scheme (e.g., see section~\ref{sec:BEADSColors}). 

This is indeed different to other finite-dimensional phase-space representations such as DROPS \cite{DROPS} which are commonly plotted such that the magnitude of a spherical function value is represented by the distance from the origin to the spherical function surface in the corresponding direction. The choice of plotting methods is in general arbitrary. However, the approach introduced for the BEADS representation is more suitable to provide a predictive visualization from all viewing perspectives as the colors can be directly assigned to specific expectation values or bit parity probabilities. As shown in Fig~\ref{fig:FigSchmidtRadial} for Schmidt form states, it is still possible to plot the BEADS representation using the magnitude as distance from the origin. Here, it is immediately clear that the zz-correlation coefficient is independent of the Schmidt angle $\theta$ by comparing the sizes of the $T\{1,2\}_{\textit{even}}$ T-Beads although the ratio of connected and compound correlation function components, and thus the color changes for different values of $\theta$. However, determining the exact correlation coefficients along arbitrary directions is not as straightforward as for the standard spherical BEADS plots.
\vspace{-.3cm}
\begin{figure}[H]
\centering
\includegraphics[width=1.\textwidth]{FigSchmidtRadial.pdf}
\caption{\label{fig:FigSchmidtRadial}BEADS representation of two-qubit Schmidt form states plotting the absolute spherical function values of beads as distance from the origin. The visualized states are equivalent to those shown in Fig.~\ref{Figure:Fig6}. Here, the \textit{red-green} and \textit{yellow-blue DROPS} color schemes and the total correlation color blending approach are applied (cf. section~\ref{sec:BEADSColors}). The representation on the left clearly shows in a visual way that the zz-correlation coefficient has a constant value of 1 for all values of $\theta$.}
\vspace{-1.cm}
\end{figure}
 
\noindent
It is also possible to plot the normalized spherical function on a sphere, that is, the original spherical function is scaled such that the maximum absolute function (expectation) value becomes~1. The norm of the (unscaled) density operator component $\rho^{\left(\ell^\prime\right)}=\sum_{j\in J\left(\ell^\prime\right)}\sum_{m=-j}^{j}{c_{j,m}^{\left(\ell^\prime\right)}T_{j,m}^{\left(\ell^\prime\right)}}$ corresponding to a bead $b^{(\ell^\prime)}$ can then be represented by the radius $r$ or volume $V$ of the sphere on which the normalized function is plotted, i.e., the sphere radius is given by
\begin{equation}\label{eq:NormRadius}
r = \left\Vert \rho^{\left(\ell^\prime\right)} \right\Vert =\sqrt{Tr\left(\rho^{\left(\ell^\prime\right)}\rho^{\left(\ell^\prime\right)}\right)}
\end{equation}%
or
\begin{equation}\label{eq:NormVolume}
r=\sqrt[3]{\frac{3}{4\pi}\left\Vert\rho^{\left(\ell^\prime\right)}\right\Vert}=\sqrt[3]{\frac{3}{4\pi}\sqrt{Tr\left(\rho^{\left(\ell^\prime\right)}\rho^{\left(\ell^\prime\right)}\right)}}.
\end{equation}%
This is advantageous in that it always ensures the appearing colors to have full brightness at the maximum absolute function value which would eventually result in darker colors in the standard BEADS plotting approach. However, this renormalization of spherical function values prevents a direct interpretation of the color, i.e., a color cannot be directly assigned to a specific correlation coefficient or bit parity probability without further considering the radius of the sphere on which the function is plotted. Exemplary BEADS visualizations using scaled spheres with radii as described in Eq.~\ref{eq:NormRadius} and~\ref{eq:NormVolume} are shown in Fig.~\ref{fig:FigSphereGrow}.
\vspace{-.3cm}
\begin{figure}[H]
\centering
\includegraphics[width=.97\textwidth]{FigSphereGrow.pdf}
\caption{\label{fig:FigSphereGrow}BEADS representation plotting variants using scaled spheres. Variants of the BEADS representation involve plotting of the spherical functions on spheres with radii corresponding to (\textbf{A}) the norm of the visualized density operator component or (\textbf{B}) chosen such that the sphere volume is equivalent to the density operator component norm which are shown for the state $\ket{\psi} = \cos{\left(\pi/8\right)}\ket{00}+\sin{\left(\pi/8\right)}\ket{11}$. Unlike in the standard plotting method (\textbf{C}), both methods involve a normalization of spherical function values such that the maximum absolute value becomes~1, thus achieving full brightness of colors. For comparison with the previously introduced plotting methods (see Fig~.~\ref{Figure:Fig5}), Schmidt form states $\ket{\psi_\theta}=\cos{\left(\theta/2\right)} \ket{00} + \sin{\left(\theta/2\right)} \ket{11}$ are visualized by using (\textbf{D}) radii calculated according to Eq.~\ref{eq:NormRadius} and (\textbf{E}) according to Eq.~\ref{eq:NormVolume}.}
\vspace{-1.cm}
\end{figure}

\subsection{High contrast BEADS visualizations}\label{app:BEADSHighContrast}
As we briefly discussed in appendix~\ref{sec:BEADSColors}, high contrast color schemes give simple clear BEADS representations due to a reduced complexity of the surface patterns of visualized beads. These color schemes are well-suited for entry level visualizations and can be easily sketched by hand.
In the following, several black and white high contrast BEADS visualizations are provided. Note that Fig.~\ref{fig:FigGatesCyclingHighContrast} and Fig.~\ref{fig:FigTeleportXHighContrast} were previously shown using the standard BEADS color schemes.
\vspace{-.5cm}
\begin{figure}[H]
\centering
\includegraphics[width=0.55\textwidth]{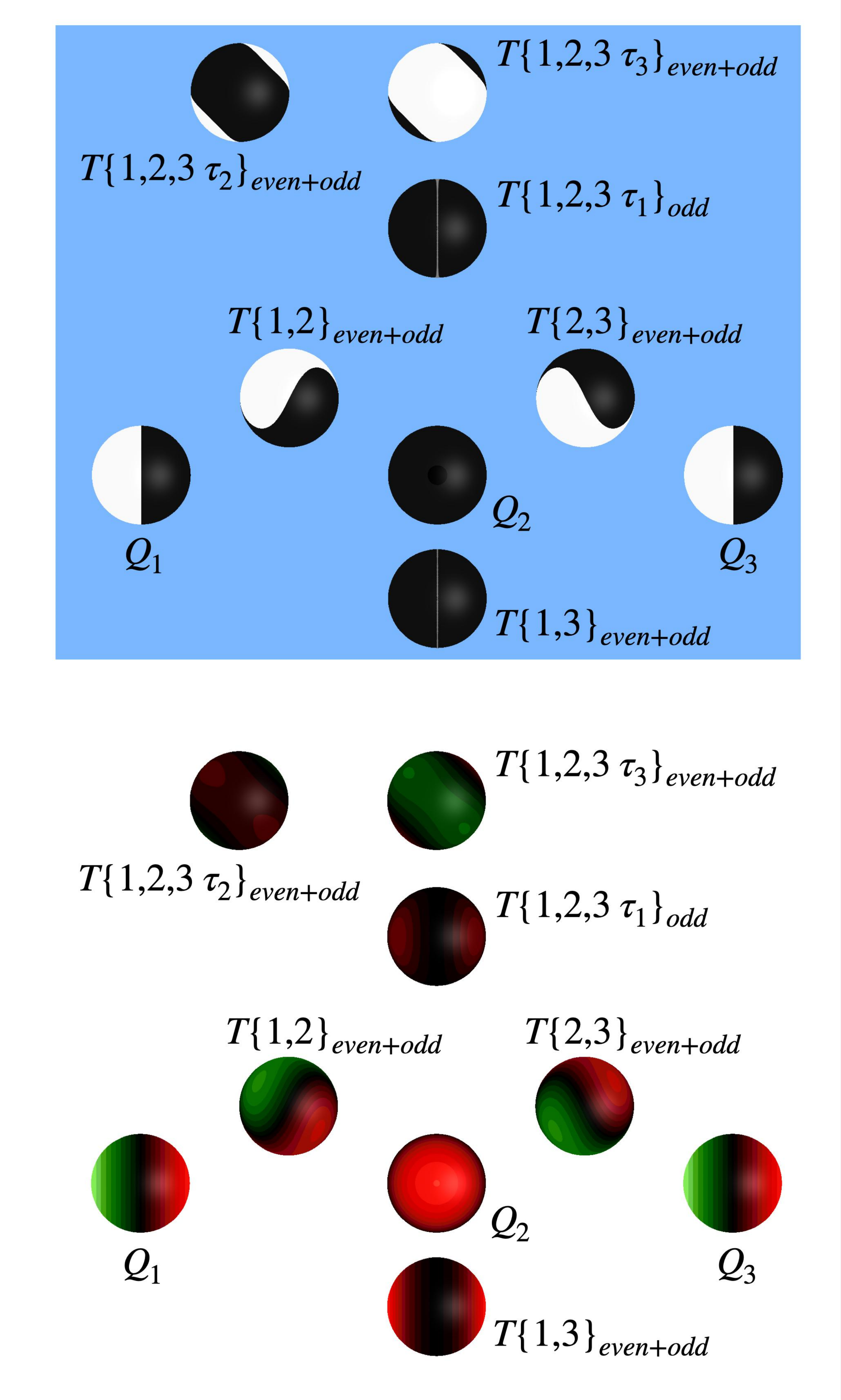}
\caption{\label{fig:FigColorHC}High contrast color scheme representation of a three-qubit state. High contrast color schemes such as the black-white high contrast scheme used in the visualization of the fully separable state $\ket{\text{+0+}}= 1/2\left(\ket{000} +\ket{001} +\ket{100} +\ket{101}\right)$ (top) are useful in providing a clear overview patterns and orientations of the involved beads. They are also easy to sketch by hand. The visualizations are given in terms of total correlation functions which only have compound contributions as the state is a product state. Unlike in the standard BEADS visualization, beads of even and odd symmetries of the same subsets are combined to reveal more elaborate surface patterns which are highly recognizable in the high contrast picture. This is possible in this case as the combined correlation expectation values still lie between $-1$ and $1$ and thus, the color schemes remain applicable. For comparison, the same state is visualized using the standard red-green discontinuous color scheme (bottom).}
\vspace{-1cm}
\end{figure}

\begin{figure}[H]
\centering
\includegraphics[width=0.8\textwidth]{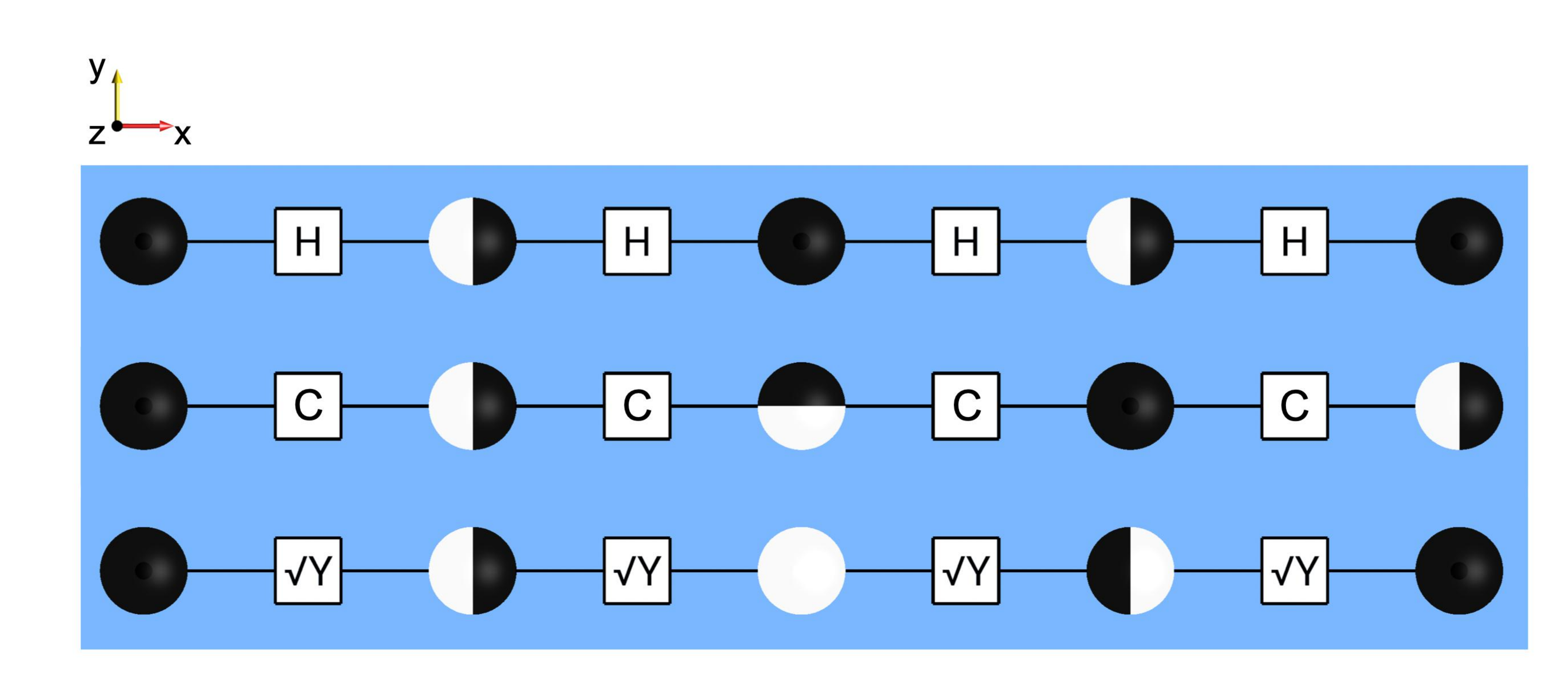}
\caption{\label{fig:FigGatesCyclingHighContrast}Comparison of Hadamard, $\sqrt{\text{Y}}$ and axis cycling gates using a high contrast BEADS representation \screen. The utilized high contrast color scheme allows to easily identify the Q-Bead orientations and thus to determine the periodicities of the visualized gates.  This figure is equivalent to Fig.~\ref{fig:FigGatesCycling}.}
\end{figure}
\vspace{-.5cm}
\begin{figure}[H]
\centering
\includegraphics[width=1.0\textwidth]{FigTeleportXHighContrast.pdf}
\caption{\label{fig:FigTeleportXHighContrast}Teleportation of the x-basis state $\ket{+}$ using the high contrast black and white BEADS color scheme \screen. The x-basis state $\ket{+}=1/\sqrt{2}\left(\ket{0}+\ket{1}\right)$ is teleported from qubit $Q_1$ to $Q_3$. For a detailed analysis see Fig.~\ref{fig:FigTeleportX}. The high contrast scheme allows for a clear overview on beads orientations and surface patterns which may serve as a starting point for deeper analysis as was performed in section~\ref{QuTel}.}
\end{figure}
\newpage

\section{Comparison and classification}
\subsection{Comparison of the BEADS representation and correlation matrix visuals}\label{sec:BEADSCMV}

\begin{figure}[H]
\centering
\includegraphics[width=.87\textwidth]{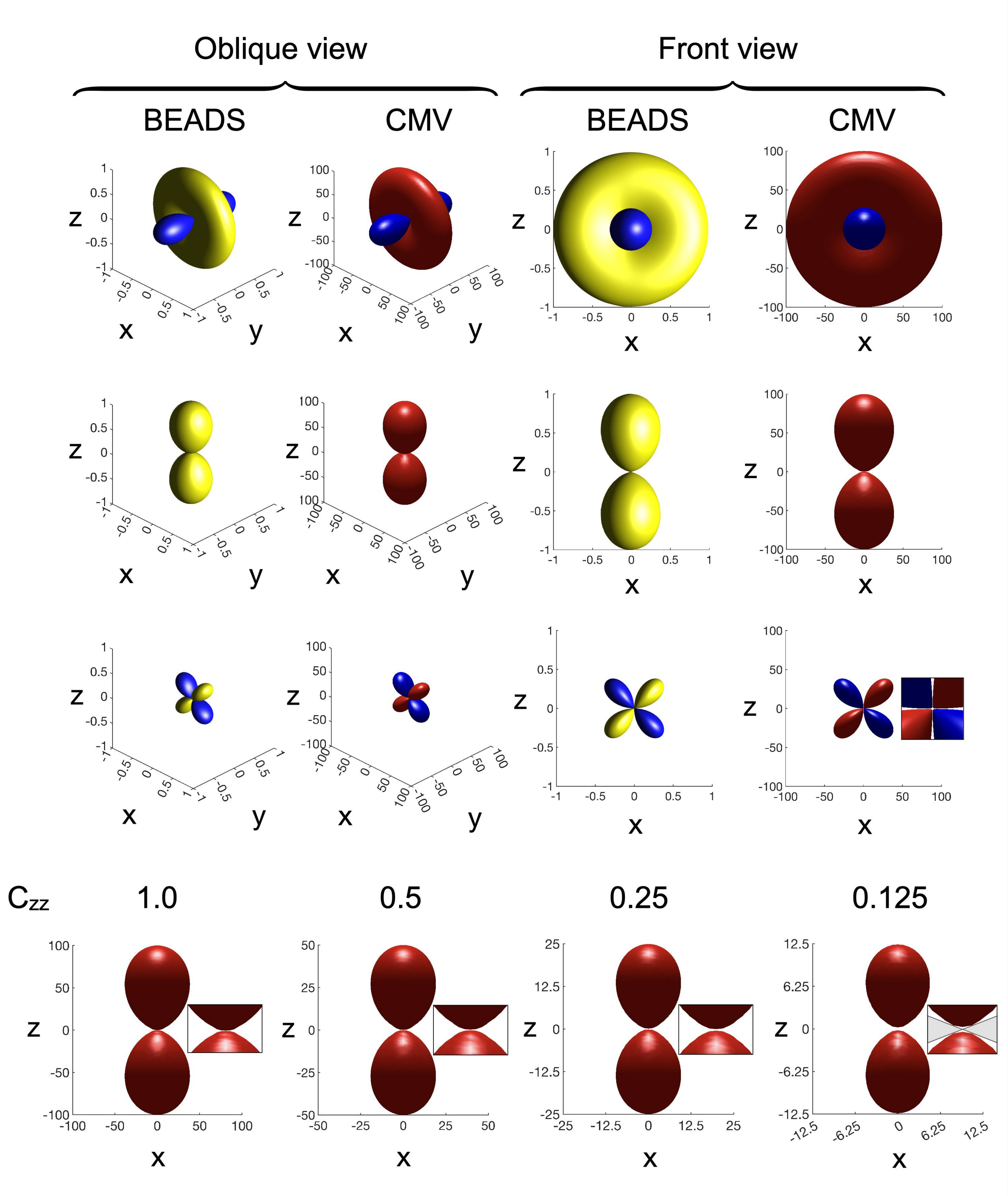}
\caption{\label{fig:FigDiscussionCMV}Comparison between BEADS and correlation matrix visuals (CMVs,  $P = \pm 0.01$). Color schemes were adopted from the original publications \cite{Hazzard1, Hazzard2} and are irrelevant for this comparison. The visualized examples (from top to bottom) comprise the bilinear connected correlation of the Bell state $\ket{\Phi^+}=1/\sqrt2\left(\ket{00}+\ket{11}\right)$, for which the corresponding correlation tensor has non-zero elements $C_{xx}=1$, $C_{yy}=-1$, $C_{zz}=1$, the bilinear correlation function of the GHZ state ($\ket{\text{GHZ}} = 1/\sqrt{2}\left(\ket{000}+\ket{111}\right)$), where only $C_{zz}=1$, and the symmetric bilinear cloverleaf-shaped correlation function component between $Q_1$ and $Q_2$ or $Q_1$ and $Q_3$ for the state $\ket{\psi}=1/2\left(\ket{000}-\ket{011}+\ket{100}+\ket{111}\right)$ with $C_{xz}=1$. While the top three examples look almost identical to the corresponding BEADS representation (plotted with magnitude of the spherical function value as distance from the origin), the sizes of the surfaces (see axes) differ greatly and the underlying correlation coefficients in arbitrary directions are only faithfully represented in the BEADS representation whereas the CMV surface is merely proportional to and an approximation of the represented correlation functions in arbitrary directions. The zoomed view shown in the third row reveals a gap between the cloverleaf lobes of the CMV at the origin which is not present in the BEADS representation and where the CMV surface and thus the correlation coefficients in a range of directions are undefined. In the bottom row, CMVs of zz-correlations with varying correlation values are shown which reveal a similar gap between CMV lobes of which the relative size increases with decreasing correlation coefficients. In the rightmost case, grey areas in the zoomed view indicate directions in which correlation coefficients are undefined.}
\end{figure}

Correlation matrix visuals (CMVs) as introduced by Mukherjee et al. \cite{Hazzard1} and applied in \cite{Hazzard2} yield visualizations of symmetric connected two-qubit correlation functions (the concept can be extended to more qubits) which, at first glance, look very similar to corresponding BEADS representations (plotted with absolute spherical function values as distance from the origin). However, the construction protocol of CMVs involves the generation of volumetric data from a connected correlation tensor \cite{Hazzard1}
\begin{equation}
Q\left(C_{ij},\mathbf{r}\right)=\mathbf{r}^TC_{ij}\mathbf{r}
\end{equation}
and the alternative form \cite{Hazzard1}
\begin{equation}
Q_f\left(C_{ij},\mathbf{r}\right)=\frac{Q\left(C_{ij},\mathbf{r}\right)}{\left(1+r^2\right)^{3/2}},
\end{equation}%
based on which the visualization is obtained as an isosurface (level set) at arbitrarily chosen isovalues $P$, i.e., the surface at which $Q_f\left(C_{ij},\mathbf{r}\right)=P$ (in \cite{Hazzard1} the proposed values of are 
$P =\pm0.01$). While this choice prohibits uniqueness of the representation in the first place, the distance of the CMV surface to the origin is only an approximation and is only proportional to permutation symmetric connected correlation function components but does not provide a direct map of the numerical correlation coefficients which would allow to directly read off the coefficients or related bit parity probabilities as is possible for fully symmetric components in the BEADS representation. In addition, CMVs commonly do not provide a faithful representation of the visualized correlation  functions in proximity to the origin, i.e., the CMV, depending on the choice of the level set and the overall size of the correlation, may not cover regions where the correlation coefficients adopt values close to zero, which leads to gaps between the CMV lobes, and thus, the correlation function represented by the CMV is undefined in certain directions. We provide a visual comparison of BEADS and CMVs in Fig.~\ref{fig:FigDiscussionCMV}.

\subsection{Comparison of different representations of quantum states}\label{sec:BEADSComp}
A detailed evaluation of various quantum mechanical representations with respect to different criteria which complements the Discussion section in the main text is given in Table~\ref{tab:BEADSDiscussion}. The assessed visualizations comprise the BEADS, DROPS \cite{DROPS}, PROPS \cite{PROPS}, Bloch vector \cite{Bloch, FeynmanBloch}, dimensional circle notation \cite{DCN}, and the Qiskit implementations of cityscape plots \cite{Qiskit}, as well as the Q-Sphere \cite{Qiskit}.

Fig.~\ref{fig:FigComparison} is an extended version of Fig.~\ref{Figure:Fig4} (see section~\ref{Preview}) which includes additional visualizations such as standard state vectors and dimensional circle notations \cite{DCN} as well as the imaginary part cityscape plots which were omitted in the main text.

\begin{figure}[H]
\centering
\includegraphics[width=1.0\textwidth]{FigComparison.pdf}
\caption{\label{fig:FigComparison}BEADS, state vector, cityscape \cite{Qiskit}, Q-Sphere \cite{Qiskit}, and dimensional circle notation (DCN) \cite{DCN} representation of several quantum states. The global phase is not visualized by the cityscape and BEADS representations since these visualizations are density-operator-based. In the published Q-Sphere representation \cite{Qiskit}, the global phase is adjusted with respect to the maximum element of the state vector (prioritizing elements closer to the end of the vector due to the little-endian convention in Qiskit) such that this element becomes positive real. The dimensional circle notation \cite{DCN} yields a faithful representation of the global phase. The visualized examples comprise a fully separable state (top), two partially entangled two-qubit states which only differ by a global $45^\circ$-rotation around the x-axis (center), and a maximally entangled three-qubit state which is the GHZ state rotated by a $-60^\circ$-rotation around the xy-bisecting axis (bottom).}
\end{figure}

\newgeometry{left=2.25cm,right=1.75cm,top=1cm,bottom=2.5cm} 
\begin{table}[h!]
\scriptsize
\centering
\caption{\label{tab:BEADSDiscussion}Evaluation of the BEADS, DROPS \cite{DROPS}, and PROPS \cite{PROPS} representations, the Cityscape plot of the density operator \cite{Qiskit}, the Q-Sphere visualization \cite{Qiskit} and the dimensional circle notation (DCN) \cite{DCN} which are specifically designed for the visualization of quantum mechanics. Representations either (\fullcirc) fully meet a criterion, (\emptycirc) partially meet a criterion or (\greycirc) do not meet a criterion. }
\vspace{.1cm}
\begin{tabular}{p{0.4\textwidth}<{\flushleft} p{0.06\textwidth}<{\centering} p{0.06\textwidth}<{\centering} p{0.06\textwidth}<{\centering} p{0.06\textwidth}<{\centering} p{0.07\textwidth}<{\centering} p{0.06\textwidth}<{\centering} p{0.06\textwidth}<{\centering}}
\hline\hline\\[-5pt]
 & BEADS & DROPS & PROPS & Bloch vector & Cityscape plot & \mbox{Q-Sphere} & Circle \mbox{notation} \\[8pt] \hline\hline
& \\[-6pt] 
\textbf{Visualization is} \\[-8pt]
-- general & \fullcirc  & \fullcirc & \fullcirc & \greycirc & \fullcirc & \fullcirc & \fullcirc \\[-8pt]
-- accurate & \fullcirc & \fullcirc & \fullcirc & \fullcirc & \fullcirc & \fullcirc & \fullcirc \\[-8pt]
-- complete & \fullcirc & \fullcirc & \fullcirc & \greycirc & \fullcirc & \fullcirc & \fullcirc \\[-8pt]
-- adaptable with respect to balance of com-\phantom{-- }pleteness vs. simplicity & \fullcirc & \emptycirc &\emptycirc & \greycirc & \fullcirc & \emptycirc & \emptycirc  \\[-8pt]
-- interactive/responsive & \fullcirc & \fullcirc & \fullcirc & \fullcirc & \fullcirc & \fullcirc & \mbox{\hspace{7pt}~\emptycirc~{(a)}}  \\[-8pt]
-- easy to interpret & \fullcirc & \greycirc & \emptycirc & \fullcirc & \emptycirc & \emptycirc & \emptycirc  \\[-6pt]\hline \\[-6pt]
\textbf{Applicable to} \\[-8pt]
-- single qubit system & \fullcirc & \fullcirc & \fullcirc & \fullcirc & \fullcirc & \fullcirc & \fullcirc  \\[-8pt]
-- two qubit system & \fullcirc & \fullcirc & \fullcirc & \greycirc & \fullcirc & \fullcirc & \fullcirc  \\[-8pt]
-- three or more qubit systems & \fullcirc & \fullcirc & \fullcirc & \greycirc & \fullcirc & \fullcirc & \fullcirc  \\[-8pt]
-- pure states & \fullcirc & \fullcirc & \fullcirc & \fullcirc & \fullcirc & \fullcirc & \fullcirc   \\[-8pt]
-- mixed states & \fullcirc & \fullcirc & \fullcirc & \fullcirc & \fullcirc & \greycirc & \greycirc   \\[-6pt]\hline \\[-6pt]
\textbf{Directly visualizes} \\[-8pt]
-- spatial orientation of qubits & \fullcirc & \fullcirc  & \greycirc & \fullcirc & \greycirc & \greycirc & \greycirc \\[-8pt]
-- spatial distribution of entanglement-based \phantom{-- }correlation coefficients & \fullcirc & \greycirc & \greycirc & \greycirc & \greycirc & \greycirc  & \greycirc \\[-8pt]
-- rotations of qubits and entanglement-based \phantom{-- }correlation functions & \fullcirc & \emptycirc & \emptycirc & \emptycirc & \greycirc & \greycirc  & \greycirc \\[-8pt]
-- direction-dependent expectation values & \fullcirc & \greycirc & \greycirc & \fullcirc & \greycirc & \greycirc  & \greycirc \\[-8pt]
-- probabilities of measurement outcomes & \fullcirc & \greycirc & \greycirc & \emptycirc & \mbox{\hspace{9.5pt}~\emptycirc~{(b)}} & \mbox{\hspace{7pt}~\emptycirc~{(b)}} & \mbox{\hspace{7pt}~\emptycirc~{(b)}} \\[-8pt]
-- reduced single qubit density operator & \fullcirc & \emptycirc & \emptycirc & \fullcirc & \greycirc & \greycirc  & \greycirc  \\[-8pt]
-- entanglement of (sub)set of qubits & \fullcirc & \greycirc & \greycirc & \greycirc & \greycirc & \greycirc  & \greycirc  \\[-8pt]
--\hspace{3.5pt}compound correlation functions of (sub)set \phantom{-- }of qubits & \fullcirc & \greycirc & \greycirc & \greycirc & \greycirc & \greycirc  & \greycirc  \\[-8pt]
-- total correlation functions of (sub)set of \phantom{-- }qubits & \fullcirc & \emptycirc & \greycirc & \greycirc & \greycirc & \greycirc  & \greycirc  \\[-8pt]
-- type and amount of entanglement/corre-\phantom{-- }lations & \fullcirc & \greycirc & \greycirc & \greycirc & \greycirc & \greycirc & \greycirc  \\[-8pt]
-- spatial symmetries of states & \fullcirc & \fullcirc & \greycirc & \emptycirc & \greycirc & \greycirc  & \greycirc  \\[-8pt]
-- permutational symmetry of states & \fullcirc & \fullcirc & \greycirc  & \greycirc & \greycirc & \greycirc  & \greycirc  \\[-8pt]
-- global phase & \greycirc & \greycirc & \greycirc  & \greycirc & \greycirc & \mbox{\hspace{7pt}~\emptycirc~{(c)}} & \fullcirc \\[-6pt]\hline \\[-6pt]
\textbf{Didactically valuable to clearly} \\[-8pt]
-- visualize a seamless transition from bit to \phantom{-- }qubit & \fullcirc & \greycirc & \greycirc & \emptycirc & \greycirc & \emptycirc & \emptycirc \\[-8pt]
-- see single-qubit gates as rotations & \fullcirc & \fullcirc & \emptycirc & \fullcirc & \greycirc & \greycirc  & \greycirc \\[-8pt]
--\hspace{3.5pt}see effect of non-selective single qubit gates \phantom{-- }on entanglement-based correlation functions & \fullcirc & \greycirc & \emptycirc & \greycirc & \greycirc & \greycirc  & \greycirc \\[-8pt]
-- see the effect of projective measurements & \fullcirc & \greycirc & \fullcirc & \emptycirc & \fullcirc & \fullcirc & \fullcirc \\[-8pt]
-- visualize simultaneously the full range of \phantom{-- }possible measurement outcomes & \fullcirc & \greycirc & \fullcirc & \emptycirc & \emptycirc & \emptycirc & \emptycirc \\[-6pt]\hline \\[-6pt]
\textbf{Is useful for} \\[-8pt]
-- understanding of quantum gates & \fullcirc & \emptycirc & \emptycirc  & \emptycirc & \emptycirc & \emptycirc & \fullcirc \\[-8pt]
-- design of quantum gates & \fullcirc & \greycirc & \greycirc & \emptycirc & \emptycirc & \emptycirc & \fullcirc  \\[-8pt]
-- analysis of quantum algorithms & \fullcirc & \greycirc & \greycirc & \greycirc & \greycirc & \emptycirc & \fullcirc \\[-8pt]
-- design of quantum algorithms & \emptycirc & \greycirc & \greycirc & \greycirc  & \greycirc & \emptycirc & \fullcirc \\[-8pt]
-- prediction of measurement probabilities & \fullcirc & \greycirc & \emptycirc & \emptycirc & \emptycirc & \emptycirc & \fullcirc \\[-6pt]\hline \\[-6pt]
\textbf{Can be augmented by }\\[-8pt]
-- also displaying numerical values & \fullcirc & \emptycirc & \emptycirc & \fullcirc & \emptycirc & \fullcirc & \emptycirc \\[-8pt]
-- creation of tangible 3D models & \fullcirc & \fullcirc & \emptycirc & \fullcirc & \emptycirc & \emptycirc & \emptycirc \\[-6pt] \hline \\[-6pt]
\textbf{Can be used for} \\[-8pt] 
-- outreach to general public & \fullcirc & \emptycirc & \greycirc & \emptycirc & \greycirc & \emptycirc  & \greycirc \\[-8pt]
-- education of high-school students & \fullcirc & \emptycirc & \greycirc & \emptycirc & \greycirc & \emptycirc & \emptycirc \\[-8pt]
-- education of bachelor and master students & \fullcirc & \emptycirc & \emptycirc & \emptycirc & \emptycirc & \emptycirc & \fullcirc \\[-8pt]
-- \mbox{research in quantum information processing} & \fullcirc & \emptycirc & \fullcirc & \emptycirc & \emptycirc & \emptycirc & \fullcirc \\[-6pt]\hline\hline
\end{tabular}
\begin{scriptsize}
\begin{enumerate}[leftmargin=.7cm,labelsep=.2cm, labelwidth=.2cm]
\setlength\itemsep{.2em}
\item[(a)] The authors are not aware of a publicly available software tool that automates the visualization of the \\dimensional circle notation representation.
\item[(b)] Limited to the probability to obtain each computational basis state.
\item[(c)] The global phase of a state in the Q-Sphere representation is eliminated for simplicity in the current Qiskit \\version but could easily be shown by adapting the program accordingly.
\end{enumerate} 
\end{scriptsize}
\end{table}%
\restoregeometry 

\subsection{Hierarchy of models, representations, and material tokens}\label{ModelsTokens}
In Fig.~\ref{fig:FigDiscussionHierarchy1} and Fig.~\ref{fig:FigDiscussionHierarchy2} we provide a hierarchical overview of models, representations and material tokens \cite{FriggHartmann, FriggNguyen, DeRegt}. The BEADS representation can be classified a visual representation but is not a physical model which can perform quantum computation on its own.
\begin{figure}[H]
\centering
\includegraphics[width=1.\textwidth]{FigDiscussionHierarchy1.pdf}
\caption{\label{fig:FigDiscussionHierarchy1}Relation between an abstract quantum state vector, its density operator, its visual quantum beads representation, and a corresponding tangible three-dimensional model (token). The left column indicates the most important involved general classes of representations and models, the middle column provides concrete concepts for the BEADS representation of quantum states, and in the right column, typical symbols and exemplary illustrations are shown. The mapping between the density operator of a (pure or mixed) state of an individual quantum system (consisting of one or more qubits) and the BEADS representation is exact in both directions (indicated by a black double-headed arrow). To visualize the state of a pure quantum system, which can always be expressed as a ket vector, the state can also be uniquely mapped to the corresponding density operator (and hence visualized) but the - in general irrelevant - global phase of the state is lost. Therefore, the inverse mapping between the density operator of a pure state and a corresponding ket state is only unique up to the global phase. This loss of the global phase information is schematically indicated by a dark grey double-headed arrow. On the one hand, this property is beneficial because it slightly simplifies the BEADS representation. On the other hand, it is important to remember that the quantum bead representation cannot be used to illustrate the effect of an operation, which only changes the global phase of a given state of interest. Depending on the used method to create material beads, the mapping between the ideal BEADS representation and material beads can be more or less exact (indicated by the light-grey double-headed arrow). However, even imperfect tangible 3D models, such as hand-colored wooden spheres or hand-made glass beads can be highly useful to understand simple quantum gates, such as the X-gate and the Hadamard gate and to roughly predict the probability of measurement outcomes. Both 2D computer graphic visualizations representations and material 3D models help to create a clear mental picture of quantum states and of their properties.}
\end{figure}
\begin{figure}[H]
\centering
\includegraphics[width=1.\textwidth]{FigDiscussionHierarchy2.pdf}
\caption{\label{fig:FigDiscussionHierarchy2}Detailed relation between an abstract quantum state vector, its density operator, its visual quantum beads representation, and a corresponding tangible three-dimensional model (token). A more detailed classification of various intermediate model and representation layers between an abstract quantum state, its density operator, its visualization, and the creation of a clear mental picture of the state, such as was shown in Fig~\ref{fig:FigDiscussionHierarchy1}, is shown.}
\end{figure}\newpage

\section{Application examples}
\subsection{Partially entangled pure states}\label{sec:PartEnt}
BEADS representations shown in the Results section of the main text are mainly based on visualizing Q-Beads and $n$-qubit connected correlation functions in the form of E-Beads. Indeed, as we discussed in appendices~\ref{sec:BEADSCorrel} and~\ref{CorrQIT}, omitting compound correlation function contributions in the visualization of quantum states does not result in any loss of information and the $n$-qubit connected correlation (E-Beads) together with linear components (Q-Beads) are sufficient to characterize an entire quantum state \cite{Mahler95}.

As we saw in the Results section, measurement outcomes of maximally entangled states can be directly predicted from E-Beads and outcomes of fully separable states can be deduced from Q-Beads only. When it comes to partially entangled states though, it is required to consider the total correlation function to forecast the correct measurement outcomes. 

Before analyzing the BEADS representation of a partially entangled state, it is thus reasonable to choose the visualized information such that the desired aspect of interest is intuitively represented: if one is interested in learning about the degree and structure (e.g., possible symmetries, locality, etc.) of manifestations of entanglement with respect to entanglement-based correlation coefficients, it is sufficient to examine E-Beads, whereas if measurement predictions are the main focus, one should view T-Beads instead.

For symmetric $n$-qubit measurements, outcomes can then be directly deduced from the spherical function value of the $n$-linear symmetric T-Bead along the chosen measurement direction. Detailed background information on the color schemes for total correlations is given in appendix~\ref{sec:BEADSColors} and Fig.~\ref{Figure:Fig6}.

A prominent three-qubit example of a partially entangled state is the W state \cite{Dur00} $\ket{\text{W}} = 1/\sqrt{3}(\ket{001} + \ket{010} + \ket{100})$ which was introduced in section~\ref{3Q}. Fig.~\ref{fig:FigW} provides visualizations of the W state with respect to different correlation functions. For a symmetric three-qubit z-measurement, considering the $T\{1,2,3\ \tau_1\}_{\textit{odd}}$ T-Bead, it can be seen that the W state is fully anticorrelated. Hence, only outcomes of odd bit parity (i.e., bit parity 1) $\ket{111}$, $\ket{001}$, $\ket{010}$, or $\ket{100}$ are expected (any outcome of bit parity 0 has zero probability) . Note that one also measures $\ket{111}$, although with zero probability, which arises from the correlations in the W state being more intricate than for GHZ states (see discussion in section~\ref{3Q}).
\begin{figure}[H]
\centering
\includegraphics[width=1.0\textwidth]{FigW.pdf}
\caption{\label{fig:FigW}Correlation functions of the three-qubit W state. The total correlation function BEADS representation of the W state \cite{Dur00} $\ket{\text{W}} = 1/\sqrt{3}(\ket{001} + \ket{010} + \ket{100})$ (top) reveals full three-qubit anticorrelation (cf. indicated expectation value at ${T\left\{1,2,3\ \tau_1\right\}}_{\textit{odd}})$ along the z-axis. Separating connected (bottom left) from compound correlation functions (bottom right) reveals that all correlation function components share the same axial symmetry for all beads. Specific correlation expectation values along the z-axis are indicated numerically in all representations.}
\end{figure}
\noindent
Of course, from the corresponding state vector expression it is straightforward to see that $\ket{111}$ cannot be measured for the W state. However, assuming that one is only provided with the BEADS representation of an arbitrary quantum state, it is still possible to determine the measurement probabilities of individual outcomes.

Hence, to see why $\ket{111}$ cannot be observed as an outcome in a three-qubit z-measurement of the W state solely from the BEADS representation, contributions represented by Q-Beads and T-Beads in the BEADS picture have to be examined. In Fig.~\ref{fig:FigW}, expectation values along the z-axis are provided for all beads and the related bit parity probabilities can be calculated according to Eq.~\ref{eq:BPP} of section~\ref{2Q}. Thus, a system of equations is obtained:\newline

\noindent
$T\{1,2,3\ \tau_1\}_{\textit{odd}}$:
\vspace{-6pt}
\begin{flalign}
\text{(I)}\qquad p(\ket{111})+p(\ket{001})+p(\ket{010})+p(\ket{100})=\frac{1-(-1)}{2}=1 &&
\end{flalign}
\noindent
and
\begin{flalign}
p(\ket{000})=p(\ket{110})=p(\ket{101})=p(\ket{011})=0, &&
\end{flalign}\vspace{2pt}

\noindent\vspace{-4pt}
$T\{k,l\}_{\textit{even}}$:
\vspace{-8pt}
\begin{subequations}
\begin{flalign}
&\text{(IIa)}\qquad p(\ket{100})+p(\ket{010})=\frac{1-\big(-\frac{1}{3}\big)}{2}=\frac{2}{3}, &&\\                                                                          
&\text{(IIb)}\qquad p(\ket{100})+p(\ket{001})=\frac{2}{3}, &&\\                                                                                                      &\text{(IIc)}\qquad p(\ket{010})+p(\ket{001})=\frac{2}{3}, &&
\end{flalign}
\vspace{2pt}
\end{subequations}

\noindent
$Q_k$:
\vspace{-8pt}
\begin{subequations}
\begin{flalign}
&\text{(IIIa)}\qquad p(\ket{100})+p(\ket{111})=\frac{1-\frac{1}{3}}{2}=\frac{1}{3}, &&\\                                                                               
&\text{(IIIb)}\qquad p(\ket{010})+p(\ket{111})=\frac{1}{3}, &&\\                                                                                                      
&\text{(IIIc)}\qquad p(\ket{001})+p(\ket{111})=\frac{1}{3}, &&
\end{flalign}
\end{subequations}
where $p\left(\bullet\right)$ is the probability to measure the state denoted in parenthesis. This system can be solved by the relation:
\begin{flalign}
&\text{(IV)}=\text{(IIa}-\text{IIIa)}\qquad p(\ket{100})+p(\ket{010})-p(\ket{100})-p(\ket{111})=\frac{2}{3}-\frac{1}{3}=\frac{1}{3}, &&\\
&\text{(IIIb}+\text{IV)}\qquad\qquad\:\quad 2p(\ket{010})=\frac{2}{3} \rightarrow p(\ket{010})=\frac{1}{3}.&&
\end{flalign}
Repeating this procedure for the remaining equations gives $p(\ket{001})=p(\ket{010})=p(\ket{100})=1/3$ and $p(\ket{111})=0$.

\subsection{Quantum gates and system dynamics}\label{sec:BEADSGates}
As introduced in section~\ref{LightInt}, quantum gates are unitary transformations acting on qubits in quantum computing algorithms and communication protocols. Thus, it is essential to understand the working principles of the most important quantum gates. While the BEADS representation is useful to visualize the \textit{outcomes} of any gate operation for arbitrary inputs, it can also be applied to monitor the \textit{dynamics} that take place during the action of a gate. 

\begin{figure}[H]
\centering
\includegraphics[width=.74\textwidth]{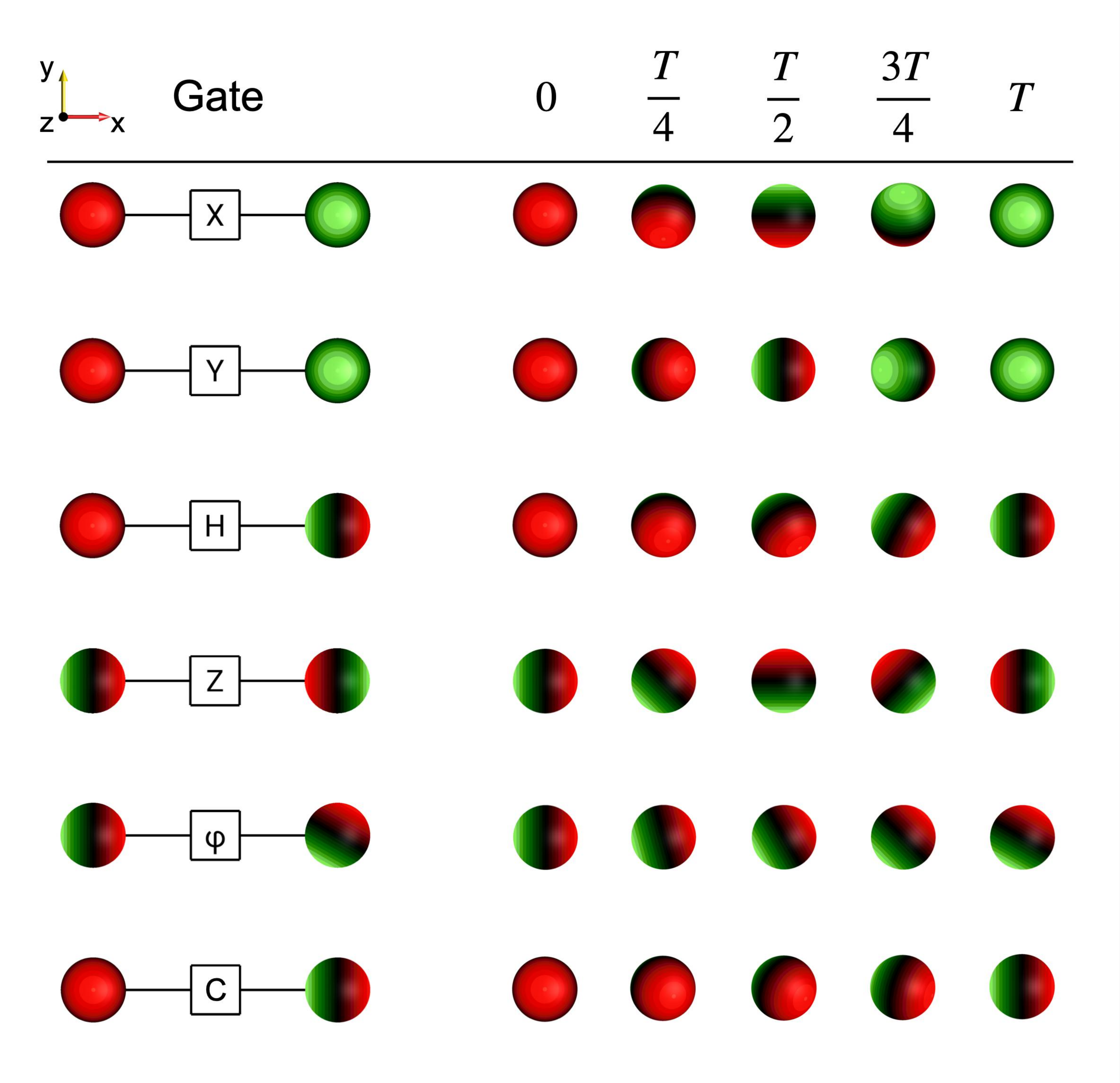}
\caption{\label{fig:FigGatesLocal}Single-qubit gates visualized by the BEADS representation \screen. Single-qubit quantum gates such as Pauli (X, Y, Z), Hadamard (H), phase ($\varphi$), or axis cycling gates (C) implement rotations which are visualized by the BEADS representation. The overall action of the gate is shown on the left whereas the dynamics which occur during the gate are outlined by BEADS representations (Q-Beads) of qubit states at different time steps during the operation (right). The Q-Beads at $T/2$ correspond to the output of the square root of each quantum gate. The symbol $\varphi$ represents a phase gate (here the phase angle is chosen to be $60^\circ$).}
\vspace{-.4cm}
\end{figure}

Indeed, quantum gates have specific durations $T$ depending on the particular physical implementation of the utilized qubits. In the simple case where a single quantum gate is implemented by a time-independent Hamiltonian $\hat{H}$, for any point in time $0 \leq t \leq T$ we thus obtain a propagator $U$ describing the time evolution of the system \cite{NielsenChuang}:
\begin{equation}
U(t)=e^{-i\hat{H}t}.
\end{equation}
\noindent
Hence, using the BEADS representation, we can not only observe the result of a quantum gate ($t=T$) but also understand the dynamics during the gate by visualizing the states $\ket{\psi_t}= U(t)\ket{\psi}$, using the initial state $\ket{\psi}$, at various points in time. As illustrated in Fig.~\ref{fig:FigGatesLocal} and discussed in section~\ref{3Q}, \textit{single-qubit gates} implement simple \textit{rotations} of the qubit state which are directly visualized by a corresponding rotation of the Q-Bead. By choosing suitable Hamiltonians $\hat{H}$, gates can be implemented such that the induced transformations are equivalent to simple rotations around desired axes for computational basis states. For instance, Pauli-X, Y, and Z gates \cite{NielsenChuang} are implemented by $180^\circ$-rotations around the corresponding axis. Phase gates (indicated by $\varphi$ in the circuit diagram), also implement z-rotations by an arbitrary phase angle $\varphi$, i.e., the Pauli-Z gate is a special case of a phase gate for which $\varphi= 180^\circ$.

The intermediate state at $t=T/2$, moreover, corresponds to the output state if one was to apply only half of the operation which is commonly denoted the square root of a quantum gate \cite{NielsenChuang}. In case of the Pauli-Y gate, the square root operation corresponds to a $90^\circ$-rotation about the y-axis. The $\sqrt{\text{Y}}$-gate transfers $\ket{0}$ to the superposition state $\ket{+}=1/\sqrt2\left(\ket{0}+\ket{1}\right)$. This state is also obtained when performing a Hadamard H \cite{NielsenChuang} or an axis cycling gate C \cite{CrooksGates, GottesmanThesis, Kubischta23, Rall17} on the same input.

\begin{figure}[H]
\centering
\includegraphics[width=.85\textwidth]{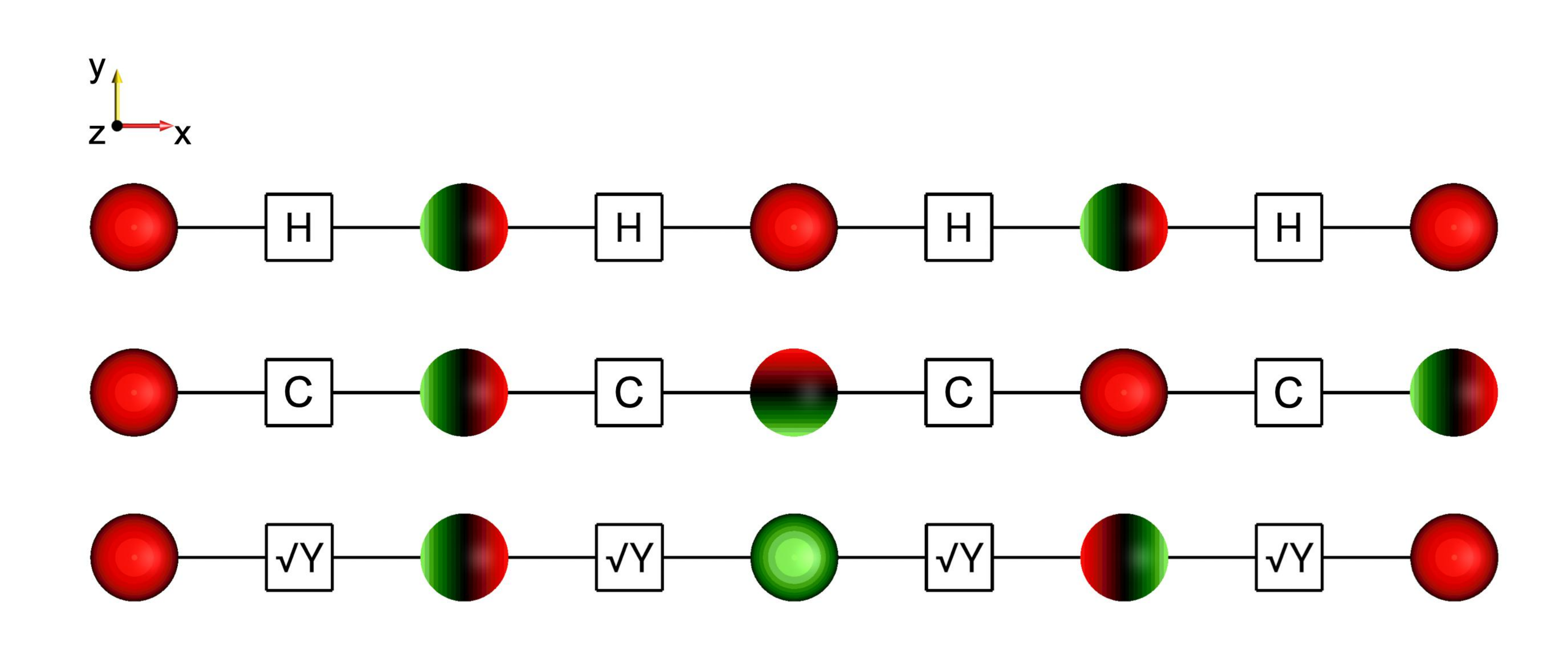}
\caption{\label{fig:FigGatesCycling}Comparison of Hadamard, $\sqrt{\text{Y}}$ and axis cycling gates \screen. H, C \cite{CrooksGates, GottesmanThesis, Kubischta23, Rall17}, and $\sqrt{\text{Y}}$-gates can be applied to bring qubits in computational basis states (here $\ket{0}$) into superposition. As visualized by the BEADS representation, the gates differ in their period which results in different (intermediate) output states when performing multiples of each gate in a row. The period $P$ of each gate can be directly read off in the BEADS picture ($P_{\text{H}}=2$, $P_{\text{C}}=3$, $P_{\sqrt{\text{Y}}}=4$). A high-contrast variant of the picture (bottom) which gives a clear overview is available in appendix~\ref{app:BEADSHighContrast}.}
\vspace{-.4cm}
\end{figure}

However, the Hadamard gate corresponds to a $180^\circ$-rotation around the xz-bisecting axis. In contrast, the axis cycling C-gate \cite{CrooksGates, GottesmanThesis, Kubischta23, Rall17} implements a $120^\circ$-rotation around the axis $R=(\text{atan}\sqrt{2},45^\circ)\approx(54.74^\circ, 45^\circ)$, corresponding to the space diagonal of the cube spanned by the x, y, and z unit vectors. When looking at the dynamics \textit{during} the gate, it is clear that the induced rotations differ greatly. Furthermore, performing multiples of the same gates in series yields differing (intermediate) outputs. Fig.~\ref{fig:FigGatesCycling} displays the discussed gates (H, C, $\sqrt{\text{Y}}$) applied four times in a row on an initial input state $\ket{0}$. One can directly see that the Hadamard gate gives the initial input state when applied two times in a row. Hence, it is said to be self-inverse. On the other hand, applying a $\sqrt{\text{Y}}$ two times outputs $\ket{1}$ and one has to perform the operation for another two times to regain the initial input state. The axis cycling gate, as its name suggests, cycles the coordinate  x-, y-, and z-axes, i.e., the computational basis (z-basis) input state is first rotated onto the x-axis, followed by a rotation onto the y-axis by the second execution of the gate, etc. Hence, this gate requires three executions until it reproduces the initial input state.

The underlying property which describes how often a gate has to be performed in order to give the original input state is called \textit{period} of a quantum gate and as shown, can be determined visually by using the BEADS representation.

It should be further pointed out, that the \textit{inverse} of a single-qubit gate simply corresponds to a rotation around the same axis but with \textit{reversed rotation direction}. As an exception, gates implemented by 180$^\circ$-rotations, e.g., the Pauli-X, Y, and Z, or the Hadamard gate are self-inverse, i.e., both rotation directions ($\pm$180$^\circ$) have the same effect for which the inverse operation is simply given by the standard gate rotation, respectively.

\begin{figure}[H]
\centering
\includegraphics[width=.85\textwidth]{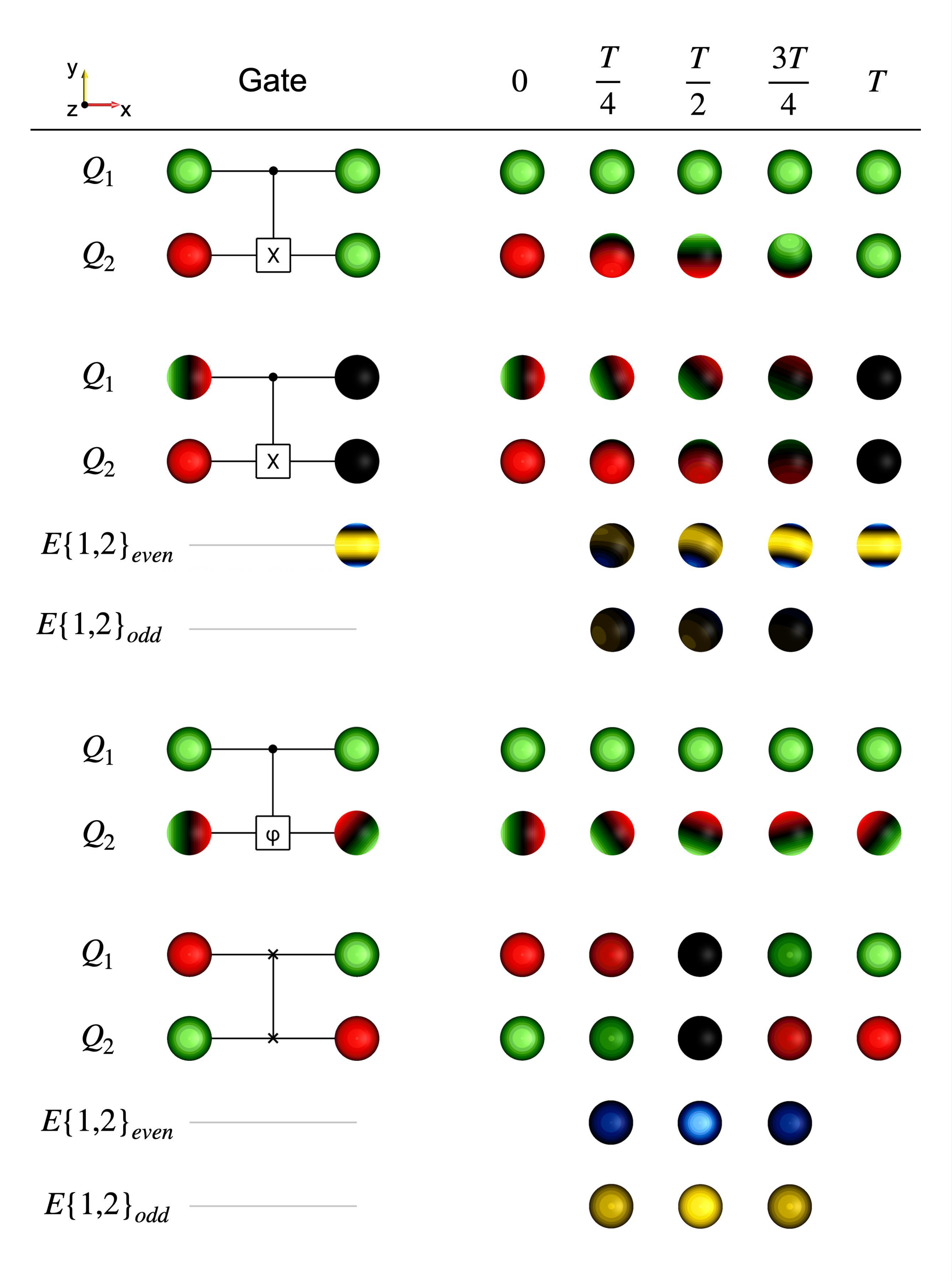}
\caption{\label{fig:FigGates2Q}Two-qubit gates visualized by the BEADS representation \screen. The dynamics of two-qubit quantum gates such as CNOT, controlled phase, or SWAP gates are not limited to simple rotations in the BEADS representation. Depending on the choice of input states, in the case of controlled gates, simple rotations of the target Q-Beads or more complicated transformations which involve rotations, scalings, and morphing when the system becomes entangled or disentangled can be observed. The SWAP gate intermittently creates entanglement which is not evident when examining only the input and output states of the operation but can be directly seen when visualizing the underlying dynamics using the BEADS representation. The overall action of the gate is shown on the left whereas the dynamics which occur during the gate are outlined on the right. The visualized states at time steps $T/2$ correspond to the output of the square root of each quantum gate.}
\vspace{-.3cm}
\end{figure}

\noindent
The effect of multiqubit quantum gates in the BEADS representation goes beyond simple rotations of individual beads. As briefly discussed in section~\ref{3Q}, multiqubit gates can induce rotations, scalings (change of brightness), and morphing. Recall that morphing describes changes in surface patterns which can only occur for correlation function beads (E-Beads, C-Beads, and T-Beads). As can be seen in Fig.~\ref{fig:FigGates2Q}, the induced effects depend on the choice of input states. For instance, when performing controlled operations, e.g. a controlled X-gate (CNOT or CX) or a controlled phase gate, with the control qubit ($Q_1$ in the examples shown in Fig.~\ref{fig:FigGates2Q}) in computational basis state $\ket{1}$, the gates lead to simple rotations of the target qubit Q-Beads $Q_2$.

On the other hand, performing a CNOT on a system with the control qubit being in an equal superposition state, e.g., $\ket{+}$ and the target being in a computational basis state creates entanglement, and thus, causes simultaneous rotations and scalings on both Q-Beads and scaling and morphing of the E-Beads.

By using the BEADS representation to visualize the dynamics of multiqubit gates, we can further observe whether entanglement is created temporarily during the action of the quantum gate, even if the input and output states are not entangled. In fact, this is a common phenomenon that, for example, occurs for SWAP gates. As can be seen in Fig.~\ref{fig:FigGates2Q}, the SWAP gate \cite{NielsenChuang} intermittently leads to a maximally entangled state at time step $T/2$ when swapping qubits in computational basis states (provided the qubits are in differing basis states).

\begin{figure}[H]
\centering
\includegraphics[width=.7\textwidth]{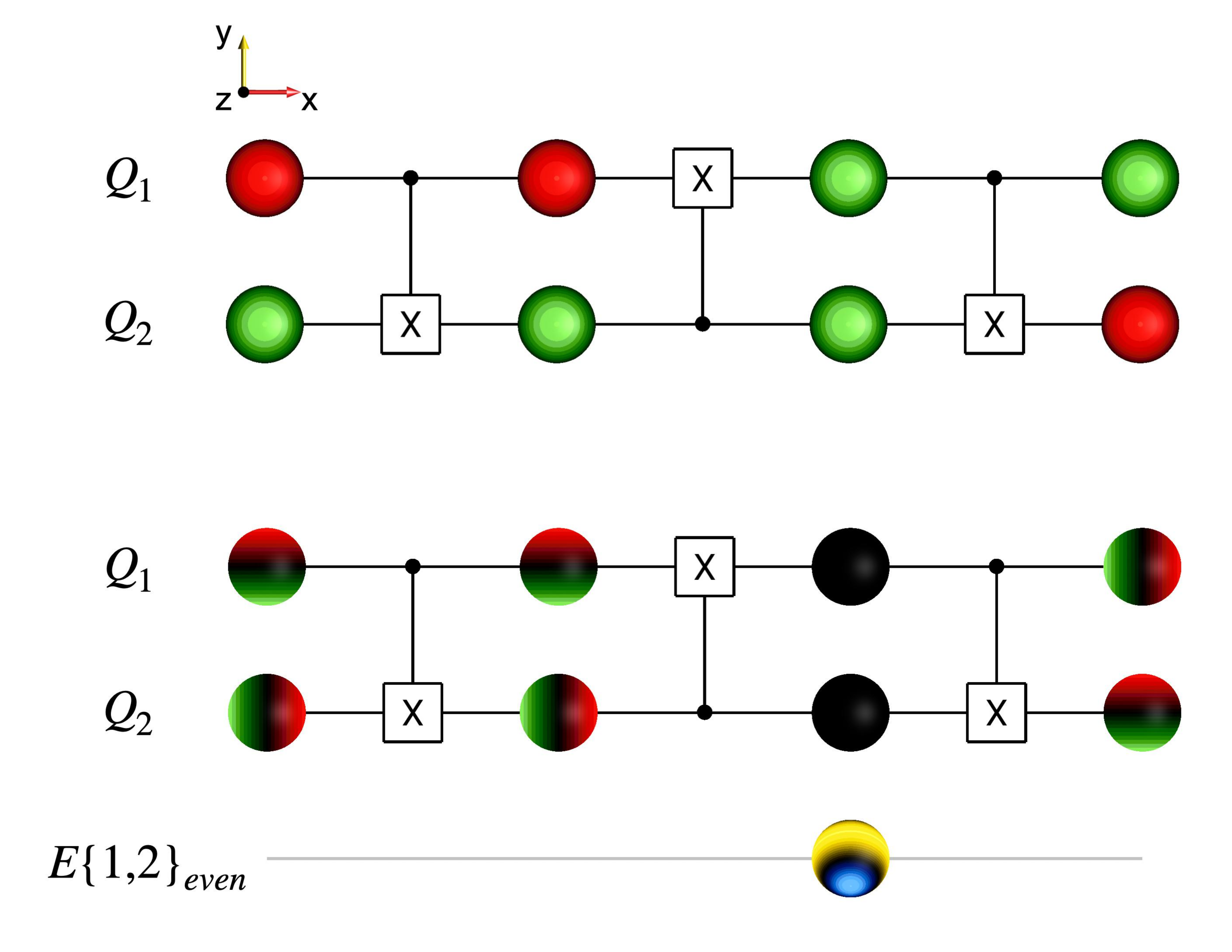}
\caption{\label{fig:FigGatesSWAPVar}Implementing a SWAP gate by a cascade of CNOT gates \screen. Performing a series CNOT$_{12}$-CNOT$_{21}$-CNOT$_{12}$ (where the first number indicates the control and the second number denotes the target qubit), implements a SWAP gate. In the visualizations above, this can be understood by comparing the initial input with the final output orientations of both Q-Beads. The input state in the top example is $\ket{01}$ whereas in the bottom example, the input is $\ket{R+} = 1/2\left(\ket{00}+\ket{01}+i\ket{10}+i\ket{11}\right)$.}
\end{figure}

\noindent
Implementations of gates such as the SWAP gate, which can be realized as a cascade of three CNOT gates (see Fig.~\ref{fig:FigGatesSWAPVar}), can be reproduced in the BEADS picture. After performing such a series of CNOT gates, the Q-Bead orientations correspond to the swapped input orientations just as in case of the SWAP gate (see Fig.~\ref{fig:FigGates2Q})

\begin{figure}[H]
\centering
\includegraphics[width=.8\textwidth]{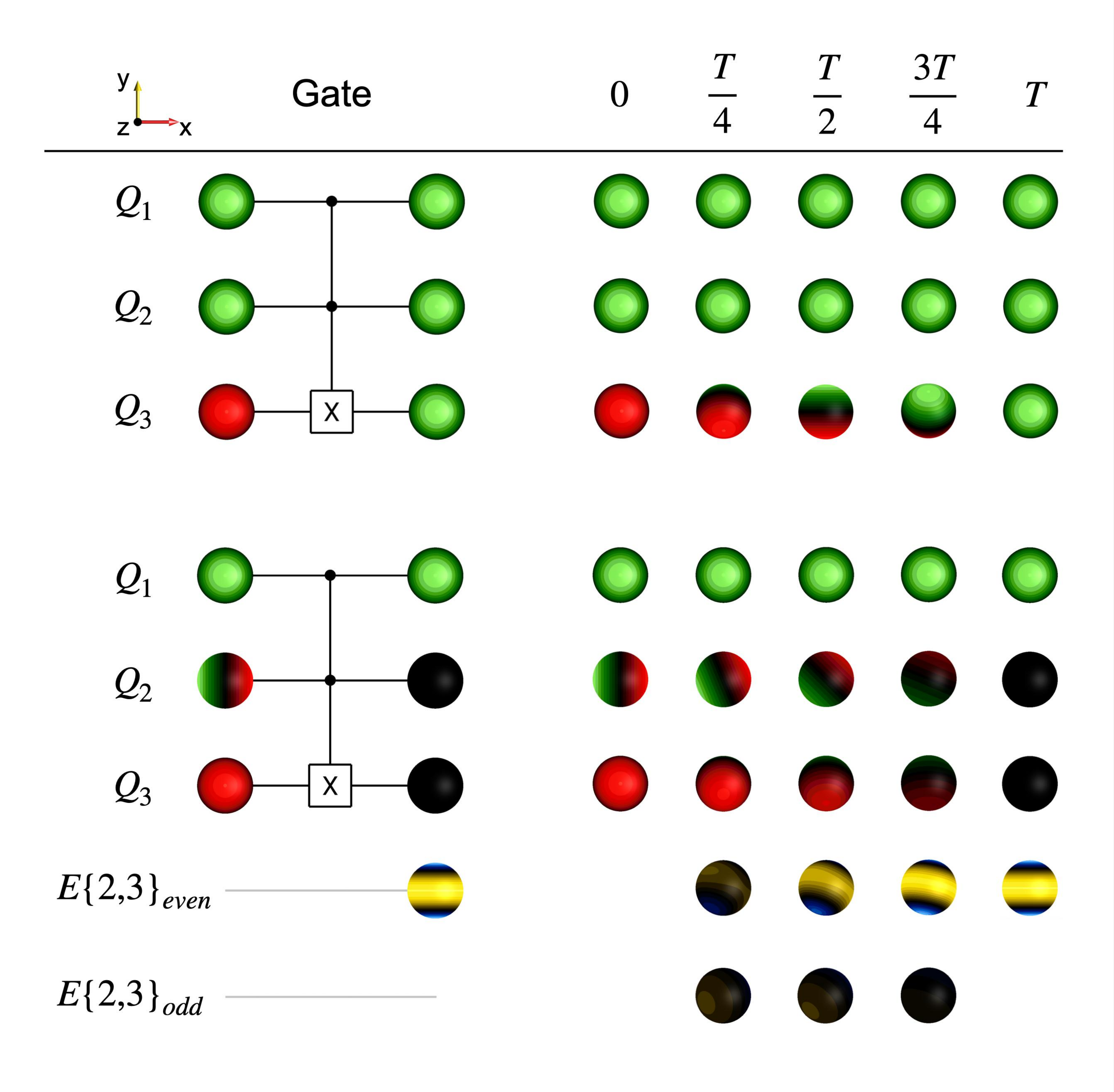}
\caption{\label{fig:FigGatesToffoli}Three-qubit Toffoli-gate visualized by the BEADS representation \screen. Different applications of the Toffoli gate (double-controlled X-gate) become clear by using the BEADS representation to visualize different outcomes. Computational basis states $\ket{1}$ of both control qubits induce simple rotations of the target Q-Bead (top). By using a computational and an x-basis input on the controls, it is possible to generate Bell pairs in a controlled manner which is clear from the characteristic E-Bead surface pattern of $E\{2,3\}_{\textit{even}}$ (bottom).}
\end{figure}

\begin{figure}[H]
\centering
\includegraphics[width=.8\textwidth]{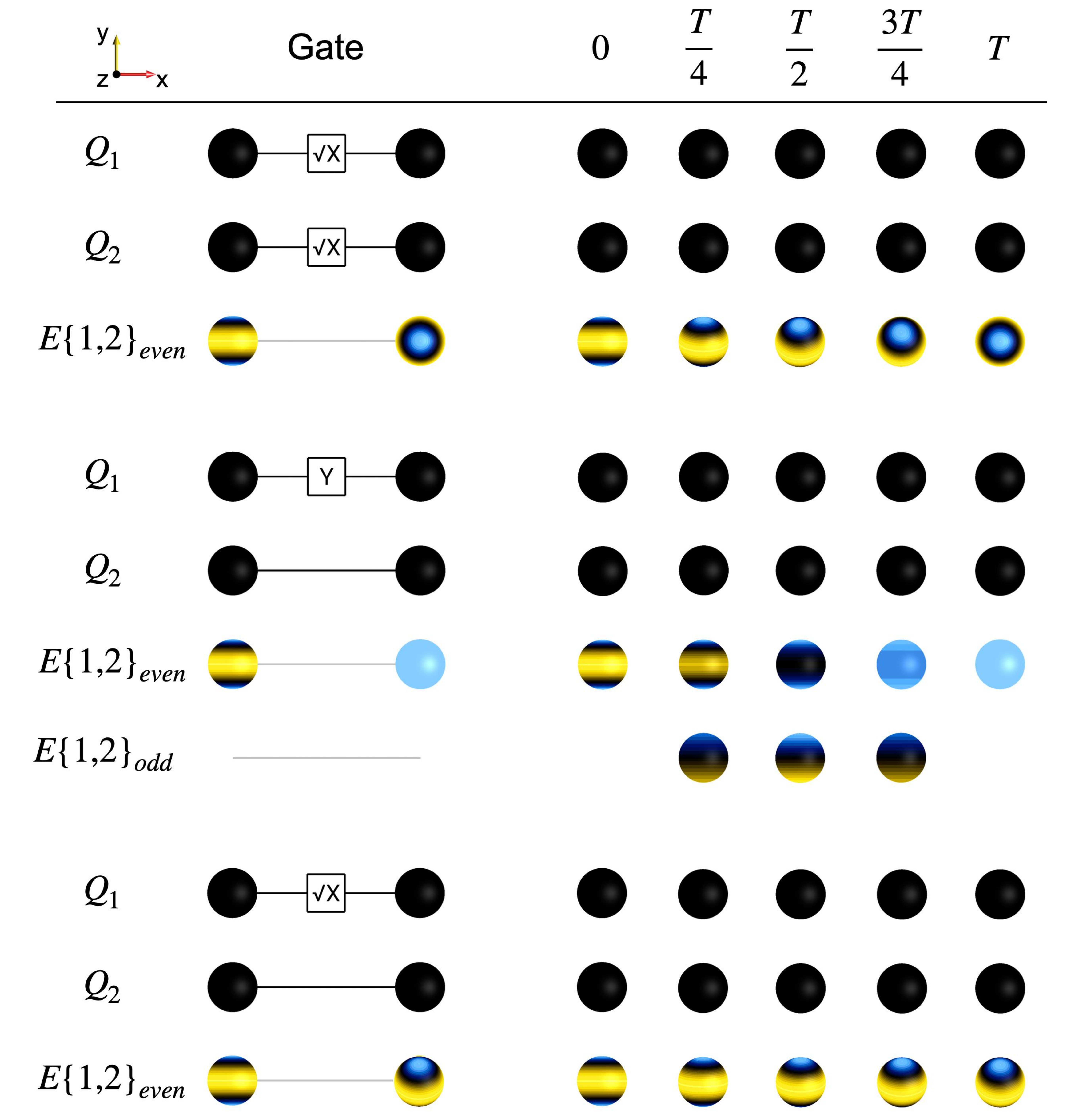}
\caption{\label{fig:FigGatesMQTransform}Effects of single-qubit gates on two-qubit entangled states \screen. Global rotations (top), i.e., rotations which are induced by the same single-qubit gate being applied to all qubits which are involved in a particular $n$-partite connected correlation function component simply rotate the corresponding E-Bead analogously. Applying single-qubit gates locally on a subset of the involved qubits in general causes morphing, i.e. a change in surface pattern (center example), yet the amount of entanglement remains unchanged. Exceptions exist for which a local rotation induces a simple rotation of the E-Bead. In the bottom example, the E-Bead representing the Bell state $\ket{\Phi^+}=1/\sqrt{2}\left(\ket{00}+\ket{11}\right)$ is rotated by $45^\circ$ (half of the gate rotation angle) around the x-axis by applying $90^\circ$-rotation around said axis ($\sqrt{\text{X}}$-gate) on qubit $Q_1$. Such exceptions can be readily identified in the BEADS picture.}
\end{figure}

\noindent
Moreover, with the BEADS representation it is possible to explore possible applications of more complicated gates such as the Toffoli gate (double-controlled NOT gate) \cite{NielsenChuang} shown in Fig.~\ref{fig:FigGatesToffoli}. Here, we can directly see that both control qubits in computational basis states implement a controlled rotation of the target qubit, whereas using an equal superposition on one control qubit while keeping the second control to a computational basis state effectively creates a Bell pair between the target qubit and the control qubit, which was input in a superposition state, in a controlled fashion, e.g., here in dependence of the state of $Q_1$.
\noindent
Extending the discussion on single-qubit operations, the effect of single-qubit gates on multiqubit entangled states shall be discussed. As an example, we visualize the transformation of the Bell pair $\ket{\Phi^+}=1/\sqrt{2}\left(\ket{00}+\ket{11}\right)$ by applying a $\sqrt{\text{X}}$-gate globally, i.e., to all qubits involved in the entanglement (or more general the correlation), in Fig.~\ref{fig:FigGatesMQTransform} for which we find that the E-Bead is simply rotated according to the rotation induced by the quantum gate on the physical qubits. This serves as a general rule and is applicable to any $n$-qubit correlation function bead (E-, T-, or C-Bead).  

If local rotations, e.g., a Y-gate, are applied on just a subset of the involved qubits (for instance, only on $Q_1$ in Fig.~\ref{fig:FigGatesMQTransform}), various effects can occur. The E-Bead is morphed if the spatial correlation function properties change. As an example, in the second circuit of Fig.~\ref{fig:FigGatesMQTransform}, the $E\{1,2\}_{\textit{even}}$ E-Bead morphs from its original appearance to a fully blue bead corresponding to the singlet state $\ket{\Psi^-}=1/\sqrt{2}\left(\ket{01}-\ket{10}\right)$. This morphing by selective local gates is quite typical. However, there are important exceptions such as the last example presented in Fig.~\ref{fig:FigGatesMQTransform} where selective gates induce simple rotations of a correlation function bead which can be readily identified in the BEADS picture. These arise due to special transformation rules which prominently occur in the analysis of multiple-quantum coherence \cite{ErnstBodenhausen} yet go beyond the scope of this work. Furthermore, E-Beads are scaled when local rotations cause a transfer to E-Beads of different permutation symmetry, e.g., from $E\{1,2\}_{\textit{even}}$ to $E\{1,2\}_{\textit{odd}}$, where the former is dimmed and the brightness of the latter E-Bead increases.

\subsection{NMR implementations of quantum gates}\label{sec:BEADSNMRDS}
In this article we mostly considers quantum gates from a theoretical perspective without focusing on specific experimental implementations, which depend on the choice of hardware. However, we can also use the BEADS representation to explore whether and how particular quantum gates can be achieved on distinct hardware.

Liquid state nuclear magnetic resonance (NMR) was one of the first platforms which allowed to realize registers of few qubits and to perform basic quantum algorithms \cite{NielsenChuang}. Coupled systems of nuclear spins-$1/2$ which constitute qubits in NMR, can, e.g., be manipulated by applying radiofrequency pulses, i.e., local rotations on individual spins and by letting the system evolve under an Ising-type interaction of the spins.

\begin{figure}[H]
\centering
\includegraphics[width=1.\textwidth]{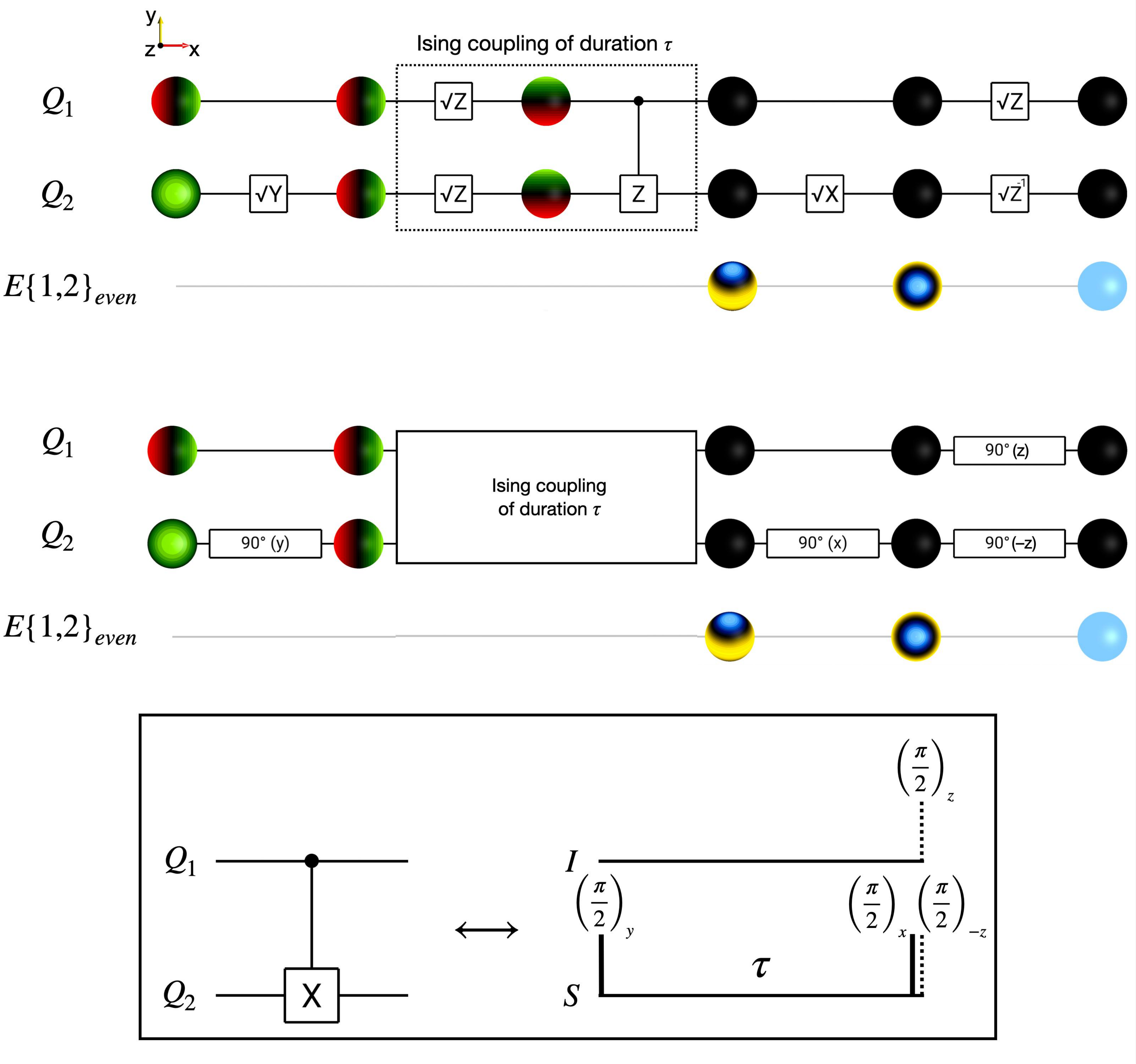}
\vspace{-.3cm}
\caption{\label{fig:FigGatesNMR}Implementation of a CNOT gate in NMR-based quantum computing \screen. A CNOT gate between two qubits $Q_1$ and $Q_2$ can be implemented as a sequence of radiofrequency pulses and an Ising-interaction between to heteronuclear spins-$1/2$ which are denoted $I$ and $S$ (bottom). Here, the delay time $\tau = 1/(2J)$, where $J$ is the coupling constant. The NMR sequence is visualized by equivalent BEADS-augmented circuits using standard gate notations (top) and NMR pulse labels (center). Applying the sequence to qubits in an initial state $\ket{-1}$ yields the maximally entangled singlet state $\ket{\Psi^-}=1/\sqrt{2}\left(\ket{01}-\ket{10}\right)$ which matches the action of the CNOT operation (up to a global phase).}
\vspace{-1cm}
\end{figure}

As shown in Fig.~\ref{fig:FigGatesNMR}, by using these building blocks, it is possible to implement a CNOT gate (up to a global phase) between two spins with zero offset frequency as a sequence of pulses and Ising interactions \cite{Marx00,NMRQCGlaser}.
\vspace{-.5cm}
\noindent
This can also be shown mathematically by calculating the total propagator $U$ of the displayed pulse sequence:
\begin{align}
U&=e^{-i\left(\frac{\sigma_{1z}}{2}\frac{\pi}{2}-\frac{\sigma_{2z}}{2}\frac{\pi}{2}\right)}\cdot\ e^{-i\frac{\sigma_{2x}}{2}\frac{\pi}{2}}\cdot\ e^{-i\frac{\sigma_{1z}\sigma_{2z}}{2}\frac{\pi}{2}}\cdot\ e^{-i\frac{\sigma_{2y}}{2}\frac{\pi}{2}} \nonumber\\
&=e^{-i\frac{\pi}{4}}\left(\begin{matrix}1&0&0&0\\0&1&0&0\\0&0&0&1\\0&0&1&0\\\end{matrix}\right)=e^{-i\frac{\pi}{4}}\text{CNOT}_{12},
\end{align}
\noindent
which is the CNOT gate up to an experimentally irrelevant global phase factor $e^{-i\frac{\pi}{4}}$.

Using the BEADS representation, we can directly verify that the action of the introduced pulse sequence implements the desired CNOT operation. In Fig.~\ref{fig:FigGatesNMR}, we chose an input state $\ket{-1}$, where the first spin is in state $\ket{-}$ and the second spin is in state $\ket{1}$, which is well-known to give the singlet state $\ket{\Psi^-}=1/\sqrt{2}\left(\ket{01}-\ket{10}\right)$ under transformation by a CNOT gate (cf.~\ref{Figure:Fig3} or supplementary section~\ref{sec:BellStates}). Indeed, the output state is $\ket{\Psi^-}$ which is represented by the characteristic uniformly colored blue E-Bead $E\{1,2\}_{\textit{even}}$ indicating full anticorrelation in any direction.

\subsection{Bell states}\label{sec:BellStates}
BEADS representations of the four Bell states (also referred to as \textit{EPR pairs}) \cite{Bell64} $\ket{\Phi^+}=1/\sqrt{2}(\ket{00}+\ket{11})$, $\ket{\Psi^+}=1/\sqrt{2}(\ket{01}+\ket{10})$, $\ket{\Phi^-}=1/\sqrt{2}(\ket{00}-\ket{11})$, and $\ket{\Psi^-}=1/\sqrt{2}(\ket{01}-\ket{10})$ and the corresponding generating circuits are shown in detail in Fig.~\ref{fig:FigBellStates}. Note that the states $\ket{\Phi^+}$, $\ket{\Phi^-}$, and $\ket{\Psi^+}$ are equivalent up to global rotations, that is, the same local rotation applied to all qubits. $\ket{\Psi^-}$, the singlet state, is fully anticorrelated in all spatial directions and thus has an entirely different correlation function BEADS visualization.
\begin{figure}[H]
\includegraphics[width=.9\textwidth]{FigBellStates.pdf}
\caption{\label{fig:FigBellStates}BEADS representation of Bell states and generating circuits \screen. Bell states are generated by choosing a two-qubit computational basis state and applying a circuit H$_1$-CNOT$_{12}$ to the system. Depending on the choice of initial state, one obtains one of four Bell states $\ket{\Phi^+}=1/\sqrt{2}(\ket{00}+\ket{11})$, $\ket{\Psi^+}=1/\sqrt{2}(\ket{01}+\ket{10})$, $\ket{\Phi^-}=1/\sqrt{2}(\ket{00}-\ket{11})$, or
$\ket{\Psi^-}=1/\sqrt{2}(\ket{01}-\ket{10})$ where all except the latter are equivalent up to global rotations.}
\end{figure}

\subsection{Graph states}\label{sec:BEADSGraphStates}
\textit{Graph states} are another important class of entangled multi-qubit states that, as their name suggests, can be described by graphs and which are used in the study of entanglement and represent a natural choice in measurement-based quantum computing \cite{HeinGraph04, HeinGraph06}. They can be transformed into each other following well-known rules. Using the BEADS representation, it is easily possible to illustrate graph state transformation rules and to get insights into the topologies and symmetries of the underlying entanglement-based connected correlations.

\begin{figure}[H]
\centering
\includegraphics[width=.8\textwidth]{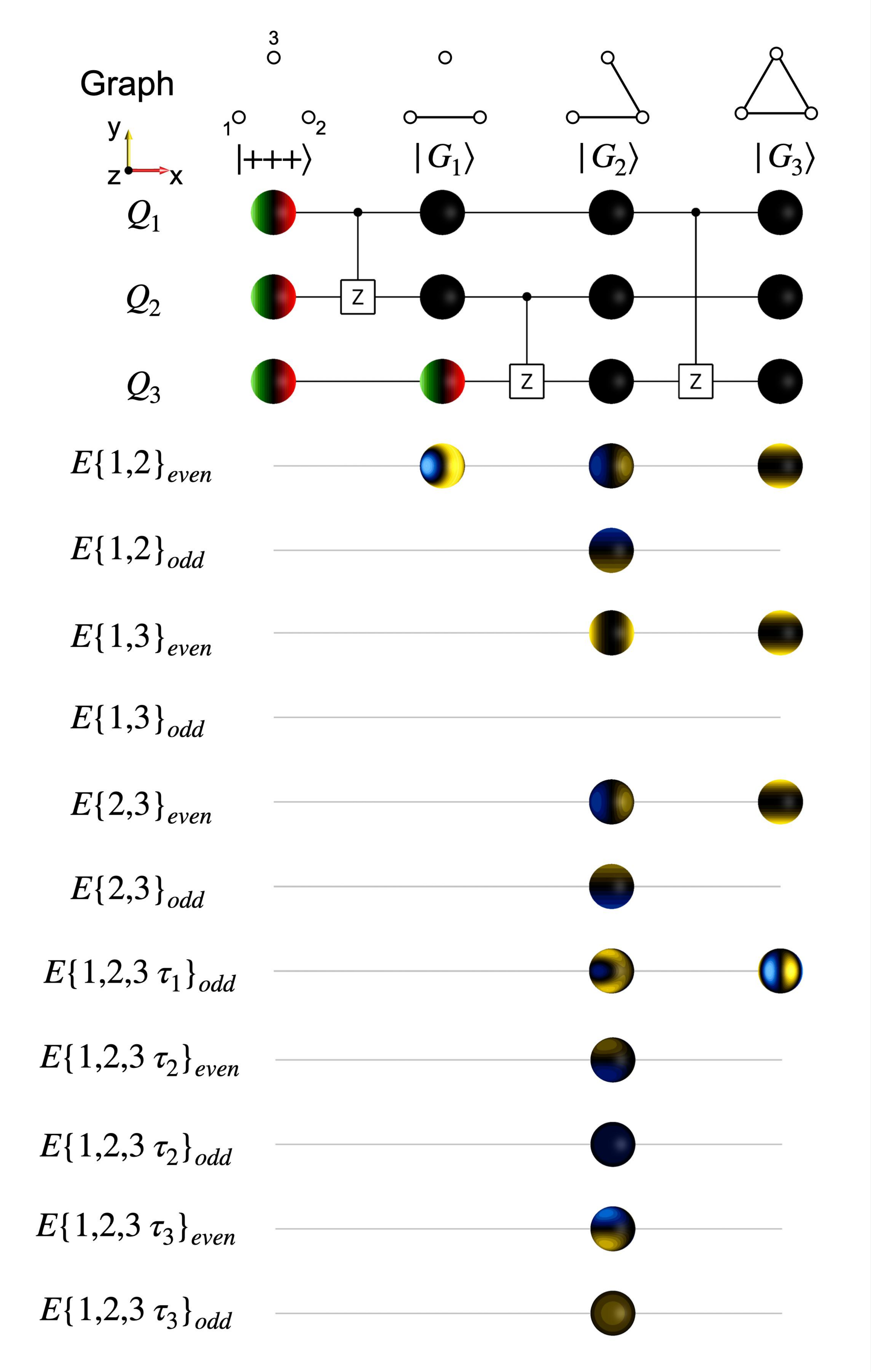}
\caption{\label{fig:FigGraph1}Creation of three-qubit graph states by performing CZ-gates \screen. Starting with an initial state  $\ket{+++}=1/\sqrt{8}\left(\ket{000}+\ket{001}+\ket{010}+\ket{011}+\ket{100}+\ket{101}+\ket{110}+\ket{111}\right)$, controlled Z-gates are performed between pairs of qubits consecutively which gives graph states $\ket{G}$ with the underlying graphs shown above the states. The graph states from left to right are: $\ket{G_1}=1/\sqrt{8}\left(\ket{000}+\ket{001}+\ket{010}+\ket{011}+\ket{100}+\ket{101}-\ket{110}-\ket{111}\right)$, 
$\ket{G_2}=1/\sqrt{8}\left(\ket{000}+\ket{001}+\ket{010}-\ket{011}+\ket{100}+\ket{101}-\ket{110}+\ket{111}\right)$, $\ket{G_3}=1/\sqrt{8}\left(\ket{000}+\ket{001}+\ket{010}-\ket{011}+\ket{100}-\ket{101}-\ket{110}-\ket{111}\right)$. Note that $\ket{G_2}$ and $\ket{G_3}$ are both maximally entangled and equivalent up to local transformations (see experiment \textbf{B} in Fig.~\ref{fig:FigGraph3}) despite having different numbers of edges.}
\end{figure}

\begin{figure}[H]
\centering
\includegraphics[width=1.\textwidth]{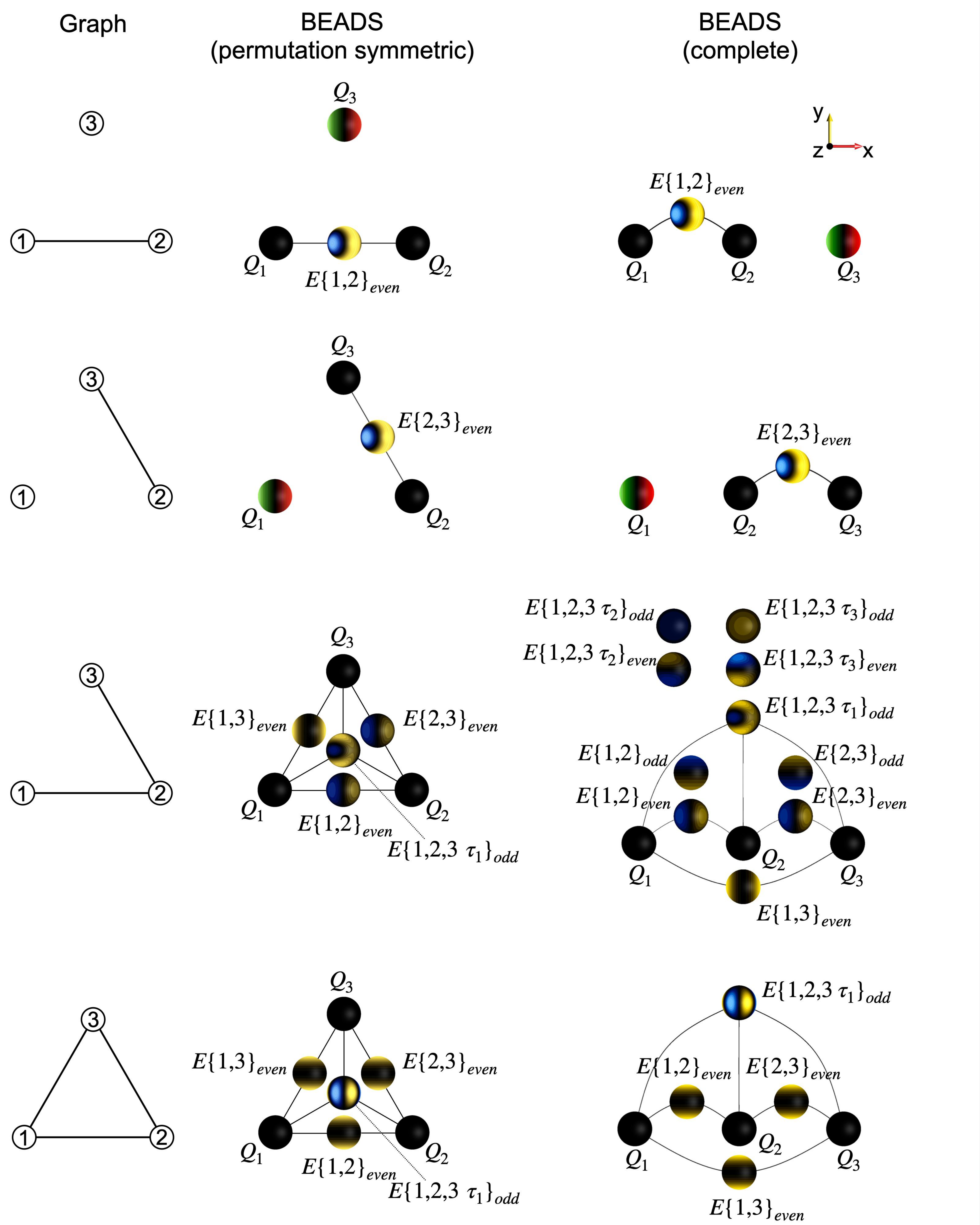}
\caption{\label{fig:FigGraph2}BEADS visualizations of graph states. Arranged in a triangular layout of beads showing only fully permutation symmetric components (center), the BEADS representations resemble the corresponding graphs. Complete BEADS visualizations of the graph states in standard layout reveal further symmetry properties.  Graph states with only one edge (top two examples) correspond to Bell pairs up to a global rotation of the entangled qubits. The chain and triangle graph states (bottom examples) are both maximally entangled but the triangle graph state is fully symmetric (only fully symmetric E-Beads are non-zero) whereas the chain graph state (second to bottom) has considerable asymmetries.}
\end{figure}

Graph states can be created by applying \textit{controlled Z-gates} (CZ) to an $N$-qubit system with all qubits in the equal superposition state $\ket{+}=1/\sqrt{2}\left(\ket{0}+\ket{1}\right)$. In the graph which represents the state, the individual qubits correspond to \textit{vertices} and each CZ$_{kl}$ operation implements an \textit{edge} between vertices $k$ and $l$. In general, a graph state $\ket{G}$ is thus defined as \cite{HeinGraph04}
\begin{equation}\label{eq:BEADSGraph}
\ket{G} = \prod_{k,l \in E} \text{CZ}_{kl} \ket{+}^\otimes,
\end{equation}
\noindent
where $E$ denotes the set of all edges. Note that all CZ-gates mutually commute and hence they can be either applied simultaneously or sequentially in any desired order to reach the graph state $\ket{G}$. Fig.~\ref{fig:FigGraph1} shows a three-qubit quantum circuit using the BEADS representation during which several graph states are generated by consecutively adding edges to the previous graph, respectively. Indeed, in the BEADS visualization we can readily see that after the first CZ operation (CZ$_{12}$) only two qubits ($Q_1$ and $Q_2$) are entangled and the resulting E-Bead surface pattern of $E\{1,2\}_{\textit{even}}$ is equivalent to that of the Bell states (aside from the singlet state, see section~\ref{2Q}) up to a global rotation. Performing a second CZ operation between qubits $Q_2$ and $Q_3$ further entangles the system. As all Q-Beads are black at this point, the state is maximally entangled, i.e., entanglement is shared between all qubits, even though in the corresponding graph not all vertices are connected by edges. This represents an important observation which can potentially lead to misconceptions when first studying graph states: The edges in a graph are indirectly connected to entanglement in the corresponding graph state, but missing edges do not strictly imply that the corresponding vertices, that is, qubits, are not entangled. We can further see this when applying a third CZ-gate which gives a complete triangle graph. The state is still maximally entangled, i.e., the amount of entanglement has not changed, however, the intrinsic structure of the entanglement has altered to fully permutation symmetric components (see E-Beads) only. This graph state is the GHZ state up to a global rotation of $90^\circ$ around the x-axis. The discussed states are shown in detail using an adapted layout, where the Q-Beads are arranged in the identical form as the vertices illustrating the graph states (in this case in a triangle) and where in addition the simplified display type H (cf. Table~\ref{tab:Variants} and Fig.~\ref{Figure:Fig10}) is applied, and using the standard BEADS layout in Fig.~\ref{fig:FigGraph2}. 

\begin{figure}[H]
\centering
\includegraphics[width=.95\textwidth]{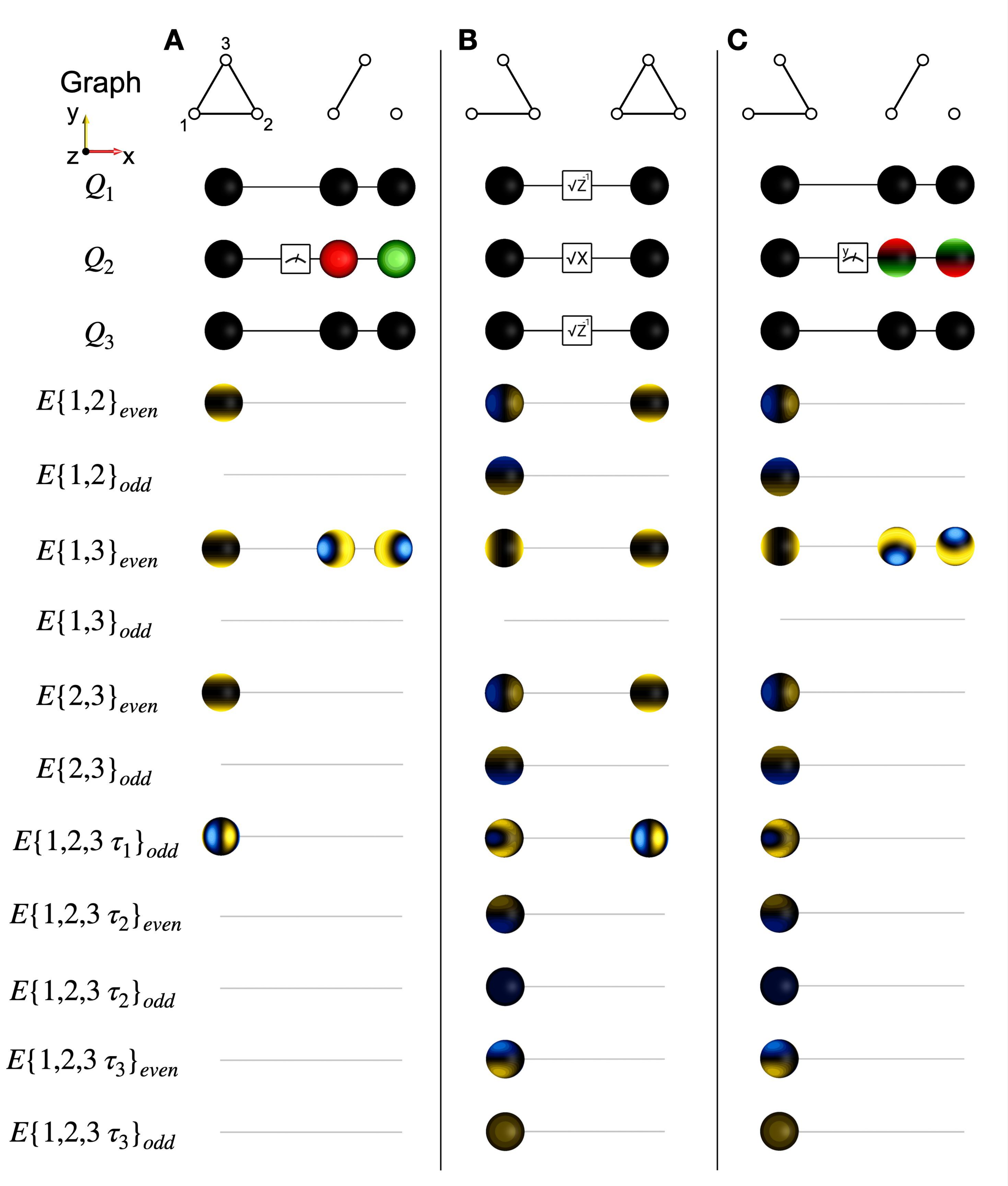}
\caption{\label{fig:FigGraph3}Transformations of graph states visualized by BEADS. When performing (\textbf{A}) a z-measurement on a qubit (vertex) in a graph state (left circuit), the vertex and its connected edges are eliminated from the graph. The resulting state must be corrected by performing a Z-gate on all neighboring vertices when the measurement outcome is $m_z=-1$ and by classically controlled rotations to bring back the measured qubit to state $\ket{+}$ which is immediately clear in the BEADS picture when comparing the resulting states with the single-edge graph states in Fig.~\ref{fig:FigGraph1} or Fig.~\ref{fig:FigGraph2}. Local complementation (\textbf{B}) creates or eliminates edges between neighboring vertices (see Eq.~\ref{eq:BEADSLocComp}). In the center circuit, local complementation is applied on vertex 2 in the graph state $\ket{G_2}$ (see Fig.~\ref{fig:FigGraph1}). Indeed, using BEADS, it is clear that the resulting state is the triangle graph state $\ket{G_3}$. Y-measurements (\textbf{C}) remove a vertex and its connected edges from a graph and additionally complement the neighboring vertices. Again, the resulting states must be corrected to represent graph states. The required corrections are obvious in the corresponding BEADS representations and amount to a $\sqrt{\text{Z}^{-1}}$-gate (if $m_y=1$) or a $\sqrt{\text{Z}}$-gate (if $m_y=-1$) on all neighboring vertices. All displayed measurement outcomes are obtained with a probability of 50\%.}
\end{figure}

Graph states further play an important role in modern quantum information science as they can be interconverted according to simple rules. As briefly discussed before, graph states are a popular choice in measurement-based quantum computing since the action of (Pauli) measurements on these states follows simple relations. For instance, a \textit{computational basis measurement} performed on one qubit corresponds to removing the dedicated vertex and all its connected edges from the graph yielding a state which is again a graph state up to outcome-dependent local corrections \cite{HeinGraph04, HeinGraph06}. This is exemplarily visualized by experiment \textbf{A} in Fig.~\ref{fig:FigGraph3} where qubit $Q_2$ is measured in the triangle graph state and we obtain a residual state which corresponds to the graph state (up to local correction) with a single edge between $Q_1$ and $Q_3$. Here, the BEADS representation allows to directly tell the \textit{correction rule} which amounts to a Z-gate on all neighbors of the measured vertex which were previously connected to the latter by an edge when the measurement outcome is $m_z=-1$ (qubit measured in state $\ket{1}$). In addition, local (classically controlled) corrections, which also can be directly inferred from the BEADS representation (e.g., $\sqrt{\text{Y}}$ if $m_z=1$ or $\sqrt{\text{Y}^{-1}}$ if $m_z=-1$), have to be applied to bring back the measured qubit to the superposition state $\ket{+}$.

Another important transformation of graph states is given by so-called \textit{local complementation} which can be achieved by applying a unitary $\sqrt{S_v}$ such that \cite{HeinGraph04, HeinGraph06}
\begin{equation}\label{eq:BEADSLocComp}
\sqrt{S_v} = e^{-i\frac{\pi}{4}\sigma_{vx}} \prod_{k\in\mathcal{N}_v} e^{i\frac{\pi}{4}\sigma_{kz}},
\end{equation}
\noindent
where $v$ denotes the target vertex and $\mathcal{N}_v$ is the set of all neighboring vertices connected to $v$ by an edge.  Note that this corresponds to performing a $\sqrt{\text{X}}$-gate on the vertex $v$ and a $\sqrt{\text{Z}^{-1}}$-gate (inverse square root of Z-gate) on all neighboring vertices. Local complementation causes an inversion of all edges between vertices in $\mathcal{N}_v$, i.e., edges are created if they were not present before the operation and vice versa. Since local complementation only involves local gates, the total amount of entanglement is not changed, yet edges are created or eliminated. In the second experiment (\textbf{B}) illustrated in Fig.~\ref{fig:FigGraph3}, we visualize how the triangle graph state can be created by local complementation of $Q_2$. By comparing the complemented state with the BEADS representations of states in Fig.~\ref{fig:FigGraph2} and Fig.~\ref{fig:FigGraph3}, we can indeed directly verify that the transformation yields the intended graph state up to obvious classically controlled corrections.
\noindent
A different way of creating edges between non-connected vertices can be achieved by \textit{y-measurements}. In contrast to z-measurements where a vertex and its connected edges are removed from a graph, y-measurements additionally lead to a \textit{complementation} of the neighboring vertices similar to what was observed for local complementation \cite{HeinGraph04, HeinGraph06}. Once again, the resulting state is a graph state up to local correction and we can directly tell the y-measurement correction rule by examining the BEADS representation of (\textbf{C}) the y-measurement circuit in Fig.~\ref{fig:FigGraph3}. Here, the following rule applies: A $\sqrt{\text{Z}^{-1}}$-gate has to be performed on all vertices in the neighborhood $\mathcal{N}_v$ of the measured qubit $Q_v$ when the measurement outcome is $m_y=1$ (corresponding to a state $\ket{R}=1/\sqrt{2}\left(\ket{0}+i\ket{1}\right)$) and a $\sqrt{\text{Z}}$-gate in case of outcome $m_y=-1$ (state $\ket{L}=1/\sqrt{2}\left(\ket{0}-i\ket{1}\right)$).

\subsection{Extended analysis of Grover's algorithm}\label{sec:BEADSGroverSupp}
Using the BEADS representation, we can directly compare the effects of different implementations of the same circuit building block of Grover's algorithm \cite{Grover} in the BEADS representation on a visual level. The remaining steps of the circuit are identical as in the initially discussed circuit (see section~\ref{Grover}).

The top circuit in Fig.~\ref{fig:FigGroverSupp} displays an alternative implementation of the two-qubit algorithm discussed in section~\ref{Grover}. Note that here we use two system qubits represented by Q-Beads $Q_1$ and $Q_2$ which are initialized in state $\ket{0}$ and an ancilla qubit denoted $Q_a$ which is initialized to $\ket{1}$. Applying a Hadamard to every qubit rotates the corresponding Q-Beads such that they become oriented parallel (state $\ket{+}$) or antiparallel (state $\ket{-}$ of the ancilla qubit) to the x-axis. We exemplarily use the same solution bitstring 01. However, unlike in the circuit in Fig.~\ref{Figure:Fig8} which uses a phase oracle $\text{U}_\omega$, this circuit applies a Boolean oracle $\text{U}_\text{f}$ which is implemented as a generalized Toffoli gate (see supplementary section~\ref{sec:BEADSGates}) on the ancilla qubit with both system qubits serving as controls or anticontrols (depending on the solution bit string being 1 or 0 at each bit position, respectively). Note that only the system qubits are affected by this operation and the oracle maximally entangles the system qubits while the ancilla qubit is left unchanged in state $\ket{-}$.

Indeed, this example nicely visualizes the \textit{phase kickback} phenomenon \cite{NielsenChuang} where the kickback from the target ancilla qubit induces exactly the same transformation on the system qubits as can be observed for the phase oracle $\text{U}_\omega$ in the previously discussed circuit, yet at the cost of an additional required qubit. The gate U$_{\text{0}}= 2\ket{00}\bra{00}-\mathds{1}$ operation, where $\mathds{1}$ is the identity matrix, fully disentangles the system. The terminal global Hadamard operation orients the Q-Beads along the z-axis. 

The three-qubit Grover search at the bottom of Fig.~\ref{fig:FigGroverSupp} displays the same circuit as was introduced in the Results section. In the single solution case, using the BEADS representation one can visually estimate the probability psolution to measure the correct solution. Indeed, this probability is given by \cite{Boyer98}
\begin{equation}\label{eq:GroverProb}
p_{\text{solution}}=\left[ \frac{1}{\sqrt{s}} \sin{\left(\left(2t+1\right)\text{asin}{\sqrt{\frac{s}{D}}}\right)}\right]^2,
\end{equation}
where $s$ is the number of solutions, $D$ is the size of the search space, and $t$ is the number of Grover iterations. After six iterations though, we almost find perfect alignment ($p_{\text{solution}} = 99.98\%$) of the Q-Beads and almost all-zero E-Beads which corresponds to measuring the solution almost with certainty.
\begin{figure}[H]
\centering
\includegraphics[width=1.0\textwidth]{FigGroverSupp.pdf}
\caption{\label{fig:FigGroverSupp}BEADS-augmented circuit representation of a three-qubit single-solution Grover's algorithm \cite{Grover} \screen. Grover's quantum search algorithm \cite{Grover} is visualized for two system qubits using a Boolean oracle implemented by using an additional ancilla qubit (top) and three qubits (bottom). Using the BEADS representation, steps, where entanglement is maximum or minimum, can be readily identified and one can directly read off the solution. For three qubits, visualizing the outcomes of each Grover iteration U$_{\text{G}}$ reveals where a solution can be measured with high probability which would otherwise have to be analyzed using analytical functions such as displayed in the bottom plot. In the three-qubit circuit, the probability of measuring the correct solution is indicated above the BEADS-augmented circuit after each Grover iteration.}
\end{figure}
 
\subsection{Quantum teleportation examples}\label{sec:BEADSTelSupp}
Here, we provide further examples of the quantum teleportation protocol \cite{Teleportation} for different initial states and BEADS representation variants.

\begin{figure}[H]
\vspace{1cm}
\centering
\includegraphics[width=1.0\textwidth]{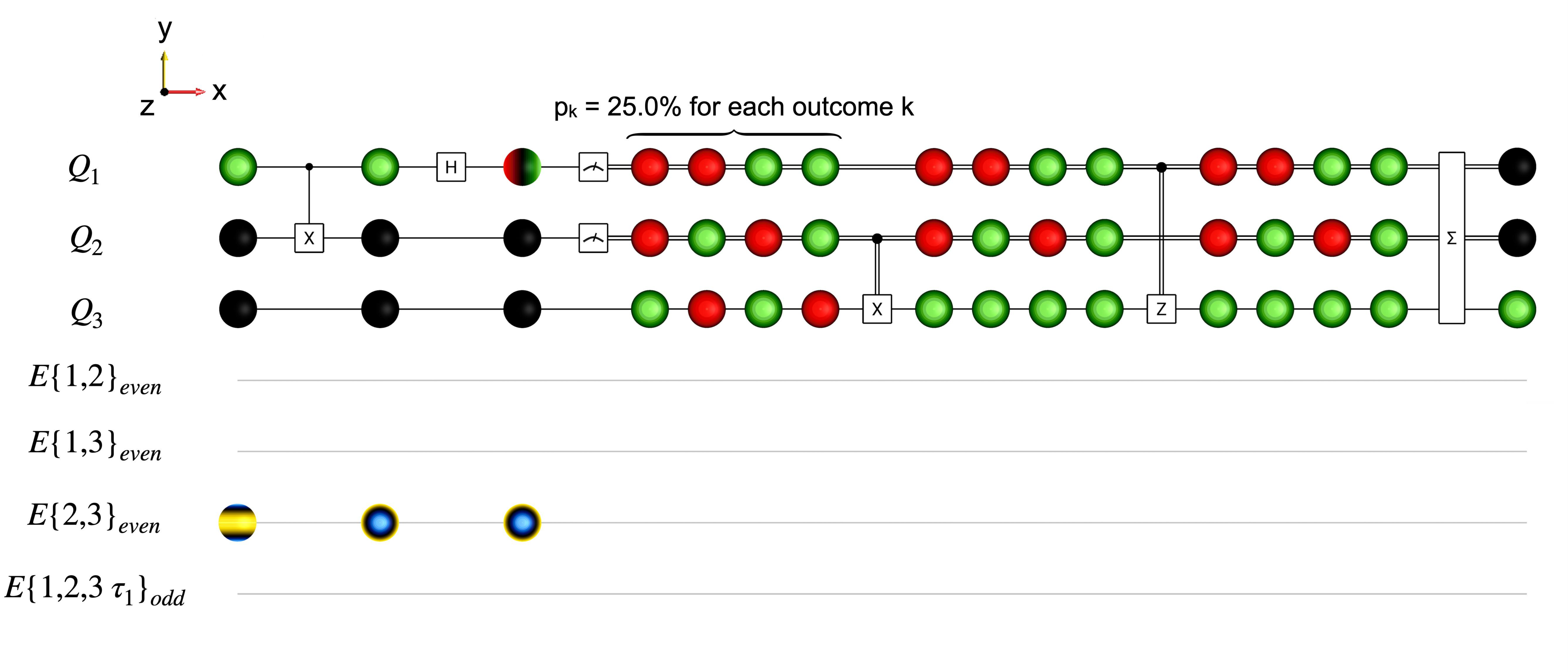}
\caption{\label{fig:FigTeleportZ}BEADS representation of the quantum teleportation of the computational basis state $\ket{1}$ which is transferred from $Q_1$ to $Q_3$ \screen. In this case, the protocol does not entangle $Q_1$ with the Bell pair formed by $Q_2$ and $Q_3$. The computational basis state $\ket{1}$ is teleported from qubit $Q_1$ to $Q_3$. Initially, qubits $Q_2$ and $Q_3$ form a Bell pair $\ket{\Phi^+}=1/\sqrt{2}\left(\ket{00}+\ket{11}\right)$, which can be created according to the standard protocol as shown in the Results section and section~\ref{sec:BellStates} of the supplementary material. The initial CNOT$_{12}$-gate causes a simple rotation (special case, see transformations induced by multiqubit gates in section~\ref{sec:BEADSGates}) of the Bell pair E-Bead which in the BEADS representation can be directly seen to be transformed into the Bell state $\ket{\Psi^+}=1/\sqrt{2}\left(\ket{01}+\ket{10}\right)$ (cf. Fig.~\ref{fig:FigBellStates} in section~\ref{sec:BellStates}). The subsequent Hadamard gate causes a rotation of $Q_1$ to give the single-qubit superposition state $\ket{-}=1/\sqrt{2}\left(\ket{0}-\ket{1}\right)$. From the E-Bead we can further tell that measurement outomes of $Q_2$ and $Q_3$ are fully anticorrelated, i.e., the qubits adopt opposite basis states when one of the qubits is measured along the z-axis. Moreover, we find Q-Bead $Q_1$ to be black along the z-axis which corresponds to a probability of $50\%$ to measure the qubit in state $\ket{1}$. Hence, upon measurement of qubits $Q_1$ and $Q_2$ we observe four possible outcomes with strictly opposite outcomes of $Q_2$ and $Q_3$ due to the discussed anticorrelation and uncorrelated results for $Q_1$. In this case, the following correction by a (classically) controlled X-gate on $Q_3$ based on the outcome of $Q_2$ already corrects all Q-Beads such that the teleportation is complete after this step ($Q_3$ is in state $\ket{1}$ in all outcome scenarios). Indeed, as can be expected, we observe no effect of the (classically) controlled Z-gate due to the axial symmetry with respect to the z-axis of Q-Beads representing computational basis states. The final step in the circuit shows the weighted sum (mixed state) of all outcome scenarios which nicely depicts that the desired state $\ket{1}$ was teleported with certainty (see $Q_3$) independent of the measurement outcomes.}
\end{figure}
\begin{figure}[H]
\centering
\includegraphics[width=1.0\textwidth]{FigTeleportX.pdf}
\caption{\label{fig:FigTeleportX}Teleportation of the x-basis state $\ket{+}=1/\sqrt{2}\left(\ket{0}+\ket{1}\right)$ \screen. Initially, qubits $Q_2$ and $Q_3$ form a Bell pair $\ket{\Psi^+}=1/\sqrt{2}\left(\ket{01}+\ket{10}\right)$. The initial CNOT$_12$-gate entangles the first qubit with the Bell pair which can be seen by the trilinear E-Beads which emerge due to genuine three-qubit entanglement. After applying the Hadamard gate to $Q_1$ we find that all permutation symmetric E-Beads ($E\{1,2\}_{\textit{even}}$, $E\{1,3\}_{\textit{even}}$, $E\{2,3\}_{\textit{even}}$, $E\{1,2,3\:\tau_1\}_{\textit{odd}}$) are black along the z-axis and hence, there will be no bilinear zz-correlation between measurement outcomes of the qubits, i.e., equal outcomes for $Q_1$ and $Q_2$ are to be expected with a probability of $50\%$. Further analysis reveals full zx-correlation of outcomes corresponding to qubits $Q_1$ and $Q_3$ (for mathematical details see supplementary section~\ref{sec:BEADSAsym}) which can also be deduced visually from the individual measurement outcomes of both qubits. Moreover, unlike in the teleportation of a y-basis state (Fig.~\ref{fig:FigTeleportYFull}), the measurement outcomes are trilinearly uncorrelated for the shown measurement. Hence, the discussed zx-correlation leads to $Q_3$ resulting in state $\ket{+}$ if the outcome of the first qubit is $m_1=1$ (corresponding to $Q_1$ in state $\ket{0}$) and in state $\ket{-}=1/\sqrt{2}\left(\ket{0}-\ket{1}\right)$ otherwise. It then can be seen that the following correction by the (classically) controlled X-gate on $Q_3$ does not have any impact as the possible states of $Q_3$ are invariant under rotations around the x-axis. All required corrections are thus implemented by the (classically) controlled Z gate and we find that $Q_3$ aligns according to the desired state in all scenarios. The final step in the circuit involves taking the weighted sum (mixed state) of all outcome scenarios which also nicely depicts that teleportation of the desired state $\ket{+}$ was achieved with certainty (see $Q_3$) independent of the measurement outcomes. A high contrast color scheme variant of this figure is available in appendix~\ref{app:BEADSHighContrast}.}
\end{figure}
\begin{figure}[H]
\centering
\includegraphics[width=1.0\textwidth]{FigTeleportYFull.pdf}
\caption{\label{fig:FigTeleportYFull}Complete BEADS representation of the teleportation of the y-basis state $\ket{R}$ \screen. The y-basis state $\ket{R}=1/\sqrt{2}\left(\ket{0}+i\ket{1}\right)$ is teleported from qubit $Q_1$ to $Q_3$. Initially, qubits $Q_2$ and $Q_3$ form a Bell pair $\ket{\Phi^+}=1/\sqrt{2}\left(\ket{00}+\ket{11}\right)$. The initial CNOT$_{12}$-gate entangles the first qubit with the Bell pair which can be seen by the trilinear E-Beads which arise due to genuine three-qubit entanglement. After applying the Hadamard gate to $Q_1$ the permutation symmetric E-Bead $E\{1,2\}_{\textit{even}}$ is black along the z-axis and hence, there is no bilinear zz-correlation between $Q_1$ and $Q_2$, i.e., equal outcomes for both qubits are to be expected with a probability of $50\%$. From the E-Bead $E\{2,3\}_{\textit{even}}$ it becomes clear that there is only symmetric bilinear xx-correlation of measurement outcomes. All non-zero odd bilinear E-Beads indicate bipartite yx-correlations between outcomes. A deeper analysis of the trilinear E-Beads (see section~\ref{sec:BEADSAsym}) reveals that the  measurement results are fully zzy-correlated (which can also be deduced by comparing the individual outcomes in the BEADS representation) for which $Q_3$ results in the state $\ket{R}$ if the product of measurement outcomes $m_1\cdot m_2=1$ and in $\ket{L}=1/\sqrt{2}\left(\ket{0}-i\ket{1}\right)$ otherwise. The following correction by the (classically) controlled X-gate on $Q_3$ based on the outcome of $Q_2$ corrects the state of $Q_3$ in the second scenario but transforms $Q_3$ in the fourth scenario such that the state of $Q_3$, undesirably, is now $\ket{L}$. However, the (classically) controlled Z-gate undoes this action as well as it corrects $Q_3$ in the third outcome case such that $Q_3$ finally aligns according to the desired state in all scenarios. The final step in the circuit is the weighted sum (mixed state) of all outcome scenarios which nicely depicts that teleportation of the desired state $\ket{L}$ was achieved with certainty (see the orientation of $Q_3$) independent of the measurement outcomes.}
\end{figure}
\begin{figure}[H]
\centering
\includegraphics[width=1.0\textwidth]{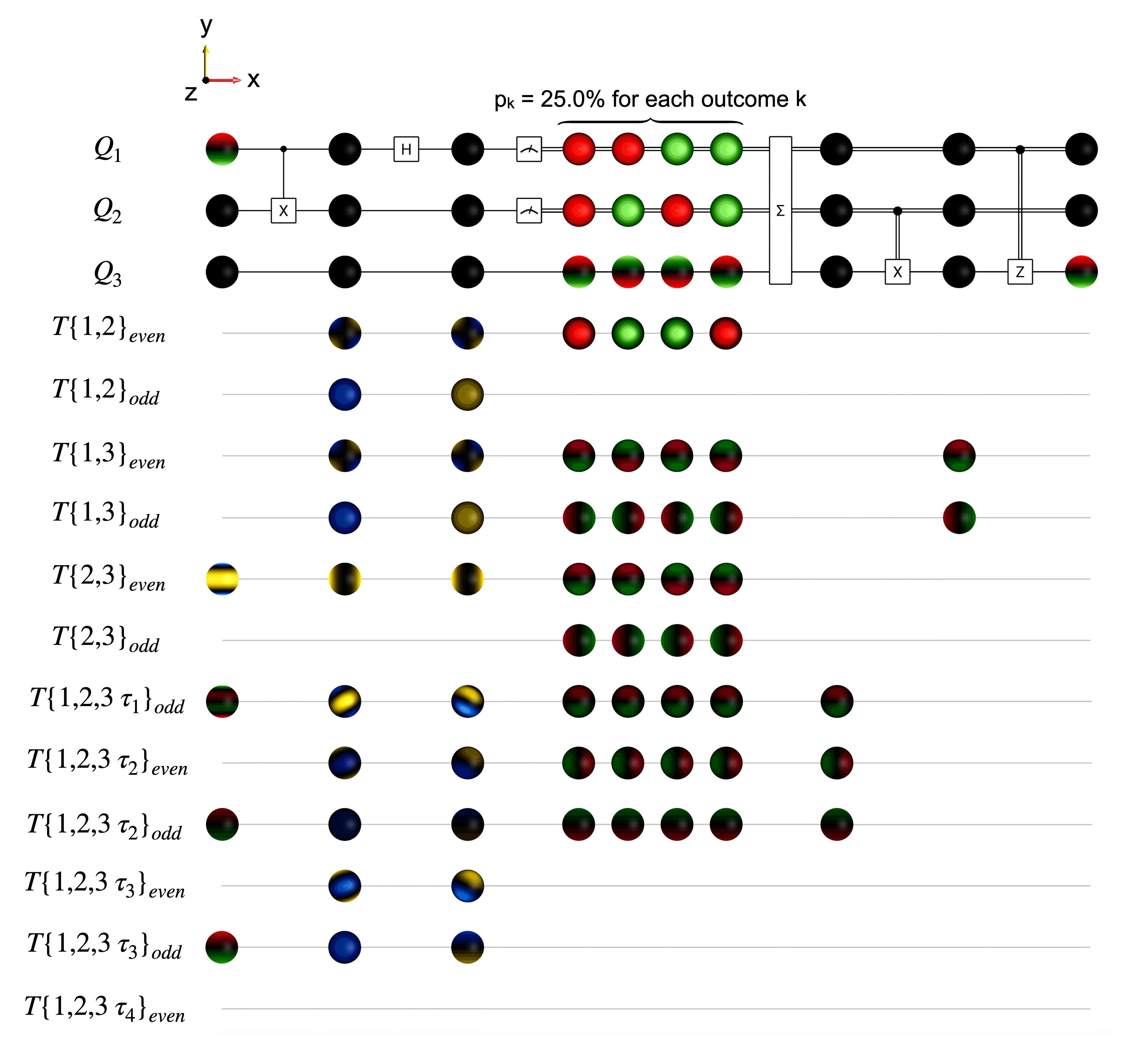}
\caption{\label{fig:FigTeleportYFullMixed}Teleportation of the y-basis $\ket{R}$ state using a mixed state BEADS representation \screen. The y-basis state $\ket{R}=1/\sqrt{2}\left(\ket{0}+i\ket{1}\right)$ is teleported from qubit $Q_1$ to $Q_3$ (for visualization variants or a detailed description see section~\ref{QuTel} or Fig.~\ref{fig:FigTeleportYFull}). Here, the different states after the measurement are combined to form a mixed state which has only non-zero correlation functions (all Q-Beads are black). It becomes clear that, despite the intriguingly different BEADS representation in comparison to the pure state equivalent (Fig.~\ref{fig:FigTeleportYFull}), performing the correction steps on this mixed state still gives the desired result where $Q_3$ is in the state $\ket{R}$.}
\end{figure}

\subsection{Entanglement swapping}\label{sec:BEADSEntanglementSwapping}
Entanglement swapping \cite{Teleportation} represents a fundamental protocol in quantum communication which is required to implement communication modules such as the quantum repeater introduced by Briegel et. al. \cite{EntanglementSwapping1}. It allows to entangle particles which are separated by large distances based on the manipulation of two Bell pairs. The basic building block hence requires four qubits and can be visualized using BEADS as is shown in Fig.~\ref{fig:FigEntSwapping}.
\begin{figure}[H]
\centering
\includegraphics[width=1.0\textwidth]{FigEntSwapping.pdf}
\caption{\label{fig:FigEntSwapping}BEADS representation of the entanglement swapping protocol. Bell pairs $\ket{\Phi^+}$ are created between qubits $Q_1$ and $Q_2$ as well as $Q_3$ and $Q_4$.  Performing a Bell measurement (the corresponding sequence of actual gates and measurements is shown at the bottom) on $Q_2$ and $Q_3$ and subsequent correction gates depending on the measurement outcomes swaps the entanglement to a new Bell pair $\ket{\Phi^+}$ between $Q_1$ and $Q_4$ \cite{EntanglementSwapping}.}
\end{figure}
\noindent
The swapping protocol is exemplarily visualized for a set of initial Bell pairs between qubits $Q_1$ and $Q_2$, as well as $Q_3$ and $Q_4$ constituting the state $\ket{\psi}=\ket{\Phi^+}\otimes\ket{\Phi^+}=1/2(\ket{0000}+\ket{0011}+$ $\ket{1100}+\ket{1111})$ where once again $\ket{\Phi^+}=1/\sqrt{2}\left(\ket{00}+\ket{11}\right)$. These Bell pairs can be created by the standard procedure shown in section~\ref{2Q}. In the entanglement swapping protocol, a Bell measurement is then performed on $Q_2$ and $Q_3$ which, after classically controlled corrections, yields a new Bell entangled pair between $Q_1$ and $Q_4$ while the entanglement between the qubits of the initial Bell pairs is destroyed. Thus, the entanglement is said to be swapped. 

It should be pointed out though, that the actual measurement operations, which are z-measurements on $Q_2$ and $Q_3$, after a basis transformation as shown at the bottom of Fig.~\ref{fig:FigEntSwapping}, as holds for any projective \textit{single-qubit} measurement, cannot create entanglement. However, the measurements can be used to \textit{localize} existing entanglement in a specific bipartite subsystem \cite{Teleportation, EntanglementSwapping1, EntanglementSwapping}. Indeed, the preceding basis transformation, more specifically the CNOT$_{23}$ gate, further entangles the system yielding the state $\ket{\psi}'=1/\left(2\sqrt{2}\right)\: (\ket{0000}+\ket{0100}+\ket{0011}+\ket{0111}-$ $\ket{1101}+\ket{1001}-\ket{1110}+\ket{1010})$ which is maximally entangled and has genuine four-qubit entanglement (i.e., the entanglement is shared between all four qubits). By measuring $Q_2$ and $Q_3$, the entanglement gets localized between $Q_1$ and $Q_4$ in form of one of the four maximally two-qubit entangled Bell states (see section~\ref{sec:BellStates}) in dependence of the measurement outcomes. The entangled state $\ket{\psi}'$ is not shown in Fig.~\ref{fig:FigEntSwapping} for reasons of compactness and clarity. In the BEADS picture, we can further see that the correction operations, which take the measurement outcomes of $Q_2$ and $Q_3$ as classical controls, yield the same Bell pair $\ket{\Phi^+}$ between qubits $Q_1$ and $Q_4$ for any combination of measurement outcomes.

\subsection{Bell's theorem}\label{sec:BEADSBellTest}
Bell's theorem \cite{Bell64} states that entanglement has implications on correlations of measurement outcomes in quantum mechanical experiments that cannot be explained by local hidden-variables. Here, we examine two possible approaches to demonstrate Bell's theorem in experimental implementations and highlight how these can be easily understood using the BEADS representation on a conceptual level especially in view of deriving correlation coefficients of asymmetric measurements.

\subsubsection*{The CHSH inequality}
The CHSH inequality which can be written as \cite{CHSH}
\begin{equation}\label{eq:CHSH}
|S|\leq2,
\end{equation}
and where
\begin{equation}\label{eq:CHSHS}
S=C\left(r_1,r_2\right)+C\left({\widetilde{r}}_1,r_2\right)+C\left({\widetilde{r}}_1,{\widetilde{r}}_2\right)-C\left(r_1,{\widetilde{r}}_2\right),
\end{equation}
\noindent
is one of the most prominent tools to reveal contradictions with local hidden-variable theories in quantum mechanics. It can be shown to always be fulfilled for local hidden-variable theories \cite{CHSH}.
Here, $C(a,b)$ denotes the correlation between measurement outcomes obtained when measuring one qubit in a pair of qubits along direction $a$ and the other one in direction $b$. The measurement directions expressed in spherical coordinates $(\theta,\phi)$ are chosen to be $r_1=(0^\circ,0^\circ)$, $\widetilde{r}_1=(90^\circ,0^\circ)$ for the first qubit and $r_2=(45^\circ,0^\circ)$, $\widetilde{r}_2=(135^\circ,0^\circ)$ for the second qubit \cite{CHSH}. Any state for which $|S|> 2$ violates the inequality and thus contradicts the concept of local hidden-variables \cite{CHSH}.

As an example, we examine the Bell state $\ket{\Phi^+}=1/\sqrt{2}\left(\ket{00}+\ket{11}\right)$. As stated for symmetric measurements in section~\ref{2Q}, we can always determine correlations of measurement outcomes $C$ \textit{numerically} as the expectation values of the corresponding observables, i.e., measurement operators:
\begin{subequations}\label{eq:CHSHExp}
\begin{align}
C\left(r_1,r_2\right)&=\left\langle\mathrm{\Phi}^+\middle|\sigma_{1z}\left(\frac{\sigma_{2x}+\sigma_{2z}}{\sqrt2}\right)\middle|\mathrm{\Phi}^+\right\rangle=\frac{1}{\sqrt2}, \\
C\left({\widetilde{r}}_1,r_2\right)&=\left\langle\mathrm{\Phi}^+\middle|\sigma_{1x}\left(\frac{\sigma_{2x}+\sigma_{2z}}{\sqrt2}\right)\middle|\mathrm{\Phi}^+\right\rangle=\frac{1}{\sqrt2},\\
C\left({\widetilde{r}}_1,{\widetilde{r}}_2\right)&=\left\langle\mathrm{\Phi}^+\middle|\sigma_{1x}\left(\frac{\sigma_{2x}-\sigma_{2z}}{\sqrt2}\right)\middle|\mathrm{\Phi}^+\right\rangle=\frac{1}{\sqrt2},\\
C\left(r_1,{\widetilde{r}}_2\right)&=\left\langle\mathrm{\Phi}^+\middle|\sigma_{1z}\left(\frac{\sigma_{2x}-\sigma_{2z}}{\sqrt2}\right)\middle|\mathrm{\Phi}^+\right\rangle=-\frac{1}{\sqrt2},
\end{align}
\end{subequations}
\noindent
which gives $S=2\sqrt{2}$ and thus violates the inequality shown in Eq.~\ref{eq:CHSH}.
Using the BEADS representation, we can choose between three different strategies to derive the required correlation coefficients \textit{graphically}. Note that Bell test experiments such as those based on finding violations of the CHSH inequality yield statistical results, i.e., must be repeated many times to show a contradiction with local-hidden variable theory.\newline

\begin{figure}[H]
\centering
\includegraphics[width=.94\textwidth]{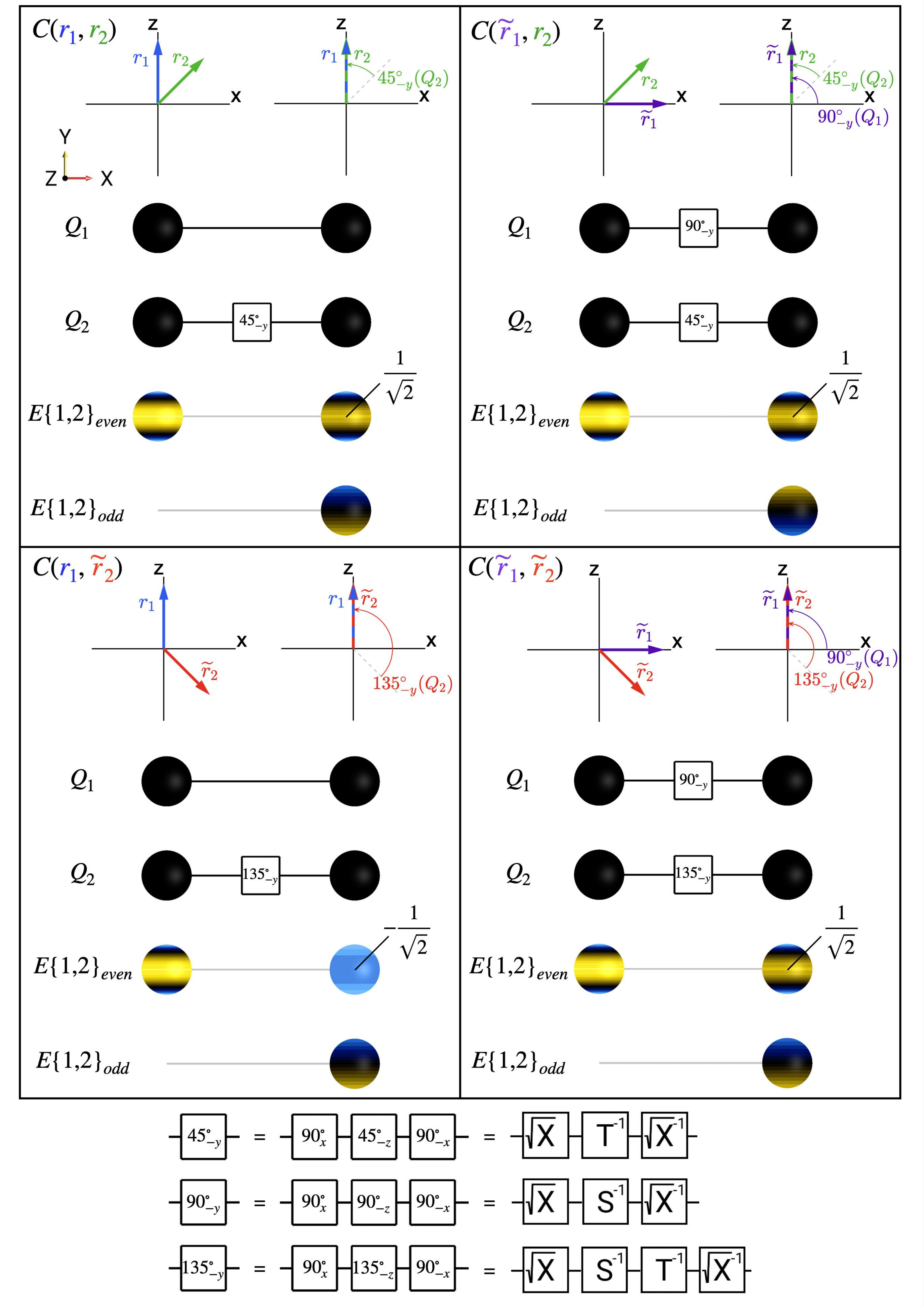}
\caption{\label{fig:FigBellCHSH1}CHSH inequality in case of a Bell state $\ket{\Phi^+}= 1/\sqrt2 \left( \ket{00} + \ket{11} \right)$ -- method \textbf{A}. The correlation coefficients of measurement outcomes (denoted in the top left corners), which are applied in the CHSH inequality, can be determined from the BEADS representation by applying local rotations to the qubits such that the measurement directions are effectively transformed to the z-axis which allows to directly read off the correlation coefficients from $E{\{1,2\}}_{\textit{even}}$ along the z-axis.}
\end{figure}

\noindent
As a first and preferred option, which is denoted \textbf{A} in Fig.~\ref{fig:FigBellCHSH1}, we can transform the Bell state by applying local rotations around the $-$y-axis by the polar angles defining the measurement direction for each of the qubits which effectively rotate the original measurement direction vectors onto the z-axis. For instance, for measurement directions $(\widetilde{r}_1,r_2)$, we apply a $90^\circ$-rotation on $Q_1$ and a $45^\circ$-rotation on $Q_2$, around the $-$y-axis, respectively. This enables us to treat the originally asymmetric measurement as a symmetric measurement in the computational basis and we can directly read off the corresponding correlation coefficient as the value of the symmetric E-Bead $E\{1,2\}_{\textit{even}}$ along the z-axis. Note that since $\ket{\Phi^+}$ is maximally entangled, $E\{1,2\}_{\textit{even}}$ is equivalent to the total correlation bead $T\{1,2\}_{\textit{even}}$. Moreover, since we now perform symmetric measurements, $E\{1,2\}_{\textit{odd}}$ is irrelevant for the interpretation. As can be seen in Fig.~\ref{fig:FigBellCHSH1}, for any combination of measurement directions except $(r_1,{\widetilde{r}}_2)$, we obtain the same positive correlation coefficient $C=1/\sqrt2$ along the z-axis. For $(r_1,{\widetilde{r}}_2)$, which corresponds to the subtracted term in Eq.~\ref{eq:CHSHS}, we get an anticorrelated state with $C=-1/\sqrt2$, and thus, reach a value $S = 4/\sqrt2  = 2\sqrt2 > 2$ in total, clearly violating inequality \ref{eq:CHSH}.\newline

\begin{figure}[H]
\centering
\includegraphics[width=1.\textwidth]{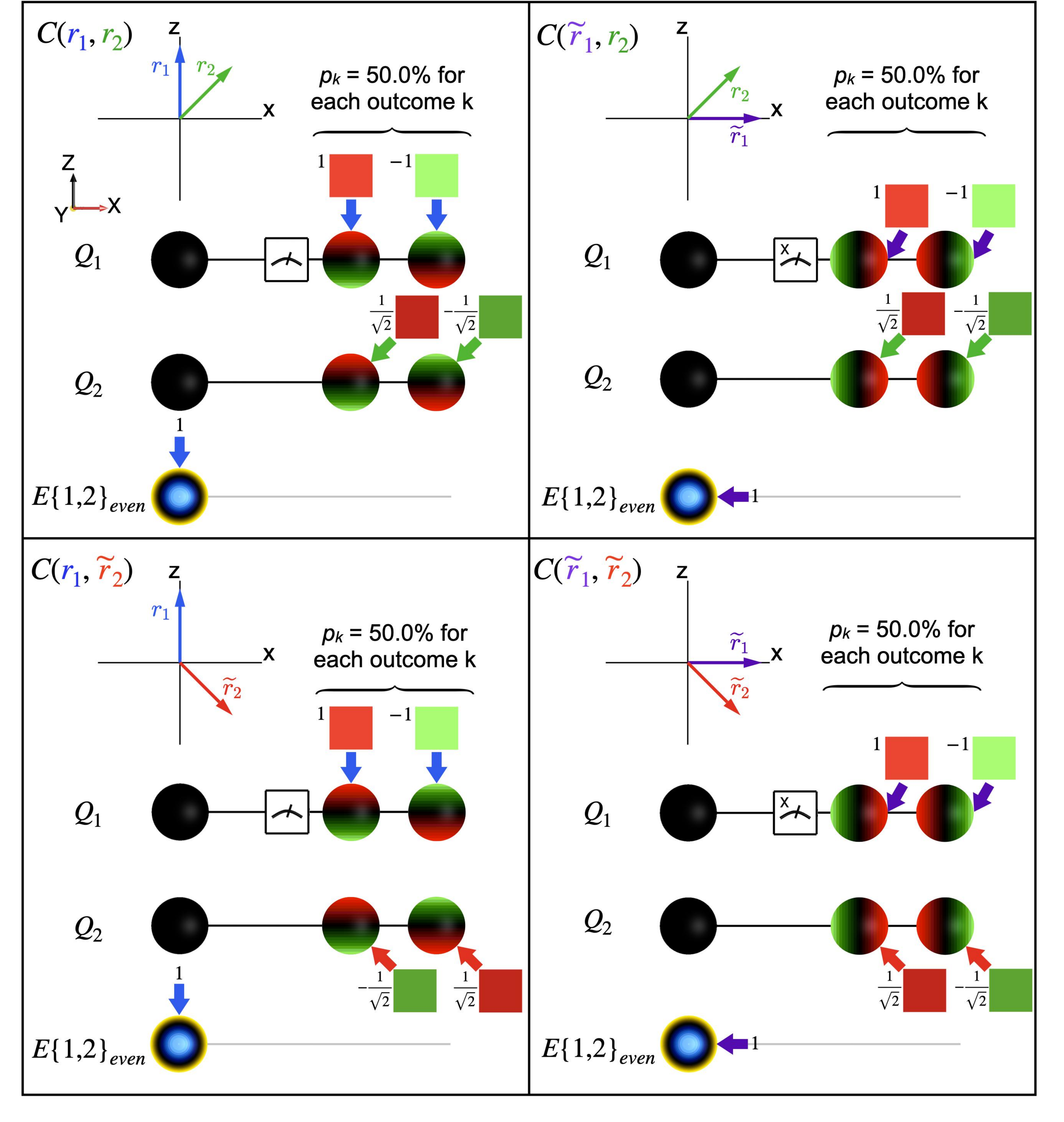}
\caption{\label{fig:FigBellCHSH2}CHSH inequality in case of a Bell state $\ket{\Phi^+}= 1/\sqrt2 \left( \ket{00} + \ket{11} \right)$ -- method \textbf{B}. By performing a local measurement on one qubit, the correlation coefficients (denoted in the top left corner of every box) which have to be applied in the CHSH inequality can be calculated as products of the single-qubit expectation values obtained from the corresponding Q-Beads. }
\end{figure}

\noindent
As a second approach \textbf{B}, we can perform a projective measurement on one of the qubits, e.g., on the first qubit $Q_1$ along direction $r_1$ (z-axis) or $\widetilde{r_1}$ (x-axis), exclusively. Note that the order of measurements is irrelevant, i.e., it is also possible to measure $Q_2$ first. However, in the following, we assume the case where $Q_1$ is measured. We can predict the post-measurement state of the second qubit $Q_2$ based on the BEADS representation of the initial Bell state which is yellow and hence fully correlated ($E\{1,2\}_{\textit{even}}=1$) in both measurement directions (see Fig.~\ref{fig:FigBellCHSH2} and recall the color scale in Fig.~\ref{Figure:Fig5} in section~\ref{2Q}). Hence, upon measuring $Q_1$, $Q_2$ will align identically as $Q_1$. The outcomes of $Q_1$ are assigned values of $\pm1$ corresponding eigenvalues of the measurement operator, e.g., in case of a z-measurement we assign 1 if $Q_1$ was measured in state $\ket{0}$ and $-1$ if we measured the qubit in state $\ket{1}$. It is now possible to directly read off the single qubit expectation values of $Q_2$ along $r_2$ and $\widetilde{r}_2$. All possible scenarios are shown in Fig.~\ref{fig:FigBellCHSH2} and the single-qubit expectation values are illustrated by corresponding color patches and numerical values. Finally, the wanted correlation coefficients can be calculated as the product of single-qubit expectation values determined for both qubits and we get:
\begin{equation}
C\left(r_1,r_2\right)=C\left({\widetilde{r}}_1,r_2\right)=C\left({\widetilde{r}}_1,{\widetilde{r}}_2\right)=1\cdot\frac{1}{\sqrt2}=\left(-1\right)\cdot\left(-\frac{1}{\sqrt2}\right)=\frac{1}{\sqrt2}
\end{equation}
and
\begin{equation}
C\left(r_1,{\widetilde{r}}_2\right)=\left(-1\right)\cdot\frac{1}{\sqrt2}=1\cdot\left(-\frac{1}{\sqrt2}\right)=-\frac{1}{\sqrt2}.
\end{equation}\newline

\noindent
Last, we put into perspective how the correlation coefficients $C$ can be derived numerically solely based on the BEADS representation of $\ket{\Phi^+}$  (method \textbf{C}). Note that the following analysis is based on the calculation of asymmetric correlations from spherical beads functions which is explained in supplementary section~\ref{sec:BEADSAsym}. Here, since all measurements are performed within the xz-half plane, we can apply Eq.~\ref{eq:MeasXZ} of section~\ref{sec:BEADSAsym}. Moreover, as the examined Bell state does not have an antisymmetric component, we can further simplify to:
\begin{align}\label{eq:CHSHnumeric}
\langle O \rangle= C(a,b) &= \cos{\Delta_\theta}\left(\left\langle T_{0,0}^{\left\{k,l\right\}_{\textit{even}}\prime}\right\rangle+\frac{\left\langle T_{2,0}^{\left\{k,l\right\}_{\textit{even}}\prime} \right\rangle}{4}+\frac{\left\langle T_{2,2}^{\left\{k,l\right\}_{\textit{even}}\prime} \right\rangle}{2}\right) \nonumber\\
&+\cos{\Sigma_\theta}\left(\frac{3 \left\langle T_{2,0}^{\left\{k,l\right\}_{\textit{even}}\prime} \right\rangle}{4}-\frac{ \left\langle T_{2,2}^{\left\{k,l\right\}_{\textit{even}}\prime} \right\rangle}{2}\right)+\sin{\Sigma_\theta}\left\langle T_{2,1}^{\left\{k,l\right\}_{\textit{even}}\prime} \right\rangle,
\end{align}
\noindent
where $\Delta_\theta = \theta_1-\theta_2$ and $\Sigma_\theta = \theta_1+\theta_2$ are the difference and sum polar angles associated with the individual measurement directions, respectively. 

\begin{figure}[H]
\centering
\includegraphics[width=1.\textwidth]{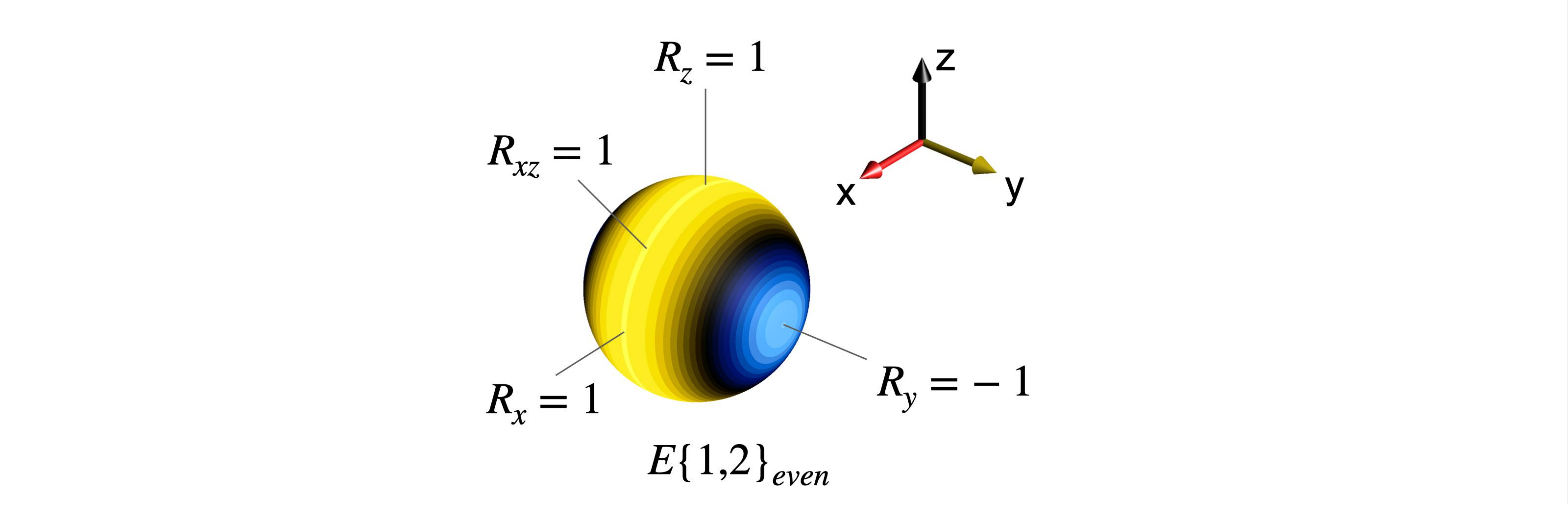}
\caption{\label{fig:FigBellCHSH3}CHSH inequality in case of a Bell state $\ket{\Phi^+}= 1/\sqrt2 \left( \ket{00} + \ket{11} \right)$ -- method \textbf{C}. By reading off the spherical function values of $E{\{1,2\}}_{\textit{even}}$ along the indicated directions, the correlation coefficients of interest can be determined numerically.}
\end{figure}

The required scaled tensor operator expectation values (cf. supplementary section~\ref{sec:BEADSAsym}) can then be obtained by reading off the E-Bead values at characteristic positions as indicated in Fig.~\ref{fig:FigBellCHSH3} and inserting them in Eq.~\ref{eq:AsymTVals2Q} which gives:
\begin{subequations}
\begin{align}
\left\langle T_{0,0}^{\left\{k,l\right\}_{\textit{even}}\prime} \right\rangle &=\frac{1}{3}\left(R_x+R_y+R_z\right)=\frac{1}{3}\left(1-1+1\right)=\frac{1}{3},\\
\left\langle T_{2,0}^{\left\{k,l\right\}_{\textit{even}}\prime} \right\rangle &=\frac{1}{3}\left(2R_z-R_x-R_y\right)=\frac{1}{3}\left(2-1+1\right)=\frac{2}{3},\\
\left\langle T_{2,1}^{\left\{k,l\right\}_{\textit{even}})\prime} \right\rangle &=R_{xz}-\frac{1}{2}\left(R_x+R_z\right)=1-\frac{1}{2}\left(1+1\right)=0,\\
\left\langle T_{2,2}^{\left\{k,l\right\}_{\textit{even}}\prime} \right\rangle &=\frac{1}{2}\left(R_x-R_y\right)=1,
\end{align}
\end{subequations}
\noindent
and inserting these values in Eq.~\ref{eq:CHSHnumeric}, as in the previously discussed approaches, we again obtain $C(r_1,r_2)=C({\widetilde{r}}_1,r_2)=C({\widetilde{r}}_1,{\widetilde{r}}_2)\ =1/\sqrt2$ and $C(r_1,{\widetilde{r}}_2)=-1/\sqrt2$.

\subsubsection*{The GHZ experiment}
Another illustrative way to prove Bell's theorem is given by the GHZ experiment \cite{GHZNoBell}. Here, we leave a numerical analysis as described in Eq.~\ref{eq:CHSHExp} to the reader.

We start with the well-known GHZ state $\ket{\text{GHZ}}=1/\sqrt2 \left(\ket{000}+\ket{111}\right)$ which is visualized in Fig.~\ref{Figure:Fig5} (cf. section~\ref{3Q}). Unlike in the previously introduced Bell test experiments, absolute contradictions between quantum mechanics and local hidden-variable theory can be revealed by using only four different measurement settings in the GHZ experiment. These settings include a symmetric measurement of all qubits along the x-axis and three asymmetric three-qubit measurements where two qubits are measured along the y-axis and one qubit is measured along the x-axis (i.e., a $y_1 y_2 x_3$, a $y_1 x_2 y_3$, and a $x_1 y_2 y_3$-measurement). It is then possible to compare the product of measurement outcomes
\begin{equation}\label{eq:GHZQuantum}
\Pi_C=\left\langle \sigma_{1x} \sigma_{2x} \sigma_{3x} \right\rangle \cdot \left\langle \sigma_{1x} \sigma_{2y} \sigma_{3y} \right\rangle \cdot \left\langle \sigma_{1y} \sigma_{2x} \sigma_{3y} \right\rangle \cdot \left\langle \sigma_{1y} \sigma_{2y} \sigma_{3x} \right\rangle
\end{equation}
\noindent
with the corresponding prediction of local hidden variable theory $\Pi_{LHV}$, where the measurement outcome of every single qubit $k$ along direction $d$ is predetermined by a hidden-variable $v_{kd}$ for which
\begin{equation}\label{eq:GHZLHV}
\Pi_{LHV}=v_{1x}v_{2x}v_{3x}\cdot\ v_{1x}v_{2y}v_{3y}\cdot\ v_{1y}v_{2x}v_{3y}\cdot\ v_{1y}v_{2y}v_{3x}=v_{1x}^2v_{1y}^2v_{2x}^2v_{2y}^2v_{3x}^2v_{3y}^2.
\end{equation}
\noindent
It is clear, that due to all hidden-variables being real numbers, $\Pi_{LHV}$ must be positive. An illustrative explanation of the GHZ-experiment is given in \cite{Maudlin11}.

To predict possible outcomes of the fully symmetric x-measurement, we analyze the $E\{1,2,3\ \tau_1\}_{\textit{odd}}$ bead ($=T\{1,2,3\ \tau_1\}_{\textit{odd}}$ due to the state being maximally entangled) which reveals a maximum positive three-qubit correlation coefficient of 1 along the x-axis. At this point, it should be recalled that positive $n$-qubit correlation implies an even number of qubits in the "down" state (eigenvalue $-1$) with respect to the corresponding measurement direction after the measurement. Hence, we expect outcomes $\ket{+++}$, $\ket{+--}$, $\ket{-+-}$, and $\ket{--+}$ with equal probability (25\%). For the different asymmetric three-qubit measurements, the corresponding correlation coefficients can once more be determined using different approaches in the BEADS representation.\newline

\begin{figure}[H]
\centering
\includegraphics[width=.9\textwidth]{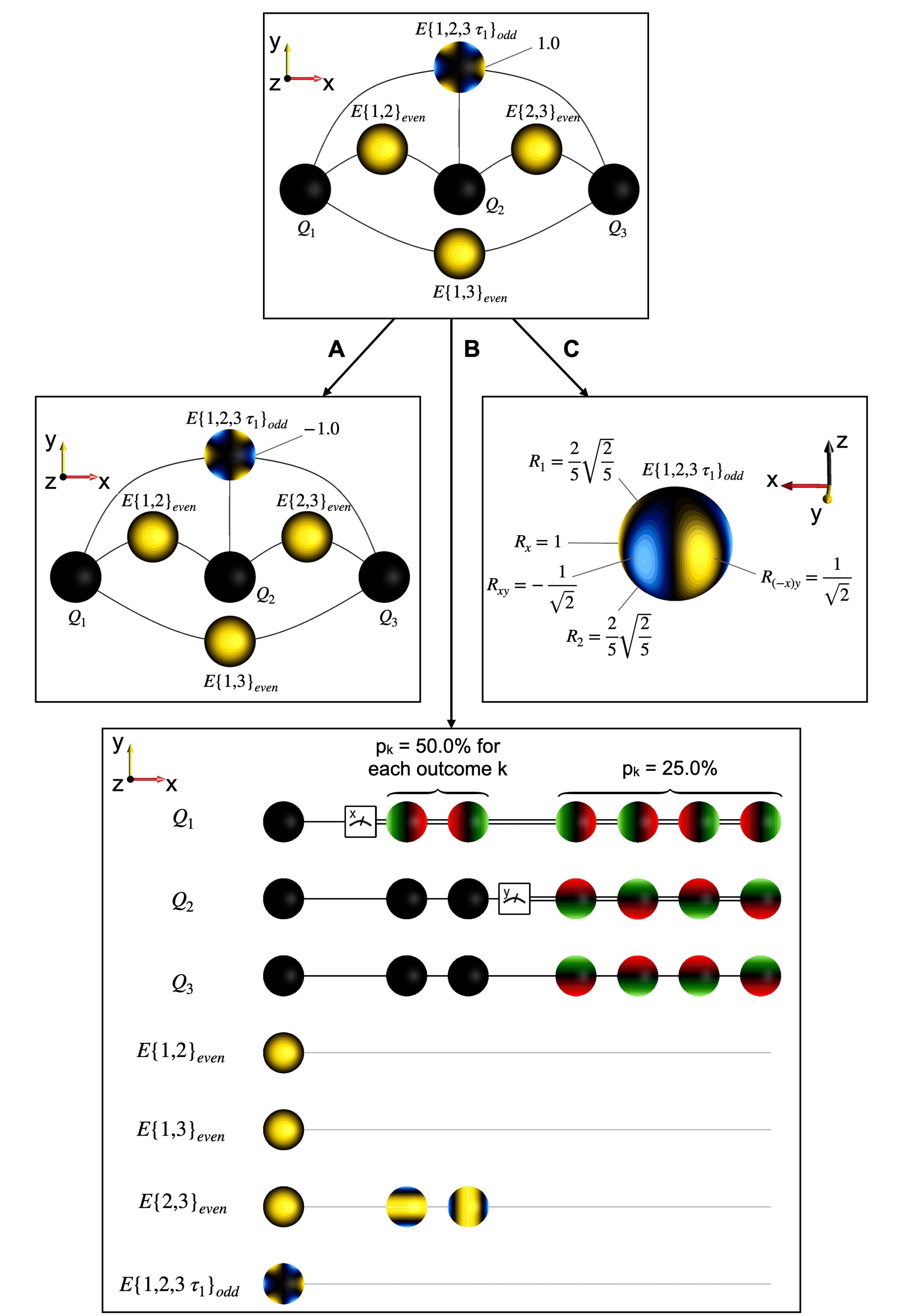}
\caption{\label{fig:FigBellGHZ}The GHZ experiment. Measurement outcomes are fully correlated when the GHZ-state is measured along the x-axis, yet the outcomes are fully anticorrelated when measuring two qubits along the y- and the remaining qubit along x-axis which contradicts a local hidden-variable theory. Using BEADS, the corresponding correlation coefficients can be determined by local transformations (\textbf{A}), local measurements (\textbf{B}) or numerically (\textbf{C}).}
\end{figure}

\noindent
As discussed for CHSH inequalities, we can first transform the GHZ state such that the measurement direction vectors are effectively co-aligned (variant \textbf{A} in Fig.~\ref{fig:FigBellGHZ}). This can be efficiently achieved by applying inverse square roots of Pauli-Z gates on the two qubits which are to be measured along the y-axis. This rotates all measurement directions to the x-axis, and thus, allows us to view the intended measurement as a fully symmetric x-measurement and to directly read off the measurement expectation value along the x-axis. Note that all bilinear E-Beads in the BEADS representation of the GHZ state are axially symmetric and $E\{1,2,3\ \tau_1\}_{\textit{odd}}$ has a threefold rotational symmetry with respect to the z-axis, the transformation results in only $E\{1,2,3\ \tau_1\}_{\textit{odd}}$ being rotated by ($\varphi_1+\varphi_2+\varphi_3)/3$ around the negative z-axis. Here, two of the rotation angles $\varphi_k$, where the index denotes the $k$-th qubit, are $\pi/2$ (square root of Z), whereas the third angle is 0, respectively, and thus, the triliniear E-Bead is rotated by $\pi/3$ around the negative z-axis. It is now possible to directly read off the correlation coefficient along the x-axis which results in being $-1$ (bright blue color), i.e., full anticorrelation, for any of the measurements. We expect measurement outcomes $\ket{---}$, $\ket{-++}$, $\ket{+-+}$, and $\ket{++-}$ with equal probability (25\%).\newline

\noindent
We obtain the same correlation values if we perform consecutive local measurements on two of three qubits (variant \textbf{B} in Fig.~\ref{fig:FigBellGHZ}). For instance, we measure the first qubit $Q_1$ along the x-axis. This results in $Q_2$ and $Q_3$ being entangled in one of two Bell states ($\ket{\Phi^+}$ or $\ket{\Phi^-}$ depending on the outcome of $Q_1$). If we then proceed to measure $Q_2$ along the y-axis, $Q_2$ and $Q_3$, due to maximum anticorrelation ($\ket{\Phi^+}$, blue color along y-axis) or correlation ($\ket{\Phi^-}$, yellow color along y-axis), adopt opposite or equal y-basis states, respectively. In Fig.~\ref{fig:FigBellGHZ}, it is clear, that possible measurement outcomes for the $x_1 y_2 y_3$-measurement are $\ket{+RL}$, $\ket{+LR}$, $\ket{-RR}$, and $\ket{-LL}$. As was discussed previously for Bell tests, we can now assign values $\pm1$ to the single-qubit outcomes and form products which here strictly result in overall correlation values of $-1$.\newline

\noindent
Alternatively (method \textbf{C} in Fig.~\ref{fig:FigBellGHZ}), we can apply the mathematical relations shown in supplementary section~\ref{sec:BEADSAsym} can be applied to calculate the correlations directly based on the BEADS representation of the GHZ state. To determine the expectation values for a particular combination of measurement directions, e.g., $x_1 y_2 y_3$, numerically based on the BEADS representation of $\ket{\text{GHZ}}$, we first identify any tensor operator component that is dependent on the desired measurement directions in Eq.~\ref{eq:AsymMaster3Q} of supplementary section~\ref{sec:BEADSAsym}. For instance, for an $ x_1 y_2 y_3$-measurement, only the $E\{1,2,3\ \tau_1\}_{\textit{odd}}$ E-Bead has non-zero expectation values for the GHZ-state. Thus, we only find components $T_{1,1}^{\{1,2,3\ \tau_1\}_{\textit{odd}}}$ and $T_{3,3}^{\{1,2,3\ \tau_1\}_{\textit{odd}}}$ to be relevant and Eq.~\ref{eq:AsymMaster3Q} simplifies to
\begin{equation}
\left\langle \sigma_{1x}\sigma_{2y}\sigma_{3y} \right\rangle =\frac{1}{3} \left\langle T_{1,1}^{{\{1,2,3\ \tau_1\}}_{\textit{odd}}\prime} \right\rangle - \left\langle T_{3,3}^{{\{1,2,3\ \tau_1\}}_{\textit{odd}}\prime} \right\rangle.
\end{equation}

\noindent
Note that this equation only includes $\tau_1$ operator components as the examined GHZ state does not show any further trilinear symmetry components.
We can calculate the expectation values $\left\langle T_{1,1}^{{\{1,2,3\ \tau_1\}}_{\textit{odd}}\prime} \right\rangle$ and $\left\langle T_{3,3}^{{\{1,2,3\ \tau_1\}}_{\textit{odd}}\prime} \right\rangle$ by reading off the spherical function values and applying them according to equations Eq.~\ref{eq:AsymTVals3Q} of supplementary section~\ref{sec:BEADSAsym} which gives:
\begin{subequations}
\begin{align}
\left\langle T_{1,1}^{{\{1,2,3\ \tau_1\}}_{\textit{odd}}\prime} \right\rangle&=\frac{4}{15}R_x\ +\ \frac{\sqrt2}{5}\left(R_{xy}-R_{\left(-x\right)y}\right)+\frac{\sqrt{10}}{12}\left(R_1+R_2\right)\nonumber\\
&=\frac{4}{15}+\frac{\sqrt2}{5}\left(-\frac{1}{\sqrt2}-\frac{1}{\sqrt2}\right)+\frac{\sqrt{10}}{12}\left(\frac{2}{5}\sqrt{\frac{2}{5}}+\frac{2}{5}\sqrt{\frac{2}{5}}\right)=0,\\
\left\langle T_{3,3}^{{\{1,2,3\ \tau_1\}}_{\textit{odd}}\prime} \right\rangle &=\frac{1}{2}R_x+\frac{1}{2\sqrt2}\left(R_{\left(-x\right)y}-R_{xy}\right)=\frac{1}{2}+\frac{1}{2\sqrt2}\left(\frac{1}{\sqrt2}+\frac{1}{\sqrt2}\right)=1.
\end{align}
\end{subequations}
Fig.~\ref{fig:FigBellGHZ} explicitly shows the spherical function values at vertices $R_x$, $R_{(-x)y}$ and $R_{xy}$, $R_1$, and $R_2$.
Repeating this procedure for all remaining measurement directions yields
\begin{equation}
\left\langle \sigma_{1x}\sigma_{1y}\sigma_{1y} \right\rangle = \left\langle \sigma_{1y}\sigma_{1x}\sigma_{1y} \right\rangle = \left\langle \sigma_{1y}\sigma_{1y}\sigma_{1x} \right\rangle =\frac{1}{3}\left\langle T_{1,1}^{{\{1,2,3\ \tau_1\}}_{\textit{odd}}\prime} \right\rangle-\left\langle T_{3,3}^{{\{1,2,3\ \tau_1\}}_{\textit{odd}}\prime} \right\rangle=-1,
\end{equation}
\noindent
which matches the results obtained with methods \textbf{A} and \textbf{B}.

Based on these findings for the GHZ state, i.e., full correlation of outcomes when measuring all qubits along the x-axis but full anticorrelation when measuring one qubit along the x-axis and the remaining qubits along the y-axis, we can now possible to examine whether these findings are compatible with the principle of locality. From the obtained correlations, it is clear that $\Pi_C=1\cdot(-1)\cdot(-1)\cdot(-1)=-1$. Indeed, if quantum mechanics would be a local theory, the outcomes observed for each individual qubit would instead multiply to 1 according to Eq.~\ref{eq:GHZLHV}.

\subsection{Supplementary video}
We provide a video which includes numerous dynamical simulations using the BEADS representation. Throughout this article, examples are marked by a screen symbol \screen$\ $if they are shown in the video. A high-resolution version of the video is available via \href{https://github.com/denhub97/QuBeads}{https://github.com/denhub97/QuBeads}.

The video is subdivided into chapters which can be used for fast navigation in most media player devices (e.g., by clicking the "$\gg$" menu or using \textit{Command} + \textit{Shift} + \textit{arrow keys} in the QuickTime player on MacOS, or by using the \textit{Playback} $\rightarrow$ \textit{Chapters} menu in VLC). The video playback can be precisely controlled by using two-finger trackpad gestures in many media players.

Recall that for dynamical simulations in the QuBeads application, which was used to create the video, the action of a gate is distributed over the entire timestep in the corresponding circuit visualization. Timesteps are separated by vertical lines in standard BEADS mode circuits (see, e.g., bottom of Fig.~\ref{fig:FigQuBeads1} of appendix~\ref{app:BEADSQuBeads}) and by beads in the BEADS-augmented circuit mode (cf. Fig.~\ref{fig:FigQuBeads1}).

\subsection{The little quantum pocket guide}
The little quantum pocket guide is the result of a collaboration between the authors and the Munich Center of Quantum Science and Technology (MCQST). This guide explains the basics of quantum information on an intuitive and easily accessible level by using the BEADS representation. It is specifically targeted at school students and non-specialists.

\end{document}